\documentclass[onecolumn]{emulateapj}

\usepackage{multirow}

\usepackage{graphics}
\usepackage{amsmath, amsthm, amssymb}
\usepackage{epsfig}

\newcommand{\sm}{{M_\odot}}

\newcommand{\beq}{\begin{equation}}
\newcommand{\eeq}{\end{equation}}
\newcommand{\beqa}{\begin{eqnarray}}
\newcommand{\eeqa}{\end{eqnarray}}

\newcommand{\gcc}         {\rm g\:cm^{-2}}

\countdef\decade=200
\advance\decade by \year
\countdef\hours=201
\advance\hours by \time
\divide\hours by 60
\countdef\mins=202
\advance\mins by \hours
\multiply\mins by 60
\multiply\hours by 100
\countdef\miltime=203
\advance\miltime by \hours
\advance\miltime by \time
\advance\miltime by -\mins

\begin{document}

\title{Mid-Infrared Extinction Mapping of Infrared Dark Clouds. II.\\ 
The Structure of Massive Starless Cores and Clumps}


\author{Michael J. Butler}
\affil{Dept. of Astronomy, University of Florida, Gainesville, FL 32611, USA}

\author{Jonathan C. Tan}
\affil{Depts. of Astronomy \& Physics, University of Florida, Gainesville, FL 32611, USA}

\begin{abstract} 
We develop the mid-infrared extinction (MIREX) mapping technique of
Butler \& Tan (2009, Paper I), presenting a new method to correct for
the Galactic foreground emission based on observed saturation in
independent cores. Using {\it Spitzer} GLIMPSE $\rm 8\:\mu m$ images,
this allows us to accurately probe mass surface densities, $\Sigma$,
up to $\simeq 0.5\:{\rm g\:cm^{-2}}$ with 2\arcsec\ resolution and
mitigate one of the main sources of uncertainty associated with
Galactic MIREX mapping. We then characterize the structure of 42
massive starless and early-stage cores and their surrounding clumps,
selected from 10 infrared dark clouds (IRDCs), measuring $\Sigma_{\rm
  cl}(r)$ from the core/clump centers. We first assess the properties
of the core/clump at a scale where the total enclosed mass as
projected on the sky is $M_{\rm cl}=60\:M_\odot$. We find these
objects have a mean radius of $R_{\rm cl}\simeq 0.1$~pc, mean
$\bar{\Sigma}_{\rm cl} = 0.3\:\gcc$ and, if fit by a power law density
profile $\rho_{\rm cl}\propto r^{-k_{\rm \rho,cl}}$, a mean value of
$k_{\rm \rho,cl}=1.1$. If we assume a core is embedded in each clump
and subtract the surrounding clump envelope to derive the core
properties, we find a mean core density power law index of $k_{\rho,c}
= 1.6$. We repeat this analysis as a function of radius and derive the
best-fitting power law plus uniform clump envelope model for each of
the 42 core/clumps. The cores have typical masses of $M_c \sim
100\:M_\odot$ and $\bar{\Sigma}_c\sim 0.1\:\gcc$, and
are embedded in clumps with comparable mass surface densities. We also
consider Bonnor-Ebert density models, but these do not fit the
observed $\Sigma$ profiles as well as power laws. We conclude massive
starless cores exist and are well-described by singular
polytropic spheres. Their relatively low values of $\Sigma$ and the
fact that they are IR dark may imply that their fragmentation is
inhibited by magnetic fields rather than radiative heating. Comparing to massive star-forming cores and clumps, there is tentative evidence for an evolution  
towards higher densities and steeper density profiles as star  
formation proceeds.
\end{abstract}


\section{Introduction}\label{S:intro}

Characterizing the initial conditions of massive star formation is
important for distinguishing between various theoretical models. For
example, McKee \& Tan (2002, 2003, hereafter MT03) have presented the
``Turbulent Core Accretion Model'', which assumes that massive stars
(including binaries and other low-order multiple systems) form from
massive starless {\it cores} in a manner that can be considered a
scaled-up version of the standard theory of low-mass star formation
(Shu, Adams, \& Lizano 1987). The core undergoes global collapse to
feed a central accretion disk. These massive starless cores are also
assumed to be near virial equilibrium and in approximate pressure
equilibrium with the surrounding {\it clump} environment, whose
pressure is likely to be set by the self-gravitating weight of the
gas, $P_{\rm cl}\simeq G\Sigma_{\rm cl}^2$, where $\Sigma_{\rm cl}$ is
the mass surface density of the clump. The clump is defined to be the
gas cloud that fragments to form a star cluster. Observed regions of
massive star formation, including revealed massive star clusters where
this activity occurred recently, have high values of $\Sigma_{\rm
  cl}\sim 1\:{\rm g\:cm^{-2}}$ (but with a dispersion of about a
factor of 10), implying large values of $P_{\rm cl}/k\gtrsim
10^8\:{\rm K\:cm^{-3}}$.

A massive, virialized core in pressure equilibrium with this
environment cannot be supported by thermal pressure, given observed
temperatures of $T\sim 10 - 20$~K, and so must be supported by some
form of nonthermal pressure, i.e. turbulence or magnetic fields. Since
massive stars are ``rare'', in the sense that they constitute only a
small mass fraction, $\sim 5-10\%$, of the observed initial stellar
mass function, then massive starless cores that will eventually form
massive stars
are also expected to be rare. Most massive structures are likely to
fragment into clusters of lower-mass stars. We anticipate that
preventing fragmentation of massive starless cores likely involves
magnetic fields being strong enough such that the core mass is approximately equal
to a magnetic critical mass and substructures are magnetically
subcritical. Krumholz \& McKee (2008) have argued fragmentation is
prevented by radiative heating from surrounding low-mass protostars,
which requires them to have high accretion rates and thus for the
clump to have a high value of $\Sigma\gtrsim 1\:\gcc$ in order to form
massive stars. There is no such constraint if magnetic fields prevent
fragmentation.

MT03 modeled massive cores as singular polytropic spheres, with a
power law density distribution of $\rho_c\propto
r^{-k_{\rho,c}}$. There were few observational constraints on this
density distribution, so MT03 assumed cores were part of a
self-similar hierarchy of structure also shared by their surrounding
clumps, where observations suggested $k_{\rm \rho,cl}\simeq 1.5$
(e.g. van der Tak et al. 2000; Beuther et al. 2002; Mueller et
al. 2002). The parameters of a fiducial massive core make it clear why
it is difficult to measure the structure observationally. For a
$60\sm$ core embedded near the center of a clump with mean
$\Sigma_{\rm cl}=1\gcc$, the radius is $R_{\rm c} = 0.057 M_{\rm
  c,60}^{1/2}(\Sigma_{\rm cl}/1\:{\rm g\:cm^{-2}})^{-1/2}$~pc. At
typical distances, $\gtrsim 2$~kpc, this radial size corresponds to
$\lesssim 5.9\arcsec$.

Until recently, measurements of mass surface densities of $\sim 1\:\gcc$
(equivalent to $N_{\rm H} = 4.27\times 10^{23}\:{\rm cm^{-2}}$ or $A_V
= 230$~mag) were based mostly on mm dust continuum measurements, which
require knowing the dust emissivity, dust temperature and dust-to-gas
ratio. In particular, the dust emissivity and temperature may be
expected to vary along the line of sight through a dense cloud. For
total power observations with single dish telescopes, resolutions have
been limited to $\sim 10\arcsec$, e.g. the diffraction limit of a 30~m
telescope at 1.2~mm. Rathborne et al. (2006) carried out a study of
this emission from 38~Galactic Infrared Dark Clouds (IRDCs) (these
types of clouds are described in more detail below). The above
considerations show that we do not expect their results to be able to
resolve the scale of individual massive cores in high pressure
environments, but are better suited to studying the properties of
clumps that might form whole star clusters. Higher angular resolution
has been achieved with interferometric observations, but these have
been possible only towards relatively limited samples of objects, many
of which are already in the process of forming a star (e.g. Bontemps
et al. 2010).

The advent of space-based, MIR, high-photometric-accuracy, imaging
surveys of the Galactic plane has opened up a new way to probe high
mass surface density structures. Indeed, these cold, high $\Sigma$
structures were first identified as ``Infrared Dark Clouds'' from
analysis of {\it Infrared Space Observatory} ({\it ISO}; P\'erault et
al. 1996) and the {\it Midcourse Space Experiment} ({\it MSX}; Egan et
al. 1998) imaging data. With the {\it Spitzer Space Telescope}, more
precise and higher angular resolution data became available. This led
Butler \& Tan (2009, hereafter BT09 or Paper I) to attempt to develop
MIR extinction (MIREX) mapping as a precision technique for probing high mass
surface density regions. 

BT09 adopted the Ossenkopf \& Henning (1994) thin ice mantle coagulated (for $10^5$~yr at densities of $10^6\:{\rm
  cm^{-3}}$ or equivalently for $\sim 10^6$~yr at densities of $\sim
10^5\:{\rm cm^{-3}}$) dust model for their fiducial analysis and a
gas-to-(refractory component)-dust mass ratio of 156 (slightly higher
than the value of 141 estimated by Draine (2011) from depletion
studies). With these values the opacity per unit gas mass in the IRAC
band 4 at $\sim 8\:{\rm \mu m}$ of a source with a spectrum typical of
Galactic diffuse MIR emission is $\kappa_{\rm 8\mu m}= 7.48\:{\rm
  cm^{2}\:g^{-1}}$. The fiducial value adopted by BT09 and in this
  paper is $\kappa_{\rm 8\mu m} = 7.5\:{\rm cm^{2}\:g^{-1}}$ so that
\begin{equation}
\label{eq:tau}
\tau_{\rm 8\mu m} = \kappa_{\rm 8\mu m} \Sigma  = 7.5 \left(\frac{\Sigma}{{\rm g\:cm^{-2}}}\right).
\end{equation}
From the variety of dust models considered by BT09, we expect $\sim
30\%$ uncertainties in the absolute value of $\kappa_{\rm 8\mu
  m}$. Within a particular IRDC, we can expect some systematic
variation in $\kappa_{\rm 8\mu m}$ due to different degrees of ice
mantle growth, but these should be at most $\sim 20\%$ (Ossenkopf \&
Henning 1994), and probably much less after averaging over conditions
on a line of sight through the cloud.

Apart from the choice of MIR opacity, there are two main sources of
uncertainty involved in MIR extinction mapping. First, the intensity
of the MIR emission behind the cloud is assumed to be smooth and must
be estimated by extrapolation from nearby regions that are assumed to
be extinction free. BT09 estimated that the extrapolation could lead
to flux uncertainties of about 10\%, corresponding to errors in
$\Sigma \simeq 0.01\:\gcc$, given fiducial dust models. This is a
minimum $\Sigma$, below which MIR extinction mapping becomes
unreliable. These problems of background estimation can be reduced by
choosing IRDCs that are in regions of the Galactic plane where the
observed surrounding emission is relatively constant and smooth around
the cloud. One systematic bias that we expect to be present is caused
by the fact that there will typically be some cloud material in the
``envelope'' region around the IRDC where extinction was assumed to be
zero. From studies of CO emission around IRDCs (Hernandez \& Tan 2011;
Hernandez et al. 2011) we estimate that this envelope typically has
$\Sigma\simeq 0.01\:\gcc$. This is an additional reason why MIR
extinction mapping becomes unreliable at low values of $\Sigma$. This
problem can be addressed by combining MIR and NIR extinction mapping
techniques (Kainulainen et al. 2011; Kainulainen \& Tan, in prep.)
The second major source of uncertainty is caused by foreground MIR
emission along our line of sight to the IRDC. Neglecting this causes
us to underestimate $\tau$ and thus $\Sigma$. The effect can be
minimized by choosing IRDCs that are relatively nearby, as was done by
BT09. BT09 also tried to estimate the expected amount of foreground
emission assuming it comes from a smoothly distributed population of
small dust grains heated by massive stars that follow an exponential
distribution in the Galaxy. For a cloud at a distance of $5$~kpc at a
Galactic longitude of $l\sim 30^\circ$, we estimate that 27\% of the
observed Galactic diffuse emission is from material in front of the
cloud. For a part of the cloud that has $\Sigma$ estimated to be $\sim
0.1\:\gcc$ in the absence of a foreground correction, applying this
correction raises $\Sigma$ by about a factor of two. If this
foreground correction is not applied then the largest values of
$\Sigma$ that can be derived are only $\sim 0.2\:\gcc$. This
model-dependent estimate of the foreground is quite uncertain and one
of the main reasons that BT09 concentrated on nearby IRDCs. Poor
understanding of the foreground emission is likely to limit the
reliability of the mass surface densities and masses of IRDCs derived
for large samples of objects (e.g. Simon et al. 2006; Peretto \&
Fuller 2009), especially for the more distant objects. Local heating
of small dust grains that then produce MIR foreground emission cannot
be accounted for in the BT09 model of foreground estimation and this
can affect even nearby IRDCs. However, choosing relatively quiescent
IRDCs can help minimize this particular source of uncertainty. One of
the main goals of this paper is to introduce a new, improved method to
measure the intensity of the foreground emission.

The structure of this paper is as follows. In \S\ref{S:method} we
introduce the method of saturation-based MIR extinction mapping. In
\S\ref{S:results} we present the results of applying this method to
study the structure of 42 massive starless and early-stage core/clumps
located in 10 IRDCs. In \S\ref{S:discussion} we discuss the
implications of these results for massive star and star cluster
formation theories.

\section{Saturation-Based MIR Extinction Mapping}\label{S:method}

The MIREX mapping technique requires knowing the intensity of
radiation directed towards the observer at a location just behind the
cloud of interest, $I_{\nu,0}$, and just in front of the cloud,
$I_{\nu,1}$. Then for negligible emission in the cloud and a
simplified 1D geometry,
\begin{equation}
\label{eq:radtrans}
I_{\nu,1}= e^{-\tau_\nu} \:I_{\nu,0},
\end{equation}
where the optical depth $\tau_\nu =\kappa_\nu \Sigma$, where $\kappa_\nu$
is the total opacity at frequency $\nu$ per unit gas mass and $\Sigma$ is the gas mass
surface density.

We cannot see $I_{\nu,0}$ directly, so it must be estimated by
interpolation from surrounding regions. BT09 tried two main ways to do
this using median filters. The Large-Scale Median Filter (LMF) method
used a square filter of size 13\arcmin\ that was much larger than the
IRDCs of interest so that the clouds did not significantly depress the
estimated median intensity. This has the advantage of not assuming any
prior knowledge about the IRDC, but the disadvantage of a coarse
angular resolution of background intensity fluctuations. For studying
specific IRDCs that can be defined as occupying a certain region of
the sky, e.g. an ellipse, we thus introduced the Small-Scale Median
Filter (SMF) method.  Here the size of the filter is set to be
one-third of the major axis of the IRDC ellipse (defined by Simon et
al. 2006), but it is only applied for background estimation outside of
the IRDC ellipse. Inside the ellipse we estimate the background by
interpolating from the surrounding background model. BT09 estimated
that the uncertainties in background estimation due to this
interpolation were at a level of $\lesssim 10\%$, which corresponds to
$\Sigma\lesssim 0.013\:\gcc$.

\begin{figure*}
\begin{center}$
\begin{array}{c}
\includegraphics[width=6.5in]{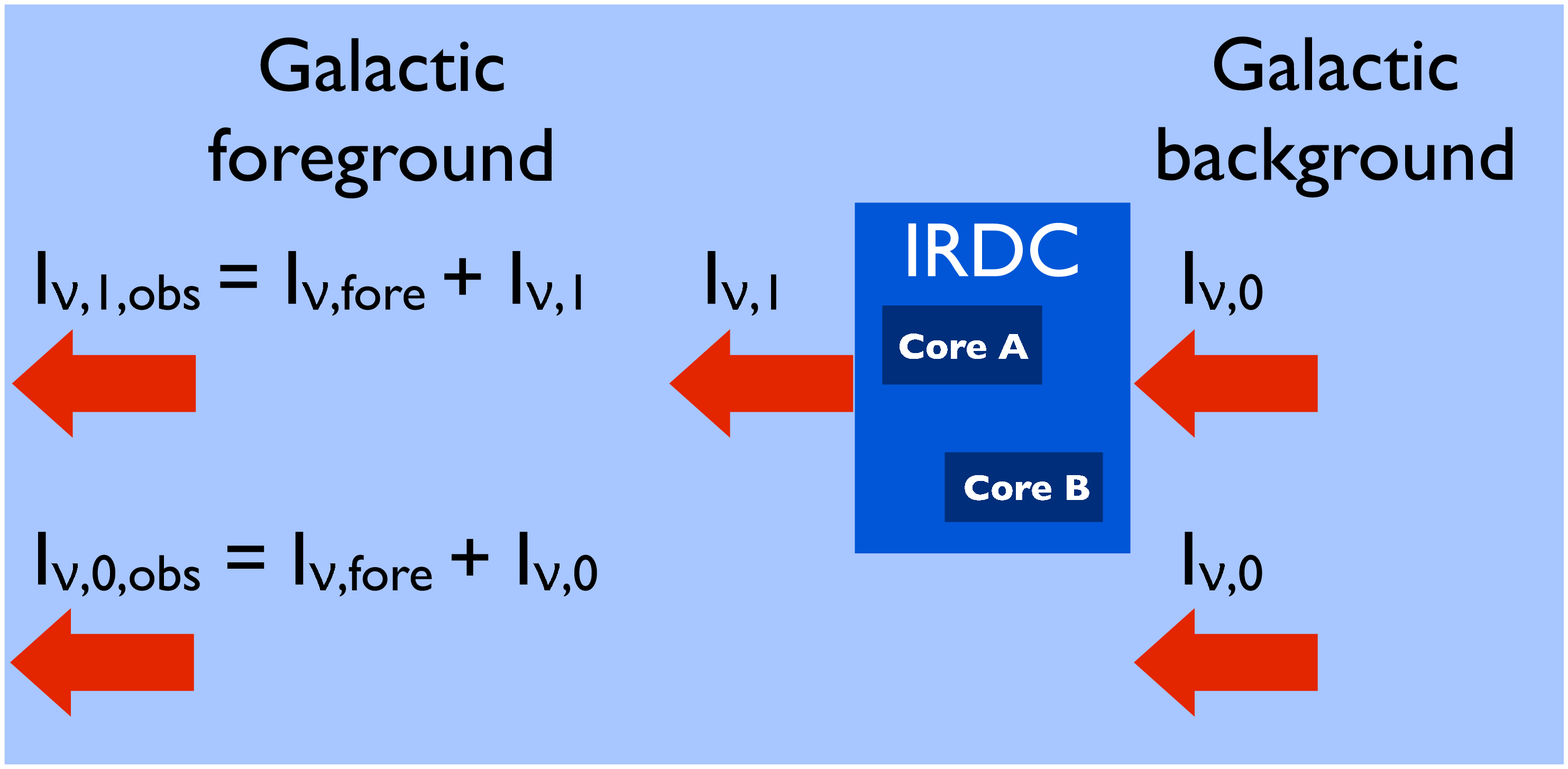}
\end{array}$
\end{center}
\caption{\footnotesize 
Schematic of simple 1D model of radiative transfer through an IRDC,
assuming negligible emission from the IRDC at frequency $\nu$.  If
independent cores (i.e. localized density maxima) A and B are both of
sufficiently high $\Sigma$, then $I_{\nu,1}\ll I_{\rm \nu,fore}\simeq
I_{\rm \nu,1,obs}(A,B)$, providing an accurate, empirical estimate of
the foreground intensity to the IRDC.
}
\label{fig:schematic}
\end{figure*}

However, because of foreground emission towards the IRDC, we actually observe (see Figure~\ref{fig:schematic})
\begin{equation}
\label{eq:radtrans2}
I_{\rm \nu,1,obs}= I_{\rm \nu,fore} + I_{\nu,1} = I_{\rm \nu,fore} + e^{-\tau_\nu} \:I_{\nu,0},
\end{equation}
and towards the IRDC surroundings, where we are trying to estimate $I_{\nu,0}$, we actually observe
\begin{equation}
\label{eq:radtrans3}
I_{\rm \nu,0,obs} = I_{\rm \nu,fore} + I_{\nu,0}.
\end{equation}
The primary uncertainty in the MIREX mapping method of BT09 for
larger values of $\Sigma$ is the estimate of the level of the
foreground contribution to the intensity, $I_{\rm \nu,fore}$.  In
order to increase the method's sensitivity to higher values of mass
surface density, we now describe a new, empirical method to estimate this
contribution.  

If a core has a high enough mass surface density, then it will block
essentially all the background emission. The observed minimum
intensity in the cloud will then be approximately equal to the
foreground emission and the angular distribution of this intensity may
appear to flatten or ``saturate''. It is difficult to be certain if an
individual dense core is saturated (as is sometimes assumed if the foreground is simply  
estimated from the darkest pixel, e.g. Ragan et al. 2009).  However, we propose that if the
minimum intensity is observed to ``be the same'' in two or more
``independent'' cores (i.e. spatially resolved peaks in $\Sigma$) in
the same cloud, then this is very likely due to saturation. In
practice, by ``be the same'' we adopt the condition to be within
$2\sigma$ of each other, where $\sigma$ is the uncertainty in the
GLIMPSE 8~$\rm \mu m$ intensities of 0.6~MJy~sr$\rm ^{-1}$ (Reach et
al. 2006). By ``independent'' we adopt an angular separation of at
least 8\arcsec\, i.e. much larger than the $2\arcsec$ FWHM of the
Spitzer IRAC $8\:{\rm \mu m}$ PSF.

The algorithm for this method is as follows (see also Figure~\ref{fig:schematic}):

\noindent
1) Define a region of the sky as the ``IRDC''. Following BT09, we use
the ellipses from the catalog of Simon et al. (2006), which were based
on MSX images.

\noindent
2) Using GLIMPSE 8 $\rm \mu m$ images, find the minimum value of
$I_{\rm \nu,1,obs}$ inside the IRDC, $I_{\rm \nu,1,obs}({\rm min})$.

\noindent
3) Search for all pixels in the IRDC with $I_{\rm \nu,1,obs}({\rm
  min})<I_{\rm \nu,1,obs}<I_{\rm \nu,1,obs}({\rm min})+2\sigma$. If
there are pixels meeting this criteria that are independent (to be
conservative we adopt $\geq 8\arcsec$ away from the IRDC minimum), then
the IRDC is defined to be saturated, all the above pixels are labeled
as ``saturated pixels'', and the following steps are carried out.

\noindent
4) The mean value of $I_{\rm \nu,1,obs}$ of the saturated pixels is
evaluated, $I_{\rm \nu,1,obs}({\rm sat})$. We set the foreground
intensity (which includes all sources of emission: Galactic, Zodiacal
and instrumental, Battersby et al. 2010) to be $I_{\rm \nu,fore} =
I_{\rm \nu,1,obs}({\rm sat}) - 2\sigma$. This subtraction is motivated
to have $I_{\rm \nu,fore}<I_{\rm \nu,1,obs}({\rm min})$ and thus give
every pixel a finite value of $\tau$ and thus $\Sigma$.

If all the ``saturated pixels'', defined above, really did have
negligible values of $I_{\nu,1}$ and had a distribution of intensities
that was relatively uniform in the above range, then
$I_{\rm \nu,fore} \simeq I_{\rm \nu,0,obs}({\rm min})+1\sigma$ and our
method would be underestimating $I_{\rm \nu,fore}$ by
$2\sigma=1.2$~MJy/sr. In fact, we do find for the $\sim 300$
``saturated pixels'' in the 10 IRDCs of our sample, a mean
value of $I_{\rm \nu,1,obs} - I_{\rm \nu,1,obs}({\rm min})\simeq
0.7$~MJy/sr.
Thus, we are likely to be
underestimating $I_{\rm \nu,fore}$ (overestimating $I_{\nu,1}$) by an
amount $\simeq 2\sigma= 1.2$~MJy/sr. This leads to a
value of $\Sigma$ where our measured values are significantly affected by saturation:
\begin{equation}
\label{eq:radtrans3}
\Sigma({\rm sat})=\frac{\tau_{\nu}({\rm sat})}{\kappa_{\nu}} = \frac{{\rm ln} (I_{\nu,0}/I_{\nu,1})}{\kappa_{\nu}},
\end{equation}
where $I_{\nu,1}/I_{\nu,0} = I_{\rm 8\mu m,1}/I_{\rm 8\mu m,0}= 
e^{-\tau_{\rm 8\mu m}({\rm sat})}$ so $\tau_{\rm 8\mu m}({\rm
  sat}) = {\rm ln} (I_{\nu,0}/I_{\nu,1})$, that is set by
$I_{\nu,1}=2\sigma\rightarrow 1.2$~MJy/sr. For a typical
IRDC with $I_{\rm \nu,0,obs}=100$~MJy/sr, $I_{\rm \nu,fore}= f_{\rm
  fore} I_{\rm \nu,0,obs}=30$~MJy/sr so that $I_{\nu,0}=70$~MJy/sr,
then $\tau_{\rm 8\mu m}({\rm sat}) = 4.07$ and $\Sigma({\rm
  sat}) = 0.544\:\gcc$. For a region of such a cloud with a
true value of $\Sigma=0.5\:\gcc$ so that $I_{\nu,1}=1.65$~MJy/sr, if
we have underestimated $I_{\rm \nu,fore}$ by $1.2$~MJy/sr, then we
would infer $\Sigma = 0.427\:\gcc$.  Similarly, for a true
$\Sigma=0.4\:\gcc$, we would infer $\Sigma = 0.361\:\gcc$.
The values of $\Sigma$(sat) calculated with $I_{\rm 8\mu
  m,1}=1.2$~MJy/sr for the 10 IRDCs in our sample are listed in
Table 1. They range from $\Sigma$(sat)=0.33 to
0.52$\:\gcc$ as one progresses along the Galactic plane towards $l=0$,
where the background is brightest.

\begin{figure*}
\begin{center}$
\begin{array}{c}
\includegraphics[width=4.7in]{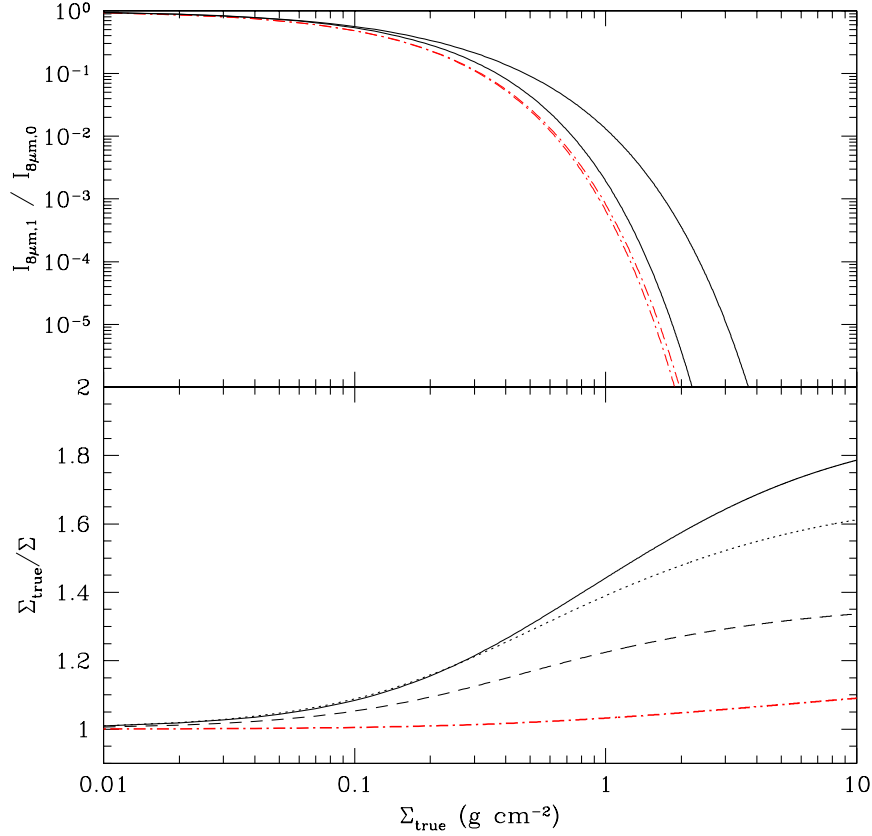}
\end{array}$
\end{center}
\caption{\footnotesize 
Effect of finite filter width on estimates of $\Sigma$ and accuracy of
an approximation using a single band-averaged opacity (see text). The
{\it Spitzer} IRAC band 4, i.e. $8{\rm \mu m}$, filter has sensitivity
from about 6.5 to 9.5~${\rm \mu m}$.
{\it Top panel:}
Ratio of transmitted to incident flux as a function of true mass
surface density, $\Sigma_{\rm true}$.  
The result for the band-average opacity for the moderately coagulated
thin ice mantle dust model of OH94 (our fiducial model) is shown by
the lower red dot-dashed line. The actual transmitted flux, calculated by
integrating the transfer equation over the bandpass, is shown by the
upper red dot-dashed line. The equivalent quantities for the Draine
(2003) $R_V=3.1$ dust model are shown by the lower and upper black solid
lines: the effect is larger here as this dust model shows larger
opacity variations across the band. 
{\it Bottom panel:} Effect on estimation of $\Sigma$. Given an
observed ratio of transmitted to incident intensities, the true mass
surface density, $\Sigma_{\rm true}$, will be greater than that
estimated using the band-averaged opacity, $\Sigma$. The ratio of
$\Sigma_{\rm true}/\Sigma$ is shown by the red dot-dashed line for the above
OH94 thin ice mantle model. The OH94 uncoagulated thin ice mantle model gives essentially the same result. The error is a few percent in the region
of interest of the IRDC cores in this study. Also shown are these effects for the Draine
(2003) $R_V=3.1$ (black solid line), $R_V=5.5$ (black dotted line) and $R_V=5.5$ Case B (black dashed line).
}
\label{fig:satcheck}
\end{figure*}

An additional uncertainty results from our use of a single effective
value of $\kappa_{\rm 8\mu m} = 7.5\:{\rm cm^{2}\:g^{-1}}$ averaged
over the {\it Spitzer} IRAC $8{\rm \mu m}$ band, weighting by the
filter response function, the spectrum of the Galactic background and
the dust opacity model (BT09). Since these functions vary over this
wavelength range (see Fig.~1 of BT09), at large optical depths the
actual transmitted intensity will be greater than that predicted,
being more dominated by the region of the spectrum with the lowest
opacity. The net effect is an underestimation of the true mass surface
density, given the observed ratio of transmitted to incident
intensities. We have investigated the size of this effect by
integrating the transfer equation~(\ref{eq:radtrans}) over the above
weighting functions (see Fig.~\ref{fig:satcheck}). For our fiducial dust
model (the moderately coagulated thin ice mantle model of OH94), which
has a relatively flat MIR opacity law, the effect is small: just a few
percent effect up to value of $\Sigma\sim 1\:\gcc$, rising to about a
10\% effect by $\Sigma=10\:\gcc$. For illustrative purposes,
Fig.~\ref{fig:satcheck} also shows the results for the Draine (2003)
$R_V=3.1$ dust model, more appropriate for the diffuse ISM, which has
bare grains and stronger variation of opacity across this wavelength
range. Now the effect leads to an underestimation of $\Sigma$ by up to
several tens of percent for $\Sigma \sim 1\:\gcc$. Other dust models
we have considered, such as the Draine (2003) $R_V=5.5$ model, have
somewhat smaller underestimation factors.


A further additional systematic uncertainty results from the fact
that the foreground intensity will vary across the IRDC, especially
due to local radiation sources. The accuracy of the $\Sigma$ values
will be higher in regions closer to the locations of saturated cores,
where $I_{\rm \nu,fore}$ has been estimated and in IRDCs with minimal
local heating sources.

\section{Results}\label{S:results}

\subsection{IRDC Properties}\label{S:results_IRDCs}

\begin{figure*}
\begin{center}$
\begin{array}{cc}
\includegraphics[width=2.48in]{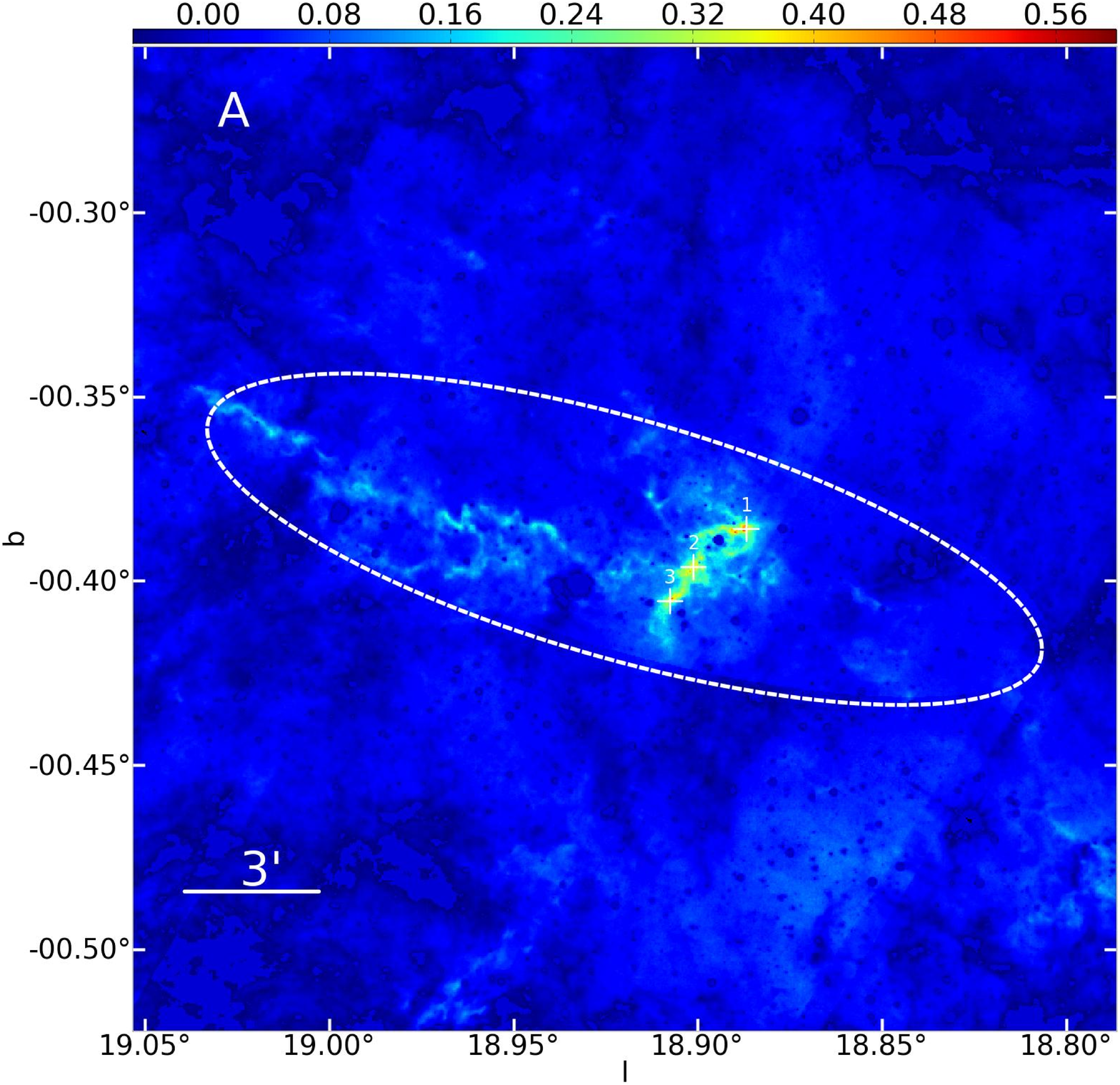} & \includegraphics[width=2.5in]{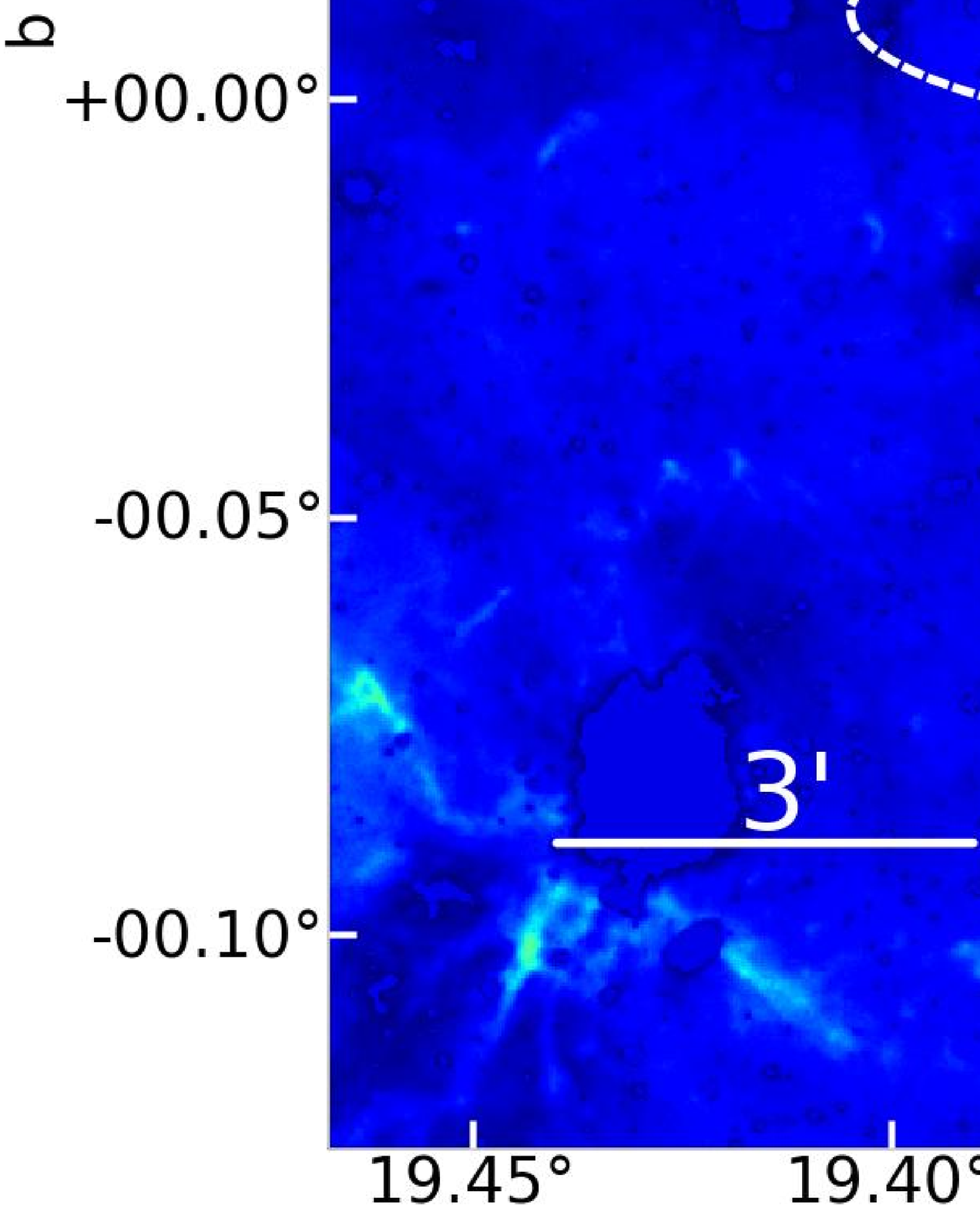} \\
\includegraphics[width=2.5in]{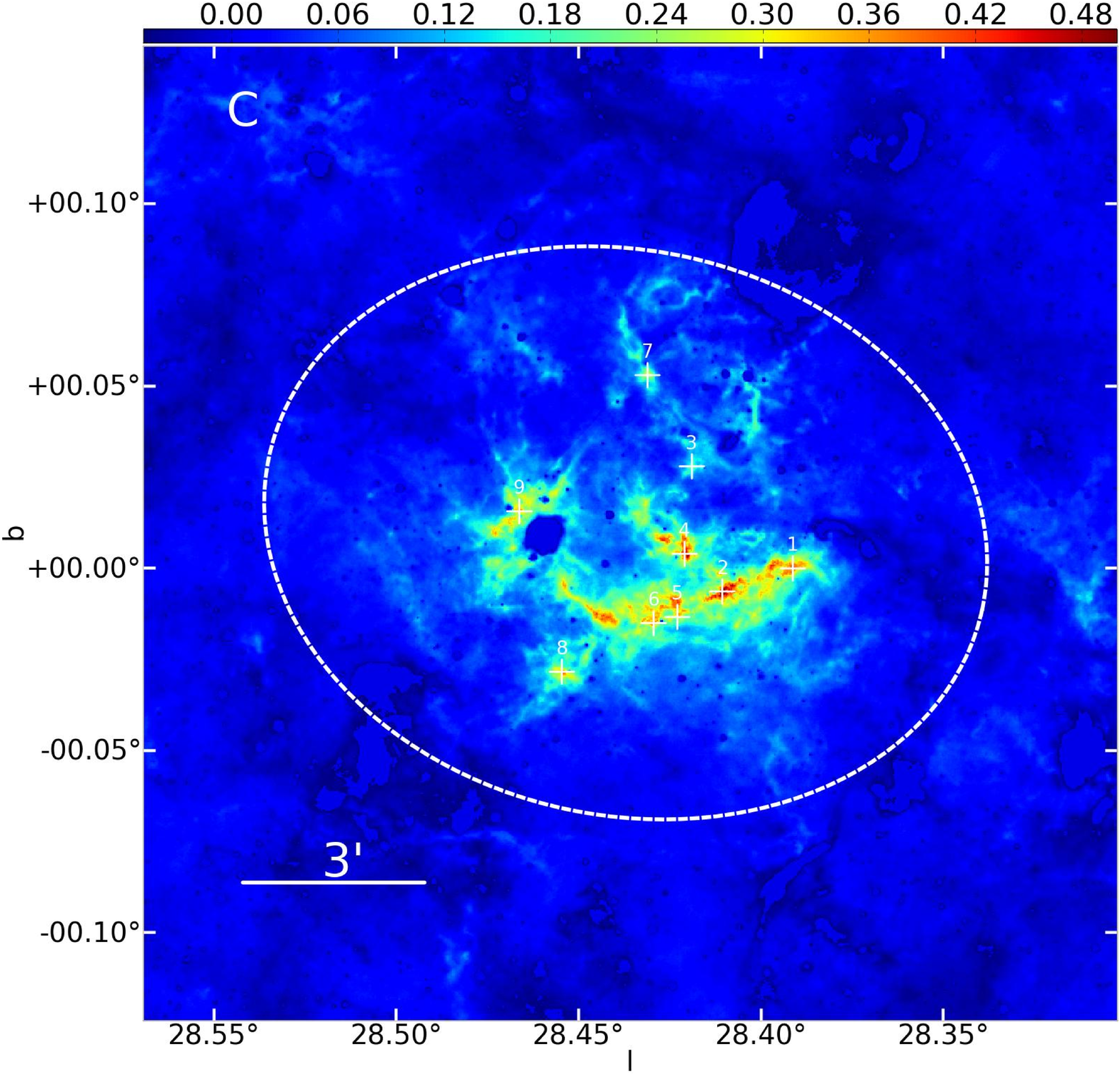} & \includegraphics[width=2.58in]{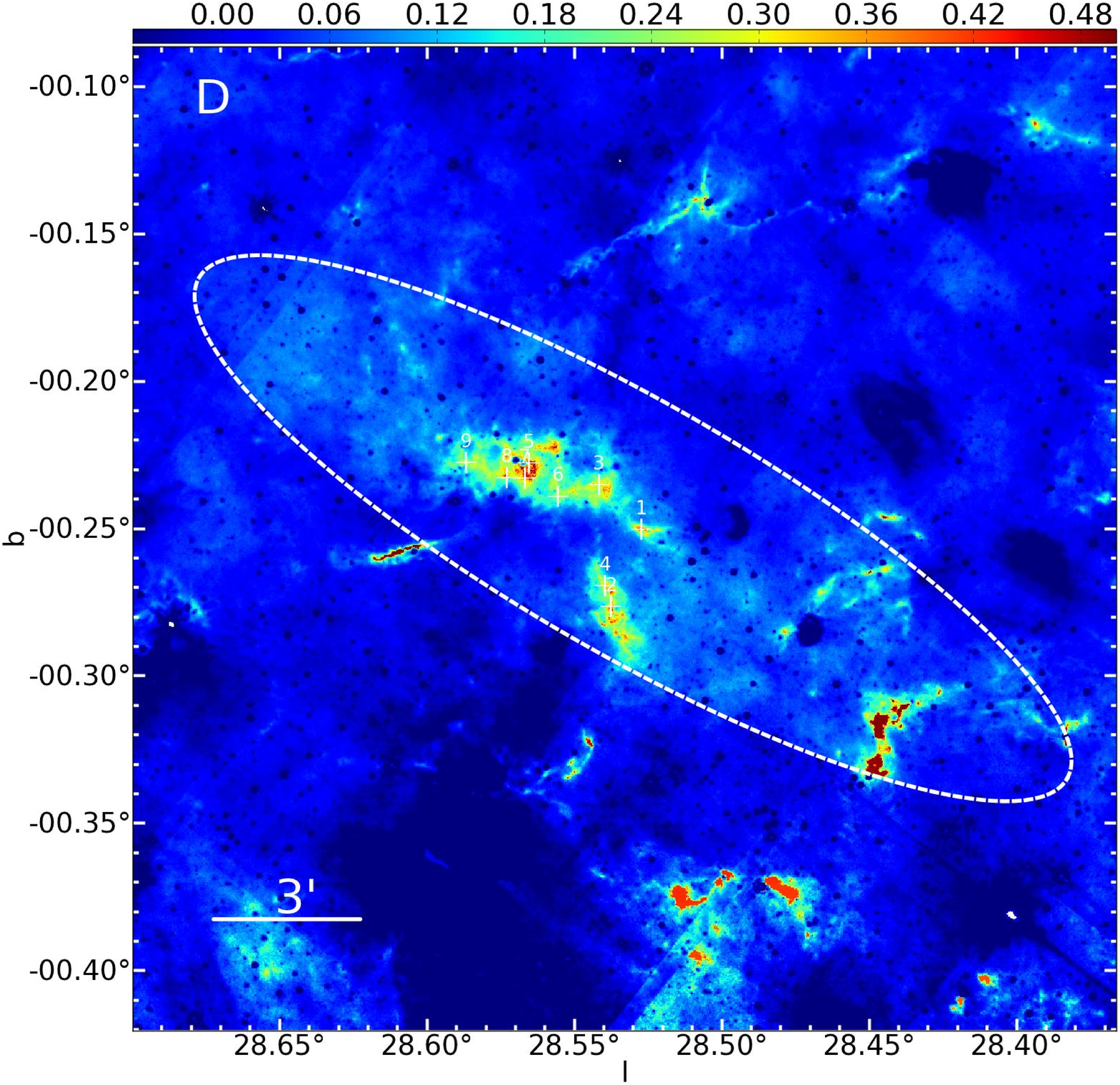} \\ 
\includegraphics[width=2.48in]{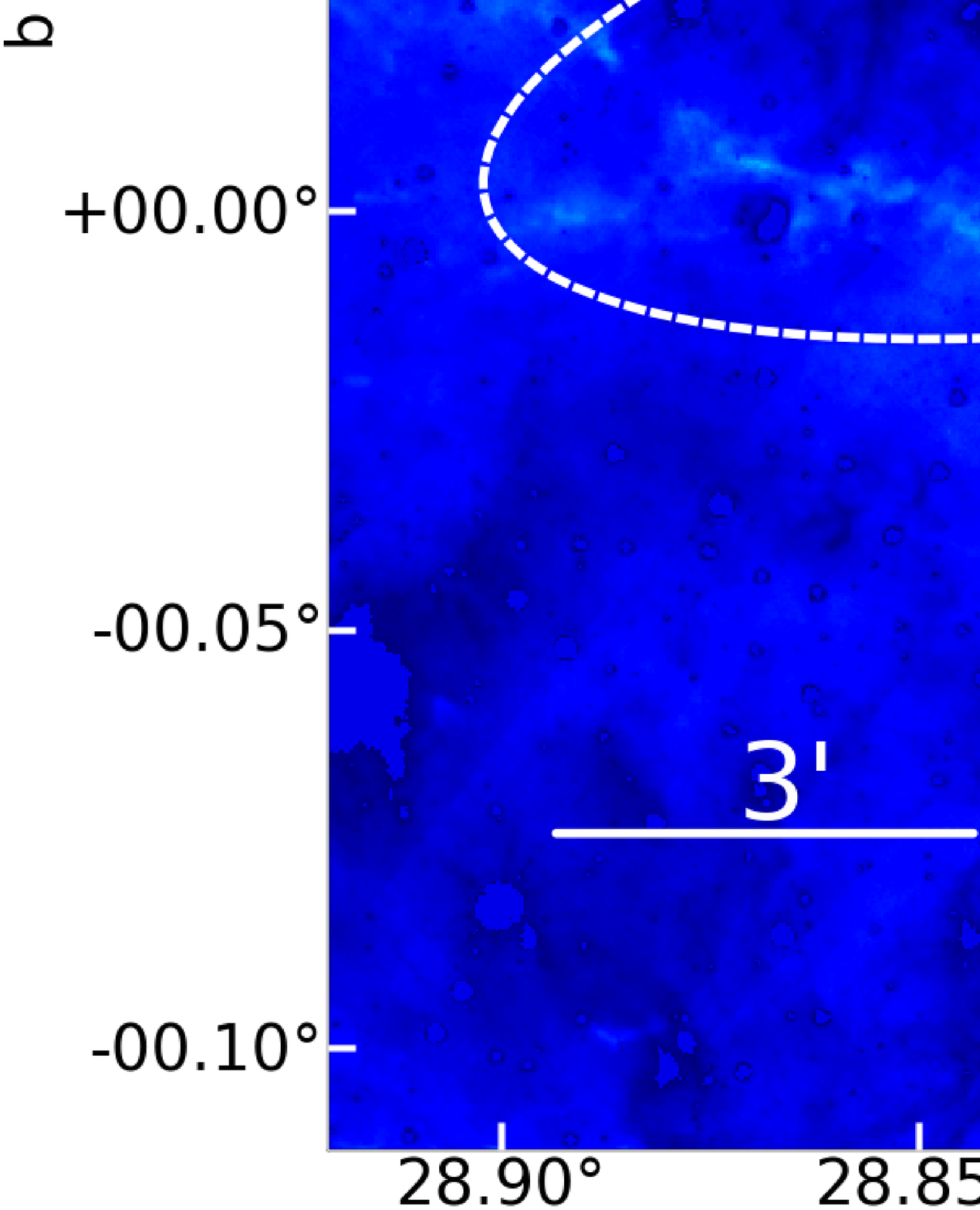}  & \includegraphics[width=2.5in]{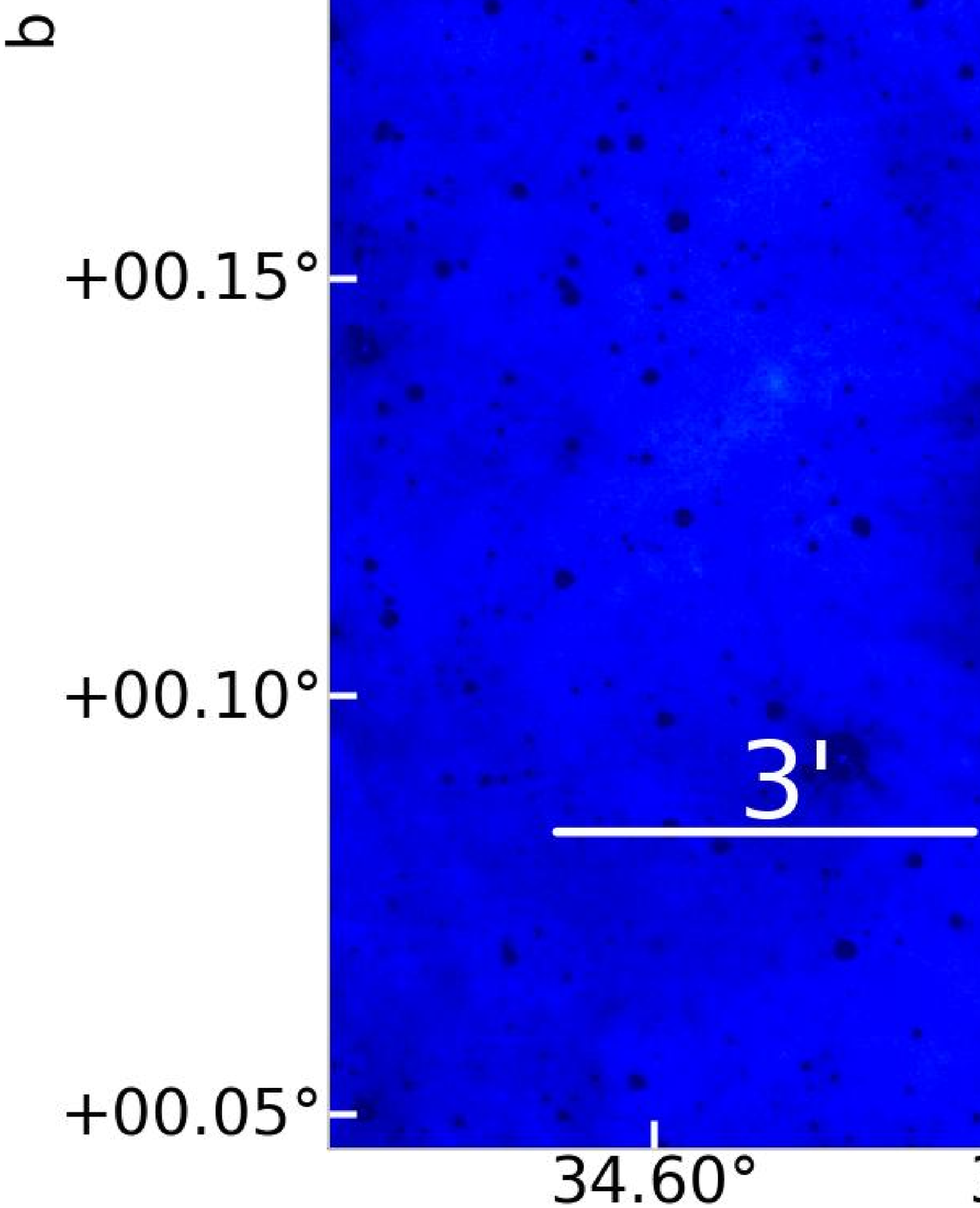} 
\end{array}$
\caption{\footnotesize
Mass surface density, $\Sigma_{\rm SMF}$, maps of IRDCs A-F derived
from MIREX mapping using {\it Spitzer} IRAC $\rm 8\:\mu m$ images with pixel scale of
1.2\arcsec\ and angular resolution of 2\arcsec\ using a saturation-based
estimate of the foreground emission (\S\ref{S:method}). The color
scale is indicated in $\gcc$. The dashed ellipse, defined by Simon et
al. (2006) based on MSX images, defines the region where the
background emission is estimated not directly from the small-scale
median filter average of the image intensity, but rather by
interpolation from nearby regions just outside the ellipse. The
locations of the massive starless cores we have selected for analysis
(\S\ref{S:results}) are marked with crosses. Bright MIR sources appear
as artificial ``holes'' in the map, where we have set the values of
$\Sigma=0\:\gcc$.
\label{fig:IRDC1}}
\end{center}
\end{figure*}

\begin{figure*}
\begin{center}$
\begin{array}{cc}
\includegraphics[width=2.5in]{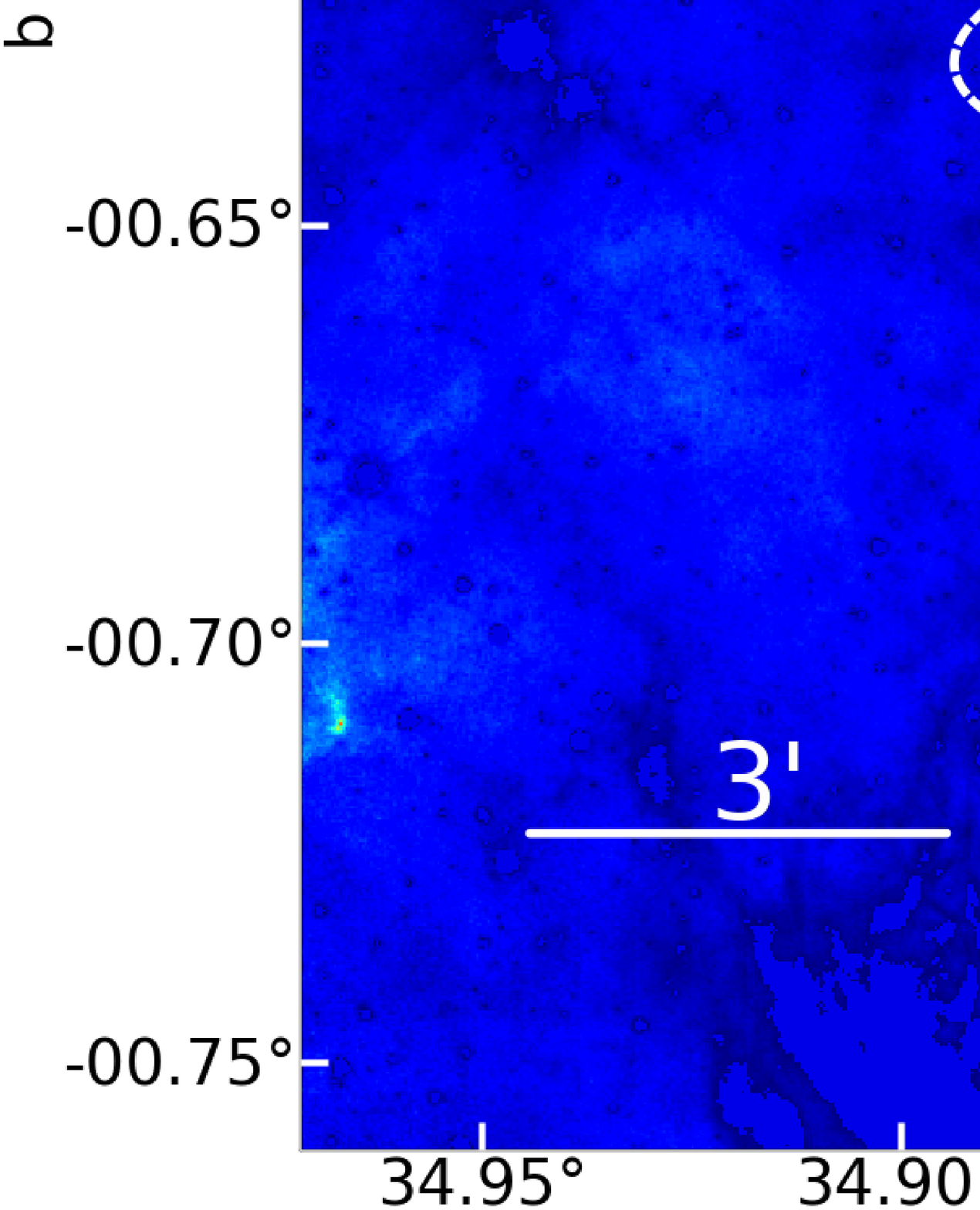} & \includegraphics[width=2.58in]{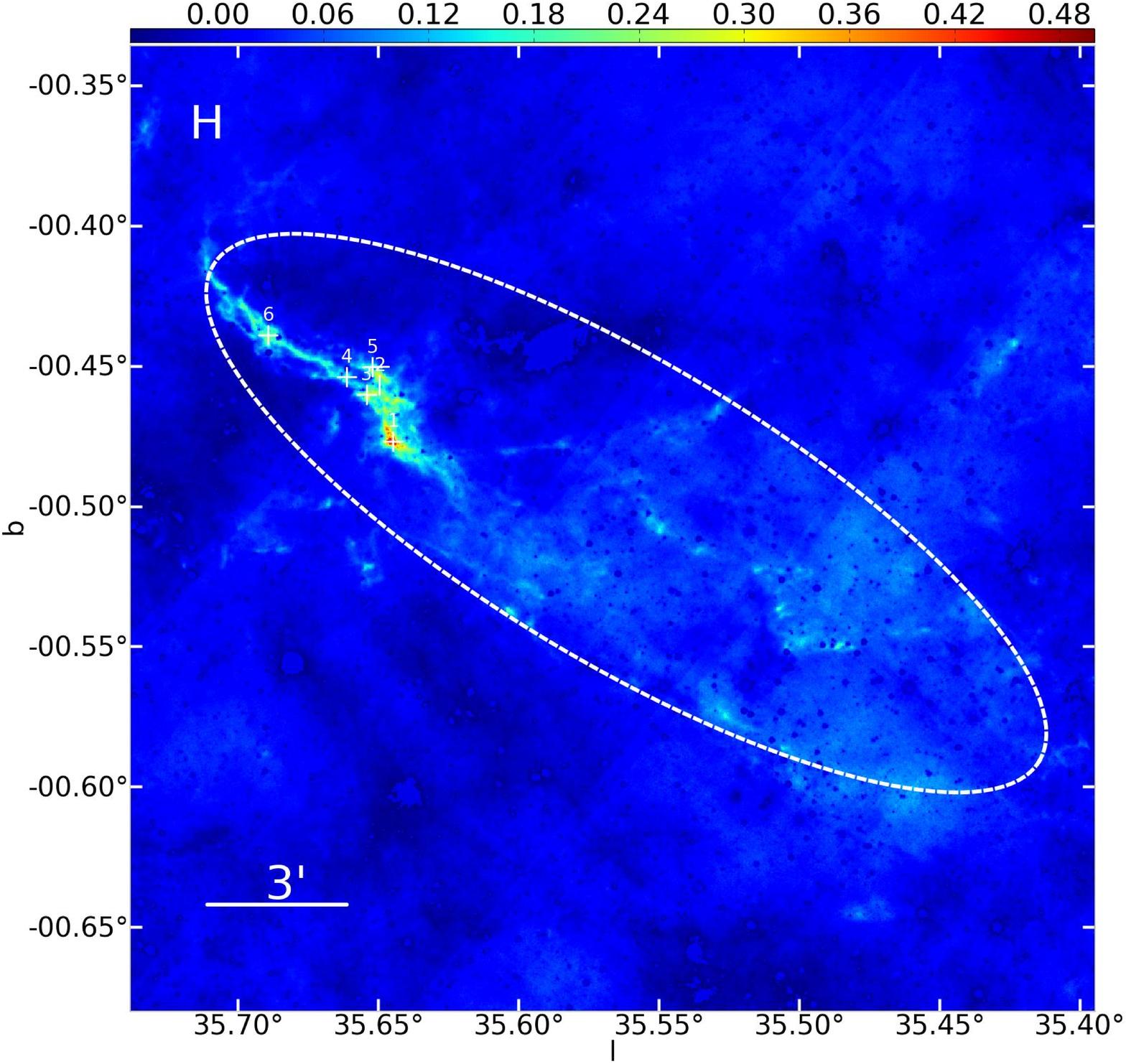} \\
\includegraphics[width=2.5in]{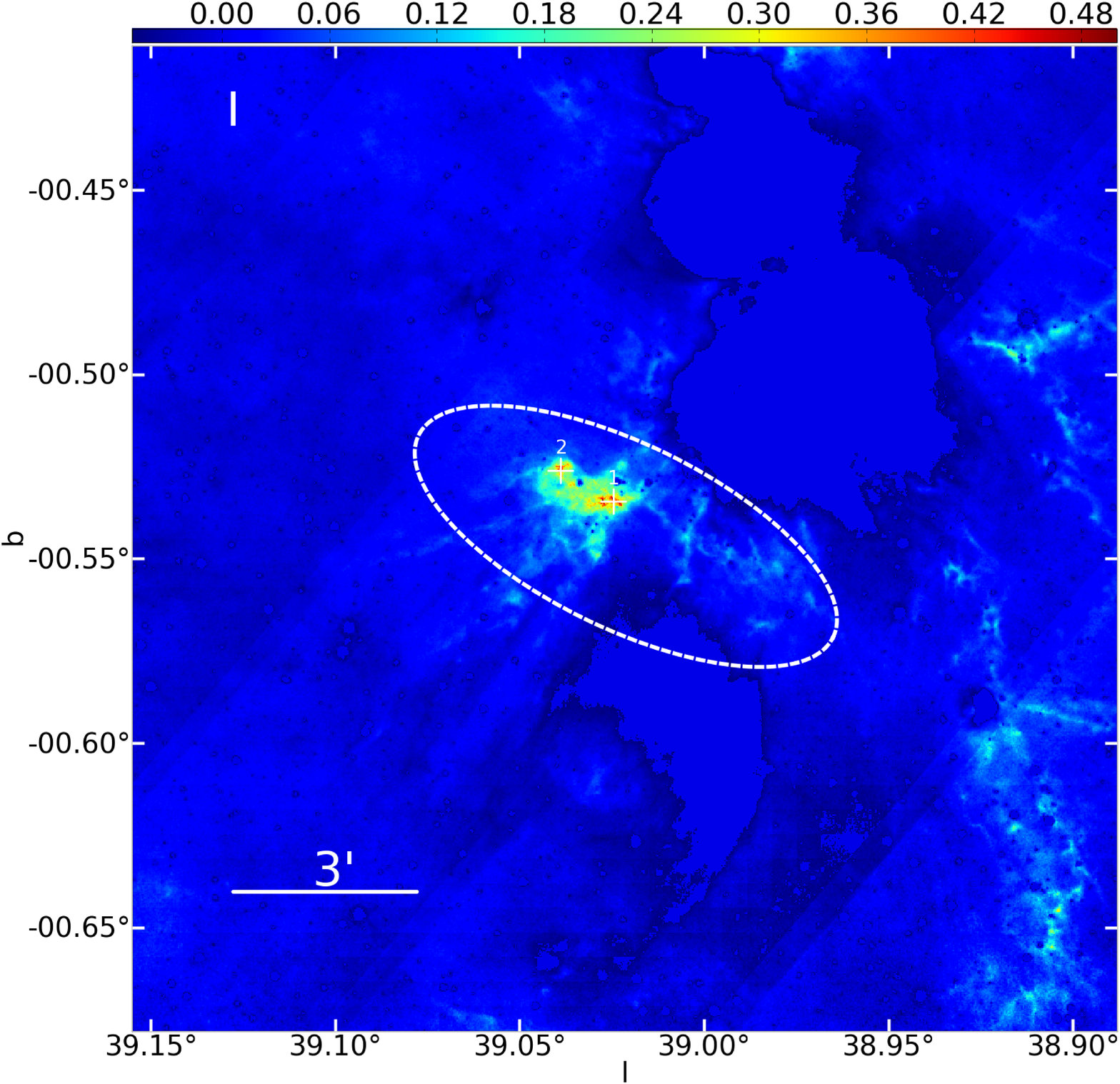} & \includegraphics[width=2.58in]{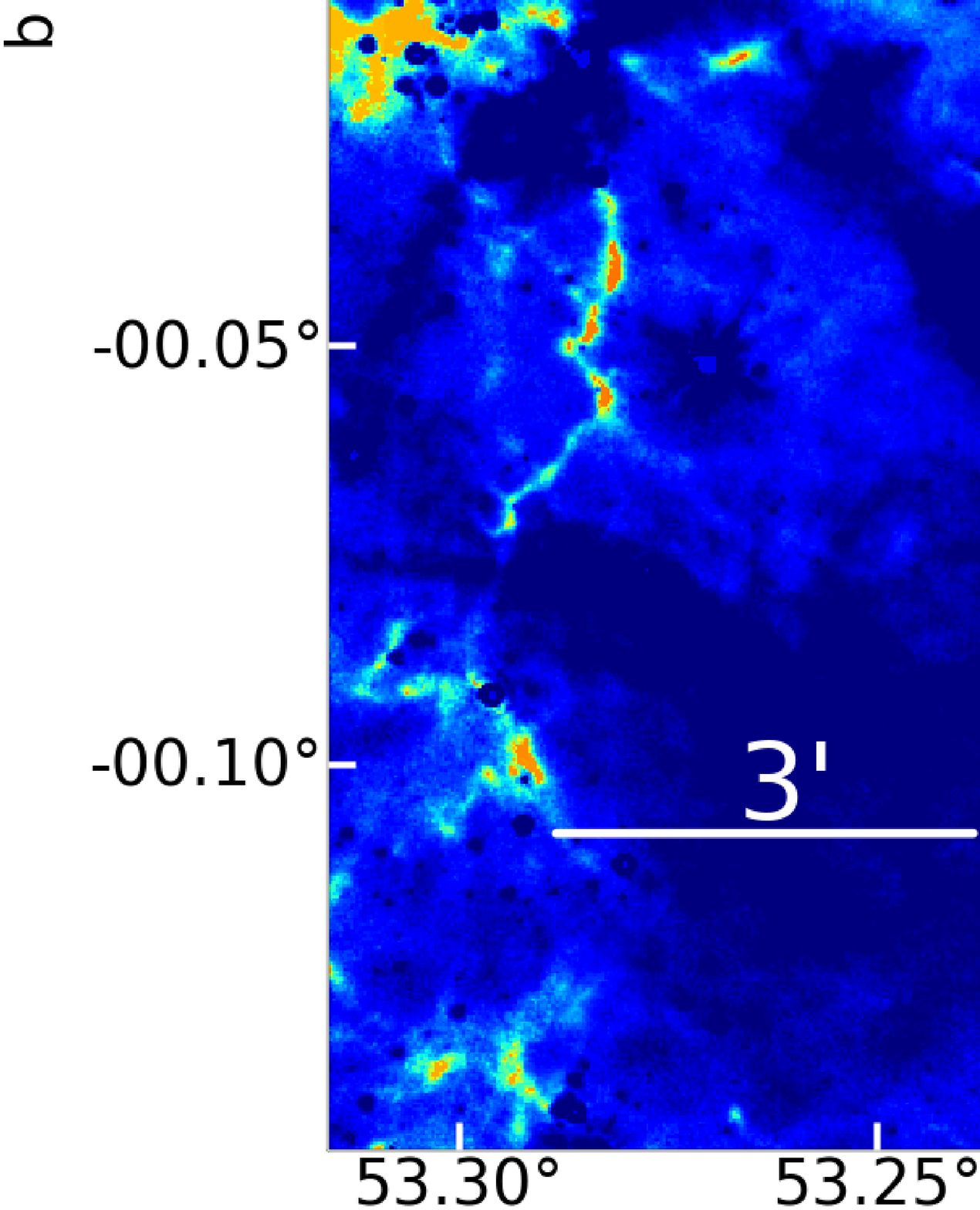}  
\end{array}$
\caption{\footnotesize
Mass surface density, $\Sigma_{\rm SMF}$, maps (in the same format as Fig.~\ref{fig:IRDC1}) of IRDCs G-J derived from
MIREX mapping using {\it Spitzer} IRAC $\rm 8\:\mu m$ images.
\label{fig:IRDC2}}
\end{center}
\end{figure*}

Following the above algorithm, we find that all 10 IRDCs of the BT09
sample exhibit the effects of saturation. In hindsight, this is not
too surprising since these clouds were selected to have relatively
high contrast against the background. The $\Sigma$ maps of the clouds
are shown in Figs.~\ref{fig:IRDC1} and \ref{fig:IRDC2}. The properties
of these clouds are listed in Table 1, where we also
compare their properties to those derived with the SMF method of BT09
with the analytic model of foreground estimation. Using the
saturation-based estimate of foreground emission, we find $I_{\rm
  \nu,fore}$ and thus $f_{\rm fore}$ has increased in all the
clouds. Thus the highest values of $\Sigma$ that we infer have risen
from $\sim 0.1-0.3\:\gcc$ in BT09 to $\sim 0.4-0.6\:\gcc$ in this paper. The
mean values, $\bar{\Sigma}_{\rm SMF}$ rise by smaller factors, so that
the total cloud masses rise by on average a factor of 2.0.
A comparison of the global properties of these IRDC with the
predictions of theoretical models of the interstellar medium will be
presented in a separate paper.

\subsection{Massive Starless Cores and Clumps}

\subsubsection{Locating the Cores}

The cores we are considering are a subset of those originally
identified by Rathborne et al. (2006) based on their mm dust continuum
emission, observed with the IRAM-30m Telescope at 11\arcsec\ FWHM
angular resolution. BT09 selected 43 cores from the Rathborne et
al. sample, excluding those with significant $\rm 8\:{\rm \mu m}$
emission and those with low-contrast against the MIR background (i.e.
with $\Sigma\lesssim 0.02\:\gcc$). Here, we have excluded one
of the BT09 cores, E4, because its GLIMPSE image suffers from a
diagonal boundary artifact where the intensity of the diffuse emission
changes abruptly.

BT09 treated the cores as circular with radii equal to half the
reported FWHM diameter of Gaussian fits that Rathborne et al. (2006)
fitted to their mm continuum images. These circles were centered at
the coordinates estimated by Rathborne et al. As discussed in
\S\ref{S:results_IRDCs}, we expect our derived values of $\Sigma$ to
be higher (and more accurate) than those of BT09. Comparing the core
masses of BT09 with those derived here for the same regions, we find
they have typically increased by a factor of about 2.2.

In this paper, we now redefine the core center to be the center of
the highest $\Sigma$ pixel inside the previous core boundary. If there
are two or more adjacent saturated pixels at the core center, then
their average position is used to define the center. In fact, 17 of
the 42 cores exhibit saturation. Occasionally, after inspecting the
$\rm 8\:{\rm \mu m}$ GLIMPSE and $\rm 24\:{\rm \mu m}$ MIPSGAL images,
we note the presence of MIR sources near ($<7.5\arcsec$) the core
center. This occurs in 9 of the 42 cores (B2, C6, C8, D5, D6, D8, E2,
E3, I1). In order to focus on massive starless and early-stage cores, we shift the
center to a new, nearby ($\lesssim 3\arcsec$) $\Sigma$ maximum to
avoid any major sources of MIR emission within a radius of
$7.5\arcsec$ of the new center. In several cases (C4, D4, F2, J1), the
$\Sigma$ map inside the Rathborne et al. core boundary does not
exhibit a well-defined high $\Sigma$ peak. In these cases we select a
new core center as close as possible to the Rathborne et al. core:
normally this is within a few arcseconds of the boundary, but for F2 it is
about 10\arcsec\ outside.


Figure \ref{fig:coreA1}a shows the $\Sigma$ map of core A1, extracted
from the larger image of IRDC A, shown in
Fig.~\ref{fig:IRDC1}a. Pixels suffering from saturation are marked
with small white squares. The core center is marked with a
cross. Similar images of all 42 cores are shown in Figs.~\ref{fig:cores1} to
\ref{fig:cores7}.

We note that 5 of the IRDCs (B, E, G, H, J) only have one core that
exhibits saturation. This is possible because the condition to
determine if an IRDC is saturated is based on independent positions
(separated by at least 8\arcsec) having the same foreground intensity
(to within $2\sigma$), rather than requiring 2 cores to meet this
condition. The 42 cores we have selected for analysis are not meant to
be a complete census of all the dense regions in these IRDCs. For
example, IRDC J only has one core selected.

The core $\Sigma$ maps exhibit complex structure. It is not easy to
define the boundary of a core from its surrounding clump, especially
when one recalls we are viewing a 3D structure in projection. $\rm
^{13}CO$(1-0) data exist for these IRDCs via the Galactic
Ring Survey (Jackson et al. 2006), but with poor angular resolution
($\sim46\arcsec$). Also, in the cores we expect CO to be highly
depleted from the gas phase due to freeze-out onto dust grain
surfaces. Widespread CO depletion has been observed in IRDC H by
comparing our $\Sigma$ map with $\rm C^{18}O$ emission observed with
the IRAM-30m Telescope (Hernandez et al. 2011). Fontani et al. (2011)
observed $\rm N_2H^+$, which does not freeze-out so readily as CO,
from 4 of our cores (C1, F1, F2, G2), but again with relatively poor
angular resolution ($>10\arcsec$). Thus, given the lack of high
angular resolution molecular line data for all the cores, here we
present a uniform analysis of core structure, based only on the
extinction maps.

For simplicity, we first make radial profiles of mean total mass
surface density, which we refer to as $\Sigma_{\rm cl}$ since it
includes contribution from the clump (see below), considering a series
of annuli extending from the core center with width equal to 1 pixel,
i.e. 1.2\arcsec\ . Fractional overlap of pixels with these annuli are
accounted for. ``Holes'' in the $\Sigma$ maps due to MIR sources are
treated as having a zero, i.e. negligible, value. In general, these
sources do not significantly affect our characterization of core
structure, at least in the inner $\sim 7.5\arcsec$, since we have
chosen cores that are relatively free of strong sources (E3 is the
worst affected, and is somewhat exceptional in this regard). Larger
annuli are minimally affected by individual MIR sources, which cover
only a small fraction of the area. We extend the radial profiles out
to a maximum angular scale equal to that reported by Rathborne et
al. (2006) based on mm dust emission, i.e. a radius equal to one FWHM
of their fitted Gaussian profile. As we will see, this is generally
larger than the scale over which the core can be considered to be a
single monolithic object.

For Core A1, Figure \ref{fig:coreA1}b shows $\Sigma_{\rm cl}(r)$ with
blue open square symbols, plotted at the radii corresponding to the
center of each annulus. The total enclosed mass, which we refer to as
the clump mass $M_{\rm cl}(r)$, is indicated by the blue long-dashed
line.

\subsubsection{Core and Clump Properties at the $60\:M_\odot$ Enclosed Mass Scale}

Before considering a more detailed analysis of the radial structure,
it is instructive to first consider the properties of these core/clump
objects at a scale where the total mass enclosed is $M_{\rm
  cl}=60\:M_\odot$. If all this mass were in a core, then such a core
has the potential to form a $\sim 30\:M_\odot$ star, given expected
star formation efficiencies of $\sim 50\%$ due to protostellar
outflows (Tan \& McKee, in prep.).  Note that there is of course no
guarantee that all our sources will collapse in this way and on
statistical grounds one would not expect them to: most are likely to
undergo fragmentation to form lower-mass stars.\footnote{One cannot
  distinguish between massive star formation Core Accretion and
  Competitive Accretion theories (Bonnell et al. 2001) simply by
  observing that a massive structure is actually composed of
  sub-fragments (c.f. Bontemps et al. 2010, their section 4.6).} For
Core A1, $60\:M_\odot$ is enclosed within $R_{\rm cl}=0.0962$~pc, so
at this scale $\bar{\Sigma}_{\rm cl}=0.431\:\gcc$. A black dashed
circle with this radius is shown in Fig.~\ref{fig:coreA1}a and these
core properties are listed in Table 2. Core A1 happens
to be one of the most extensively saturated cores at the scale of an
enclosed mass of $60\:M_\odot$ (along with C2, H1, I1, I2, J1), so these
numbers are likely to be significantly affected by saturation (which
causes us to underestimate $\Sigma$), so actually the radius enclosing
$60\:M_\odot$ would be smaller and $\bar{\Sigma}$ larger.

The distributions of the radii, $R_{\rm cl}$, and mean mass surface
densities, $\bar{\Sigma}_{\rm cl}$, of the 42 core/clumps at the
$M_{\rm cl}=60\:M_\odot$ scale are shown in Figure~\ref{fig:hist}a
(with the 6 highly saturated cores --- A1, C2, H1, I1, I2, J1 --- shown as
a shaded subset). The mean/median/RMS dispersion-about-the-mean of
$R_{\rm cl}(M_{\rm cl}=60\:M_\odot)= 0.121/0.114/0.0238$~pc. The
mean/median/RMS dispersion values of $\bar{\Sigma}_{\rm cl}(M_{\rm
  cl}=60\:M_\odot)$ $ = 0.296/0.318/0.0952 \:\gcc$ (see also
Table 3).

We next fit a power law density distribution, 
\begin{equation}
\label{eq:PL}
\rho_{\rm cl}(r) = \rho_{\rm s,cl} \left(\frac{r}{R_{\rm cl}}\right)^{-k_{\rm \rho,cl}}, 
\end{equation}
where $\rho_{\rm s,cl} = \mu_{\rm H} n_{\rm H,s,cl}$ (with $\mu_{\rm
  H}=2.34\times 10^{-24}\:{\rm g}$) is the density at the surface of
the clump, $R_{\rm cl}$. We project the above distribution to derive
$\Sigma_{\rm cl}(r)$, which we then convolve with a Gaussian with a
FWHM of 2\arcsec\ (to allow for the {\it Spitzer} IRAC $\rm 8\:\mu m$
PSF). We then fit this model to the observed $\Sigma_{\rm cl}(r)$
profile, excluding annuli that are significantly ($>50\%$) affected by saturated
pixels.  For Core A1, $k_{\rm \rho,cl}=1.40$ and $n_{\rm H,s,cl} =
2.47 \times 10^5\:{\rm cm^{-3}}$.  For the whole sample, the
mean/median/dispersion values of $k_{\rm \rho,cl} = 1.09/1.10/0.236$
and $n_{\rm H,s,cl}=(1.76/1.85/0.852)\times 10^5\:{\rm cm^{-3}}$. These
distributions are shown in Fig.~\ref{fig:hist}a with the blue dotted
histograms. The values for individual cores are listed in Table 2.

The above analysis is somewhat simplistic in that it has assumed the
structure exists in isolation. In reality, we see that these high
$\Sigma$ objects are surrounded by regions that also have significant
mass surface densities. Thus next we model the cores with a similar
power law density structure, 
\begin{equation}
\label{eq:PLcore}
\rho_{c}(r) = \rho_{\rm s,c} \left(\frac{r}{R_{\rm c}}\right)^{-k_{\rho,c}}, 
\end{equation}
but now when comparing to the observed $\Sigma$ maps we account for
the mass surface density of the surrounding clump medium, $\Sigma_{\rm
  cl,env}$. We estimate $\Sigma_{\rm cl,env}$ using the
observed value in the annular region from $R_c$ to $2R_c$. This choice
is motivated by the desire to sample a region of the clump that has a
scale comparable to the core in both size and mass\footnote{We have
  also tried measuring $\Sigma_{\rm cl,env}$ from a thin,
  1.2\arcsec\-wide annulus just outside $R_c$, which generally leads
  to larger estimated values of $\Sigma_{\rm cl,env}$. However,
  we consider that this thin-shell annulus does not sample a large
  enough region and mass of the clump that, via self-gravity, would be
  responsible for setting core's surrounding pressure (see discussion
  in \S\ref{S:intro}).}. We assume this same value of $\Sigma_{\rm
  cl,env}$ covers the area of the core, and so subtract it from
the interior $\Sigma_{\rm cl}(r)$ profile to derive the mass surface
density profile of the core, $\Sigma_c(r)$. Thus note that
$\Sigma_{\rm cl,env}= \bar{\Sigma}_{\rm cl} - \bar{\Sigma}_c$.

For the same $M_{\rm cl}=60\:M_\odot$ enclosed mass scale as defined
above, we set $R_c = R_{\rm cl}(M_{\rm cl}=60\:M_\odot)$. The masses
contained in the cores, based on integrating the $\Sigma_c(r)$
profile, are of course less than the $60\:M_\odot$ we previously
identified with the clump. For Core A1, we derive a core mass of
$M_c=37.9\:M_\odot$ with $\bar{\Sigma}_c=0.316\:\gcc$. Its value of
$k_{\rho,c}=2.04$ and $n_{\rm H,s,c}=1.21 \times 10^5\:{\rm cm^{-3}}$,
i.e. a steeper density profile with a lower value of the volume
density at the surface than was derived previously. For the 42 cores
we find the mean/median/dispersion values of $M_c = 30.4/30.4/12.7
\:M_\odot$, $\bar{\Sigma}_c = 0.139/0.160/0.0738\:\gcc$,
$k_{\rho,c}=1.64/1.67/0.271$ and $n_{\rm H,s,c}=
(0.639/0.750/0.394)\times 10^5\:{\rm cm^{-3}}$ (See red solid line
histograms in Fig.~\ref{fig:hist}a and Tables 2 \&
3). Compared to the clump results (i.e. derived
from the total $\Sigma$ profiles), above, for the envelope-subtracted
core properties we necessarily find smaller surface densities, steeper
density profiles and smaller volume densities.

The power law fits ignore annuli affected by significant saturation, where
$\Sigma$ is underestimated. Thus we also estimate a core mass, $M_{\rm
  c,PL}$, based on extrapolation of the power law fits to the center
of the core:
\begin{equation}
\label{eq:MPL}
M_{\rm c,PL} = \frac{4\pi}{3-k_{\rho,c}}\rho_s R_c^3 = \frac{43.5}{3-k_{\rho,c}} \frac{n_{\rm H,s,c}}{10^5\:{\rm cm^{-3}}} \left(\frac{R_c}{0.1\:{\rm pc}}\right)^3\:M_\odot  \:\: (k_{\rho,c}<3)
\end{equation}
There are no cores where the derived $k_\rho>3$ for which the inner
boundary condition would have be considered.  If there were, then in
these cases we would expect to truncate the power law at the Jeans
scale in the core. For Core A1, $M_{\rm c,PL} = 49.2\:M_\odot$, about
30\% times larger than $M_c$. Such an increase is expected since this
is one of the most extensively saturated cores. For the rest of the 42
cores the change is typically much smaller. The mean/median/dispersion
values of $M_{\rm c,PL} = 31.1/31.0/13.5\:M_\odot$ and
$\bar{\Sigma}_{\rm c,PL}=0.154/0.171/0.0899$ (See red solid line
histograms in Fig.~\ref{fig:hist}a and Tables 2 \&
3).

\subsubsection{Best-fit Power Law Cores}

We now repeat the power law core plus clump envelope fitting procedure
as a function of radius, starting at the inner region with 3
unsaturated annuli for the core. An annulus twice as large in radius
is used to estimate the value of $\Sigma$ of the clump envelope,
$\Sigma_{\rm cl,env}$. We assess the relative goodness of
fit of this model as a function of $r$ by finding the minimum of the
reduced $\chi^{2}$ parameter, defined by
\begin{equation}
\chi^{2} \equiv \sum_{i=1,N} \frac{1}{\nu} \frac{[\Sigma_{\rm c,PL}(r)-\Sigma_{c,i}(r)]^{2}}{\sigma_i^2}
\end{equation}
where $N$ is the number of annuli, $\nu=N-2$ is the number of degrees
of freedom and $\sigma$ is the error for each annulus, which we take
to be $\sigma = 0.01\:\gcc + 0.2\Sigma_c$. Note, that because of the
2\arcsec\ angular resolution of {\it Spitzer} IRAC, adjacent annuli
are not completely independent. However, the relative values of
$\chi^2$ should still give a measure of the best-fitting
model. 

We place some additional constraints on the fitting. First, we do not
allow the best-fit core to extend beyond neighboring core centers
(from our sample of 42 cores). Second, to prevent independent discrete
structures that are not part of our core sample from influencing the
fitting, we check for a $3 \sigma$ rise in the $\Sigma_{c}$ profile by
comparing the annulus before any rise begins to the following local
maximum. If this occurs, we ignore fits beyond the pre-rise annulus,
and search inward for a local maximum in $\chi_{\nu}^{2}$ and define
that to be the best-fit radius (for example, this occurs in Core
A3). Third, if more than $25 \%$ of an annulus is composed of MIR
emission pixels, we do not extend the fit any further.  In these
cases, a prior unaffected annulus with a local maximum in
$\chi_{\nu}^{2}$ is chosen as the best-fit radius (this
circumstance only arises in Core E3). As a result of the above
constraints, it is possible that the global minimum of $\chi_{\nu}^{2}$
will not be chosen as the ``best-fit''.

\begin{figure*}
\begin{center}$
\begin{array}{cc}
\includegraphics[width=2.8in]{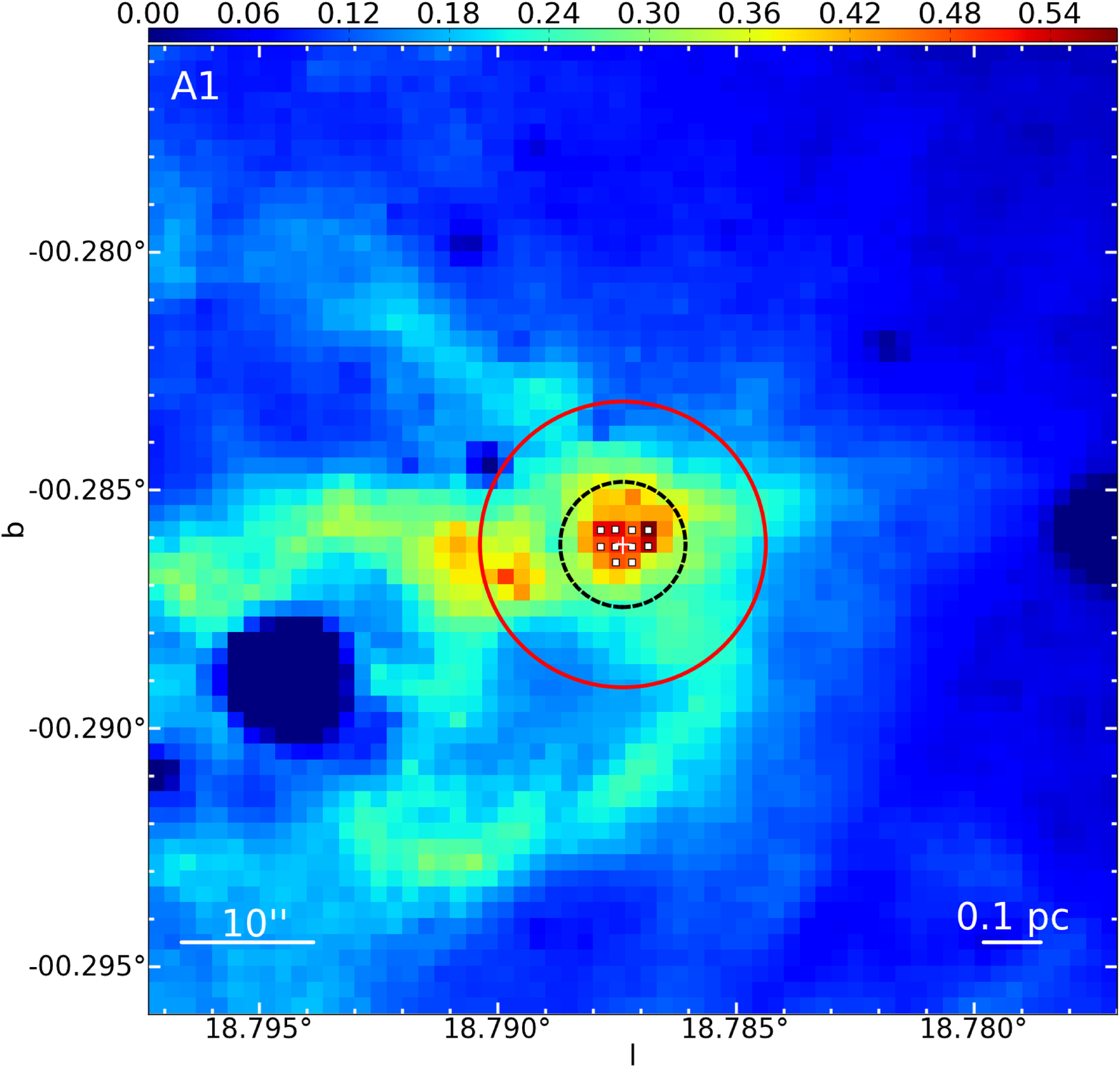} & \includegraphics[width=2.8in]{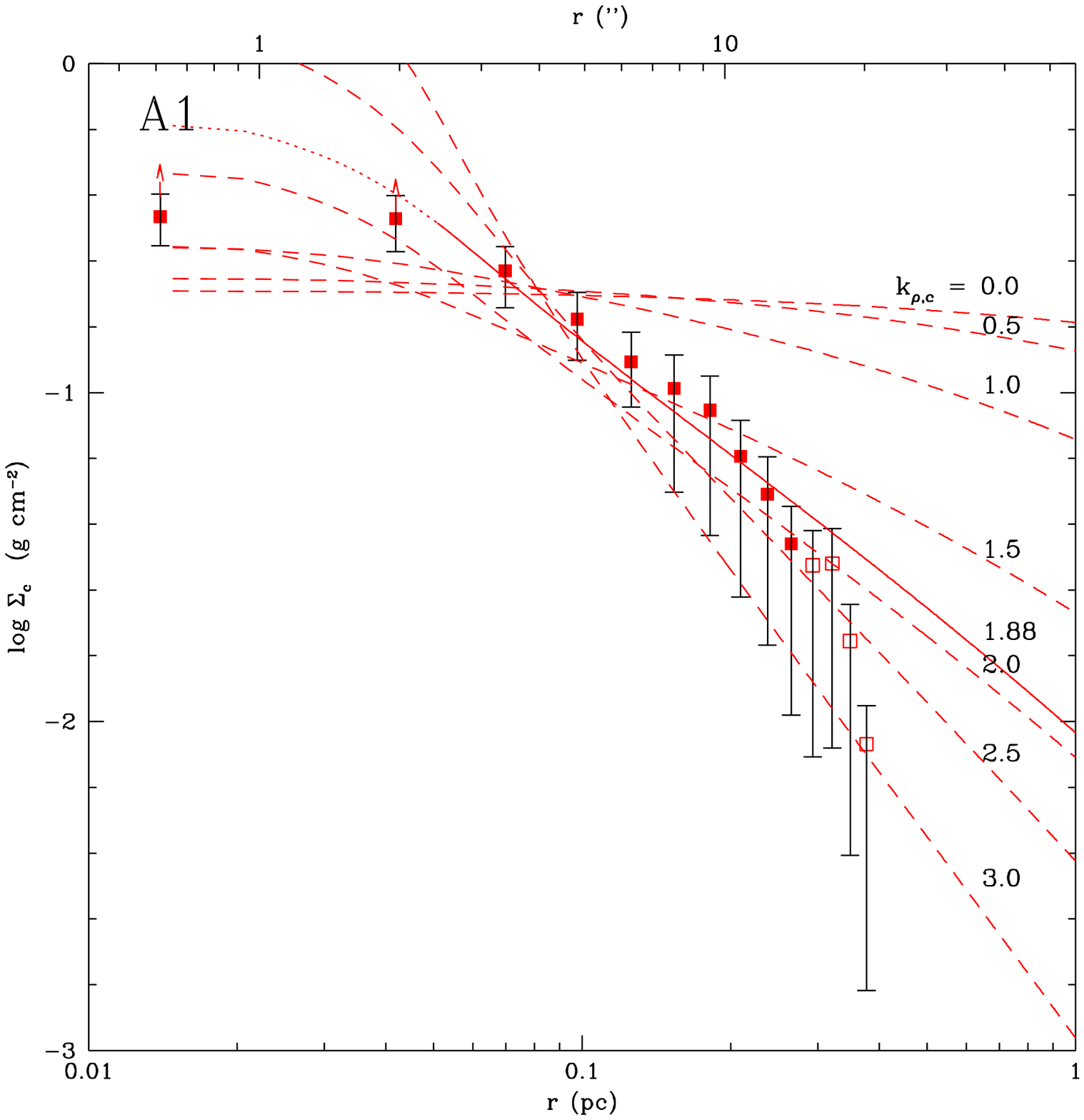}  \\
\includegraphics[width=2.8in]{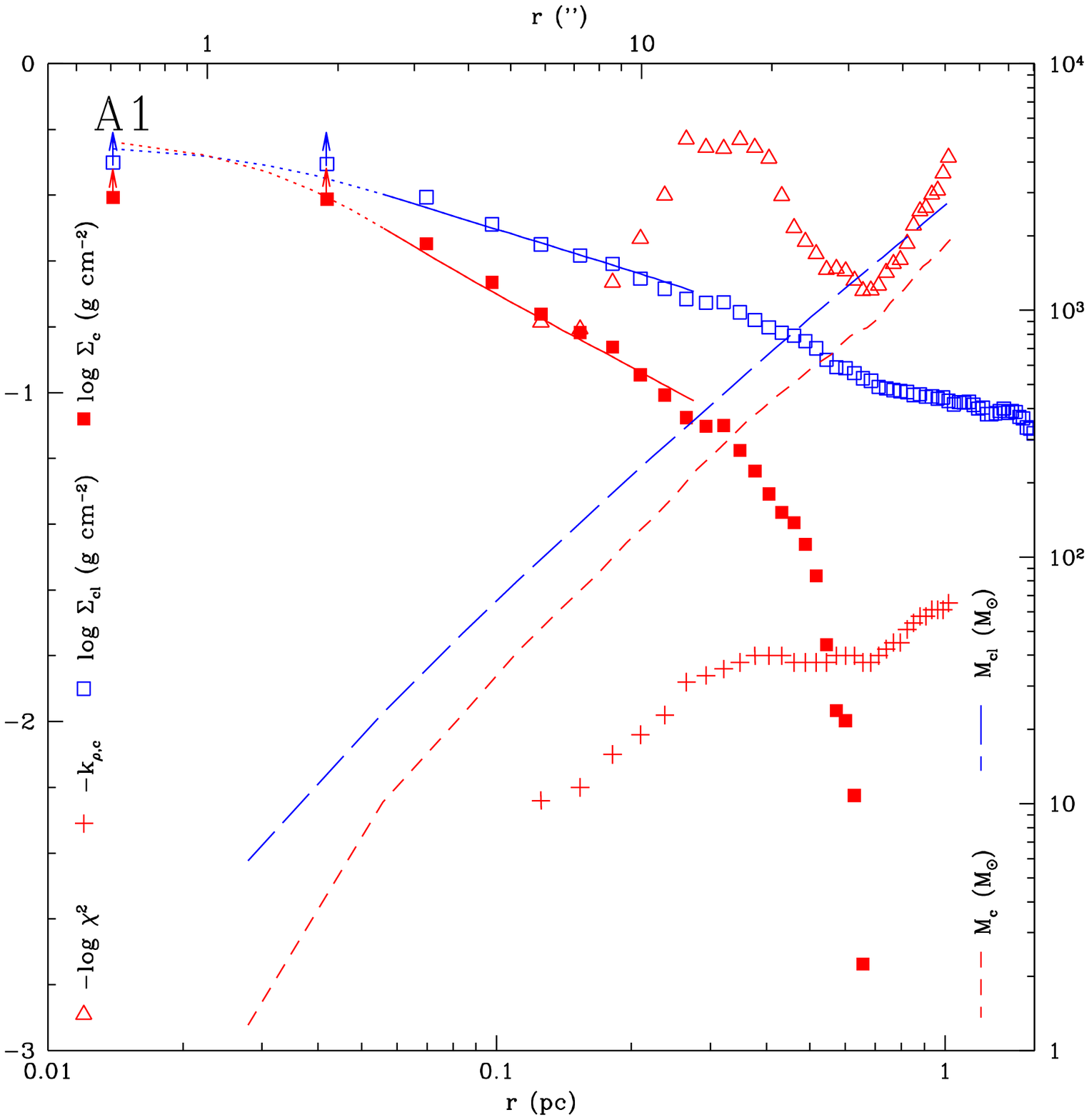} & \includegraphics[width=2.8in]{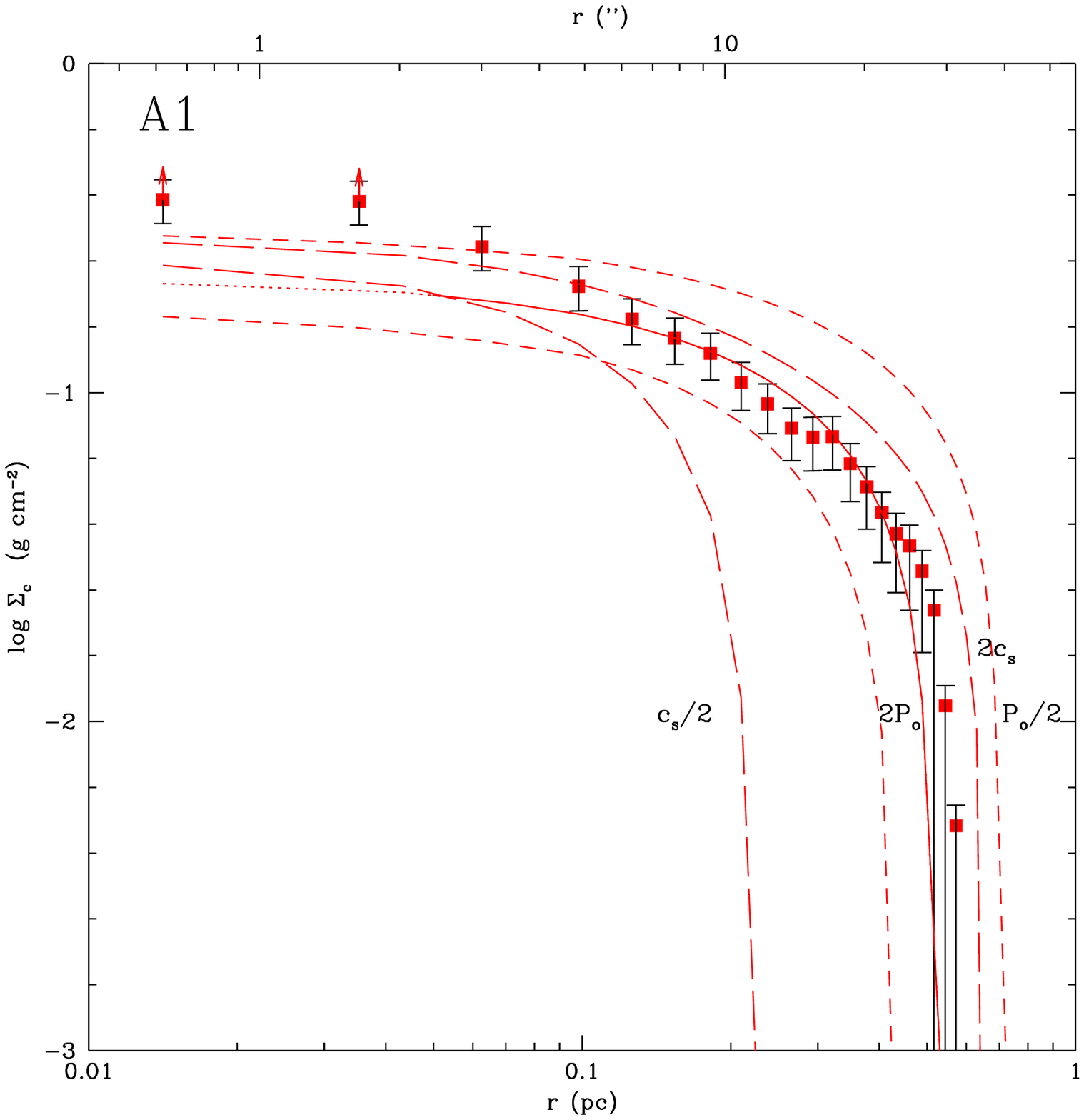}
\end{array}$
\caption{\footnotesize
{\it (a) Top Left:} Mass surface density, $\Sigma_{\rm SMF}$, map in
$\gcc$ of IRDC Core A1, extracted from the map of IRDC A
(Fig.~\ref{fig:IRDC1}). The core center is marked with a
cross. Saturated pixels, for which $\Sigma$ is a lower-limit of the
true value, are marked with small white squares. The black dashed
circle shows the radius enclosing a total mass of $60\:M_\odot$. The
red solid circle shows the extent of the core derived from the
best-fit power law (PL) core plus envelope model (see text). {\it (b)
  Bottom Left:} Radial profiles of Core A1: observed log~$\Sigma_{\rm
  cl}/({\rm g\:cm^{-2}})$ (blue open squares, plotted at annuli
centers) derived from the map shown in (a); total projected enclosed
mass, $M_{\rm cl}$, (blue long-dashed line [see right axis]); core
mass, $M_c$ after clump envelope subtraction (red dashed line [see
  right axis]); index of core PL density profile, $k_{\rho,c}$, (red
crosses); $-{\rm log}\:\chi^2$ (red triangles) of the PL plus envelope
fit (best-ﬁt has a maximum or local maximum value [see text]); the best-fit PL plus envelope
model (blue solid line; dotted line shows range affected by saturation
that was not used in the fitting); log~$\Sigma_{\rm c}/({\rm
  g\:cm^{-2}})$ of best-fit core after envelope subtraction (red solid
squares) and PL fit (red solid line; dotted line shows range affected
by saturation that was not used in the fitting). {\it (c) Top Right:}
$\Sigma_c(r)$, i.e. after clump envelope subtraction for the best-fit
model (red solid squares; open squares show residual, post-subtraction 
envelope material). PL models with various values of
$k_{\rho,c}$ are indicated (dashed lines), including the best-fit
model with $k_\rho=1.88$ (solid line).  {\it (d) Bottom Right:} As for
(c) but for Bonnor-Ebert (BE) plus envelope fitting. $\Sigma_{c}(r)$,
i.e. after clump envelope subtraction for the best-fit model (red
solid squares). Best-fit BE model (solid line) and models varying
$c_s$ (long-dashed lines) and $P_0$ (dashed lines) by factors of 2
from this are shown (see text).
\label{fig:coreA1}
}
\end{center}
\end{figure*}

\begin{figure*}
\begin{center}$
\begin{array}{ccc}
\hspace{-0.0in} \includegraphics[width=2.15in]{A1.eps} & \hspace{-0.2in} \includegraphics[width=2.15in]{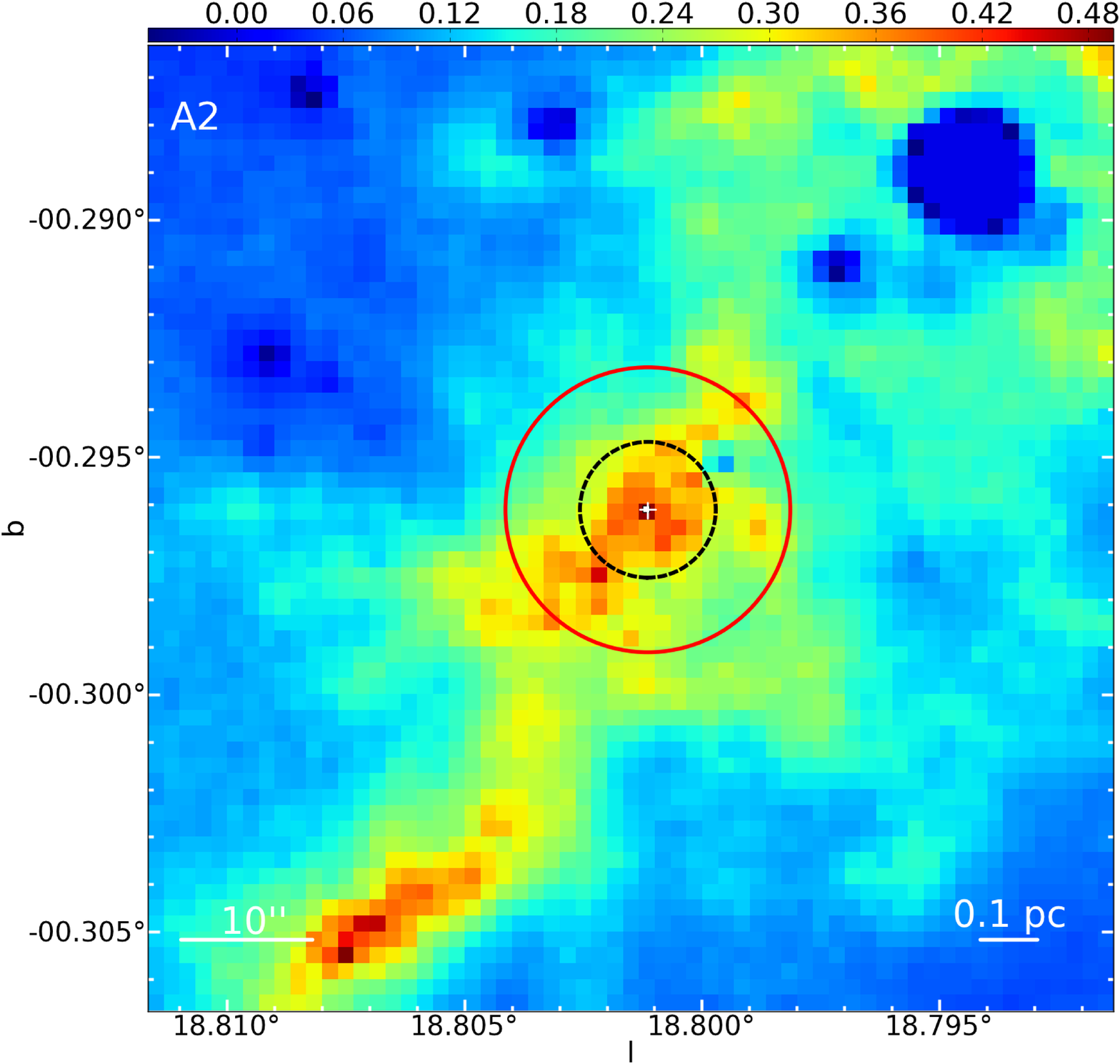} & \hspace{-0.3in} \includegraphics[width=2.15in]{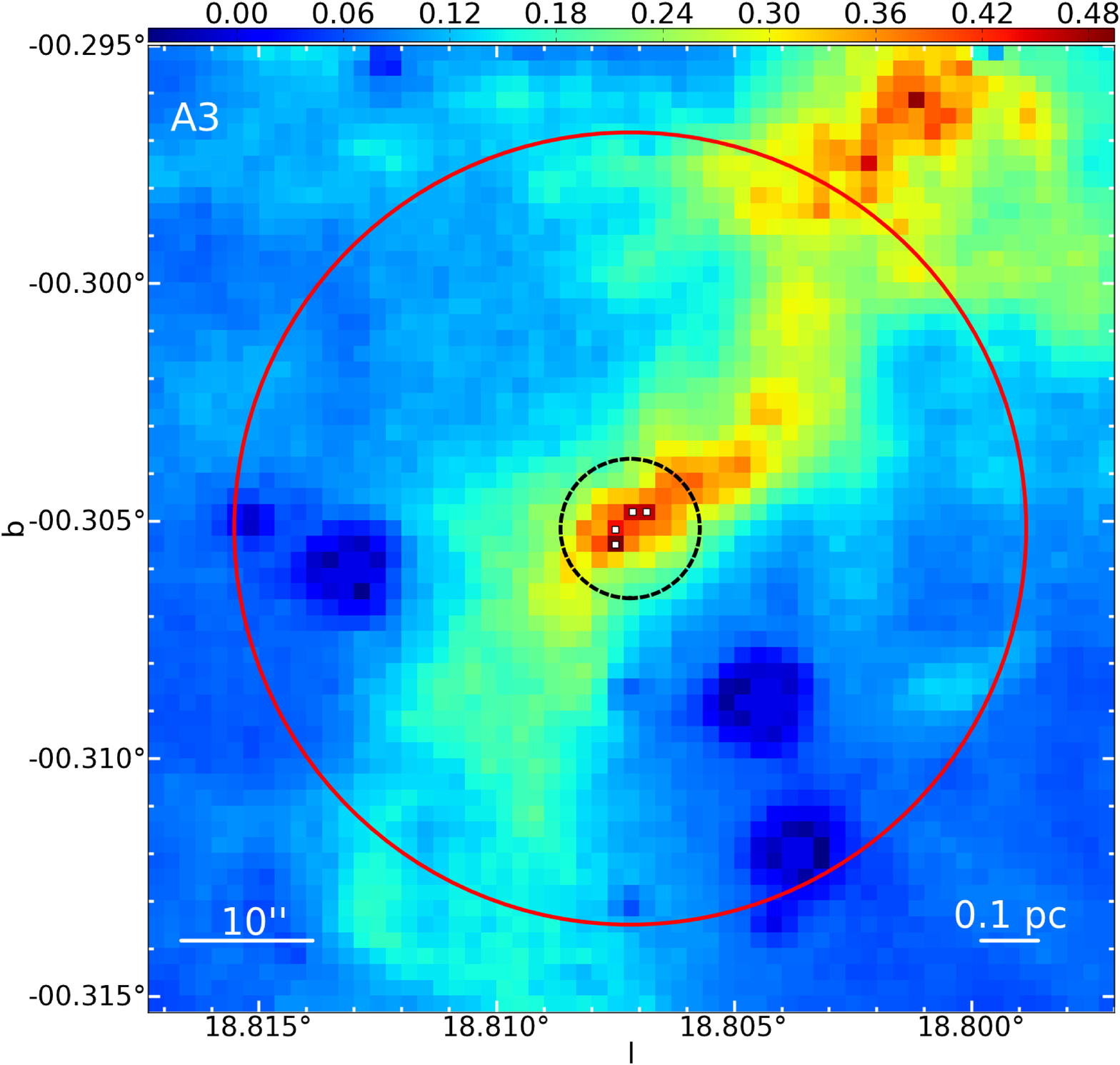} \\
\includegraphics[width=2.2in]{A1b.eps} & \includegraphics[width=2.2in]{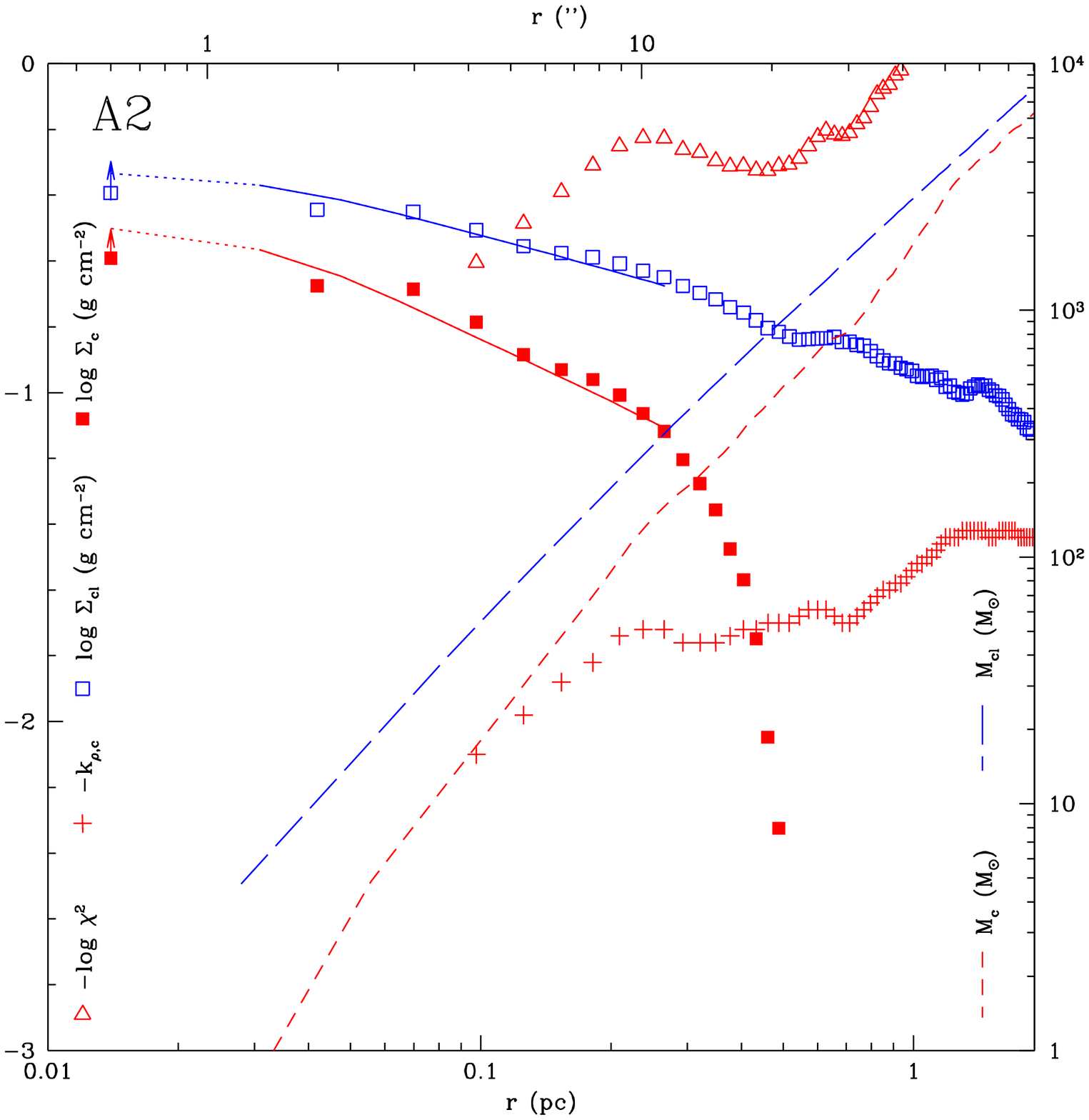} & \hspace{-0.1in} \includegraphics[width=2.2in]{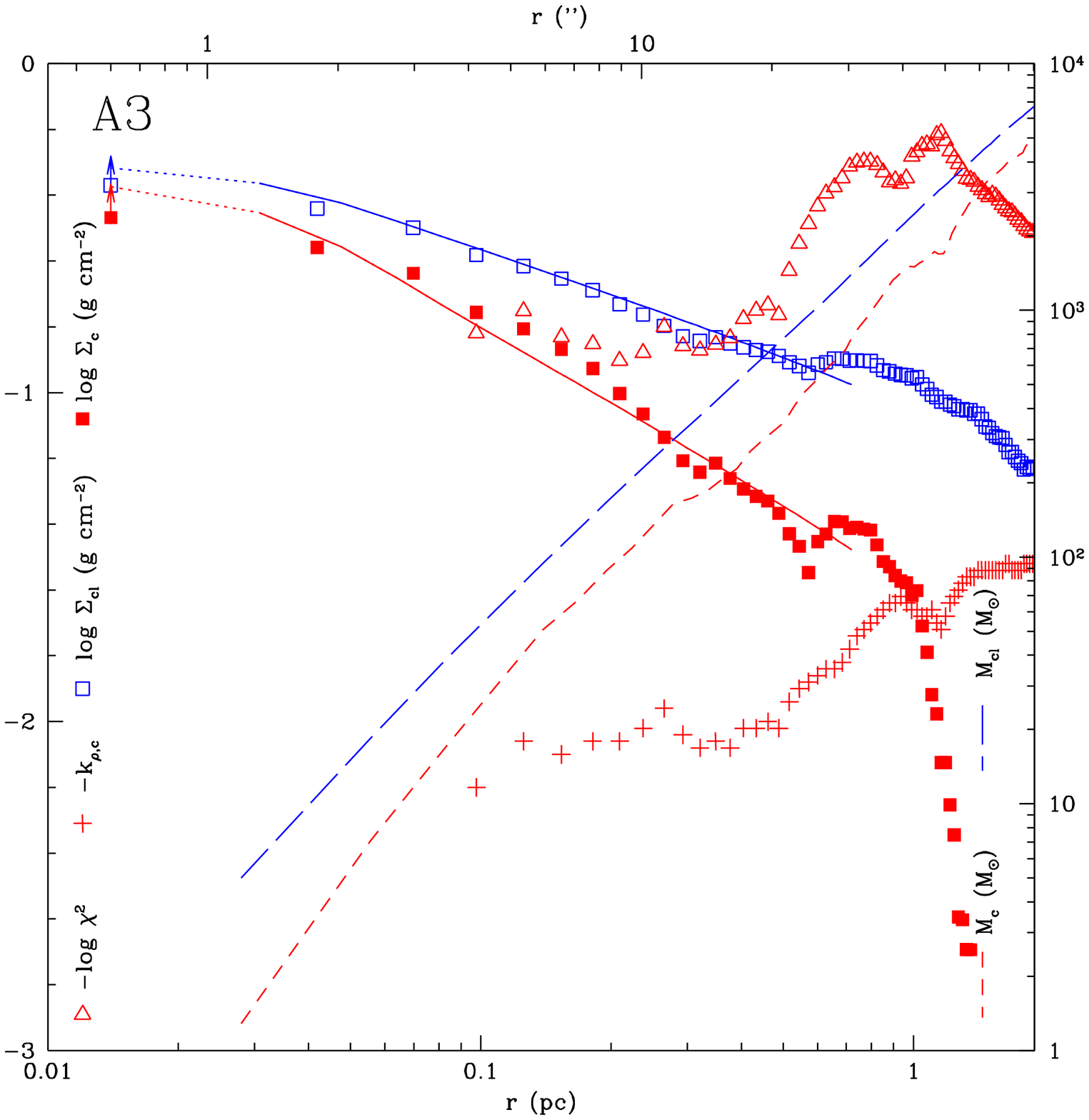} \\
\hspace{-0.0in} \includegraphics[width=2.15in]{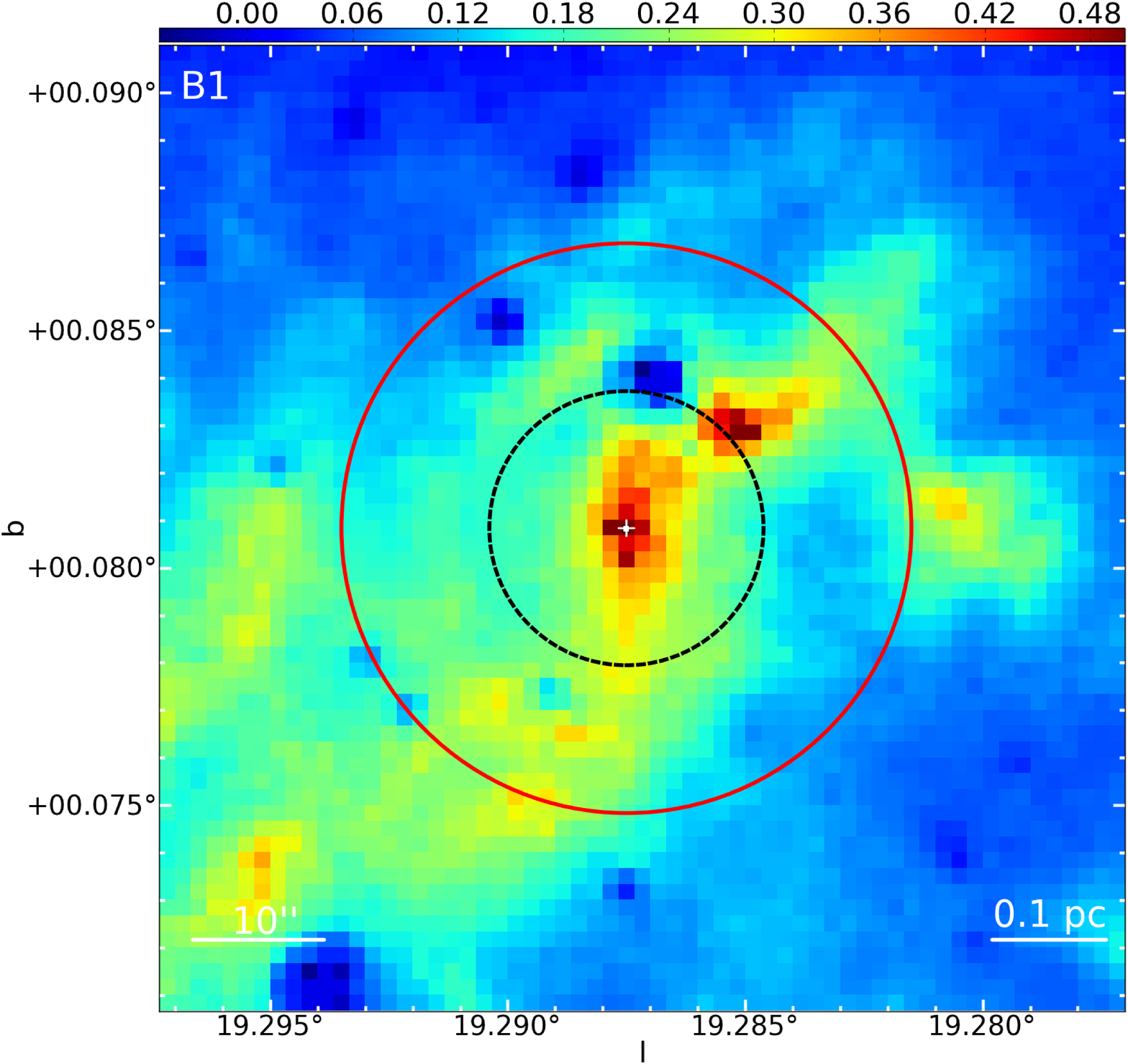} & \hspace{-0.2in} \includegraphics[width=2.15in]{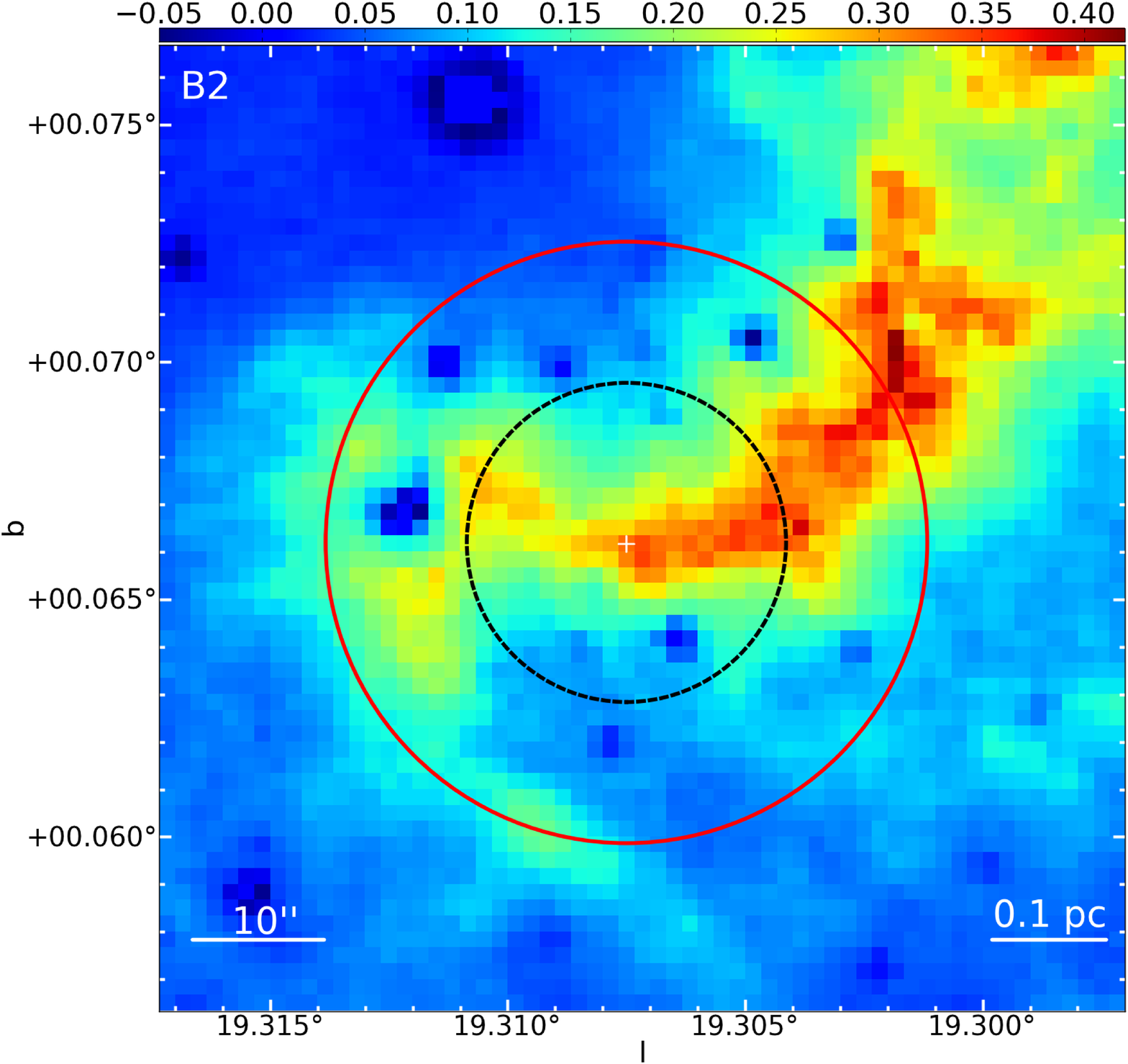} & \hspace{-0.3in} \includegraphics[width=2.15in]{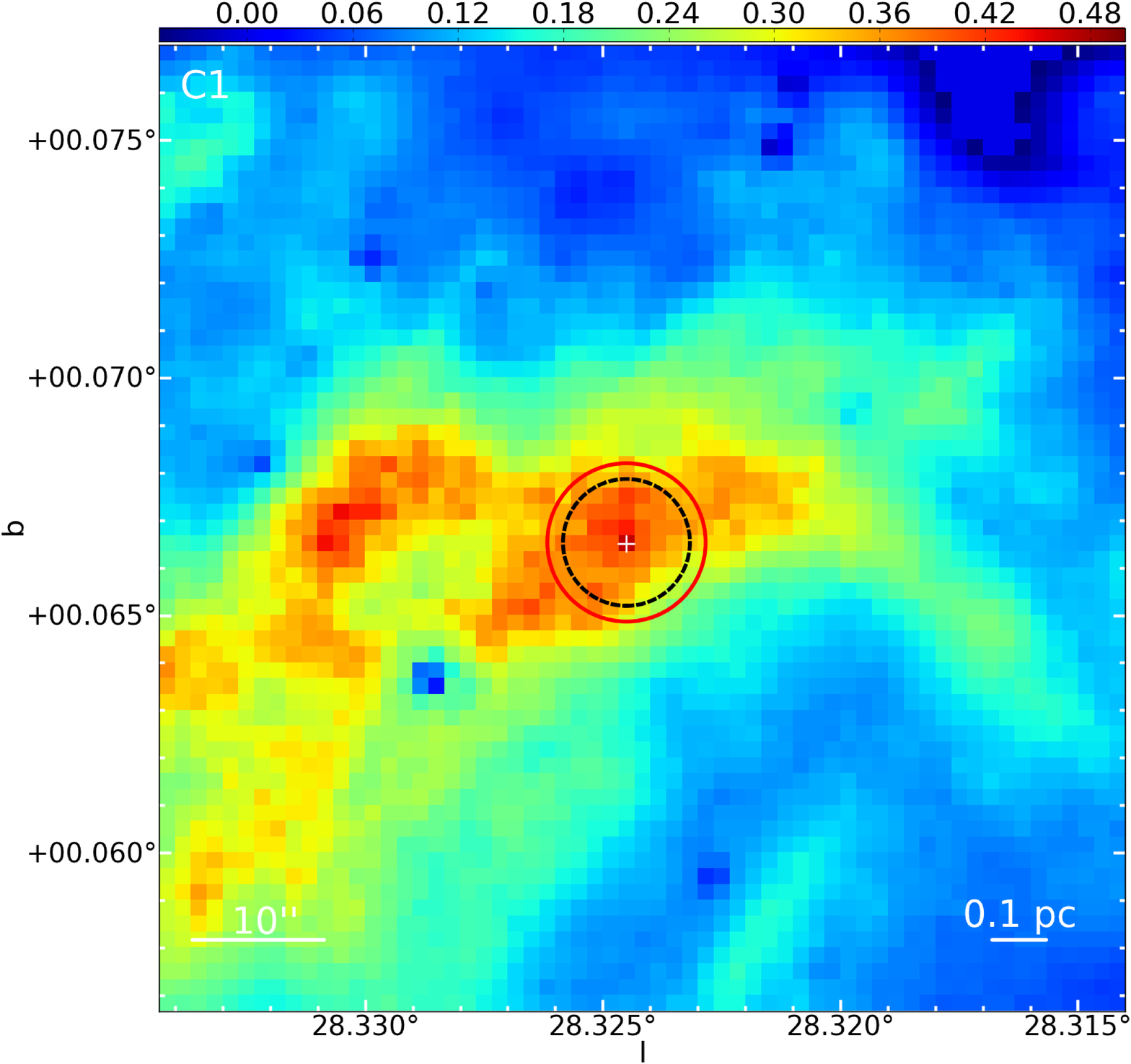} \\
\includegraphics[width=2.2in]{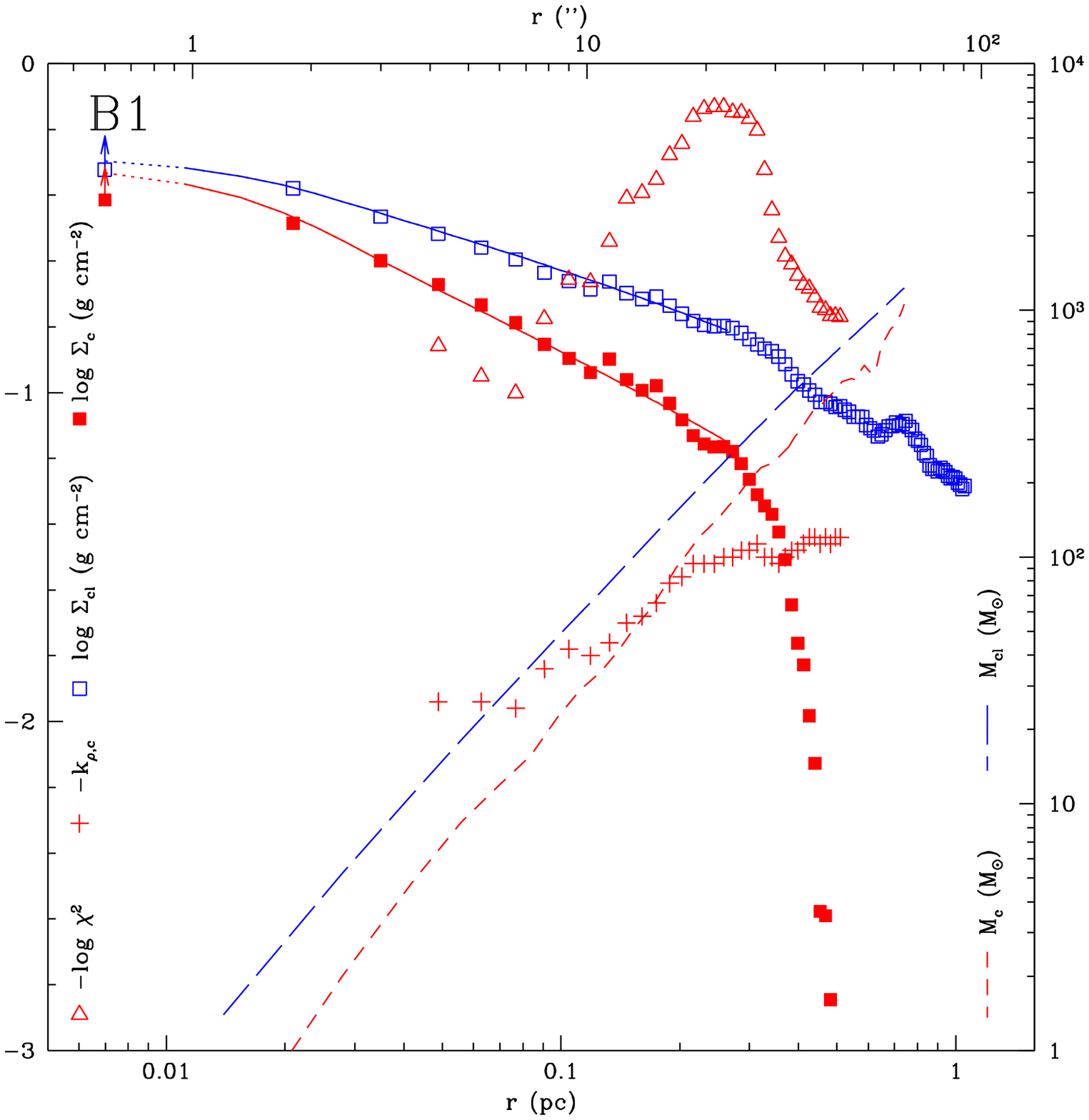} & \includegraphics[width=2.2in]{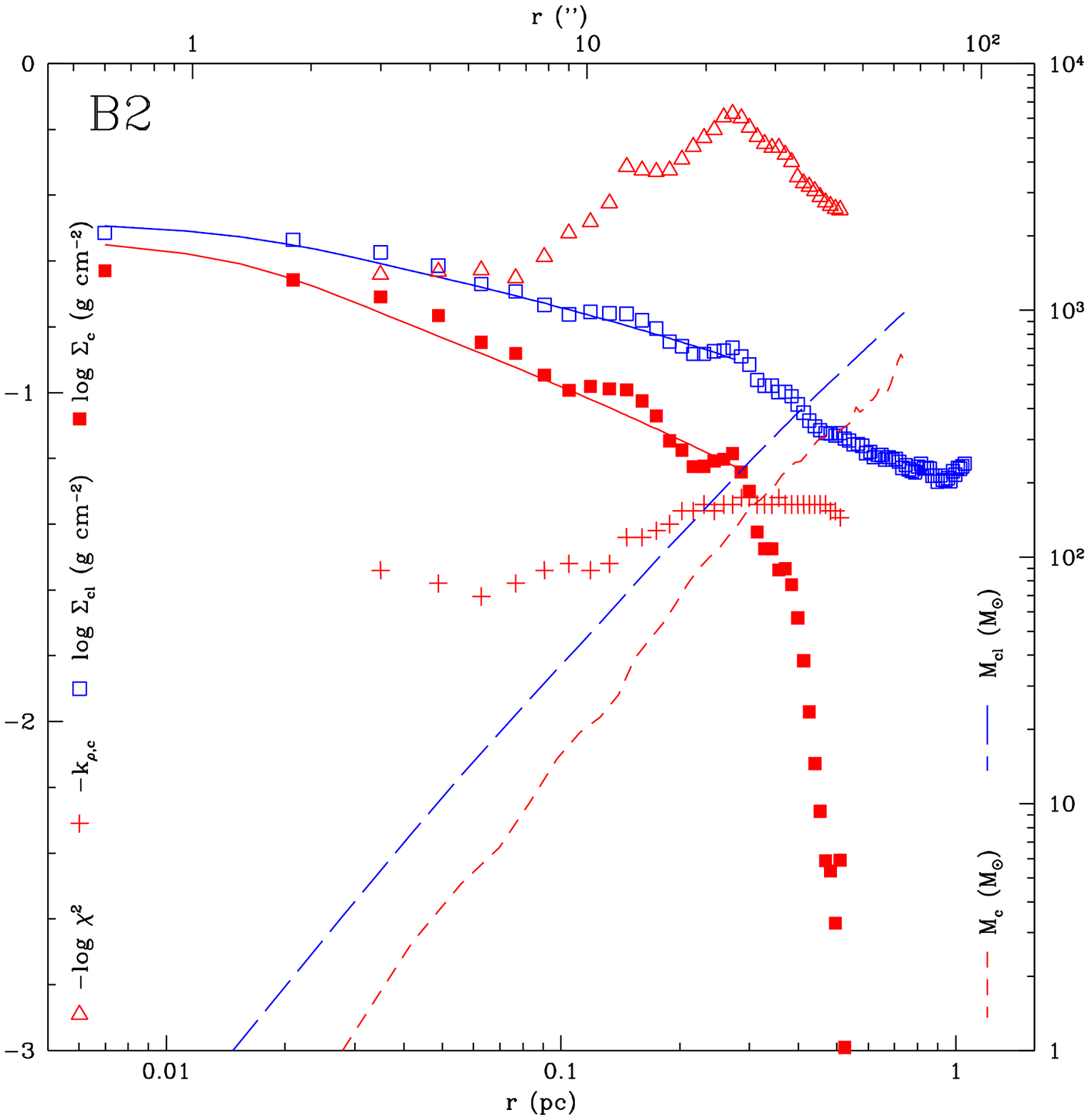} & \hspace{-0.1in} \includegraphics[width=2.2in]{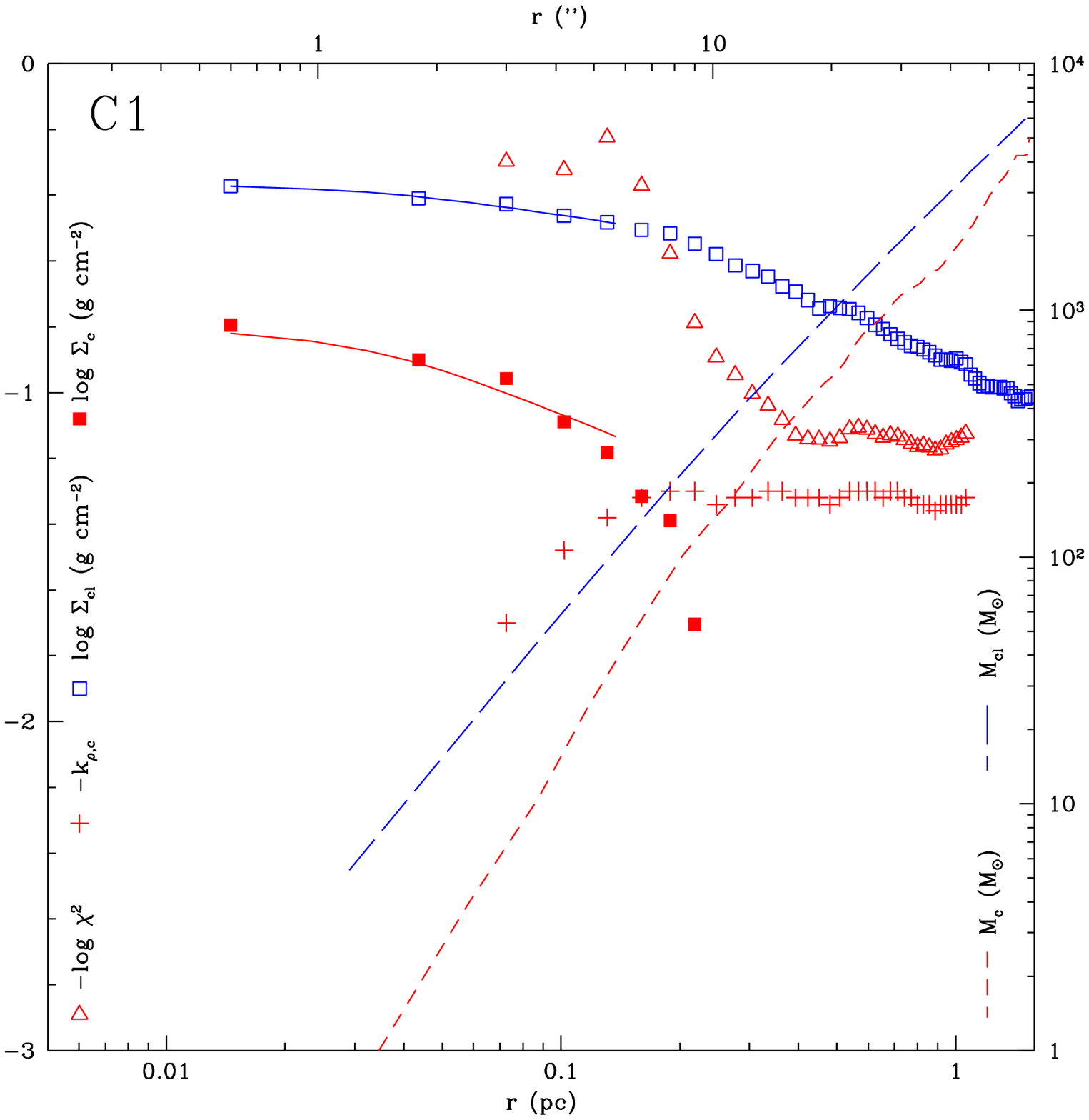}
\end{array}$
\end{center}
\vspace{-0.3in}
\caption{\footnotesize
Core A1, A2, A3, B1, B2, C1 $\Sigma$ maps (notation as Fig.~\ref{fig:coreA1}a) and azimuthally averaged radial profile figures (notation as Fig.~\ref{fig:coreA1}b).
\label{fig:cores1}
}
\end{figure*}

\begin{figure*}
\begin{center}$
\begin{array}{ccc}
\hspace{-0.0in} \includegraphics[width=2.15in]{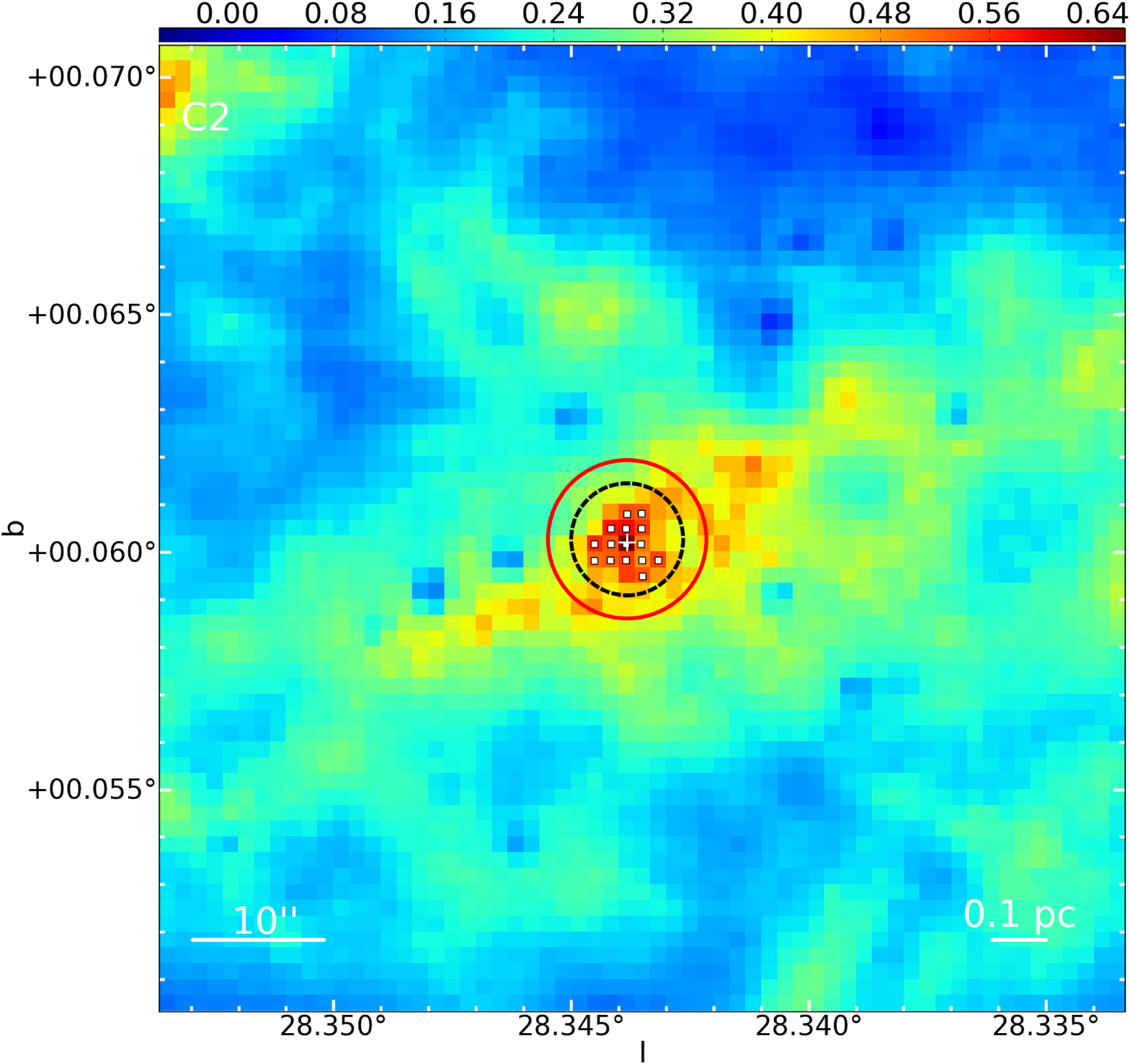} & \hspace{-0.2in} \includegraphics[width=2.15in]{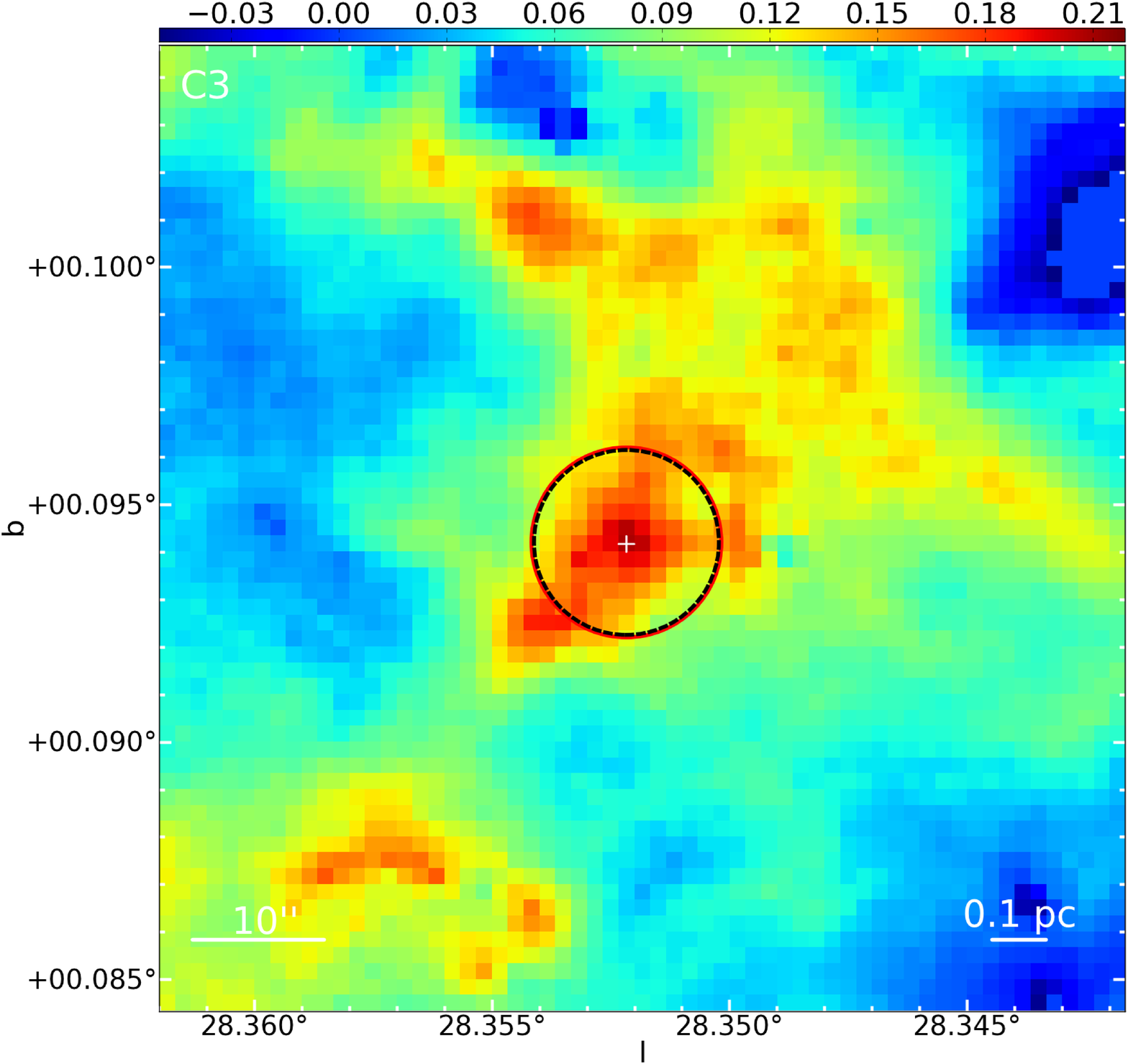} & \hspace{-0.3in} \includegraphics[width=2.15in]{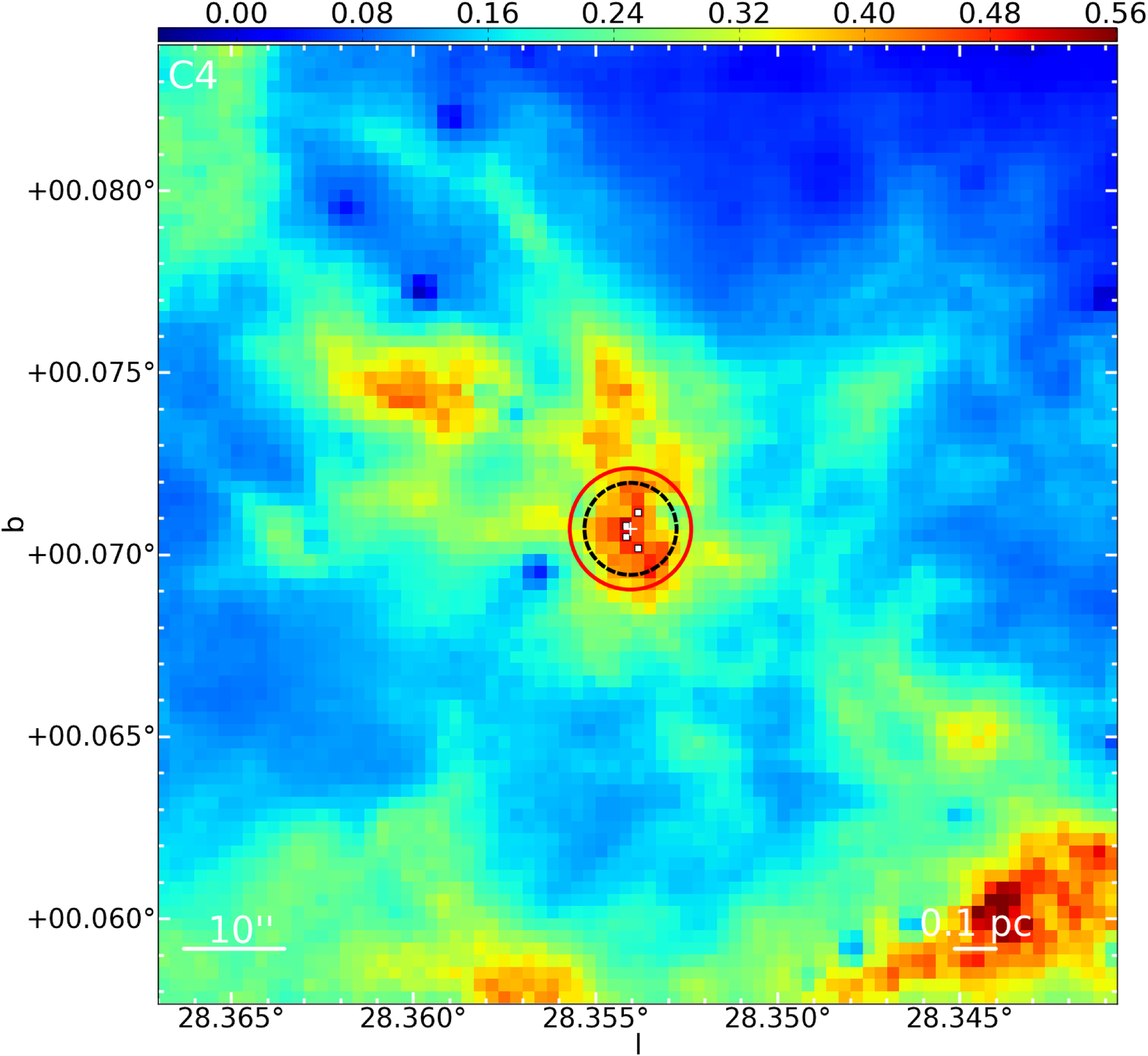} \\
\includegraphics[width=2.2in]{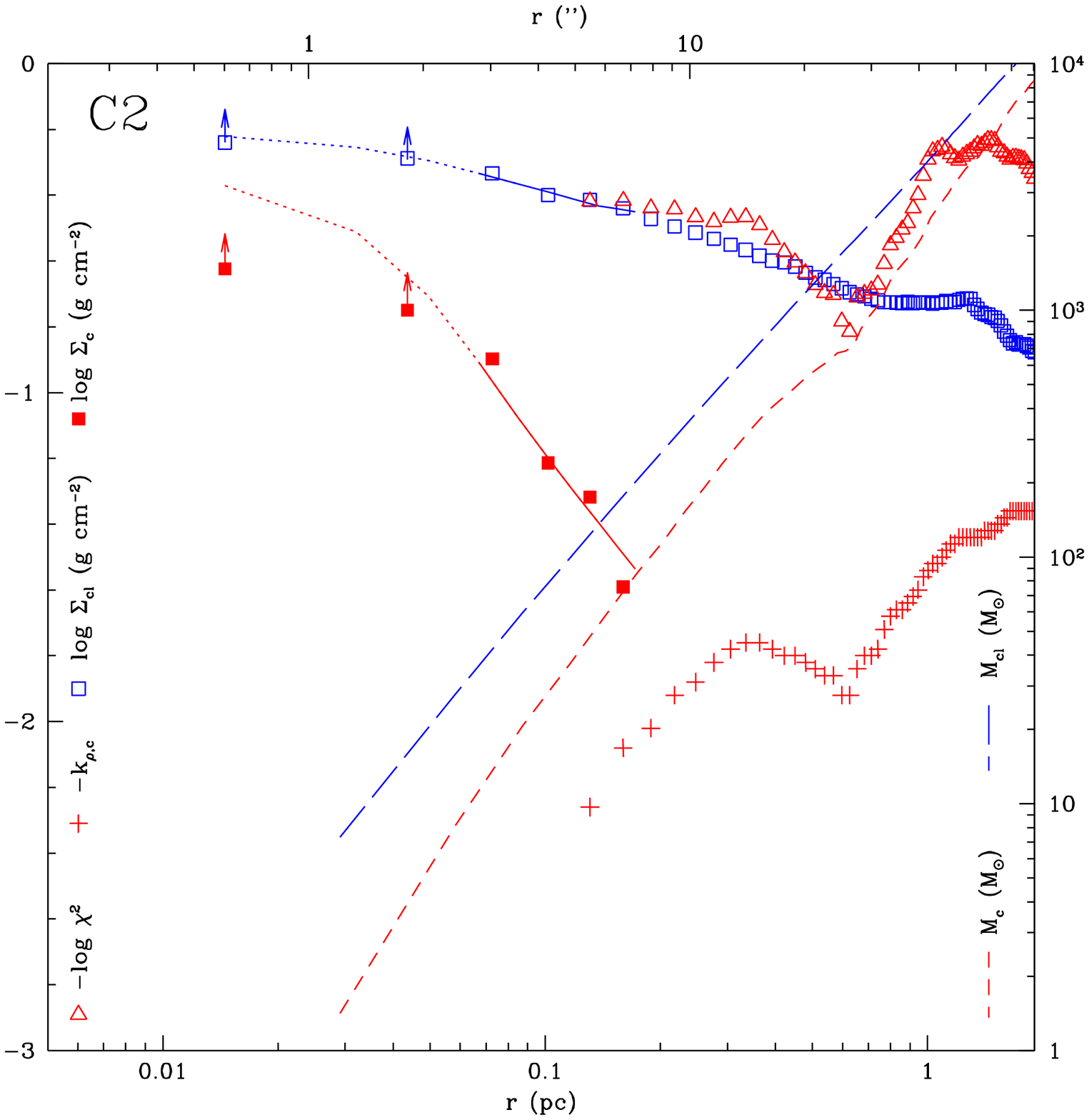} & \includegraphics[width=2.2in]{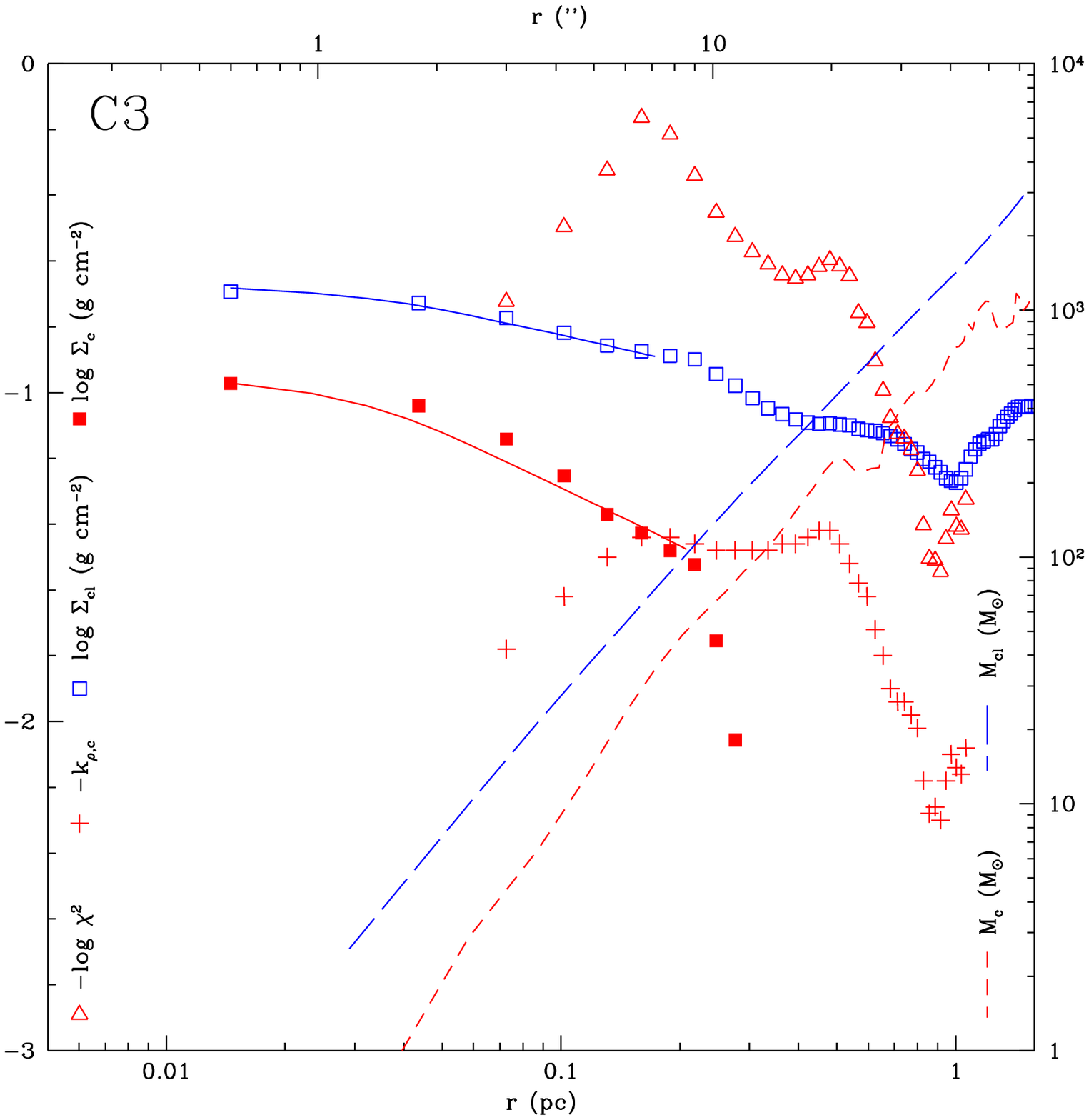} & \hspace{-0.1in} \includegraphics[width=2.2in]{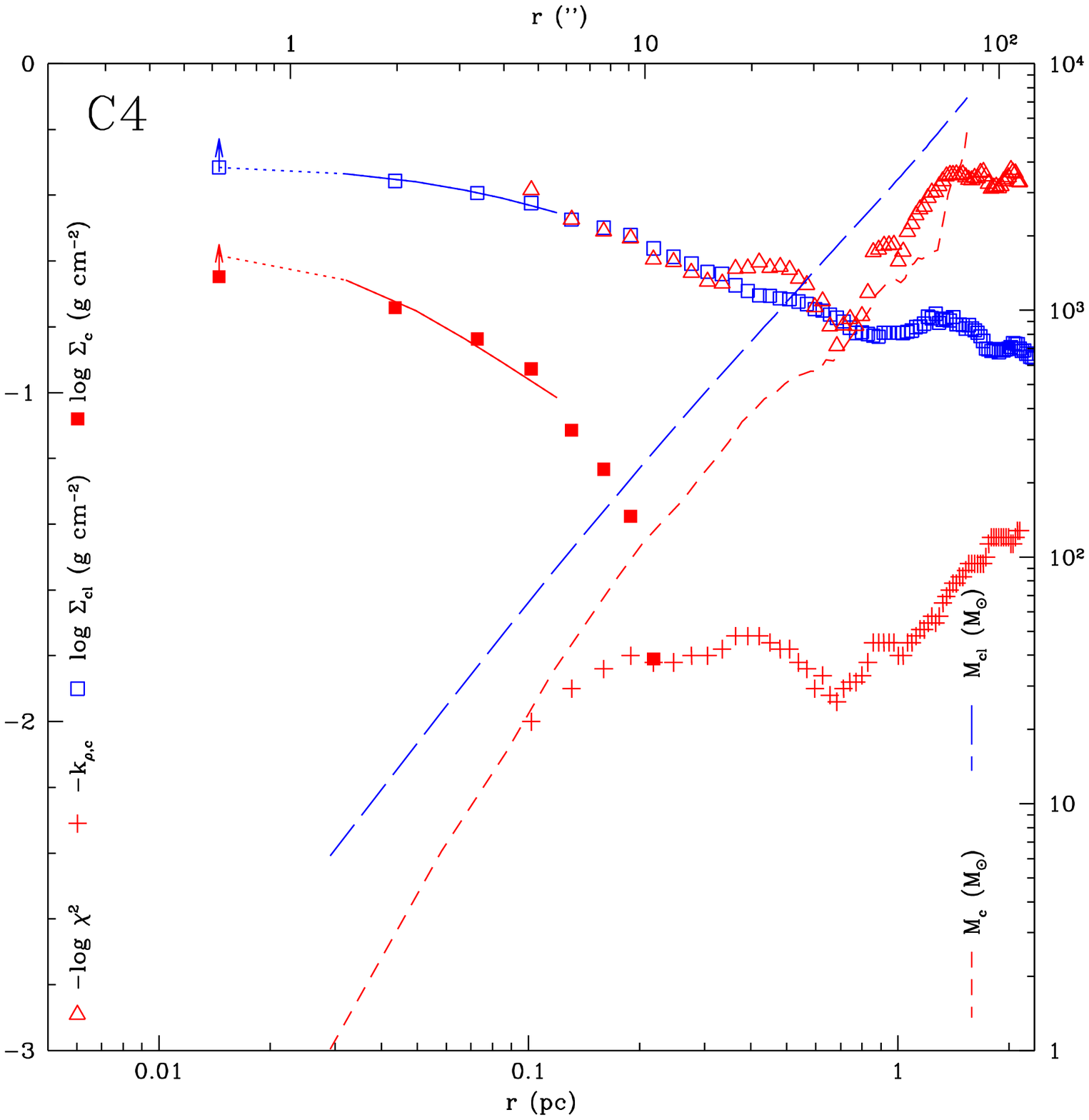} \\
\hspace{-0.0in} \includegraphics[width=2.15in]{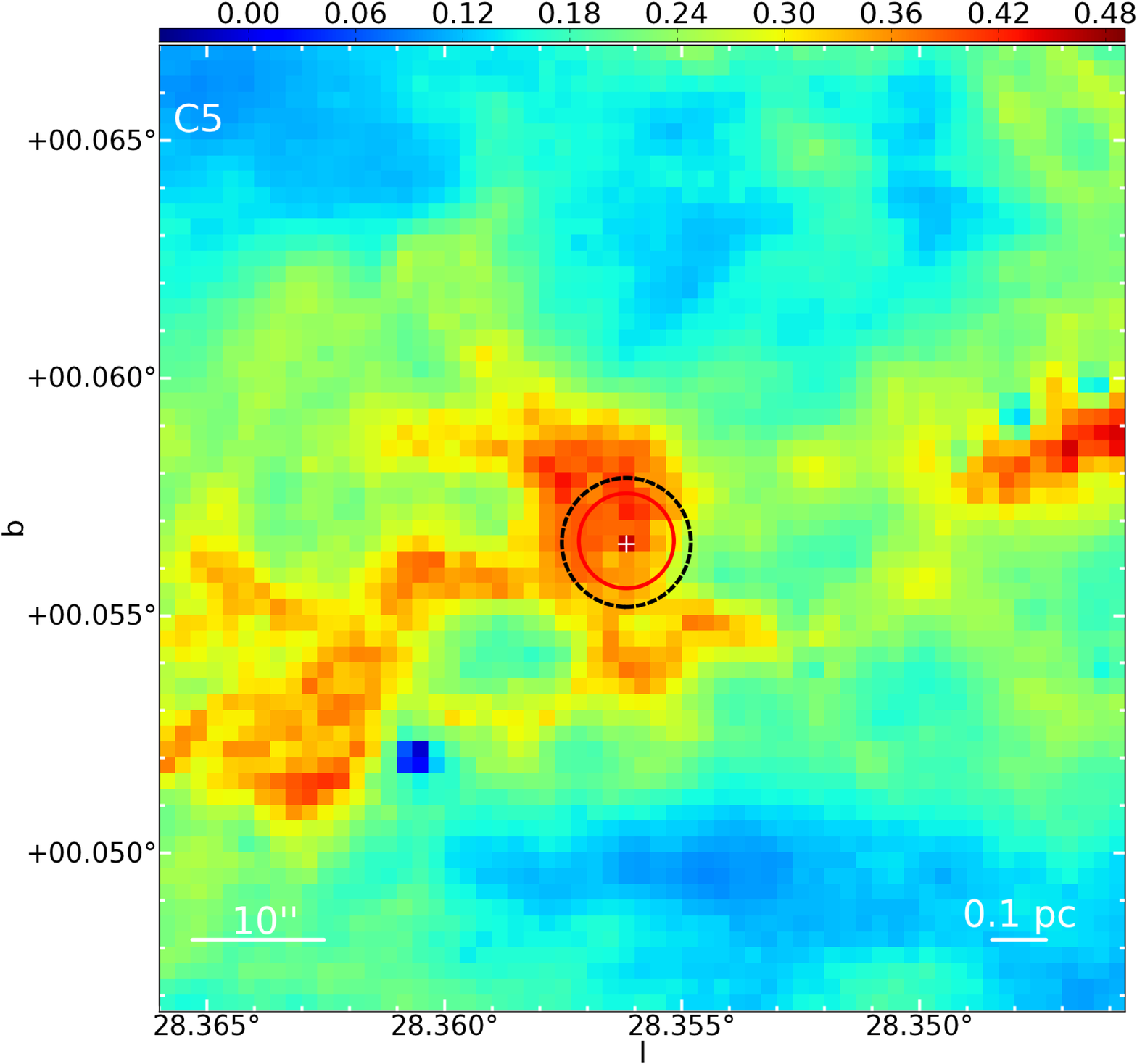} & \hspace{-0.2in} \includegraphics[width=2.15in]{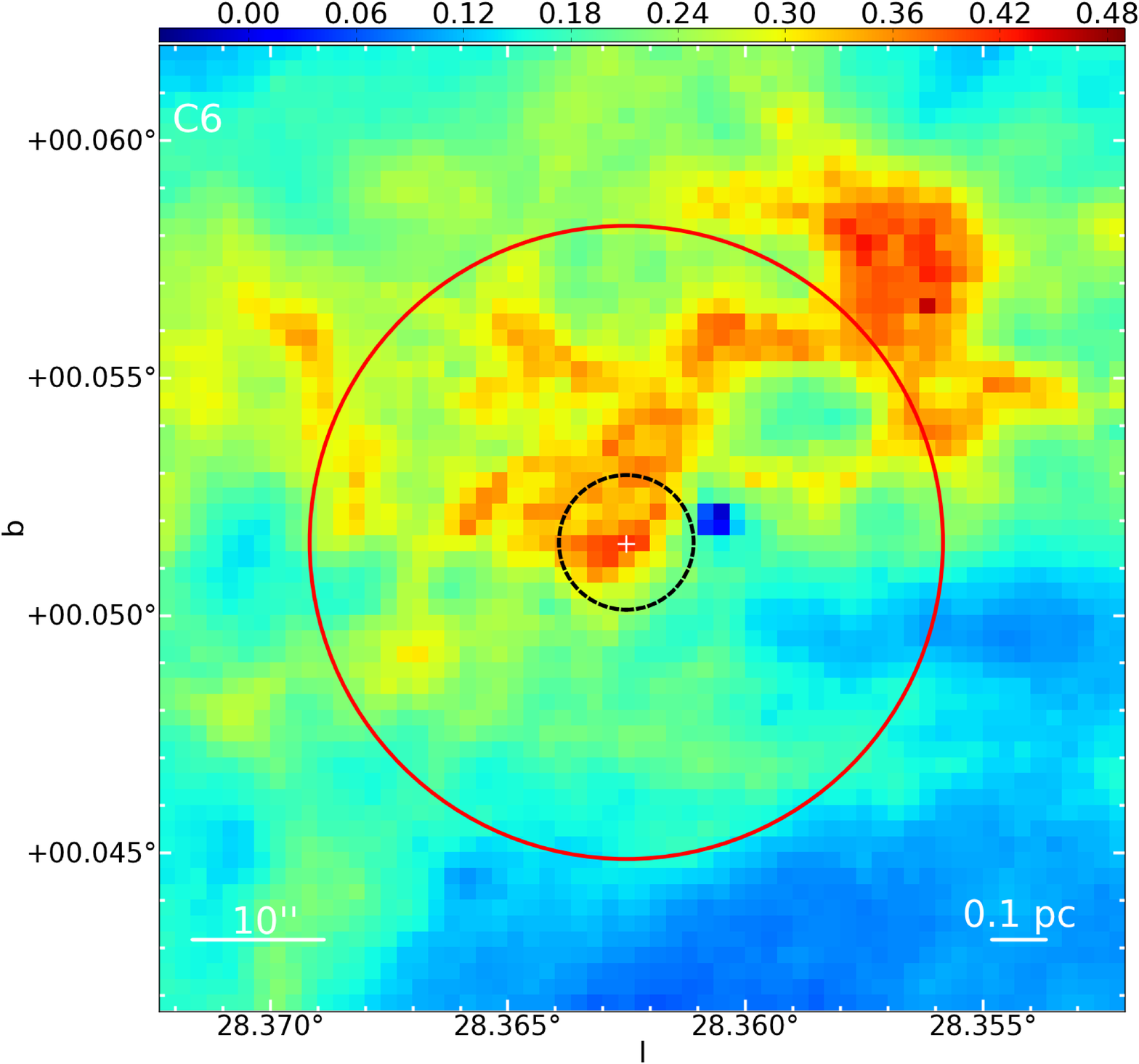} & \hspace{-0.3in} \includegraphics[width=2.15in]{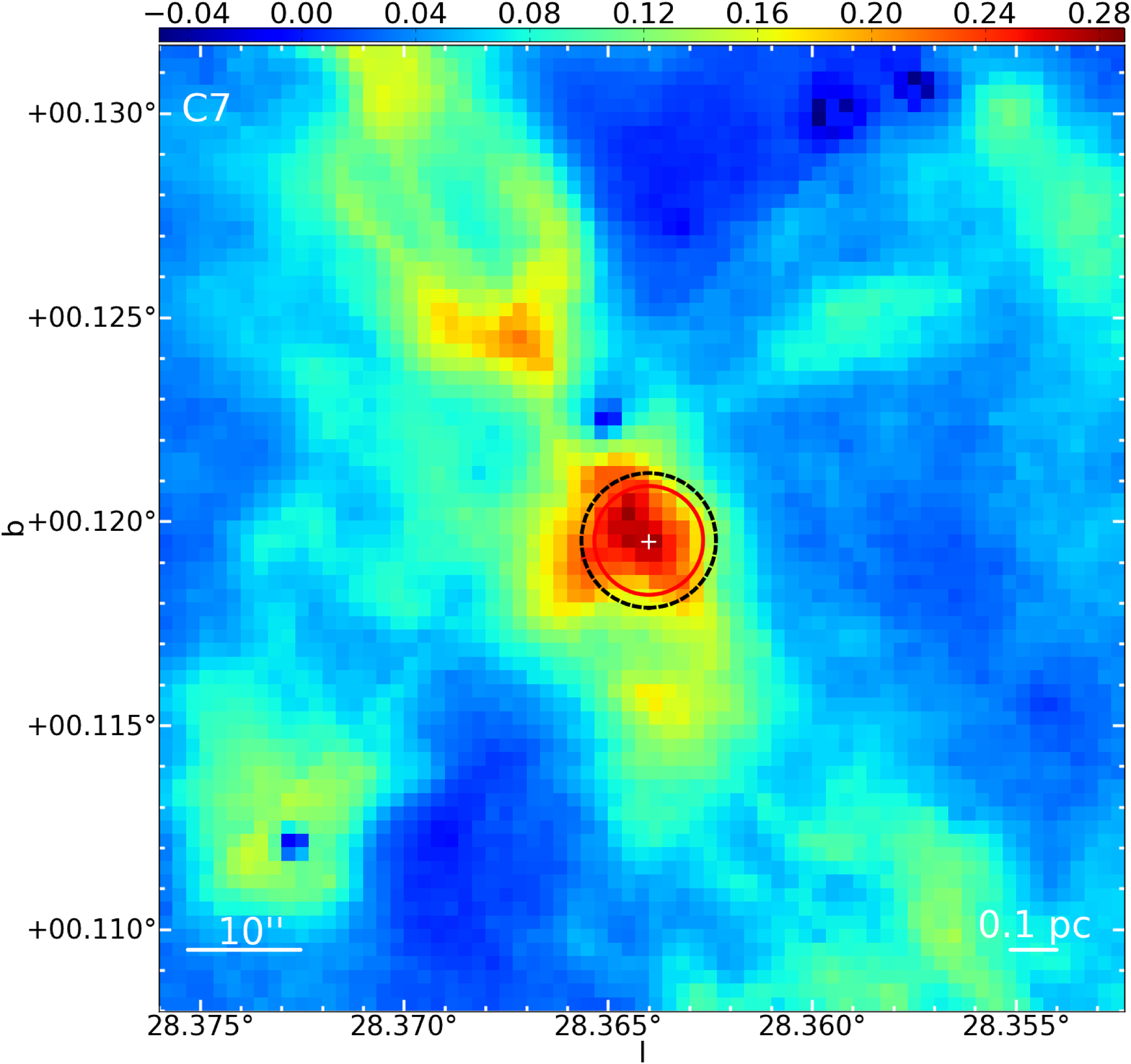} \\
\includegraphics[width=2.2in]{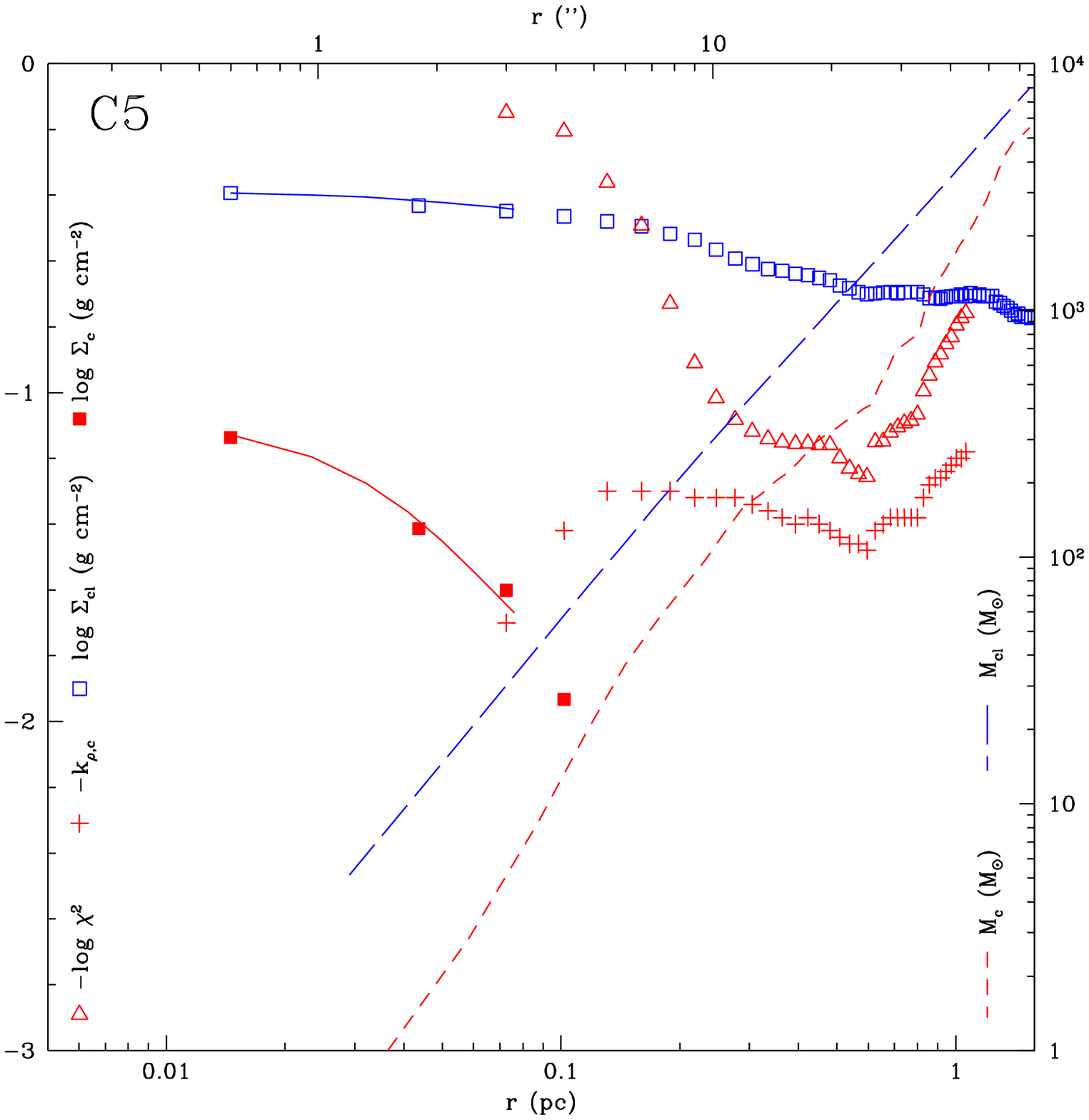} & \includegraphics[width=2.2in]{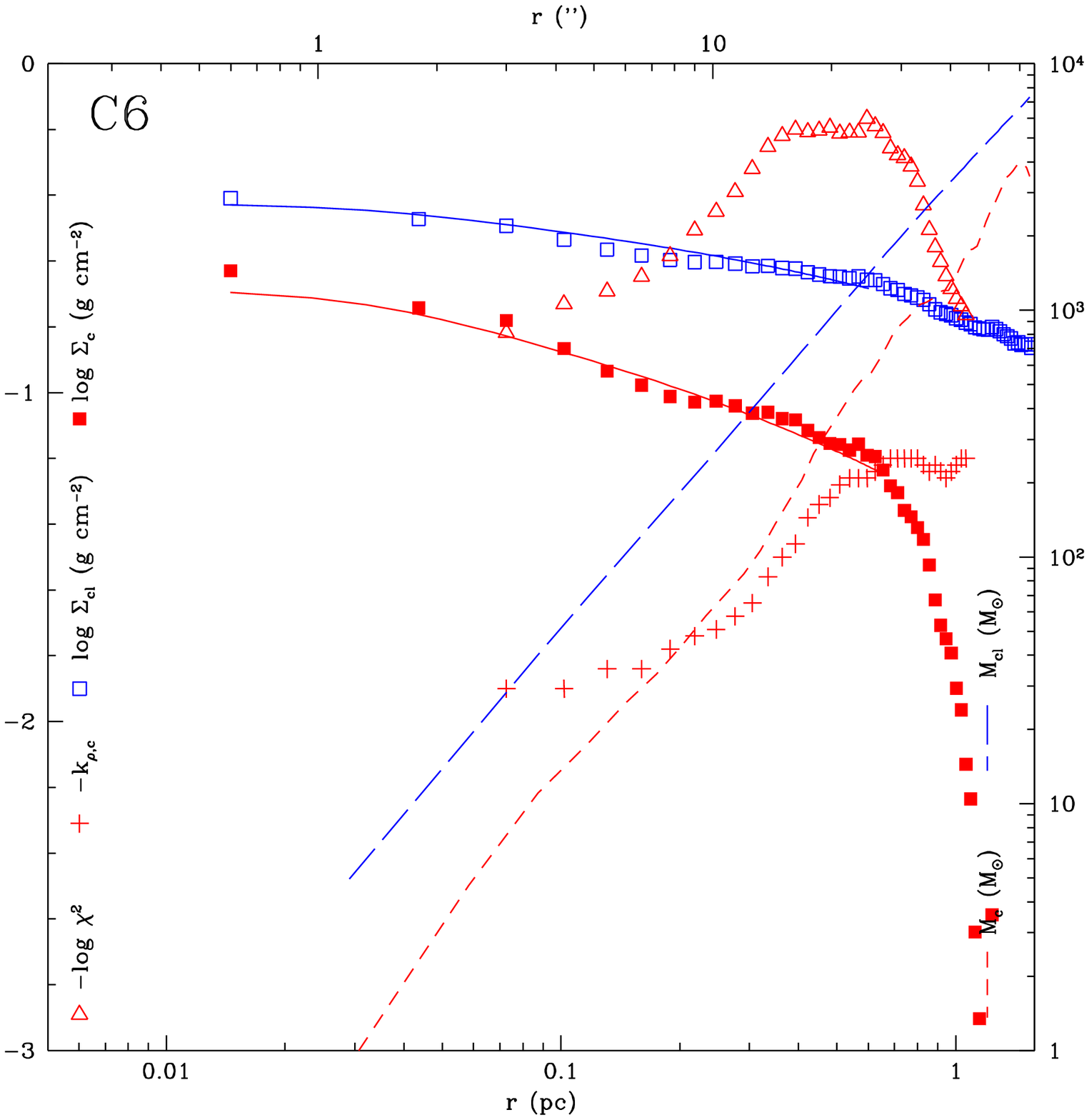} & \hspace{-0.1in} \includegraphics[width=2.2in]{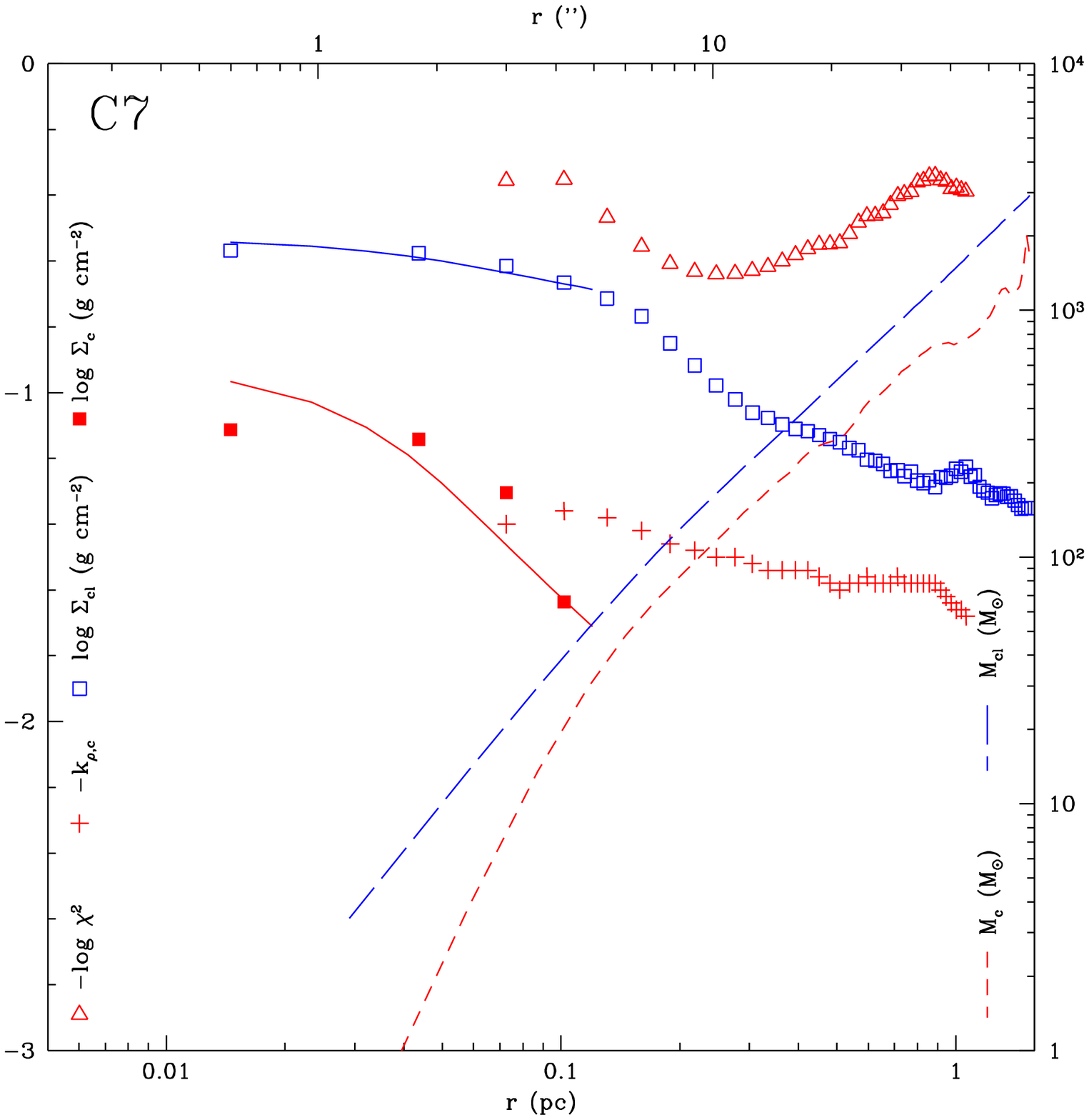}

\end{array}$
\end{center}
\vspace{-0.3in}
\caption{\footnotesize
Core C2, C3, C4, C5, C6, C7 $\Sigma$ maps (notation as Fig.~\ref{fig:coreA1}a) and azimuthally averaged radial profile figures (notation as Fig.~\ref{fig:coreA1}b).
\label{fig:cores2}
}
\end{figure*}

\begin{figure*}
\begin{center}$
\begin{array}{ccc}
\hspace{-0.0in} \includegraphics[width=2.15in]{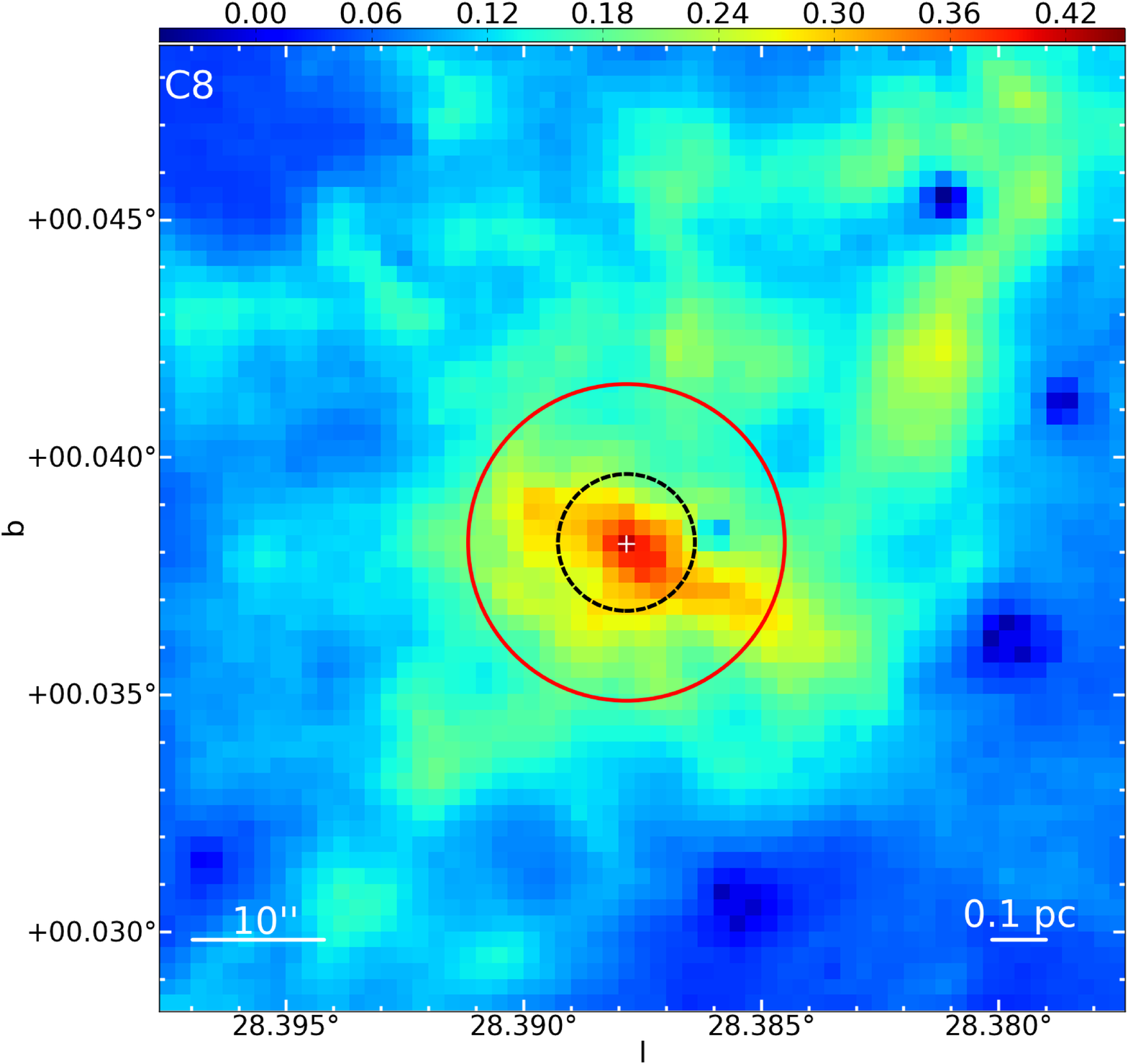} & \hspace{-0.2in} \includegraphics[width=2.15in]{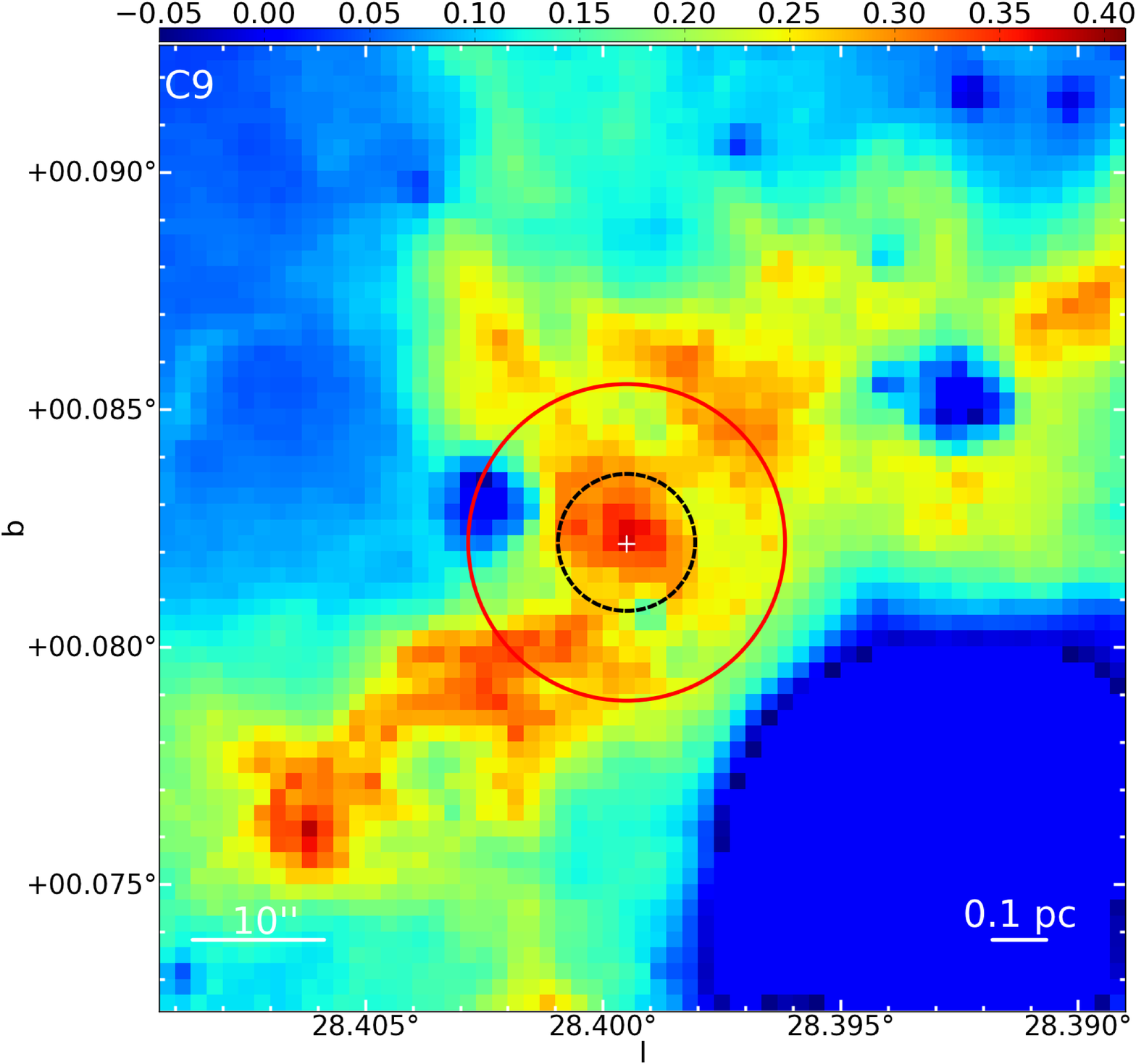} & \hspace{-0.3in} \includegraphics[width=2.15in]{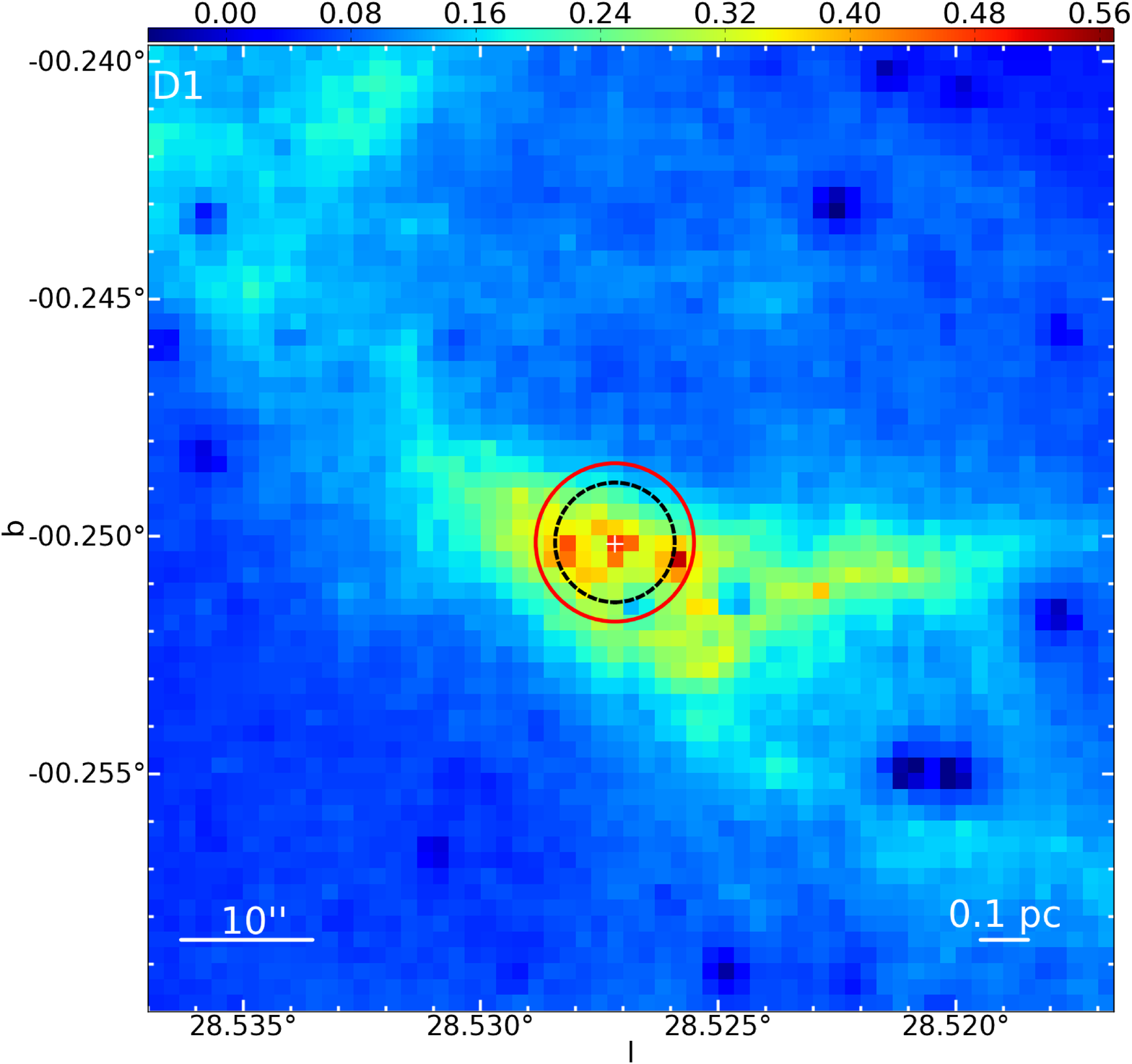} \\
\includegraphics[width=2.2in]{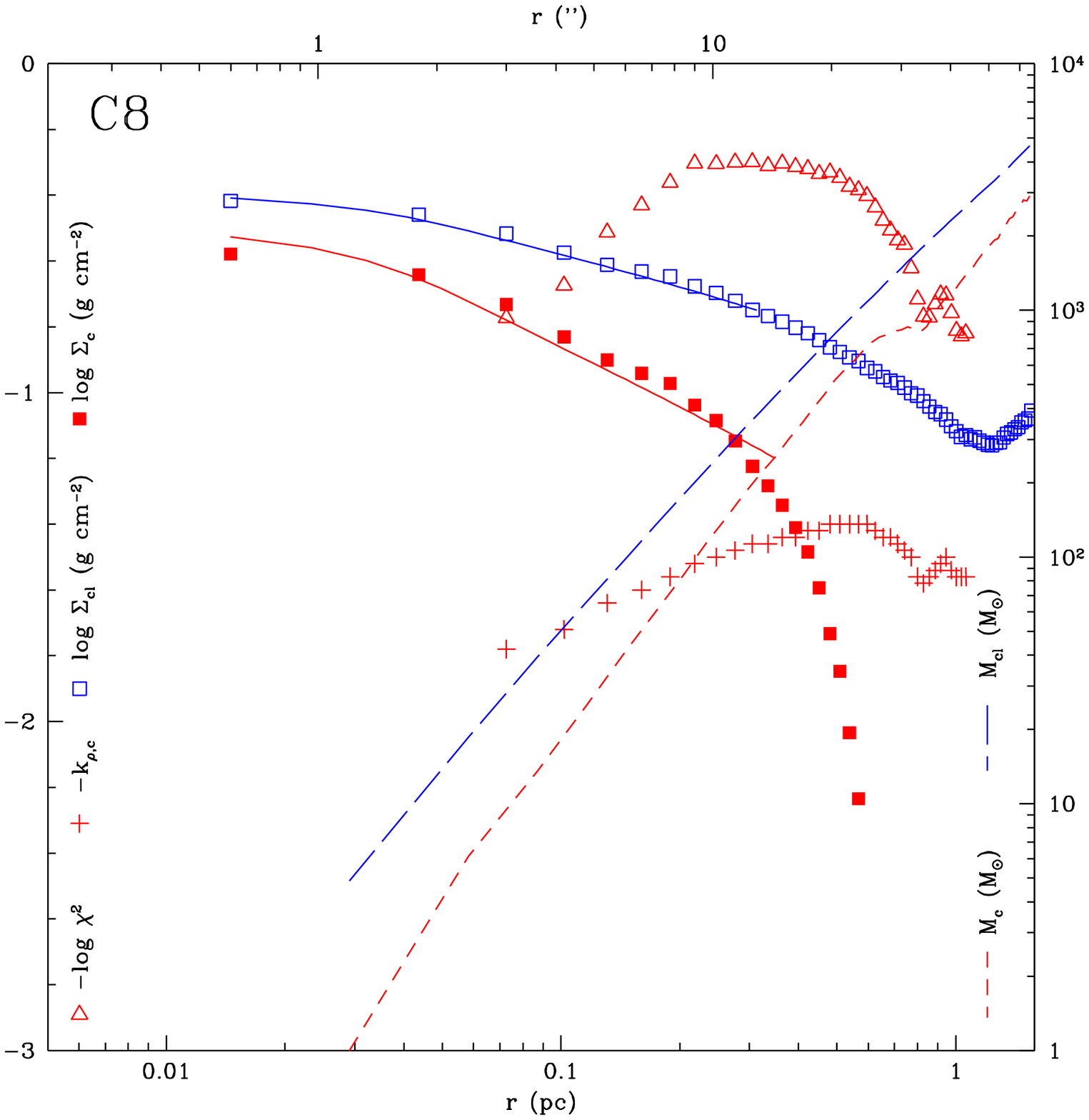} & \includegraphics[width=2.2in]{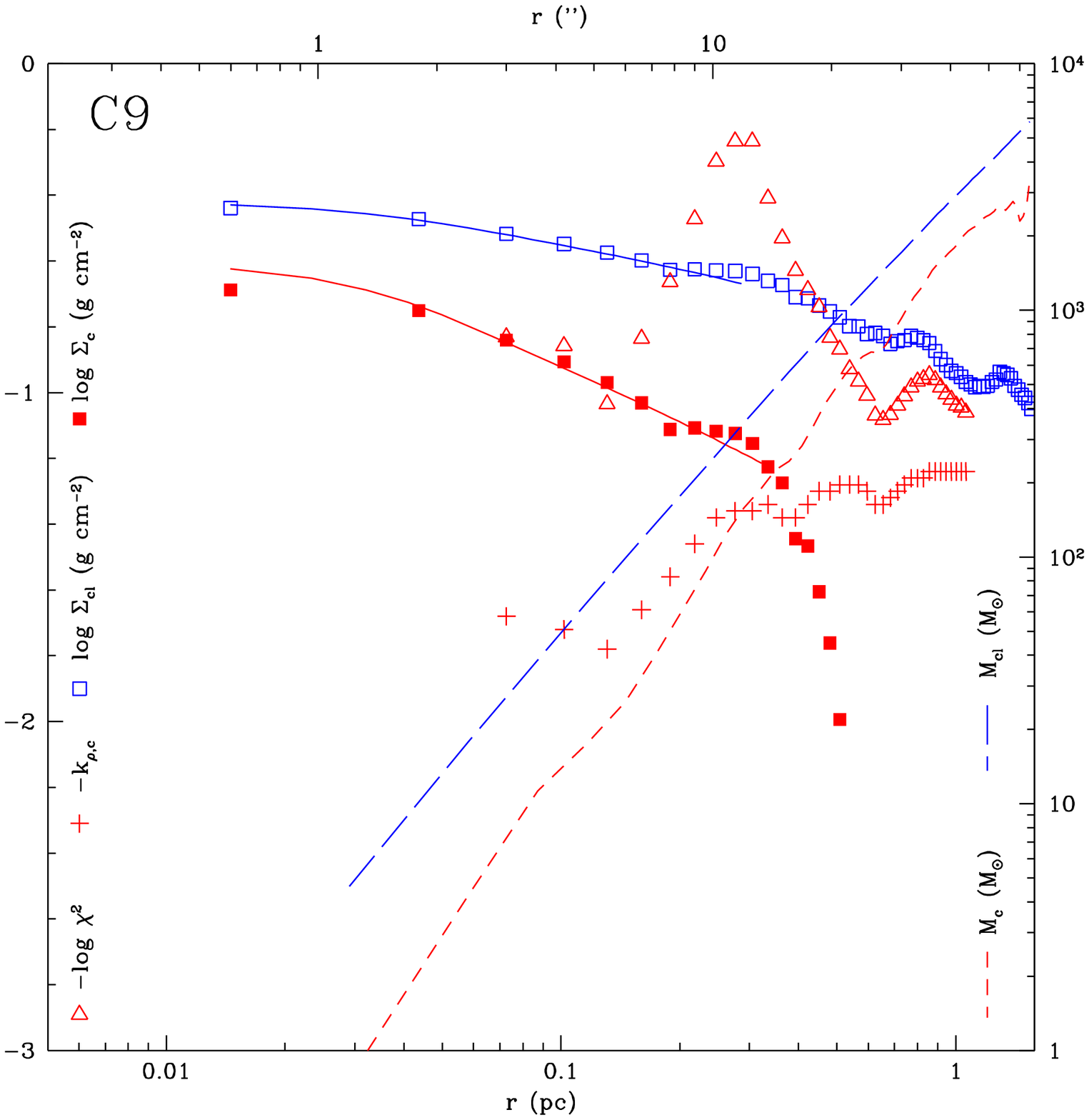} & \hspace{-0.1in} \includegraphics[width=2.2in]{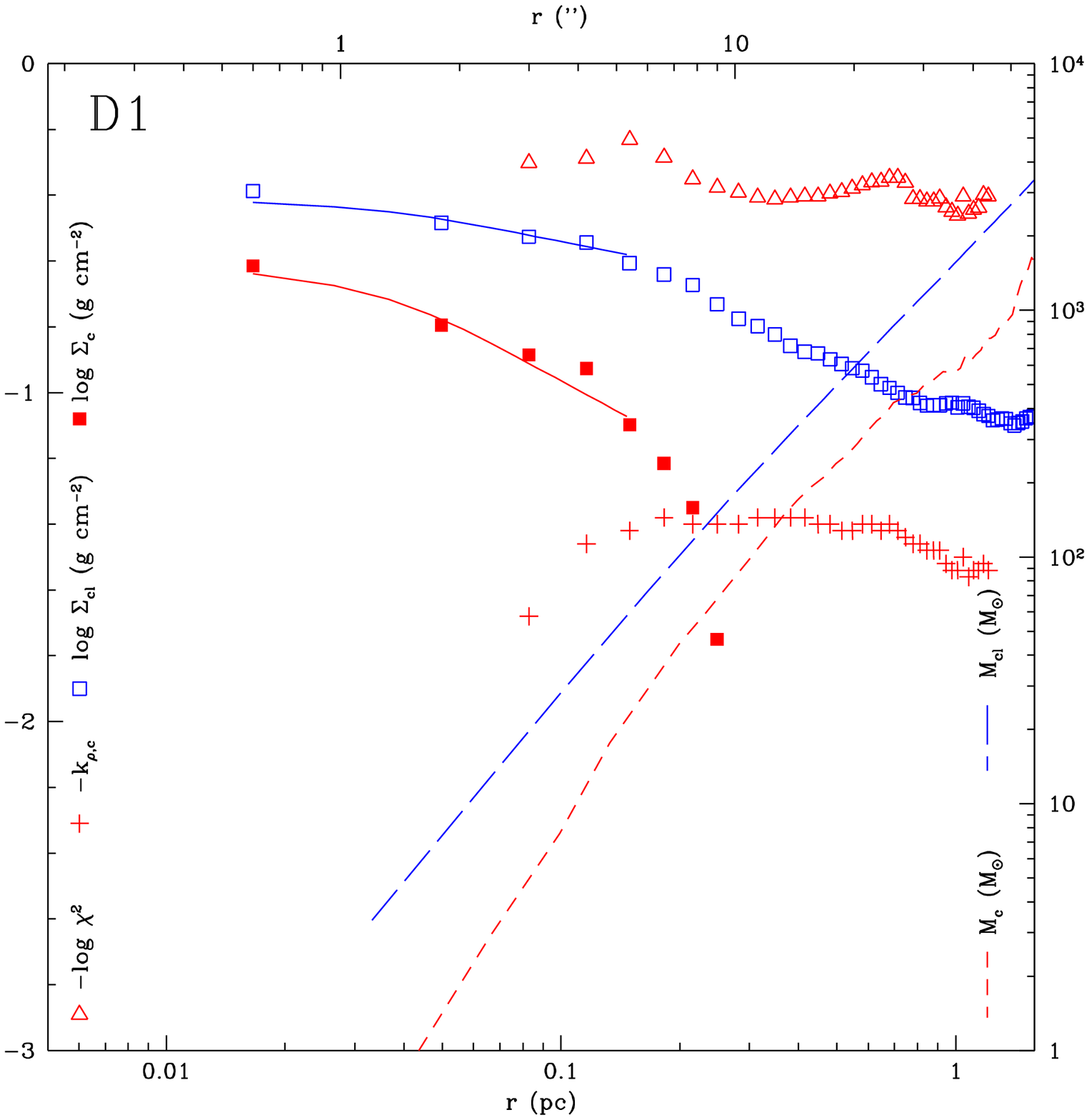} \\
\hspace{-0.0in} \includegraphics[width=2.15in]{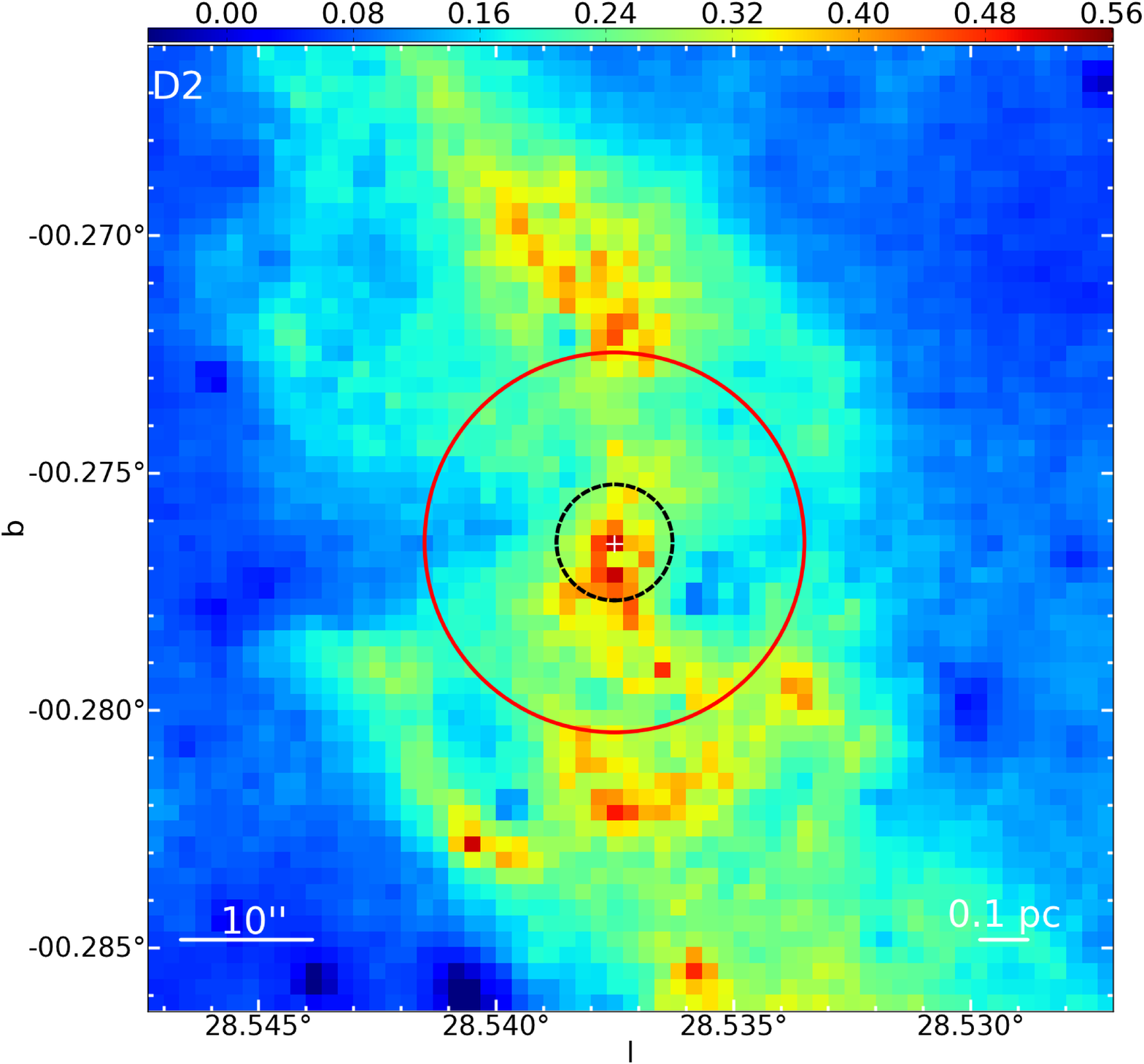} & \hspace{-0.2in} \includegraphics[width=2.15in]{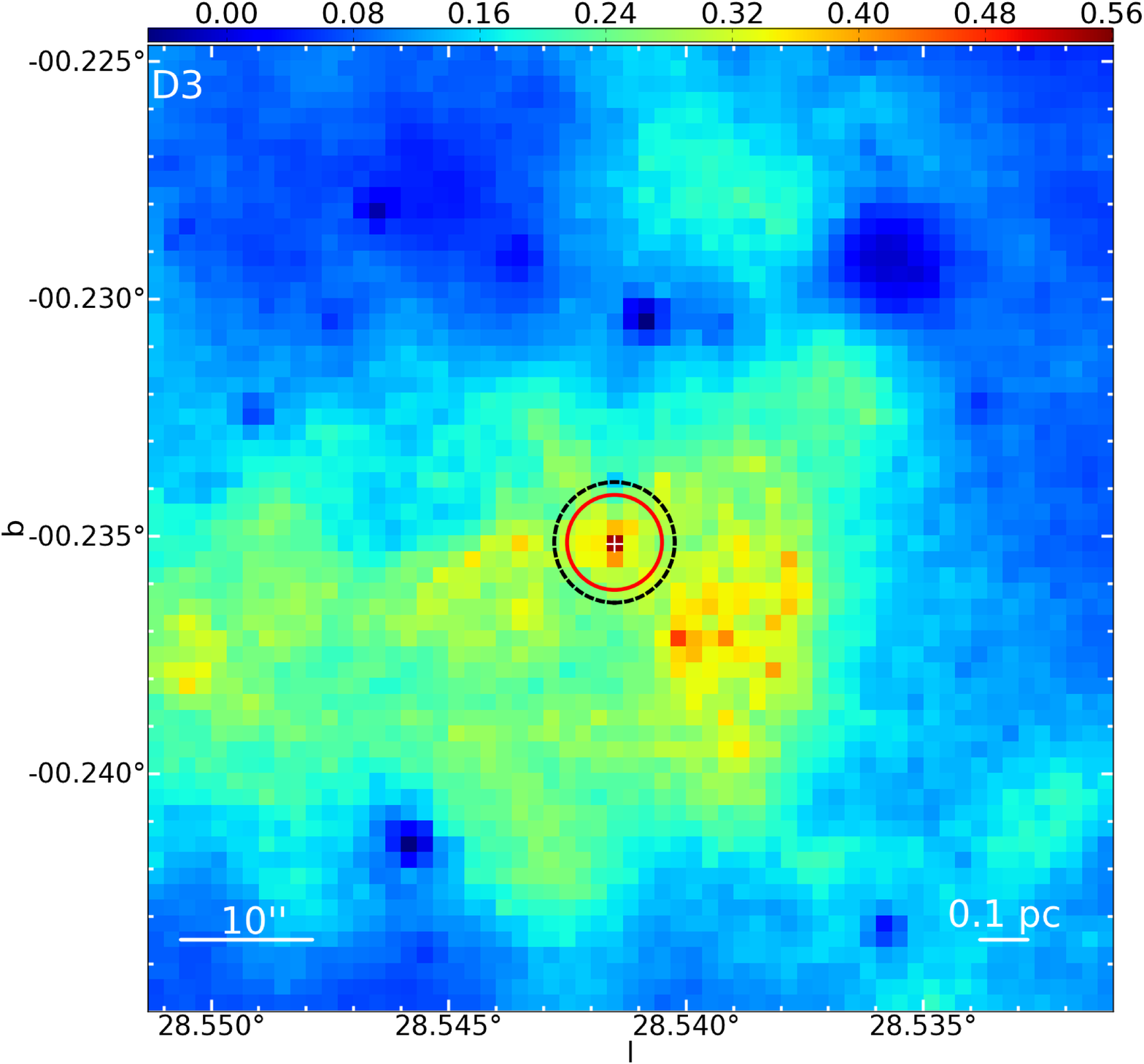} & \hspace{-0.3in} \includegraphics[width=2.15in]{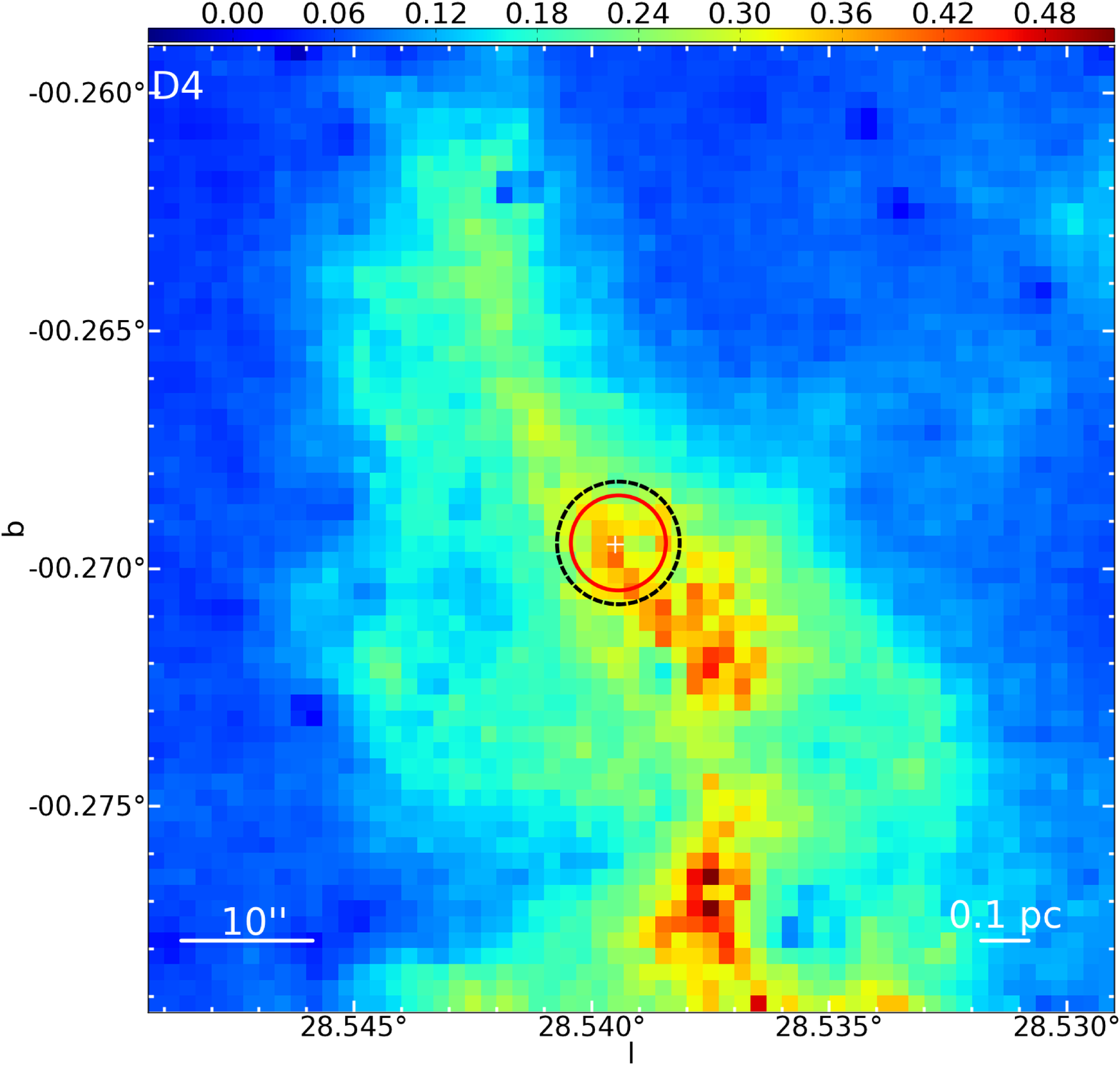} \\
\includegraphics[width=2.2in]{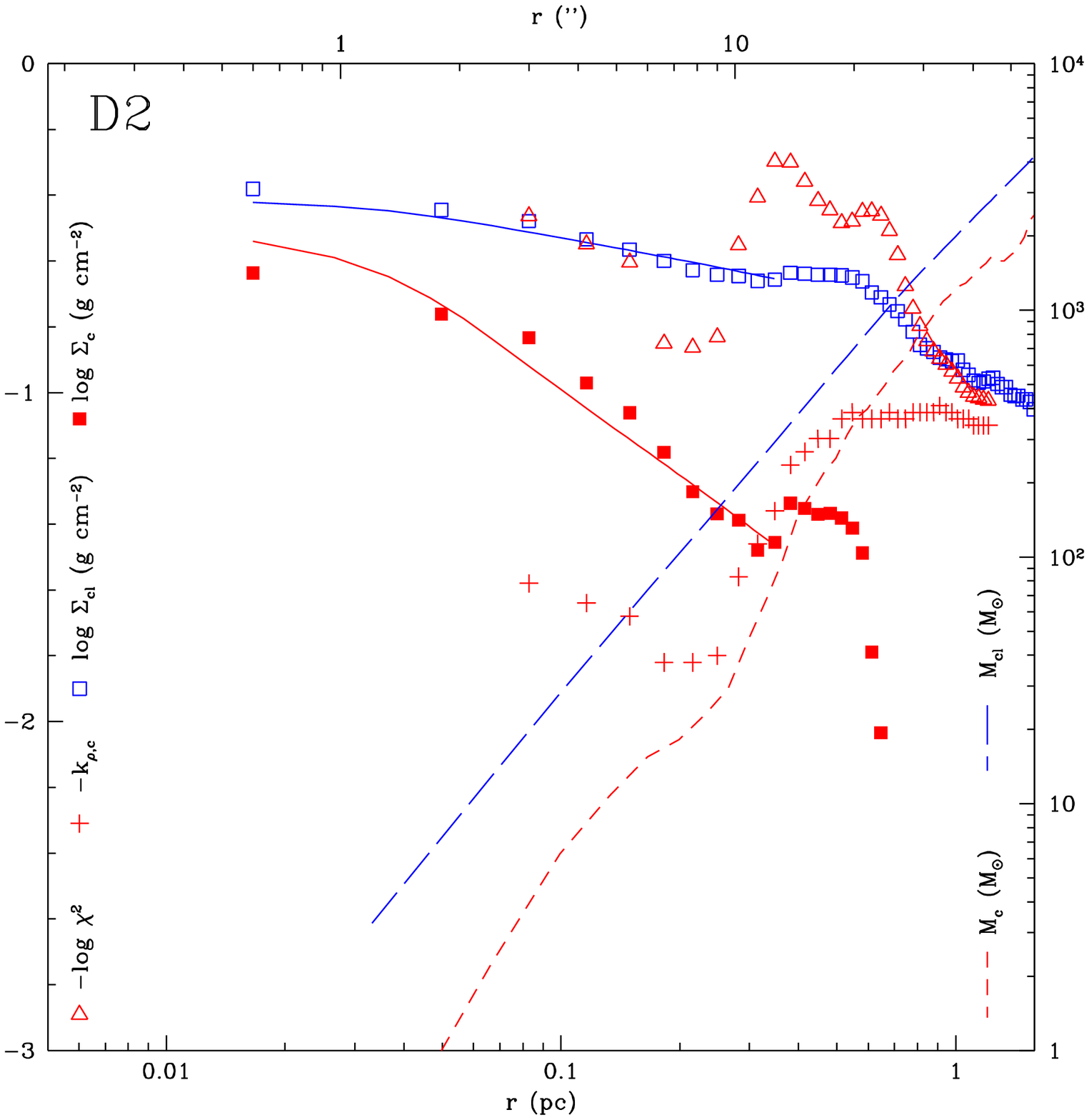} & \includegraphics[width=2.2in]{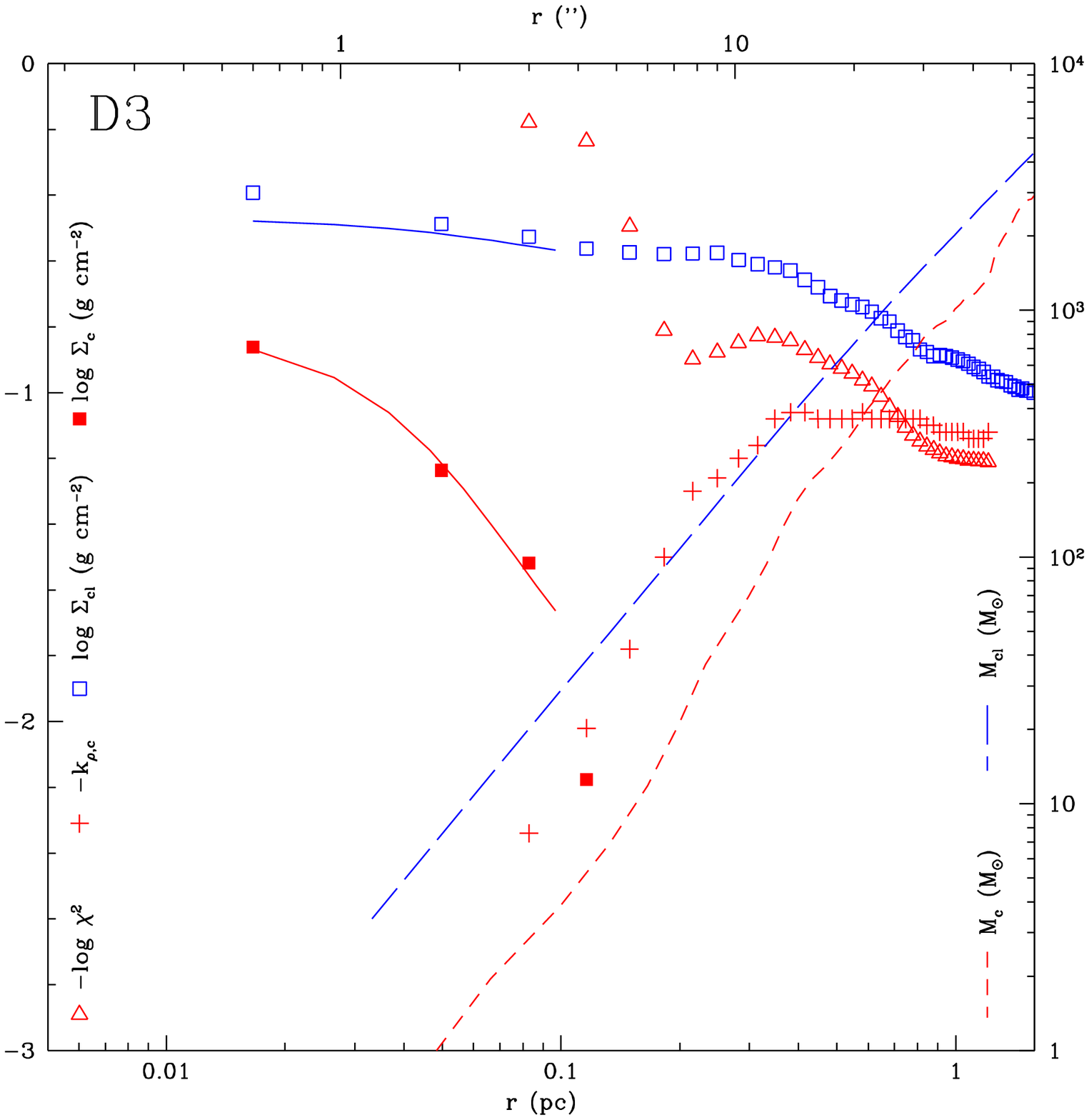} & \hspace{-0.1in} \includegraphics[width=2.2in]{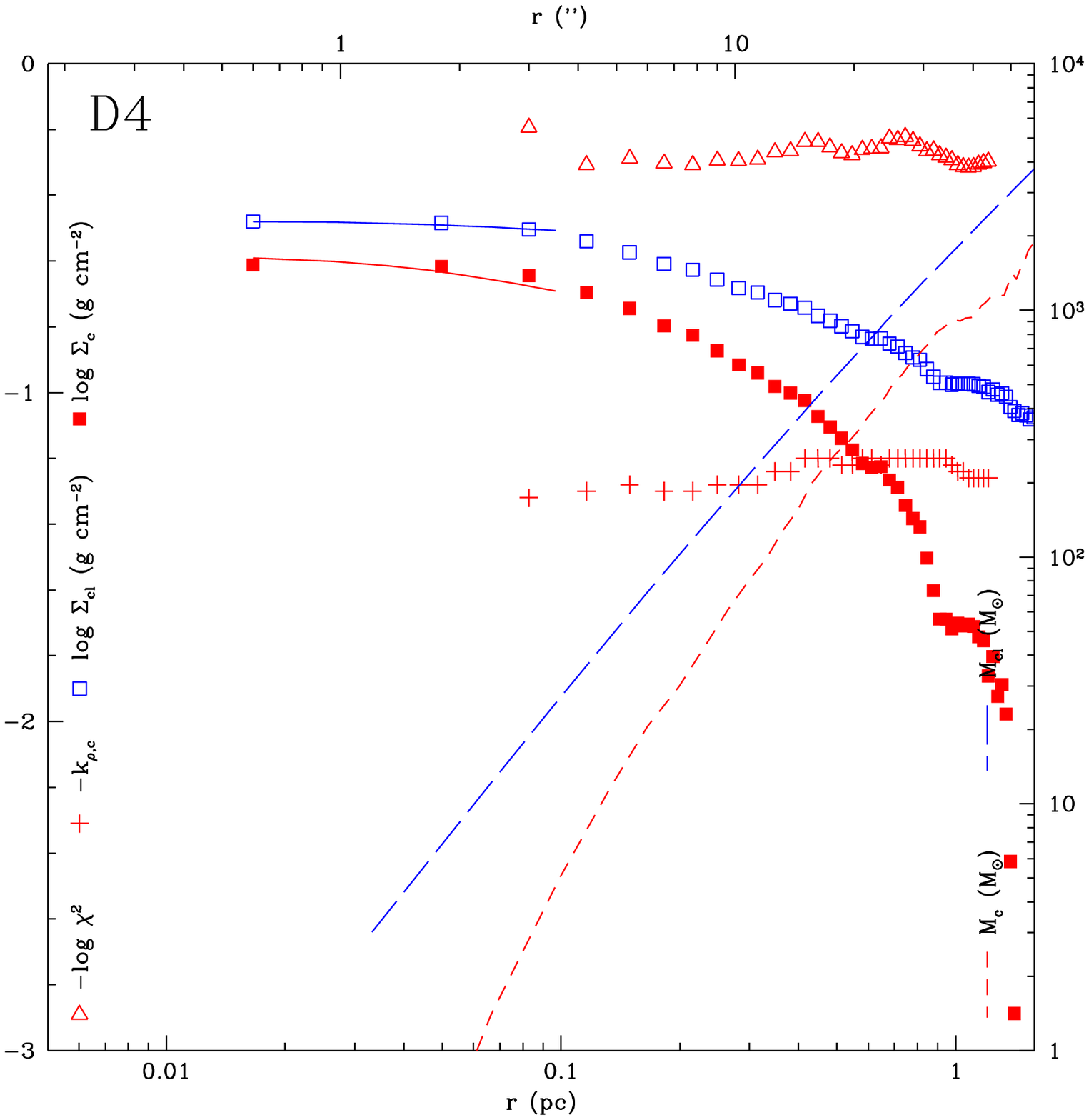}
\end{array}$
\end{center}
\vspace{-0.3in}
\caption{\footnotesize
Core C8, C9, D1, D2, D3, D4 $\Sigma$ maps (notation as Fig.~\ref{fig:coreA1}a) and azimuthally averaged radial profile figures (notation as Fig.~\ref{fig:coreA1}b).
\label{fig:cores3}
}
\end{figure*}

\begin{figure*}
\begin{center}$
\begin{array}{ccc}
\hspace{-0.0in} \includegraphics[width=2.15in]{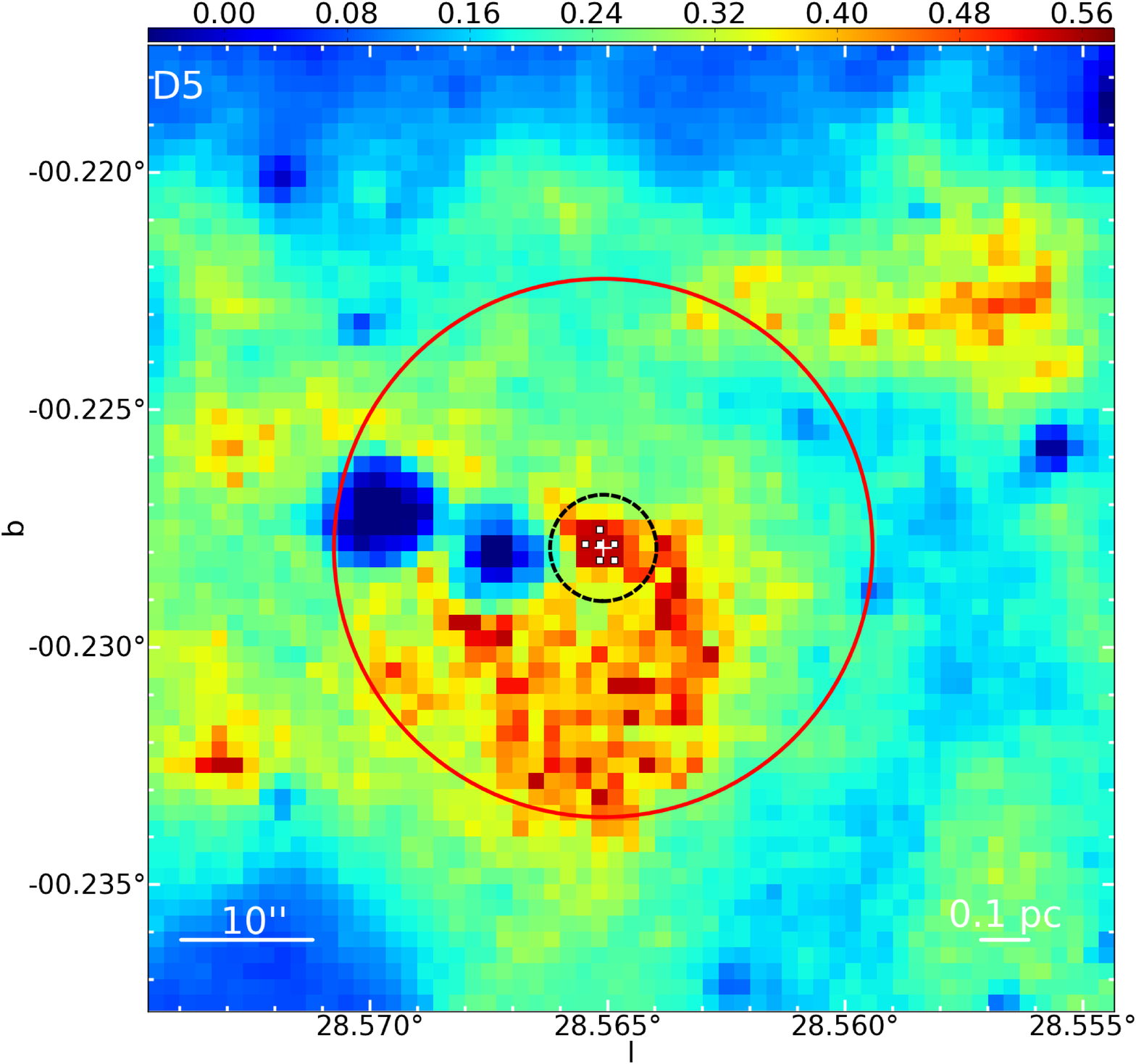} & \hspace{-0.2in} \includegraphics[width=2.15in]{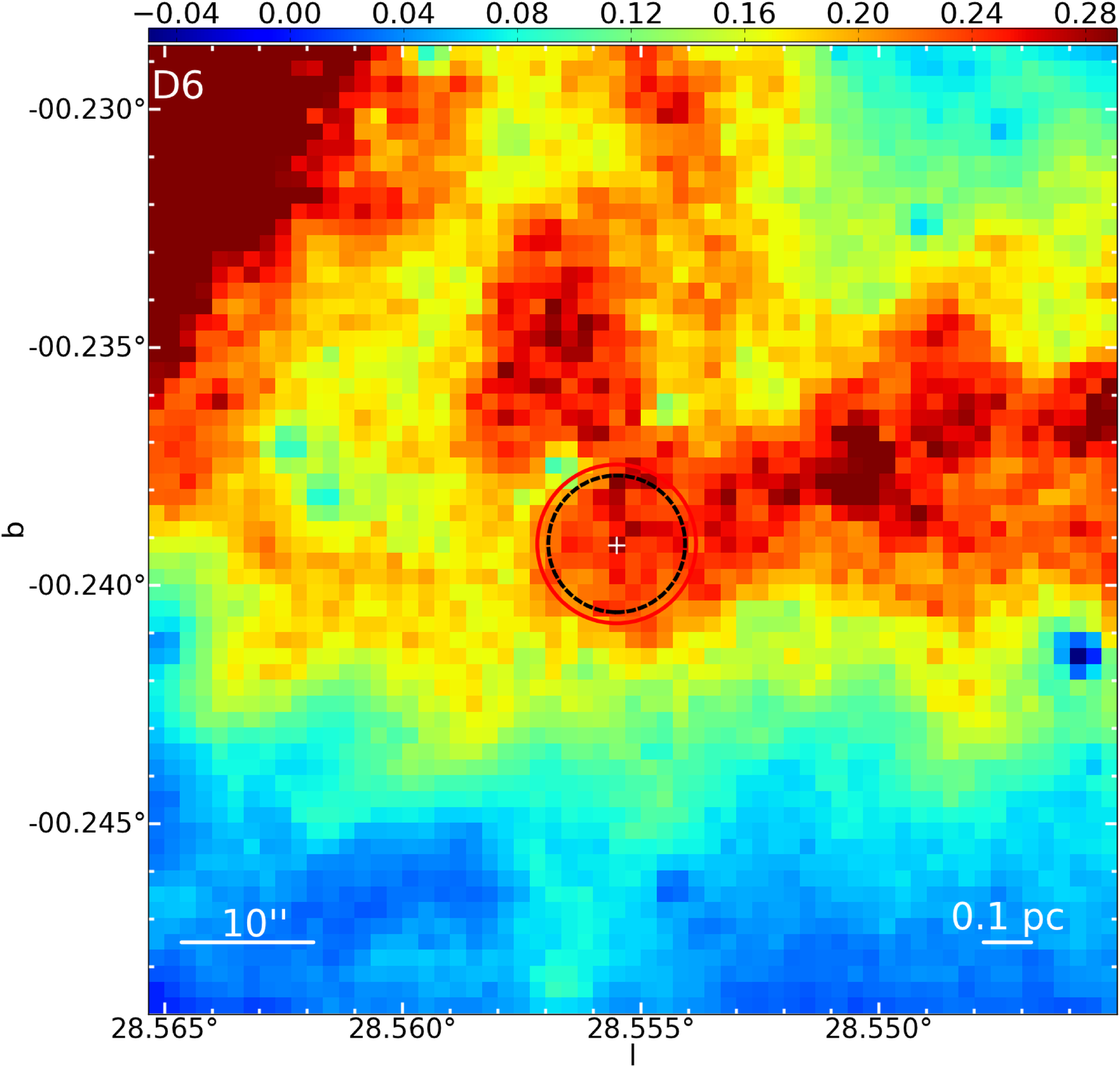} & \hspace{-0.3in} \includegraphics[width=2.15in]{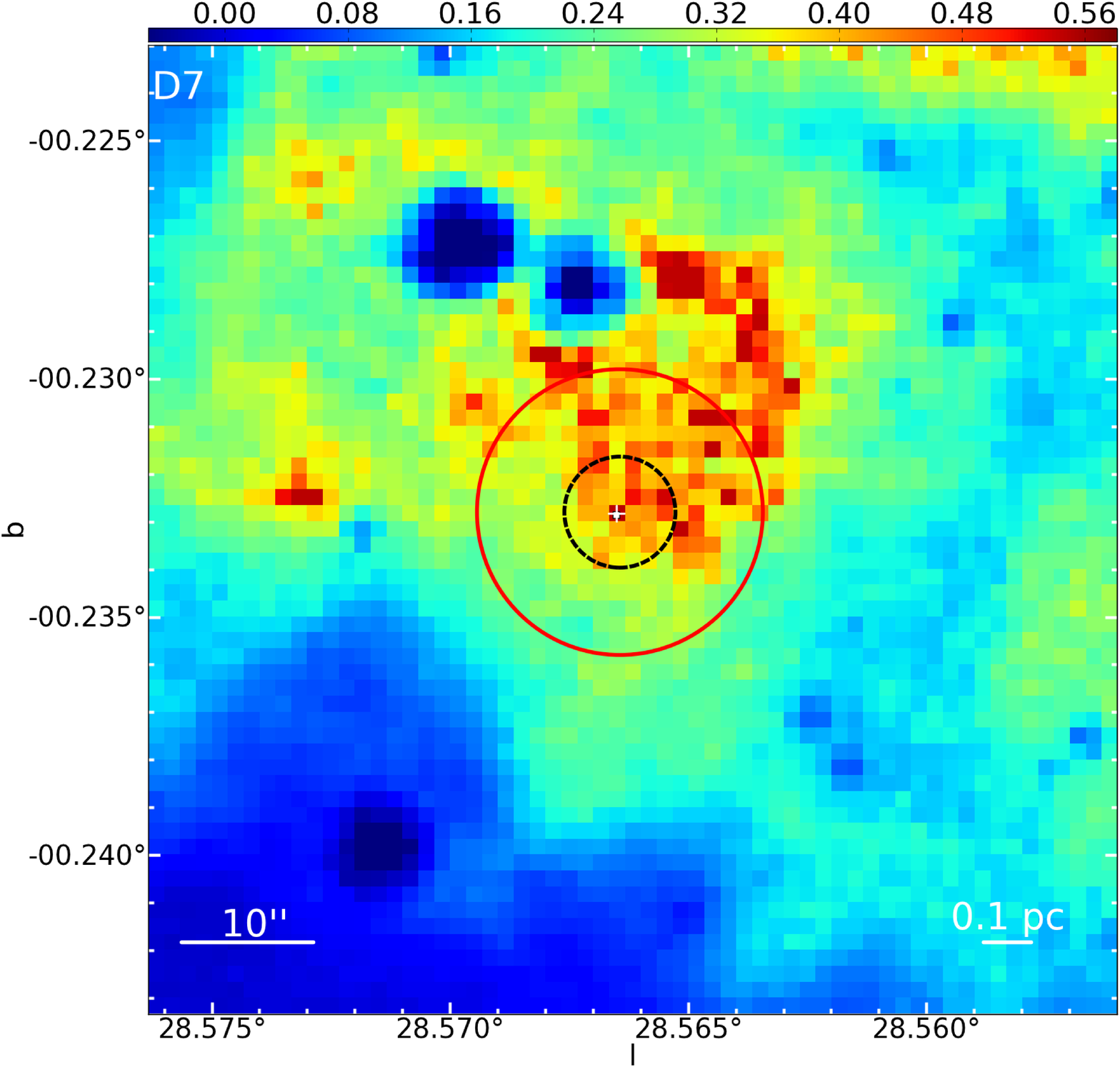} \\
\includegraphics[width=2.2in]{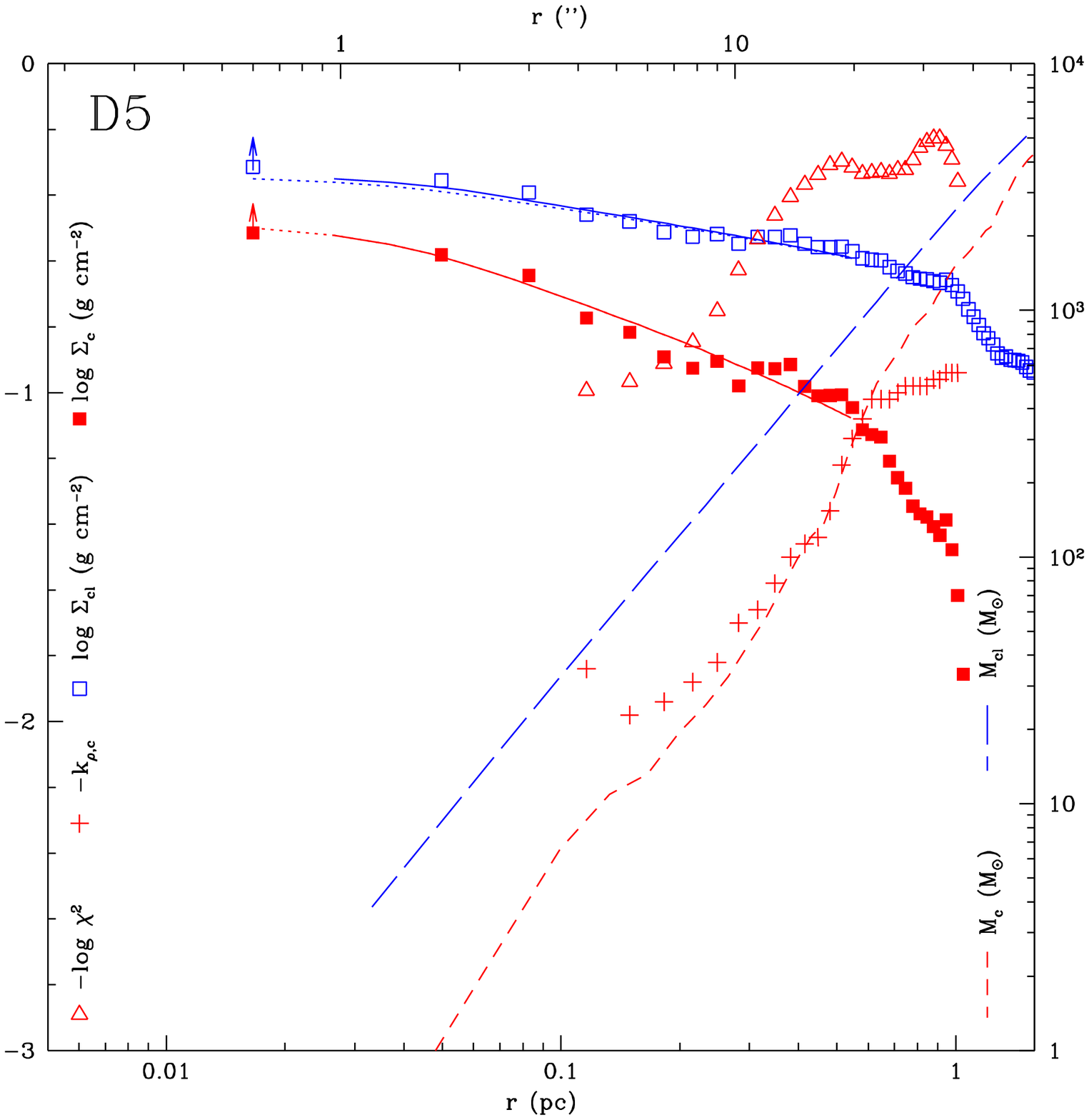} & \includegraphics[width=2.2in]{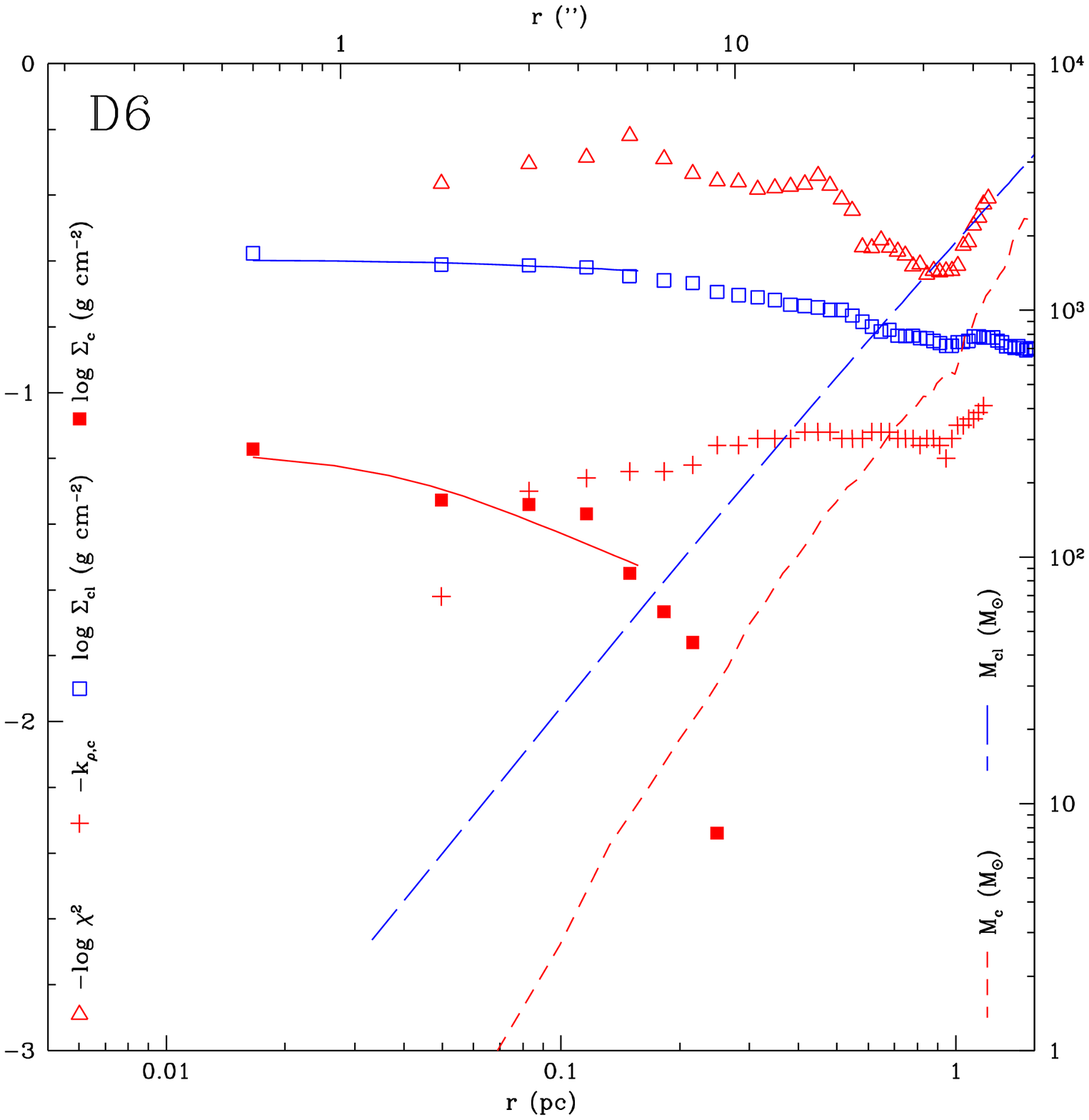} & \hspace{-0.1in} \includegraphics[width=2.2in]{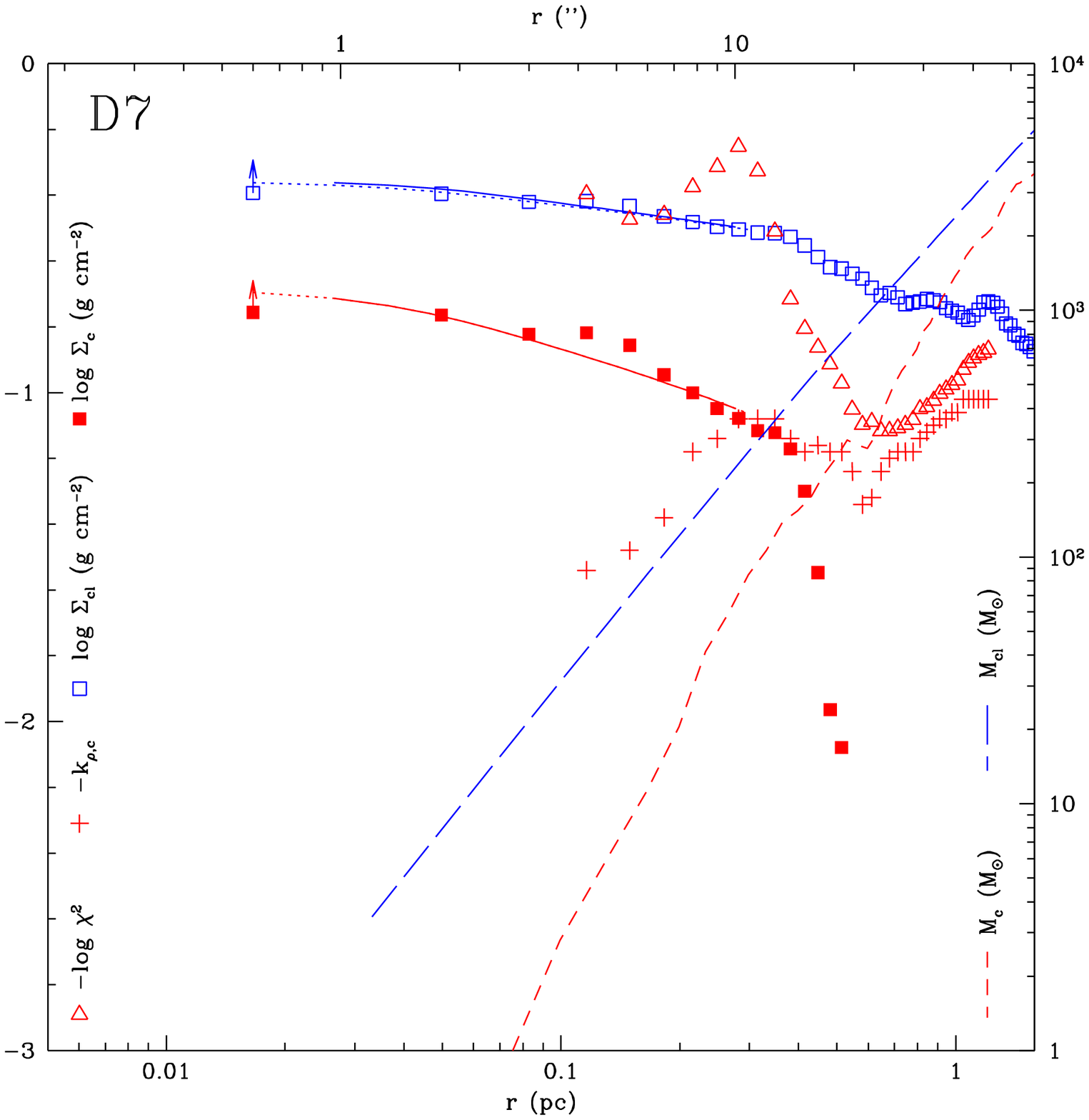} \\
\hspace{-0.0in} \includegraphics[width=2.15in]{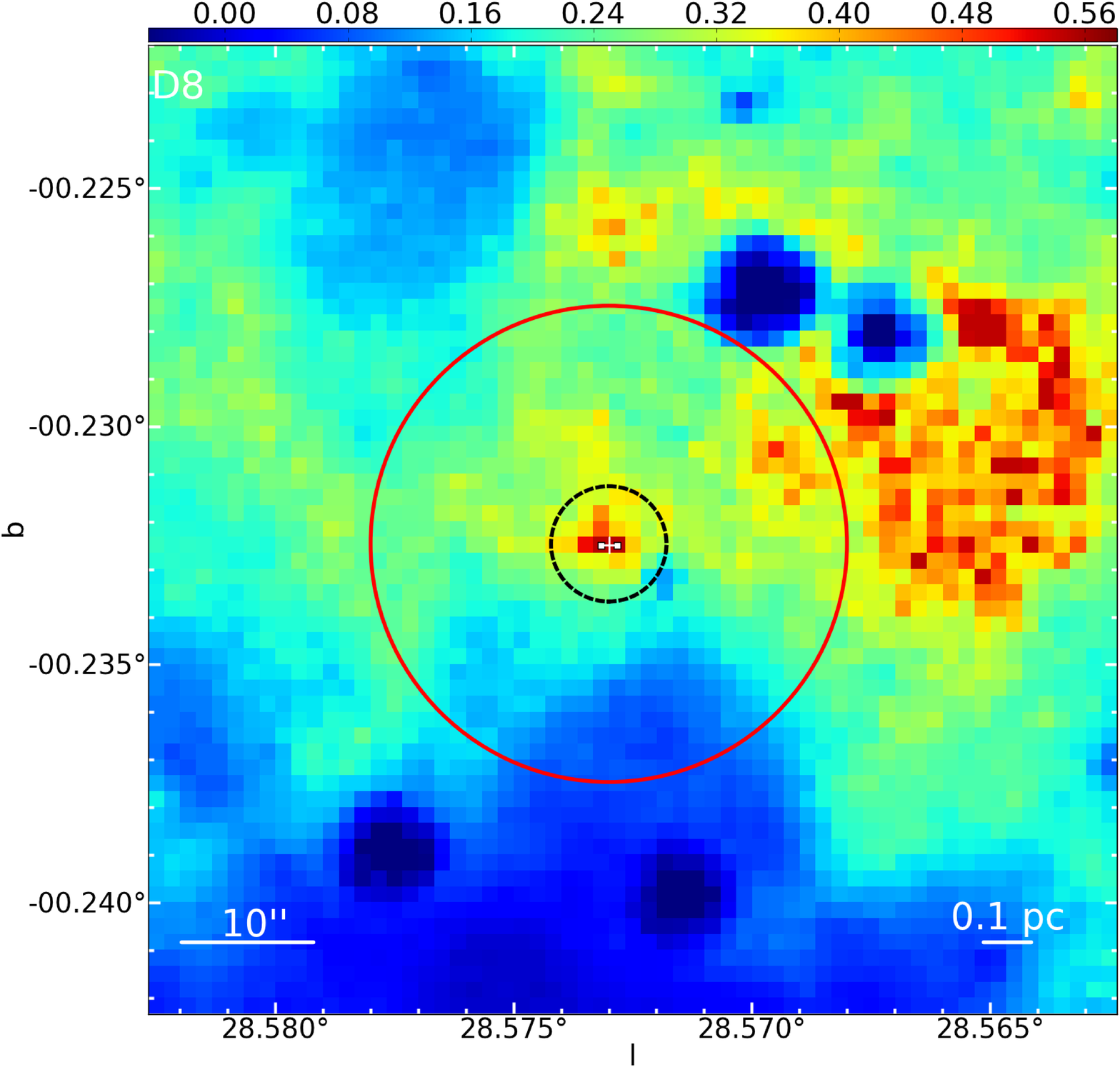} & \hspace{-0.2in} \includegraphics[width=2.15in]{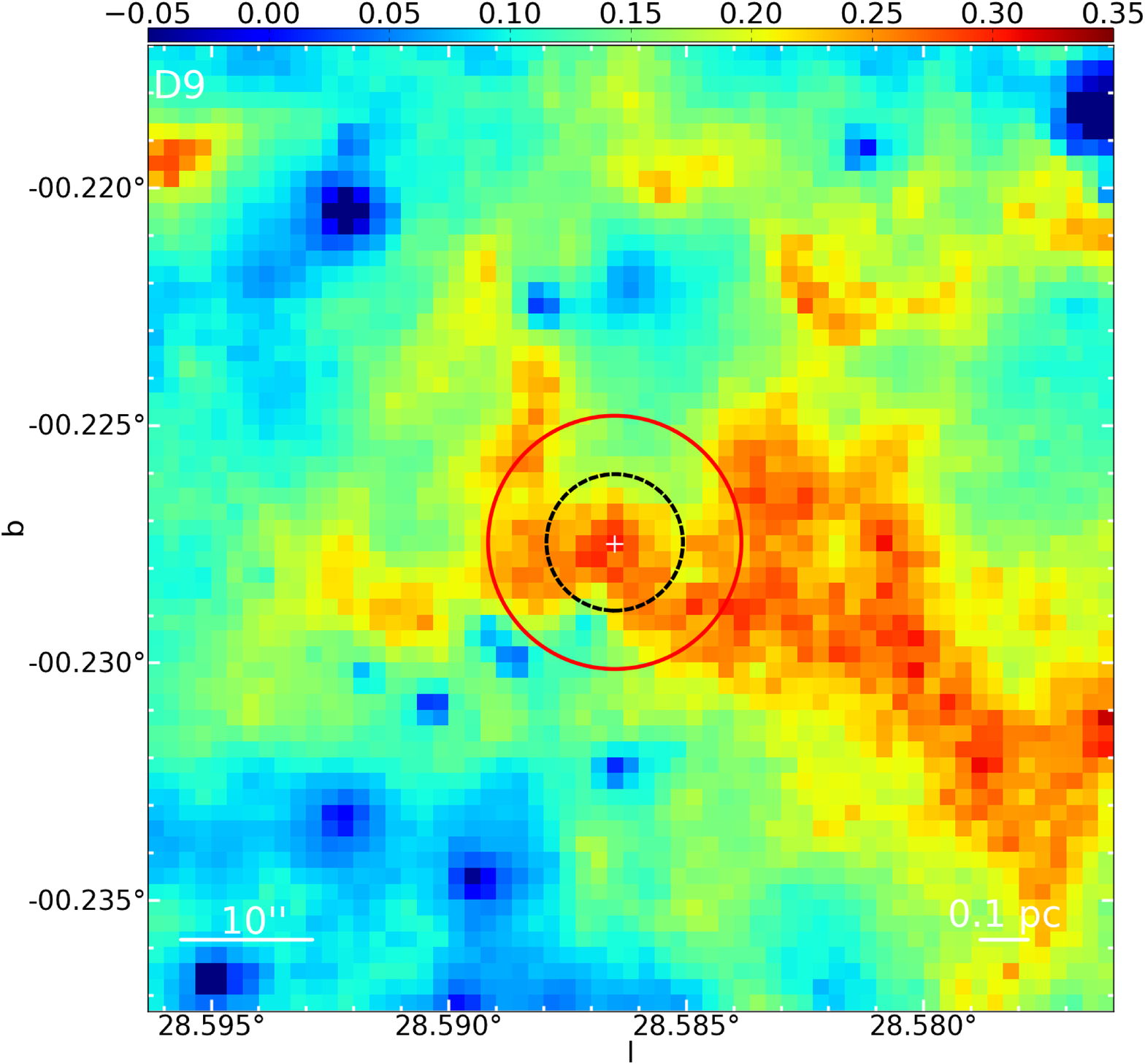} & \hspace{-0.3in} \includegraphics[width=2.15in]{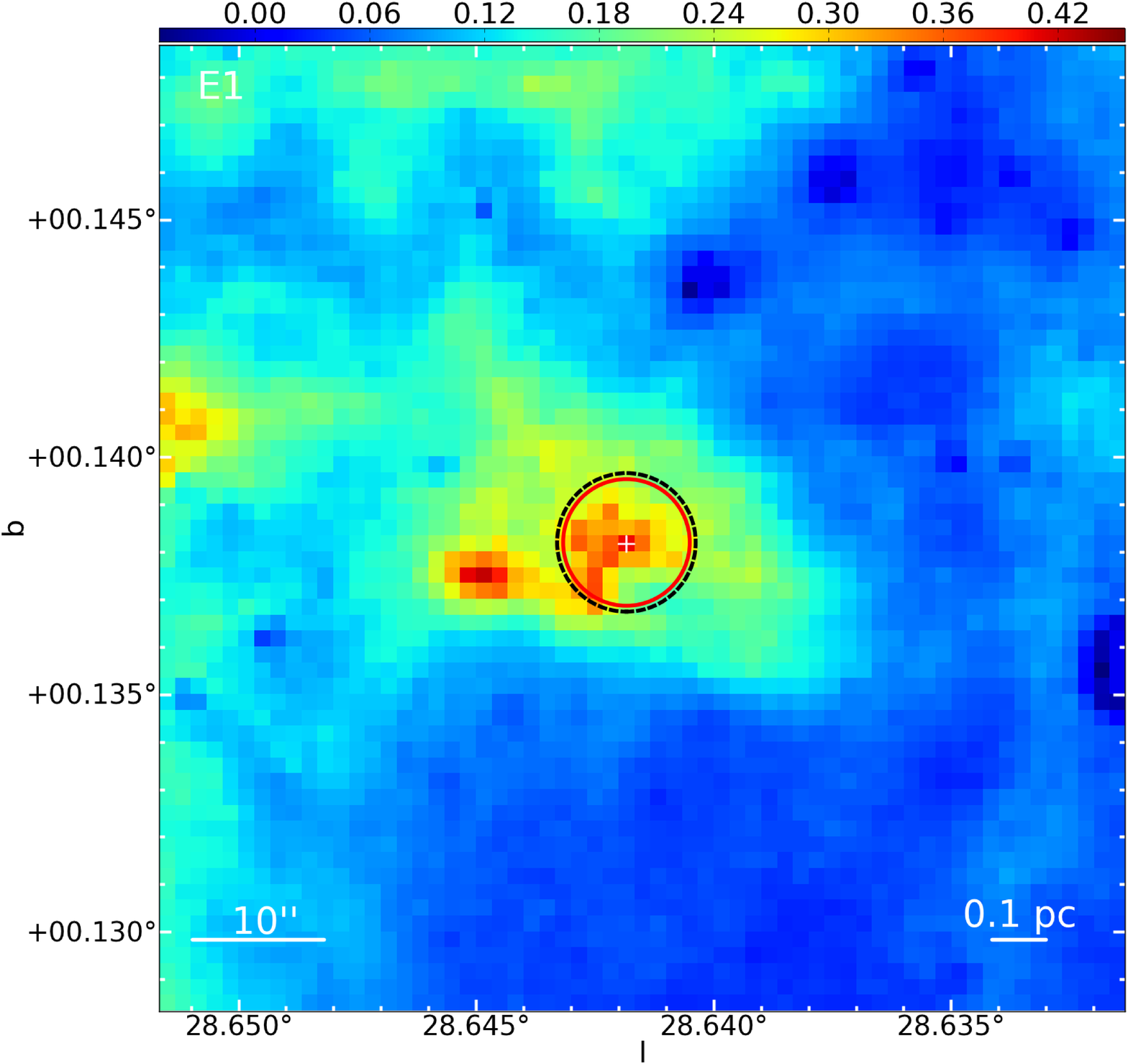} \\
\includegraphics[width=2.2in]{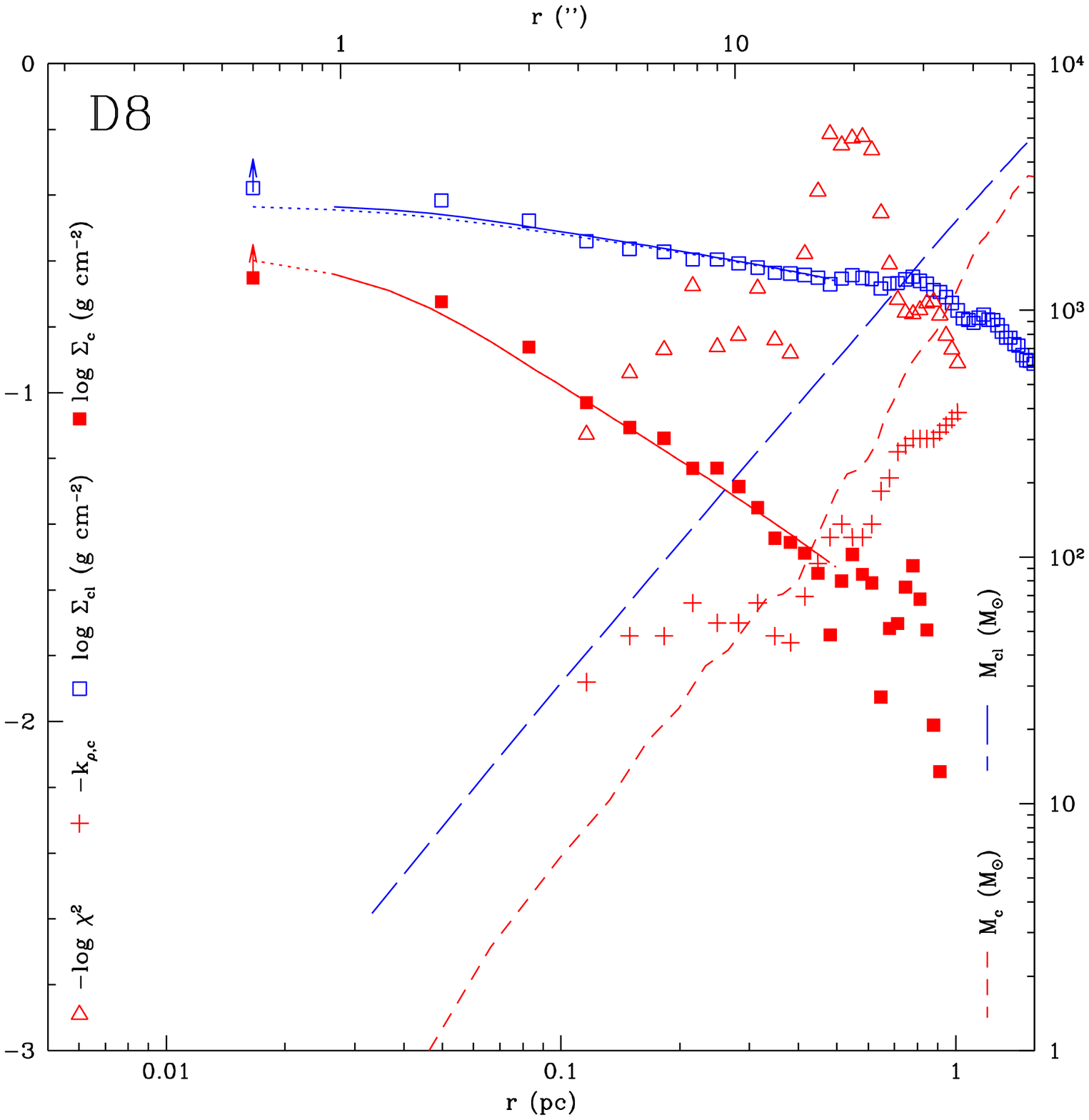} & \includegraphics[width=2.2in]{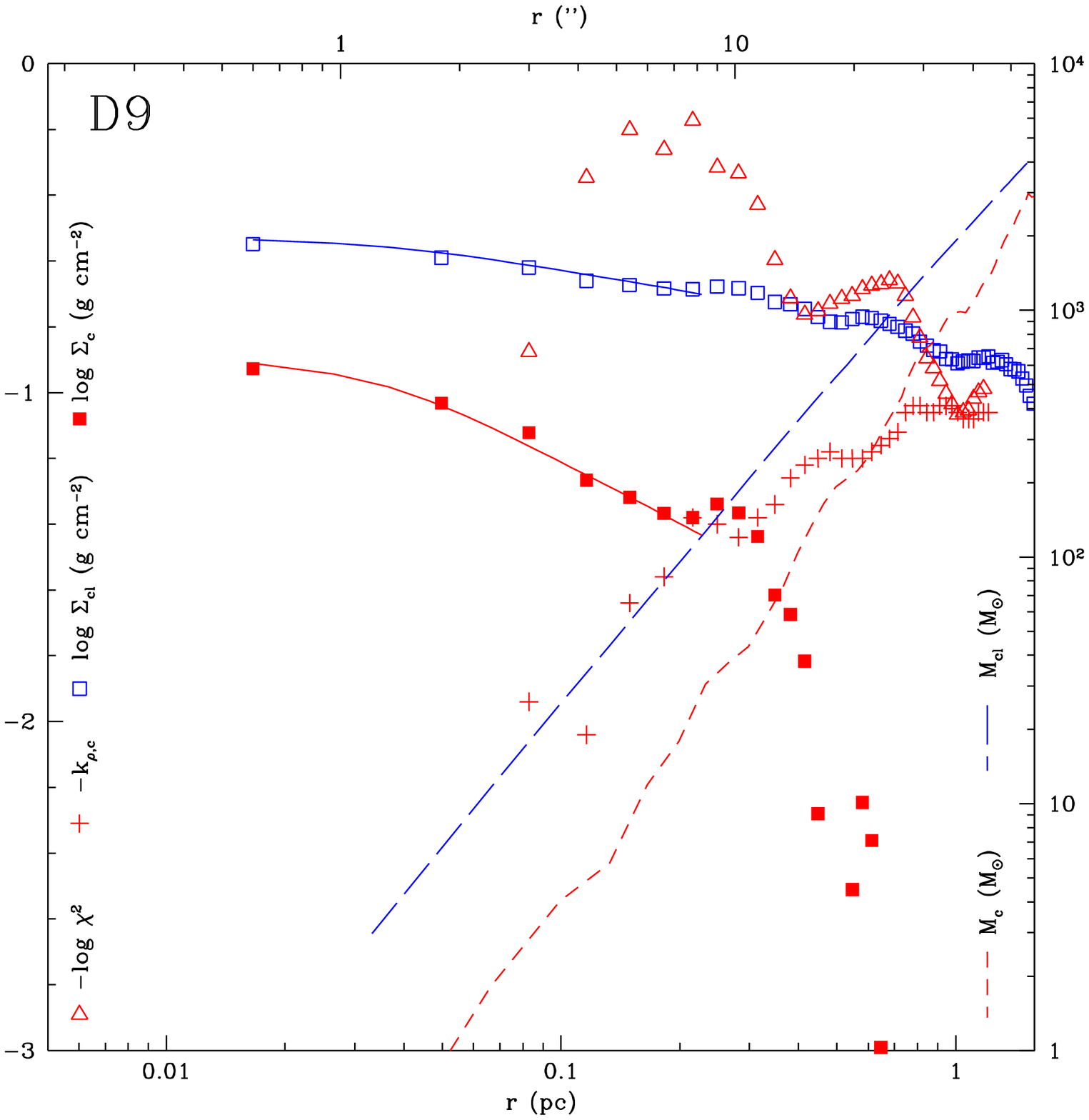} & \hspace{-0.1in} \includegraphics[width=2.2in]{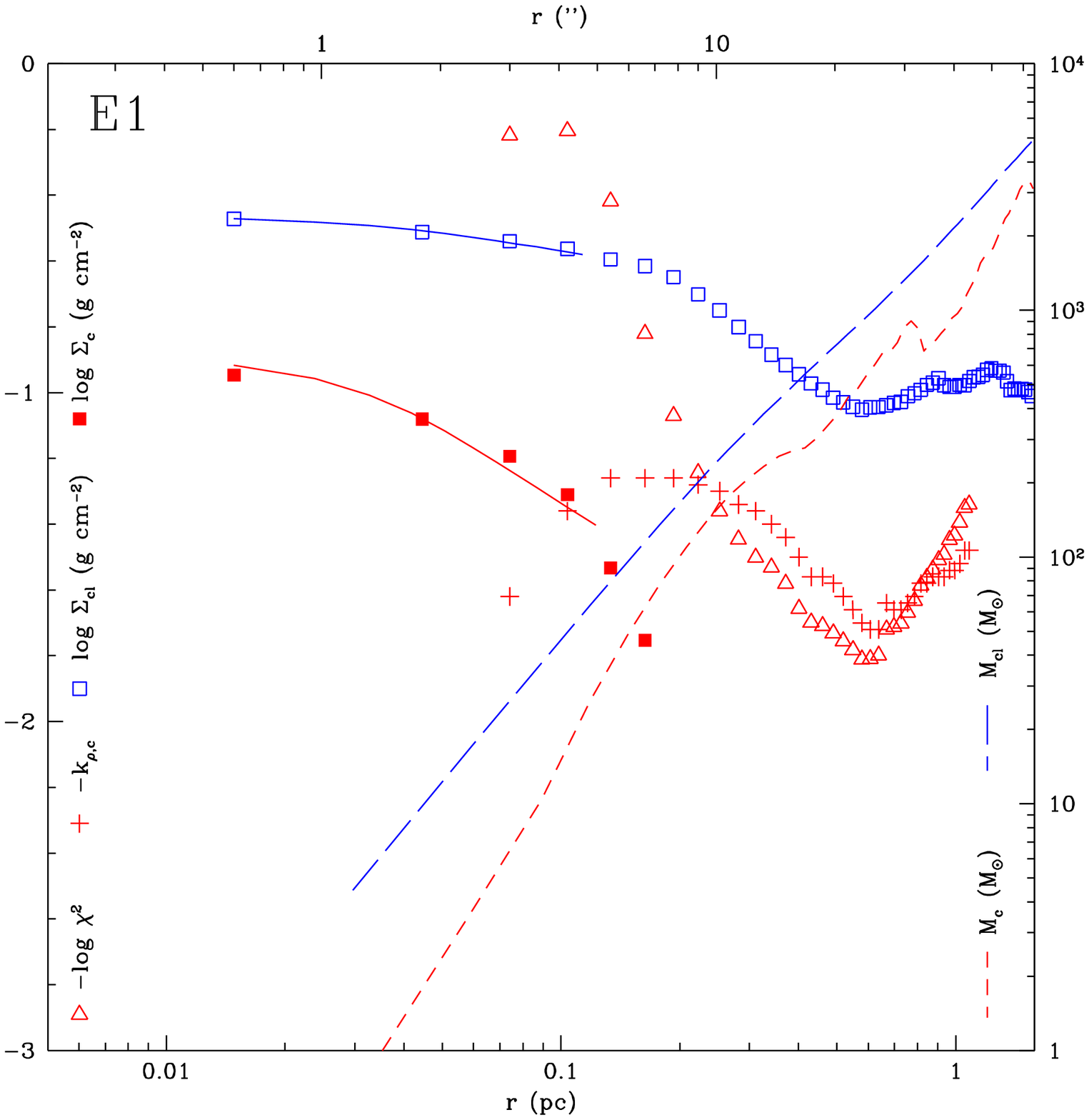}
\end{array}$
\end{center}
\vspace{-0.3in}
\caption{\footnotesize
Core D5, D6, D7, D8, D9, E1 $\Sigma$ maps (notation as Fig.~\ref{fig:coreA1}a) and azimuthally averaged radial profile figures (notation as Fig.~\ref{fig:coreA1}b).
\label{fig:cores4}
}
\end{figure*}

\begin{figure*}
\begin{center}$
\begin{array}{ccc}
\hspace{-0.0in} \includegraphics[width=2.15in]{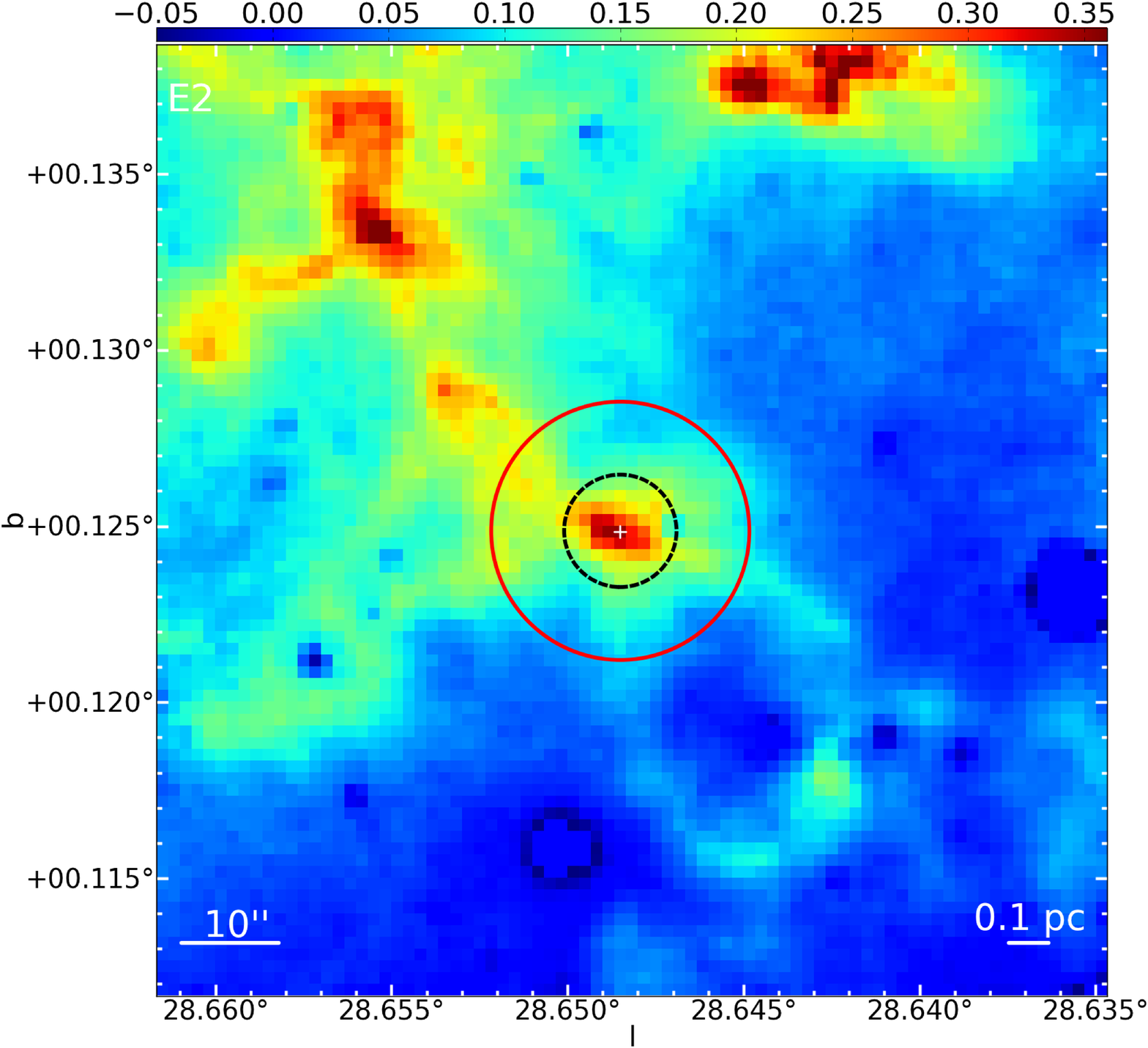} & \hspace{-0.2in} \includegraphics[width=2.15in]{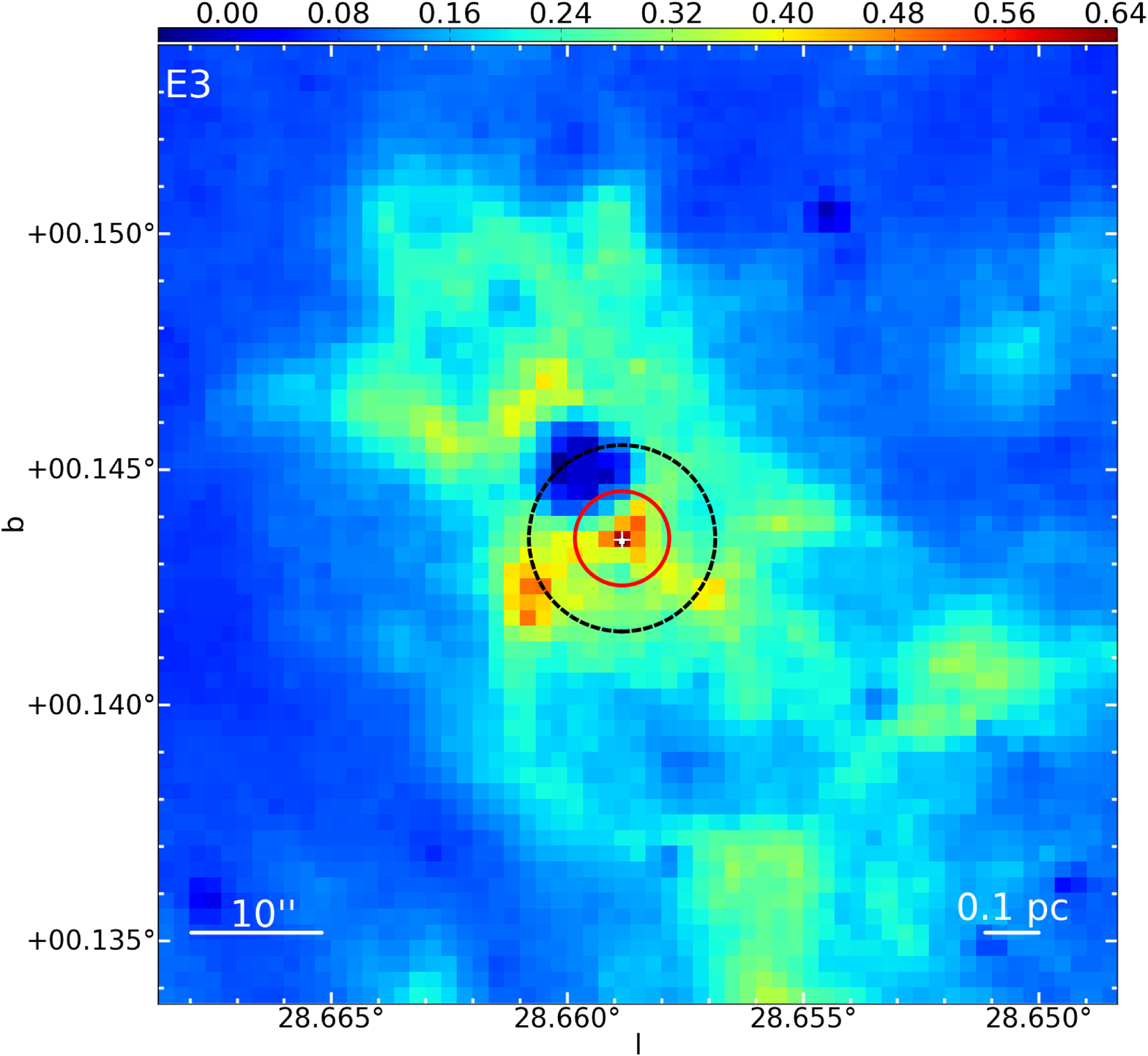} & \hspace{-0.3in} \includegraphics[width=2.15in]{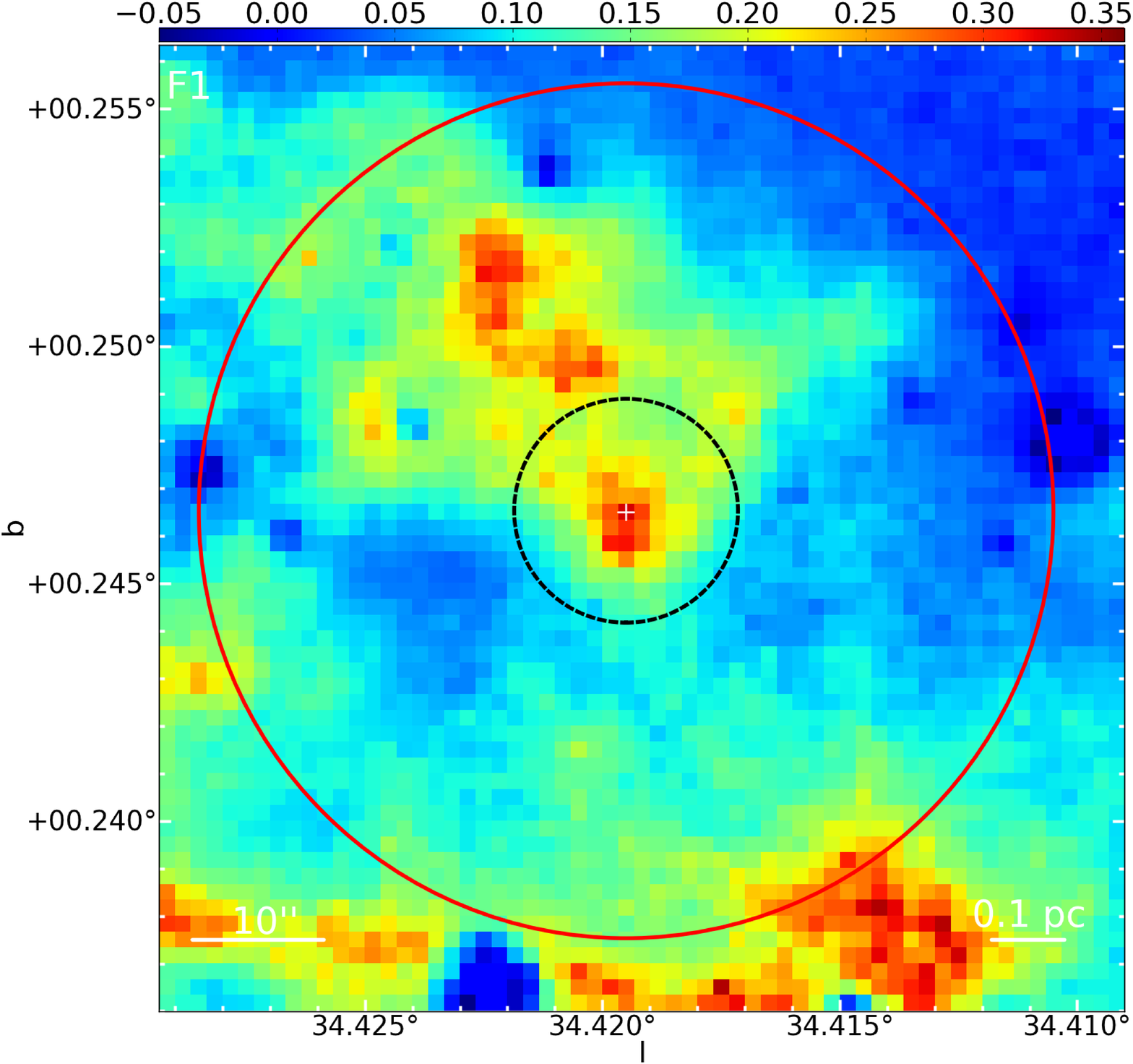} \\
\includegraphics[width=2.2in]{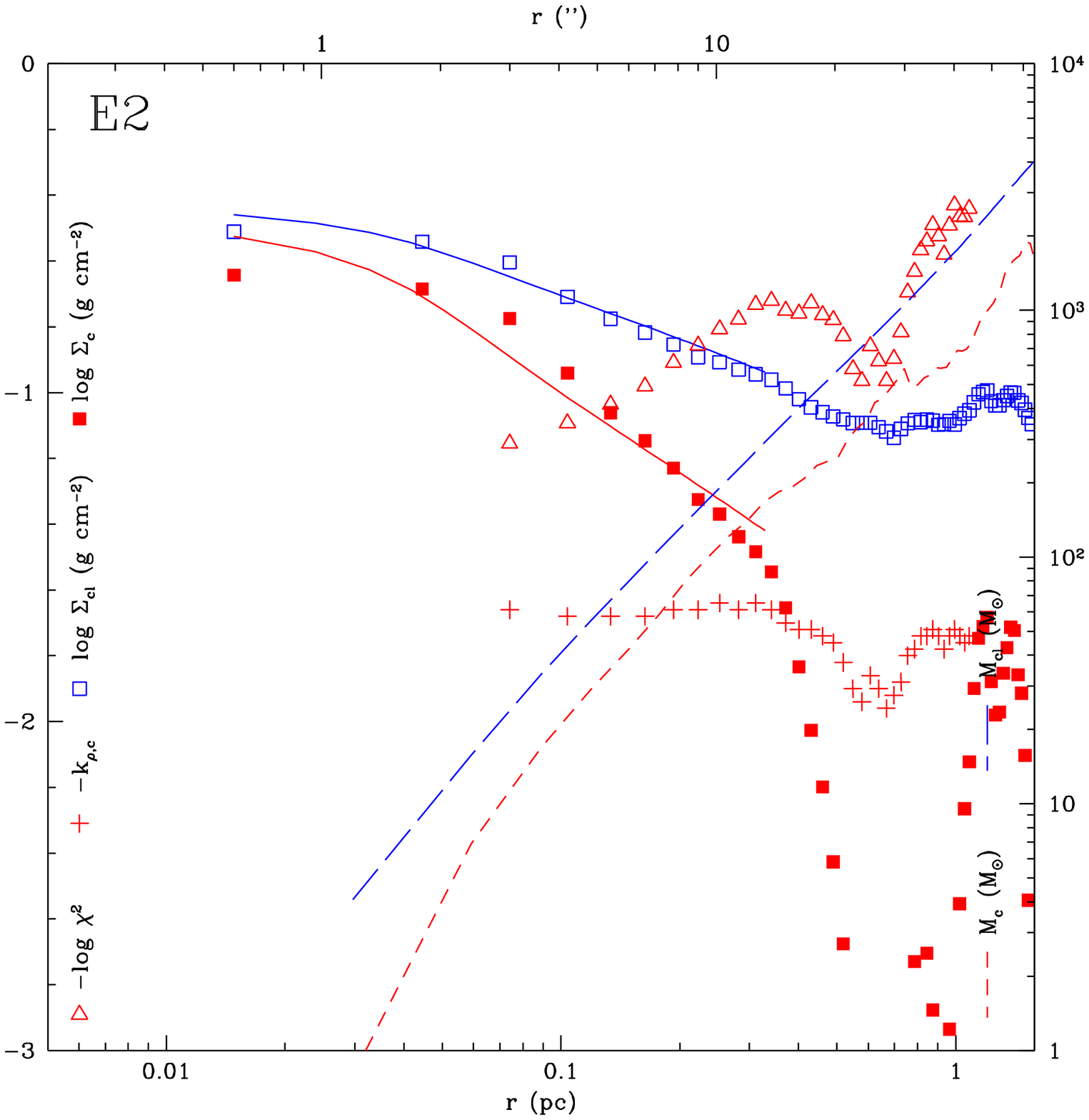} & \includegraphics[width=2.2in]{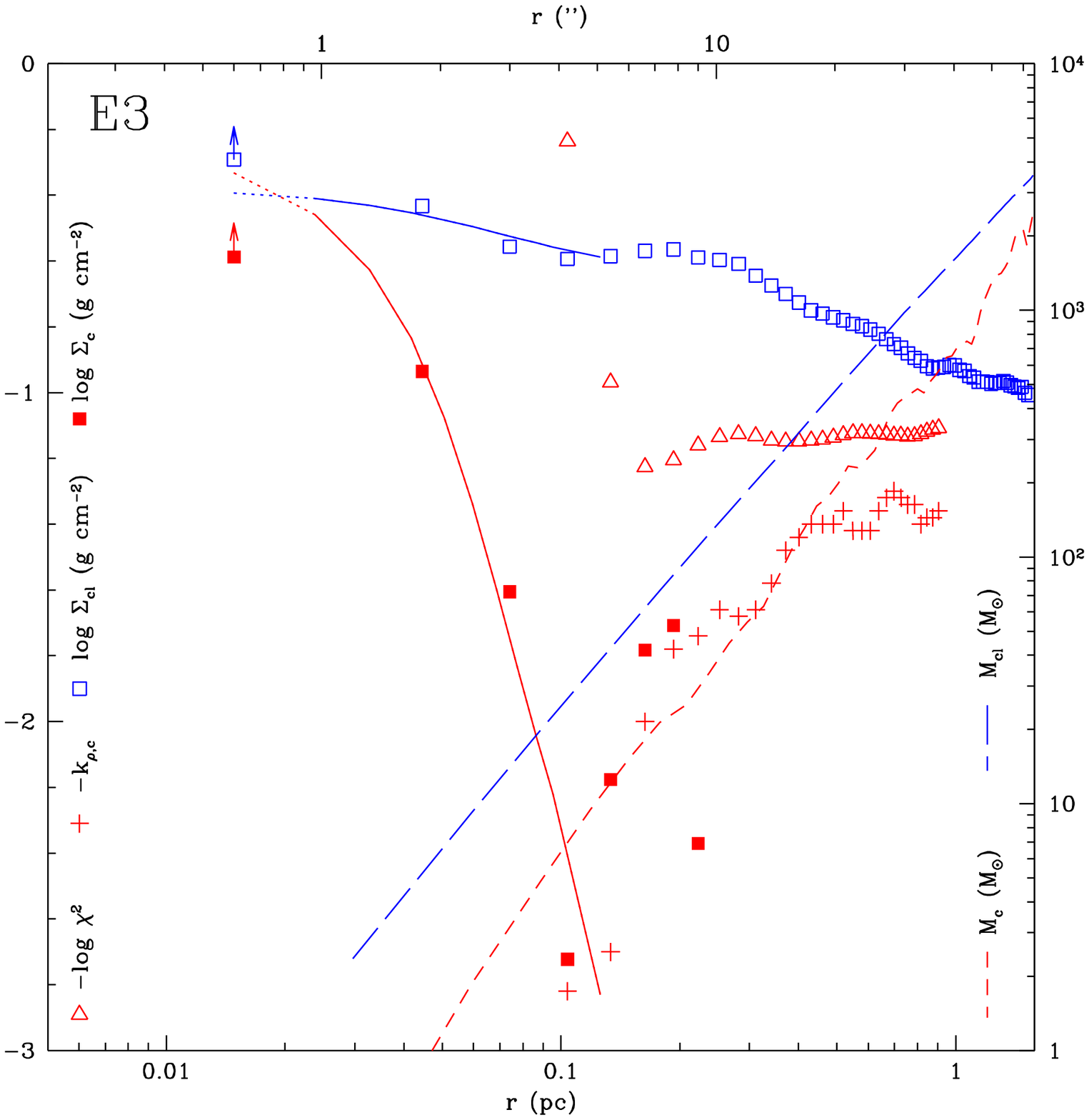} & \hspace{-0.1in} \includegraphics[width=2.2in]{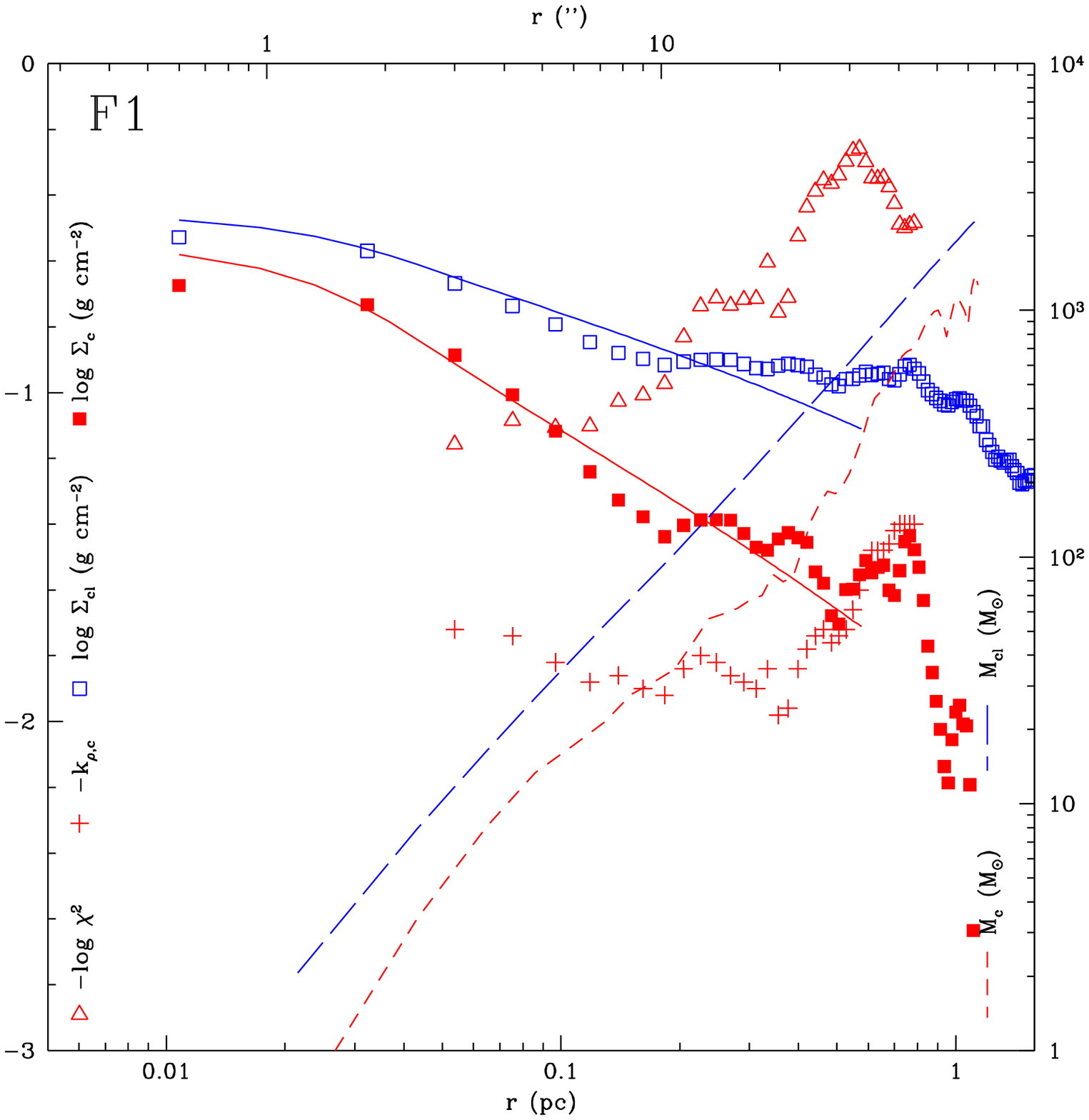} \\
\hspace{-0.0in} \includegraphics[width=2.15in]{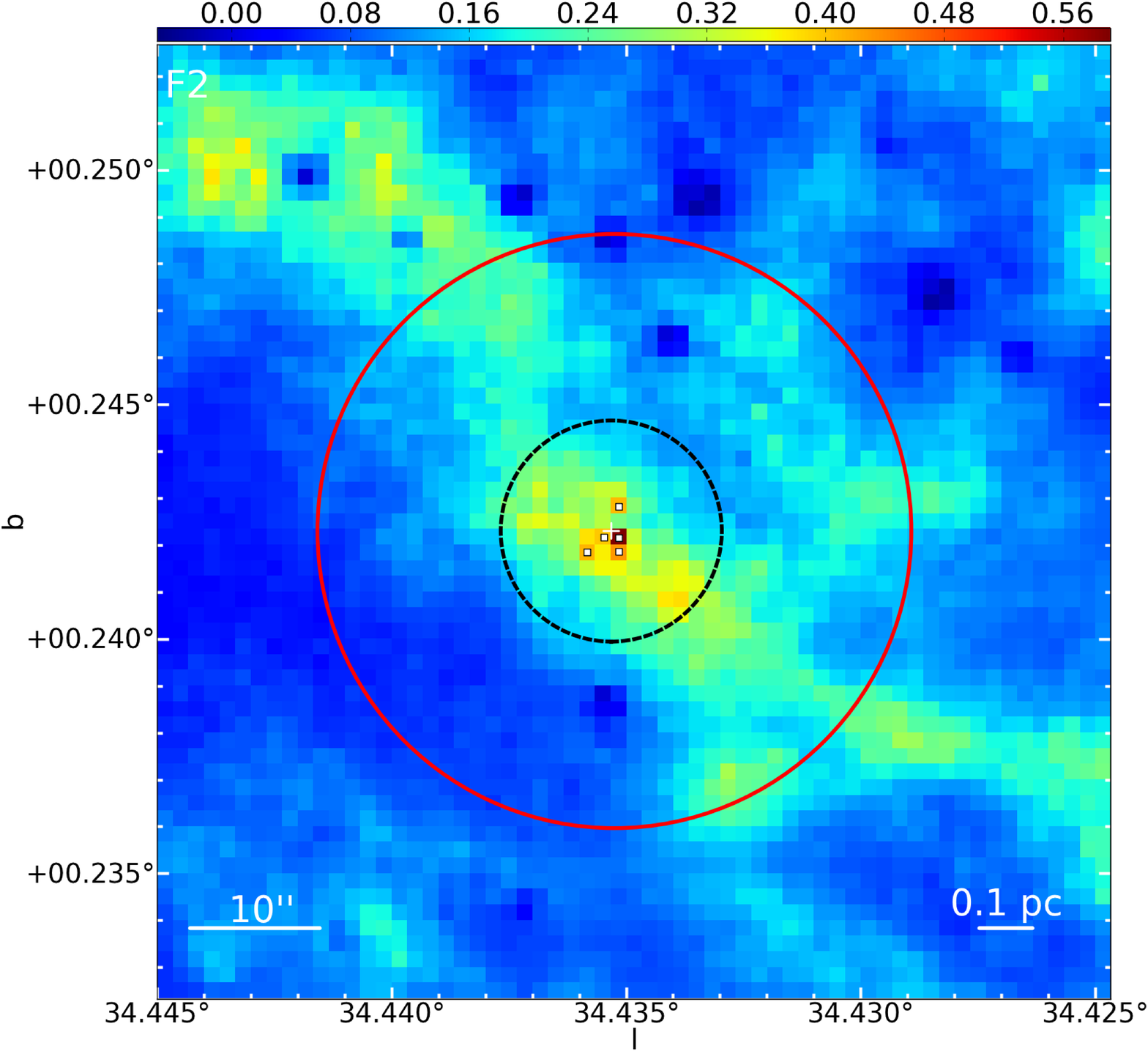} & \hspace{-0.2in} \includegraphics[width=2.15in]{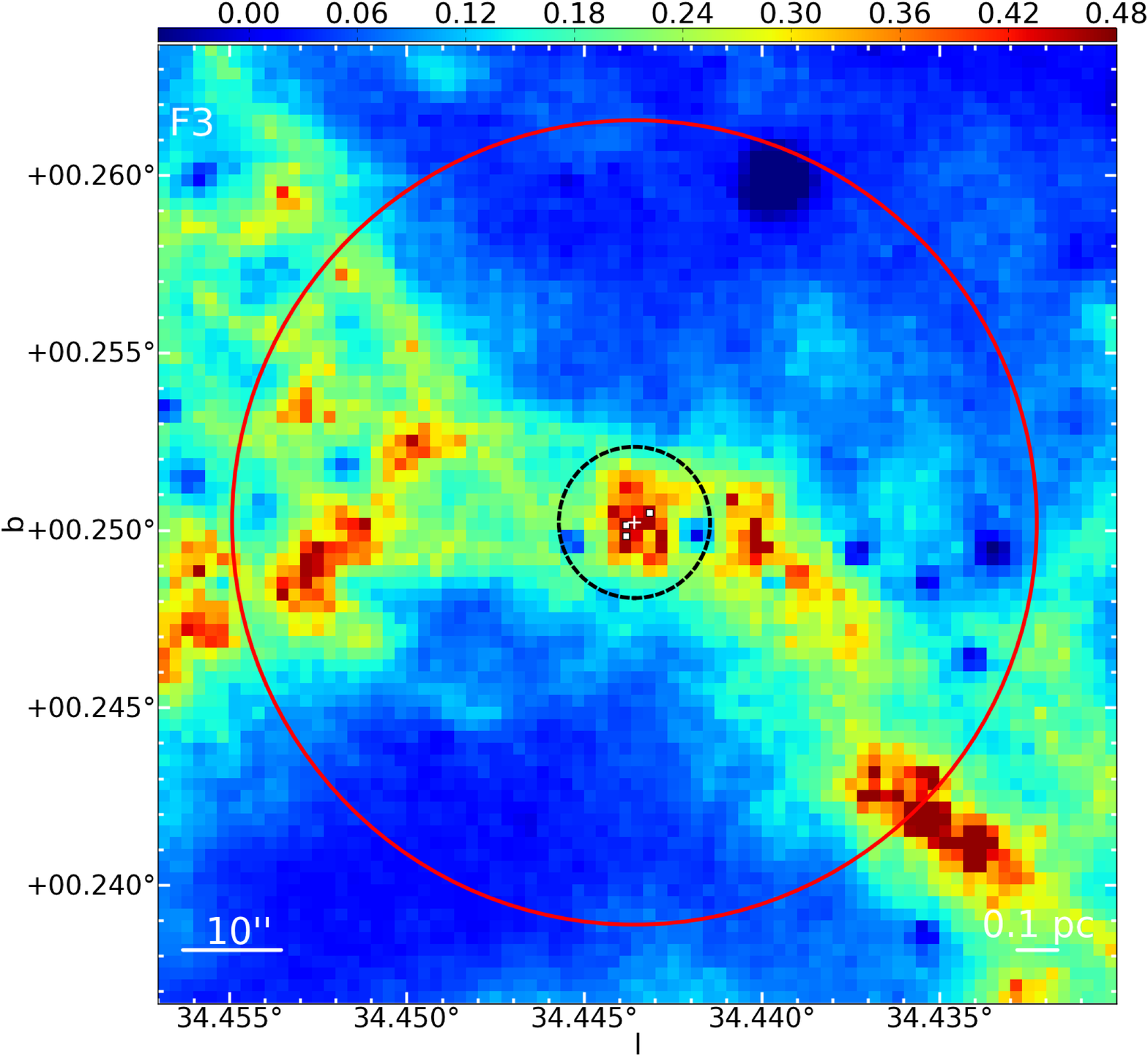} & \hspace{-0.3in} \includegraphics[width=2.15in]{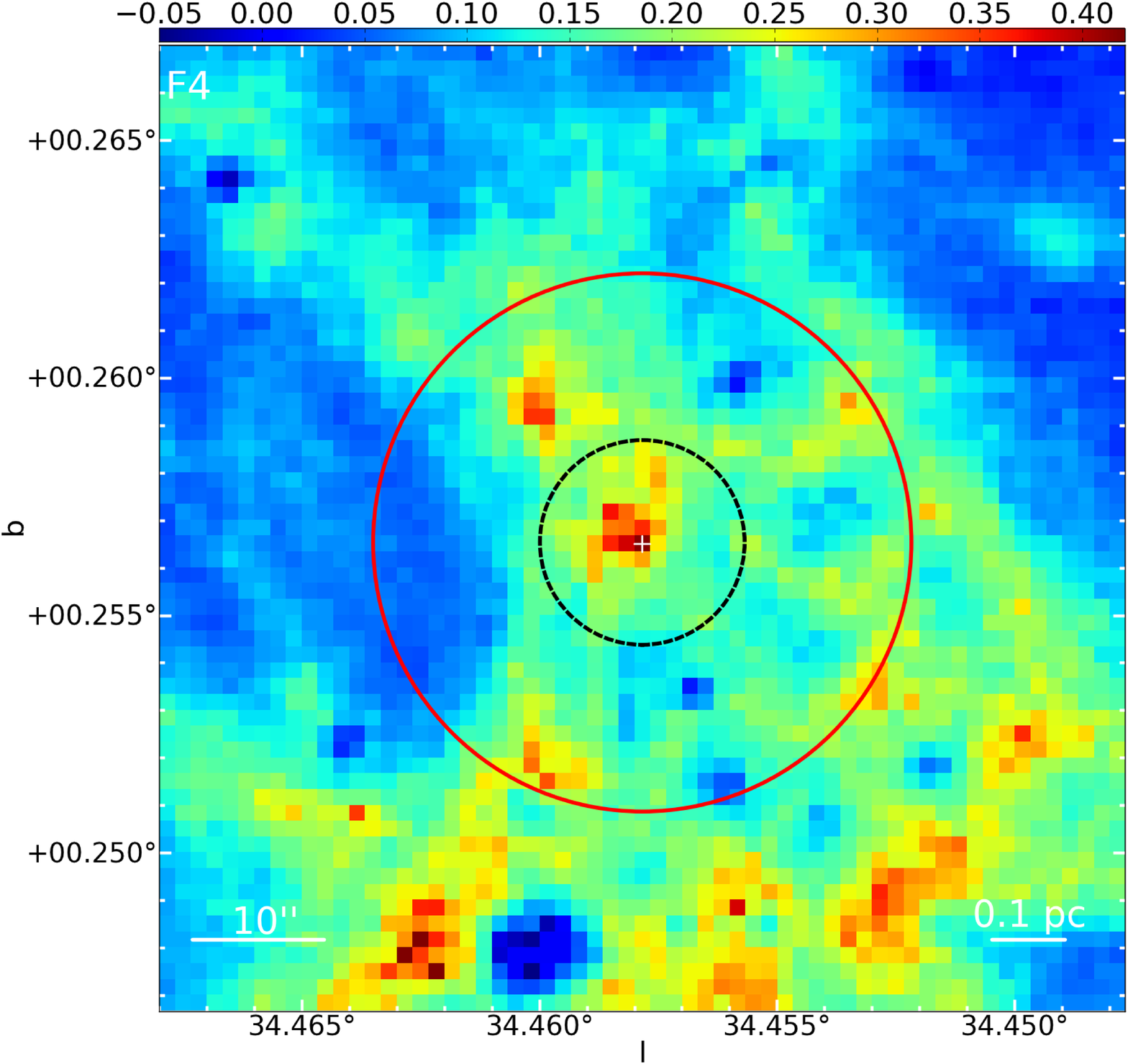} \\
\includegraphics[width=2.2in]{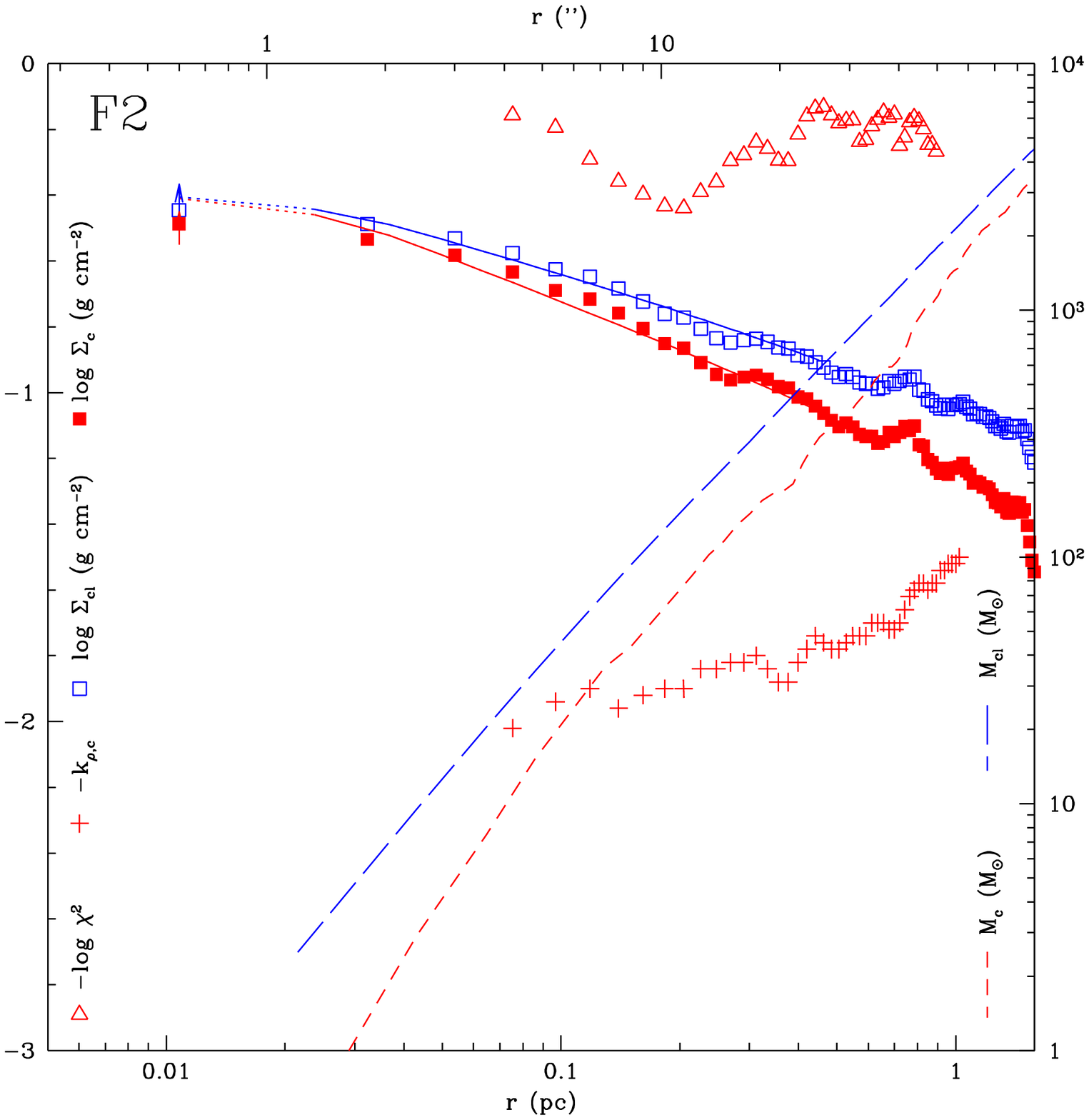} & \includegraphics[width=2.2in]{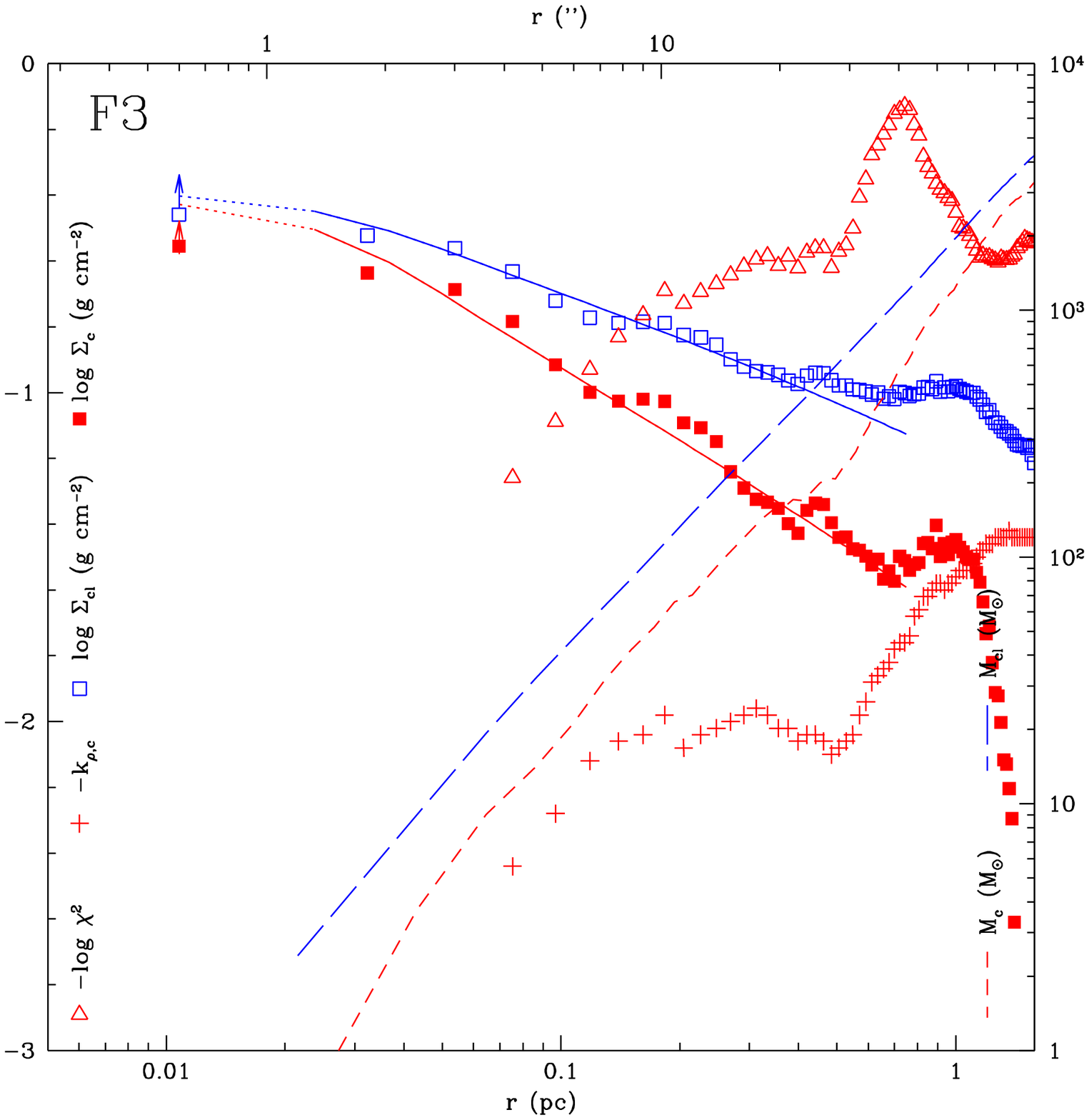} & \hspace{-0.1in} \includegraphics[width=2.2in]{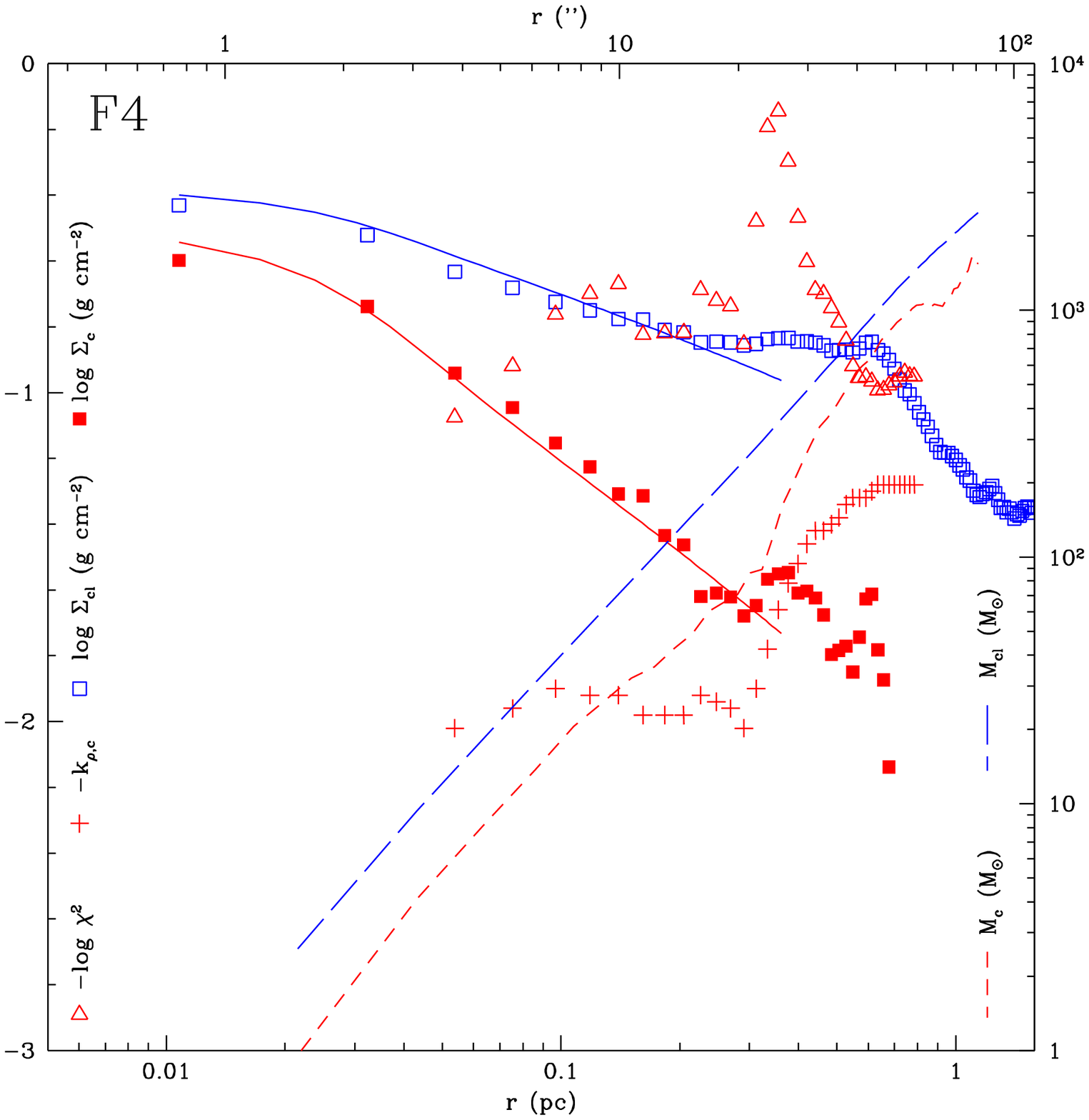}
\end{array}$
\end{center}
\vspace{-0.3in}
\caption{\footnotesize
Core E2, E3, F1, F2, F3, F4 $\Sigma$ maps (notation as Fig.~\ref{fig:coreA1}a) and azimuthally averaged radial profile figures (notation as Fig.~\ref{fig:coreA1}b).
\label{fig:cores5}
}
\end{figure*}

\begin{figure*}
\begin{center}$
\begin{array}{ccc}
\hspace{-0.0in} \includegraphics[width=2.15in]{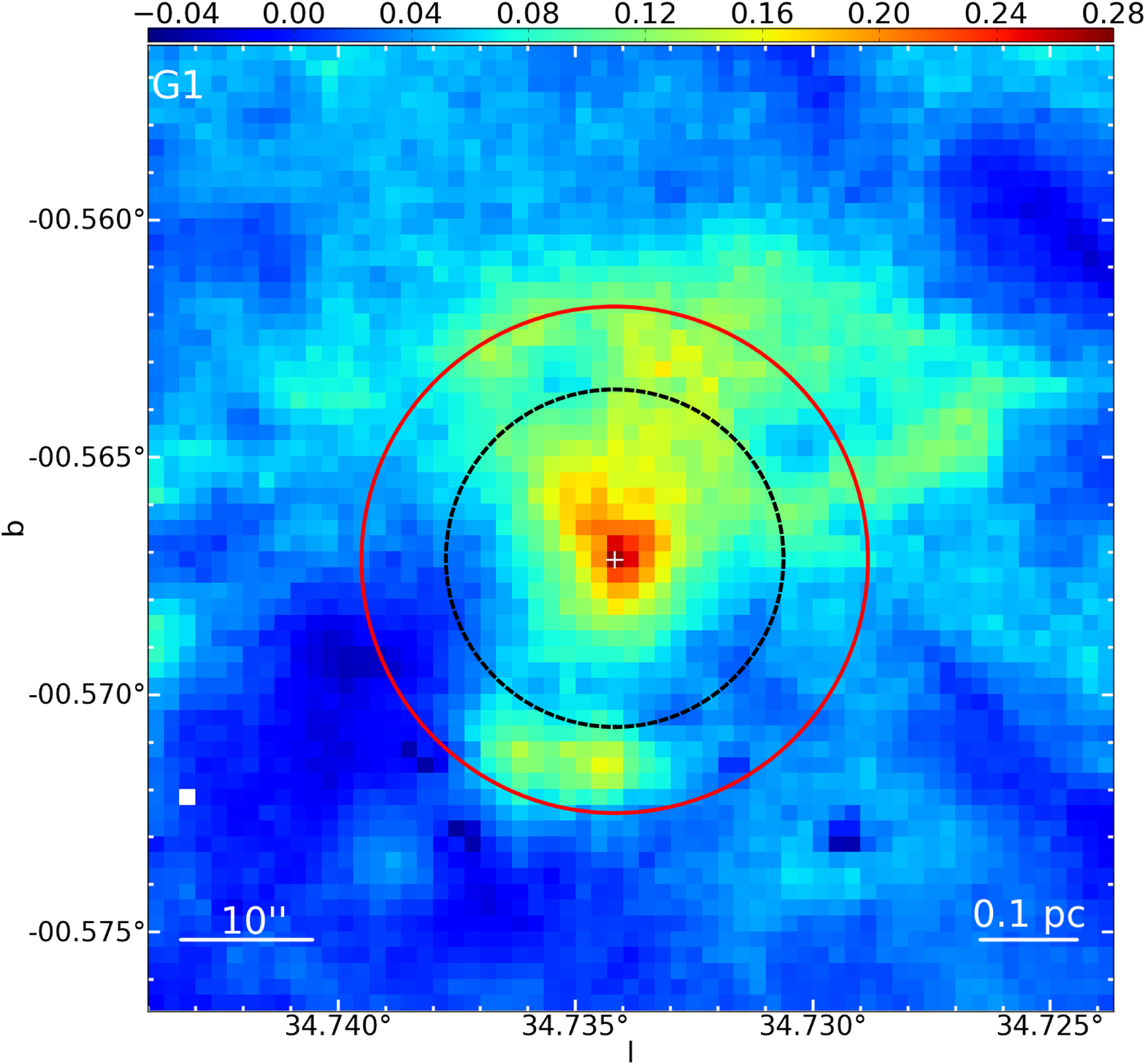} & \hspace{-0.2in} \includegraphics[width=2.15in]{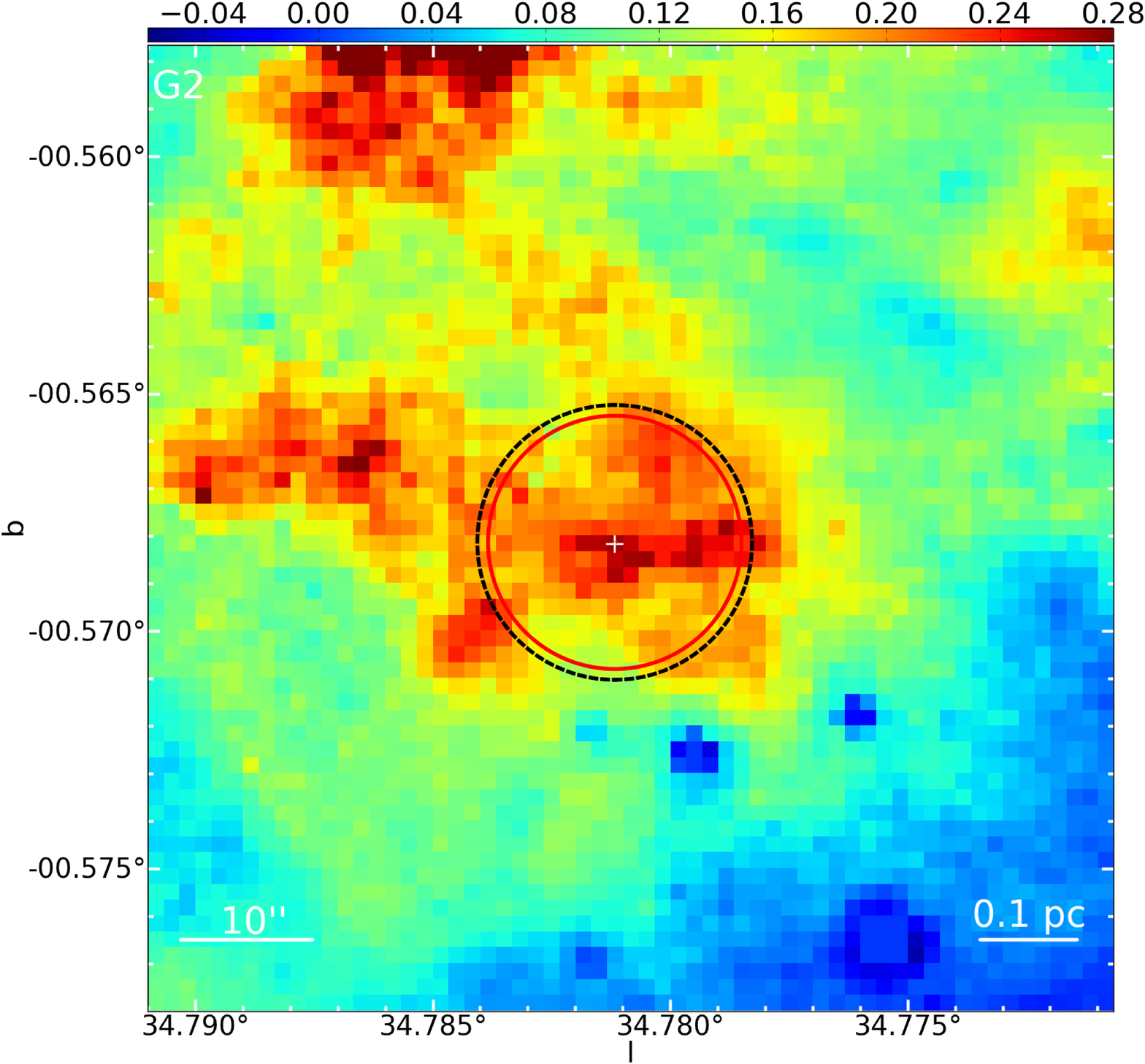} & \hspace{-0.3in} \includegraphics[width=2.15in]{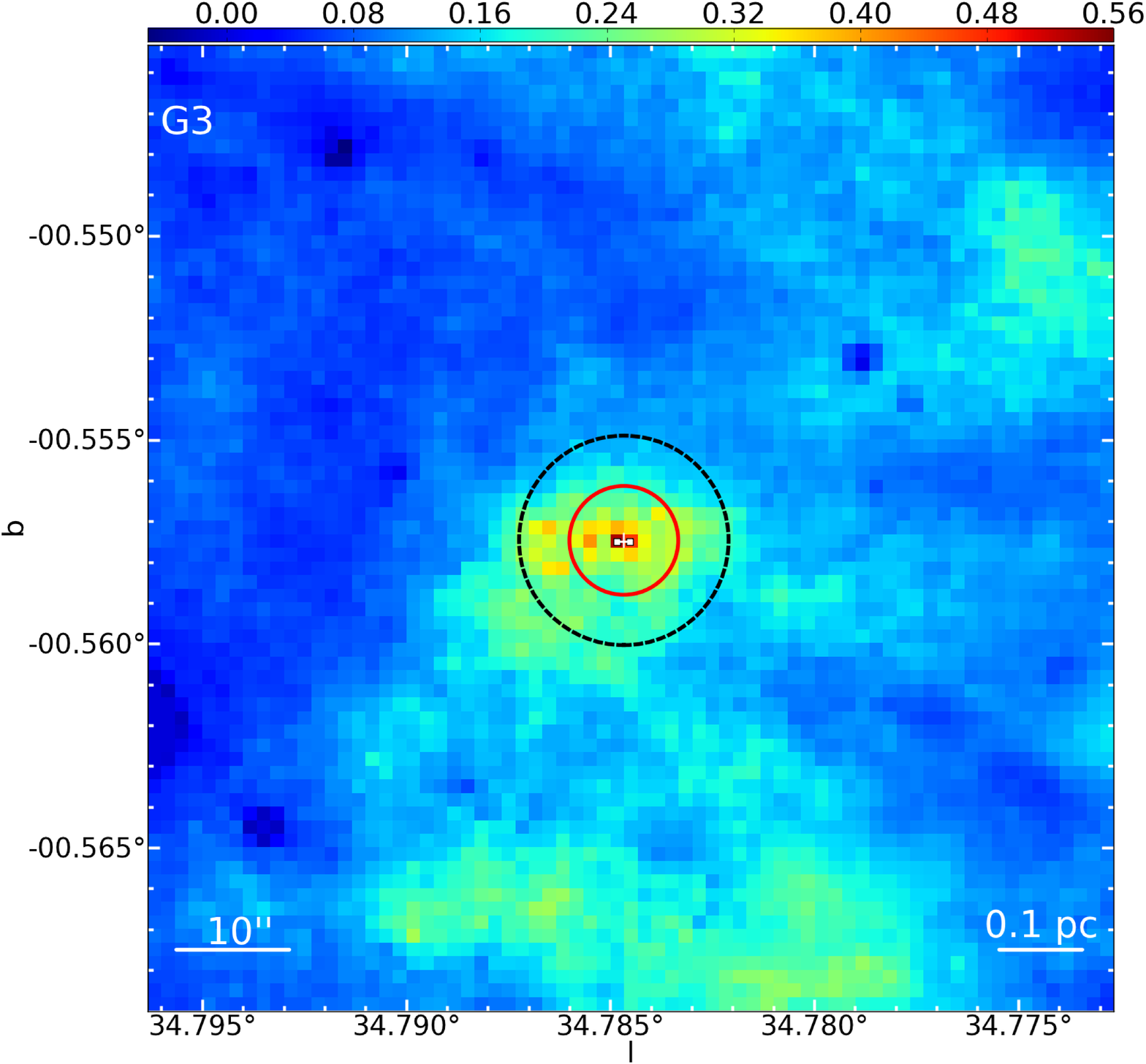} \\
\includegraphics[width=2.2in]{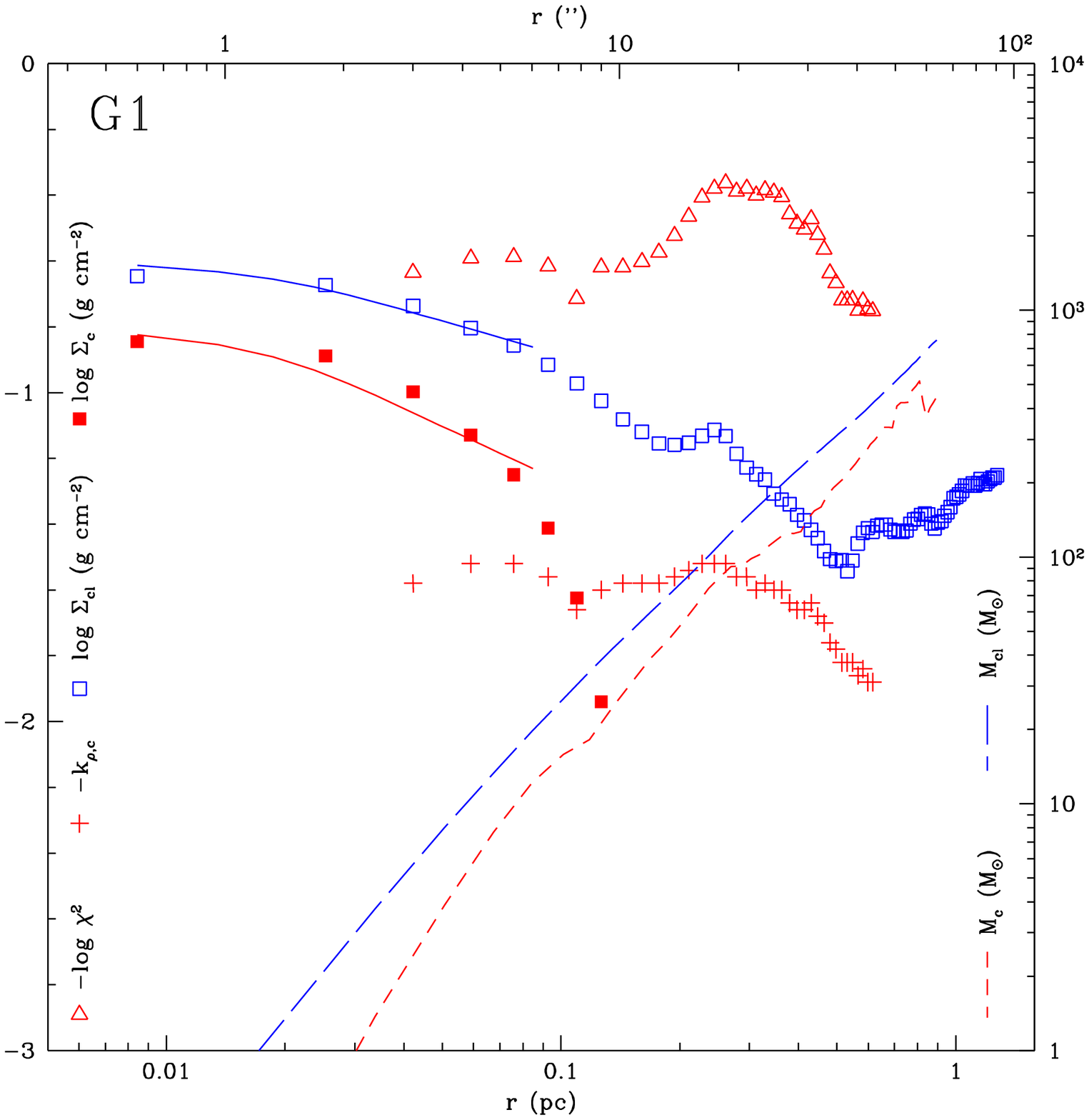} & \includegraphics[width=2.2in]{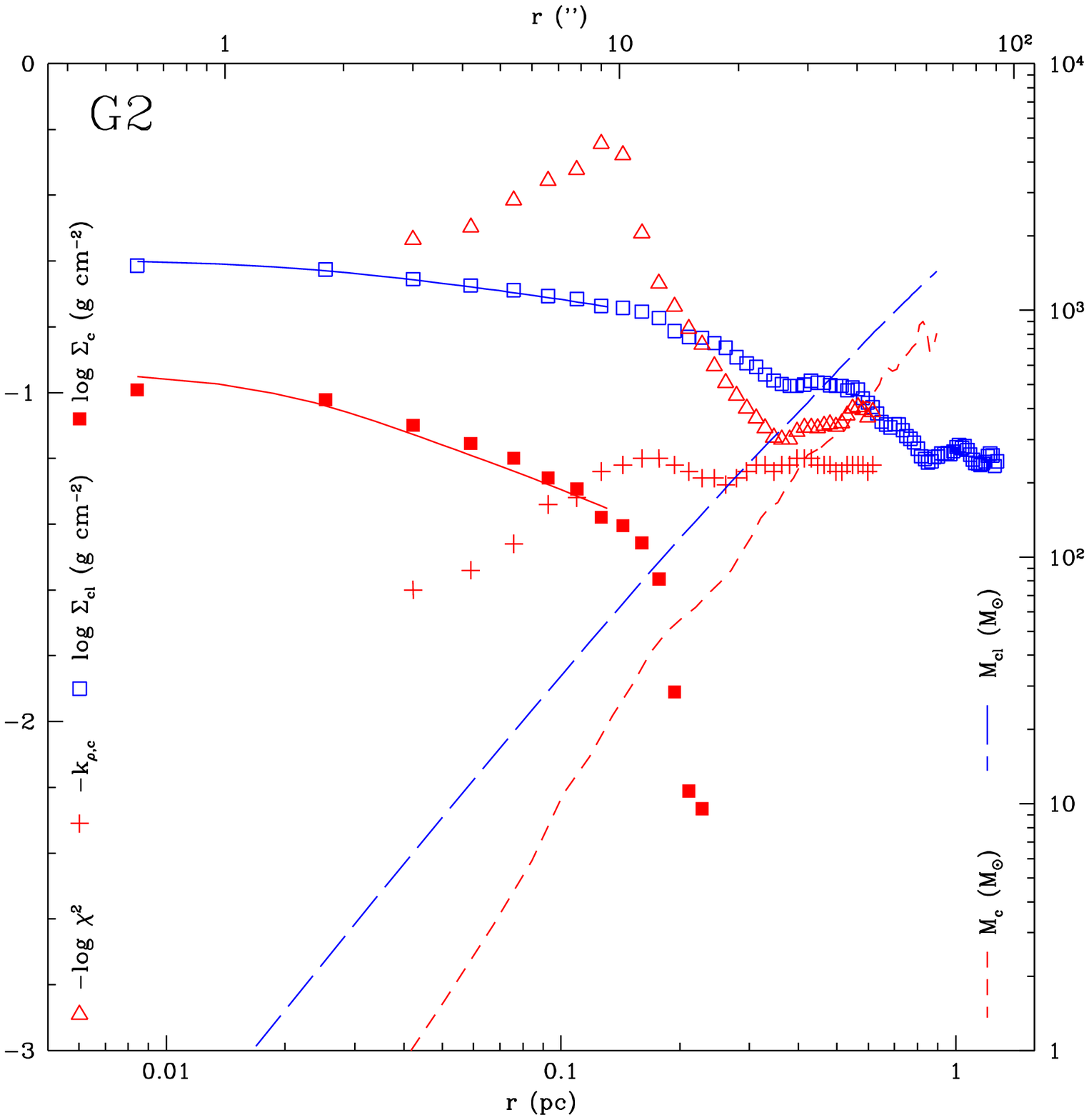} & \hspace{-0.1in} \includegraphics[width=2.2in]{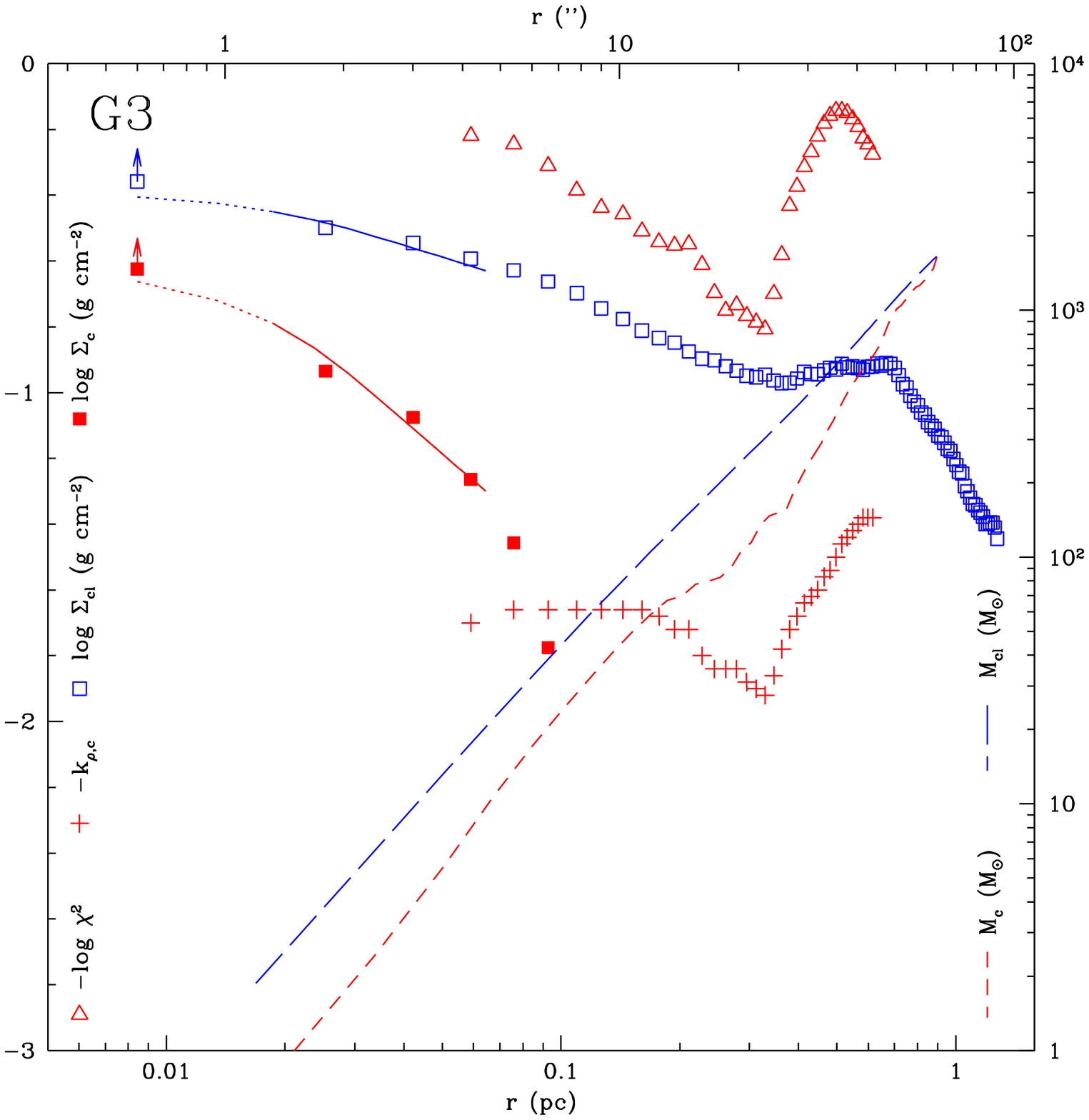} \\
\hspace{-0.0in} \includegraphics[width=2.15in]{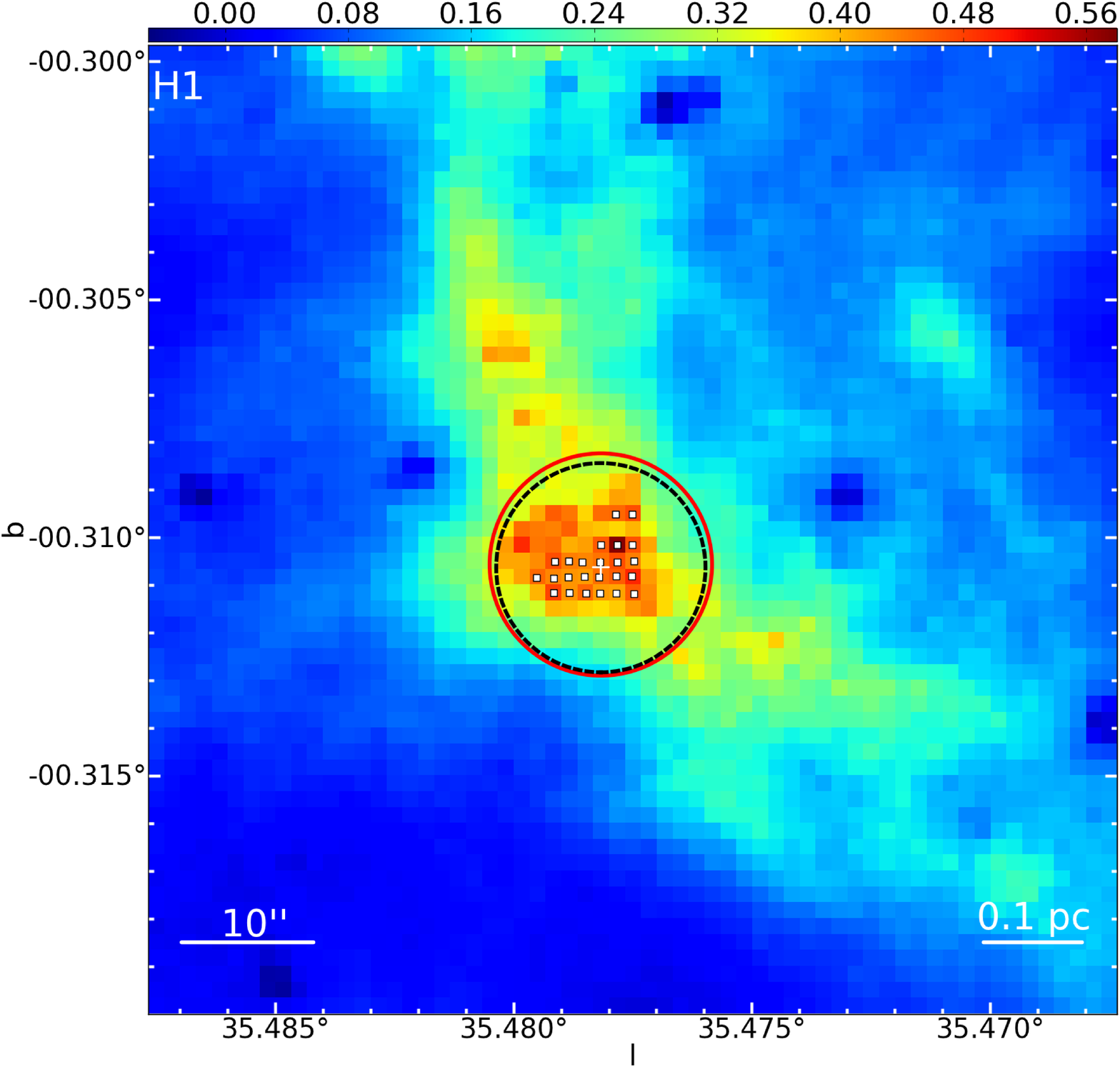} & \hspace{-0.2in} \includegraphics[width=2.15in]{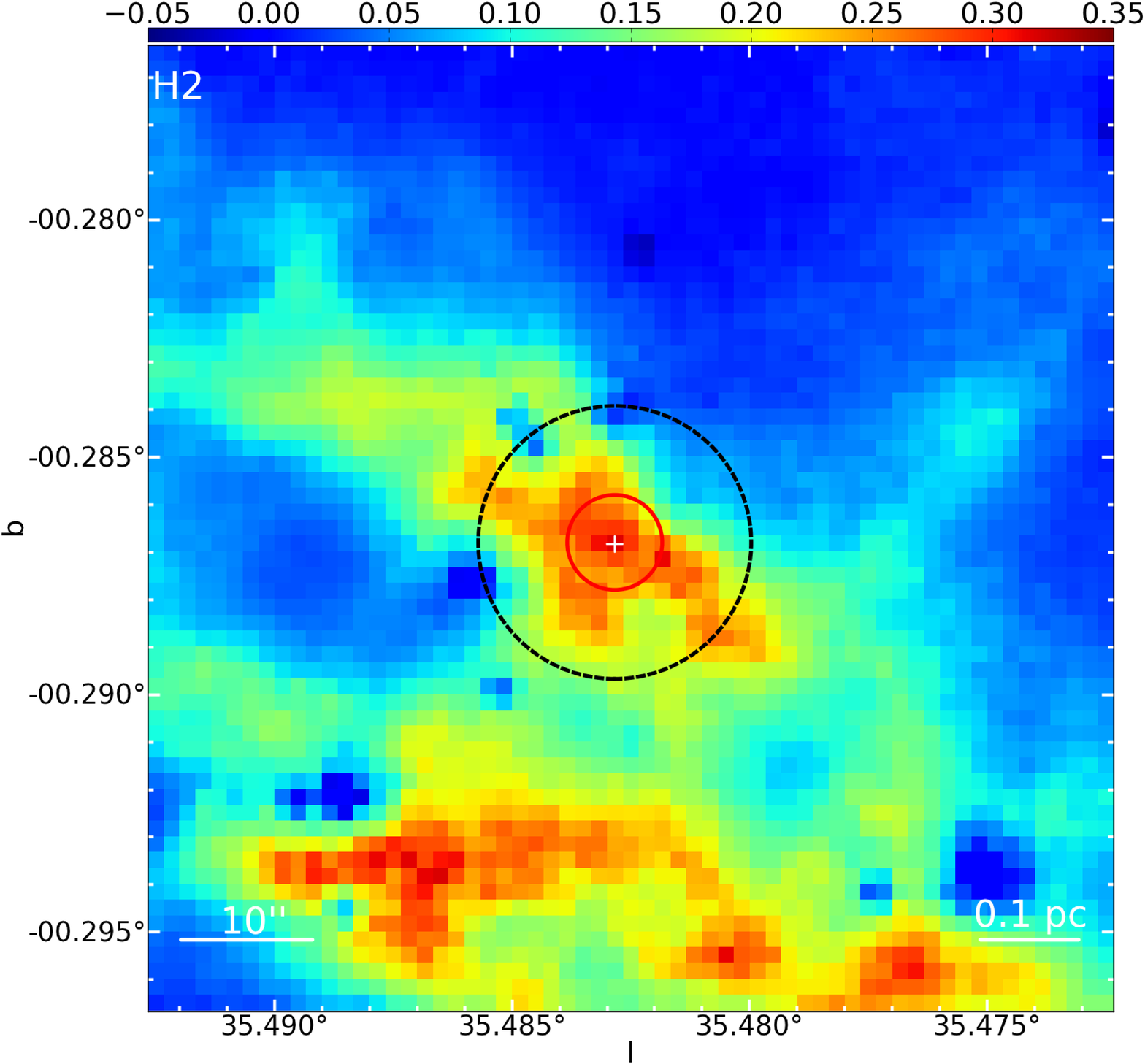} & \hspace{-0.3in} \includegraphics[width=2.15in]{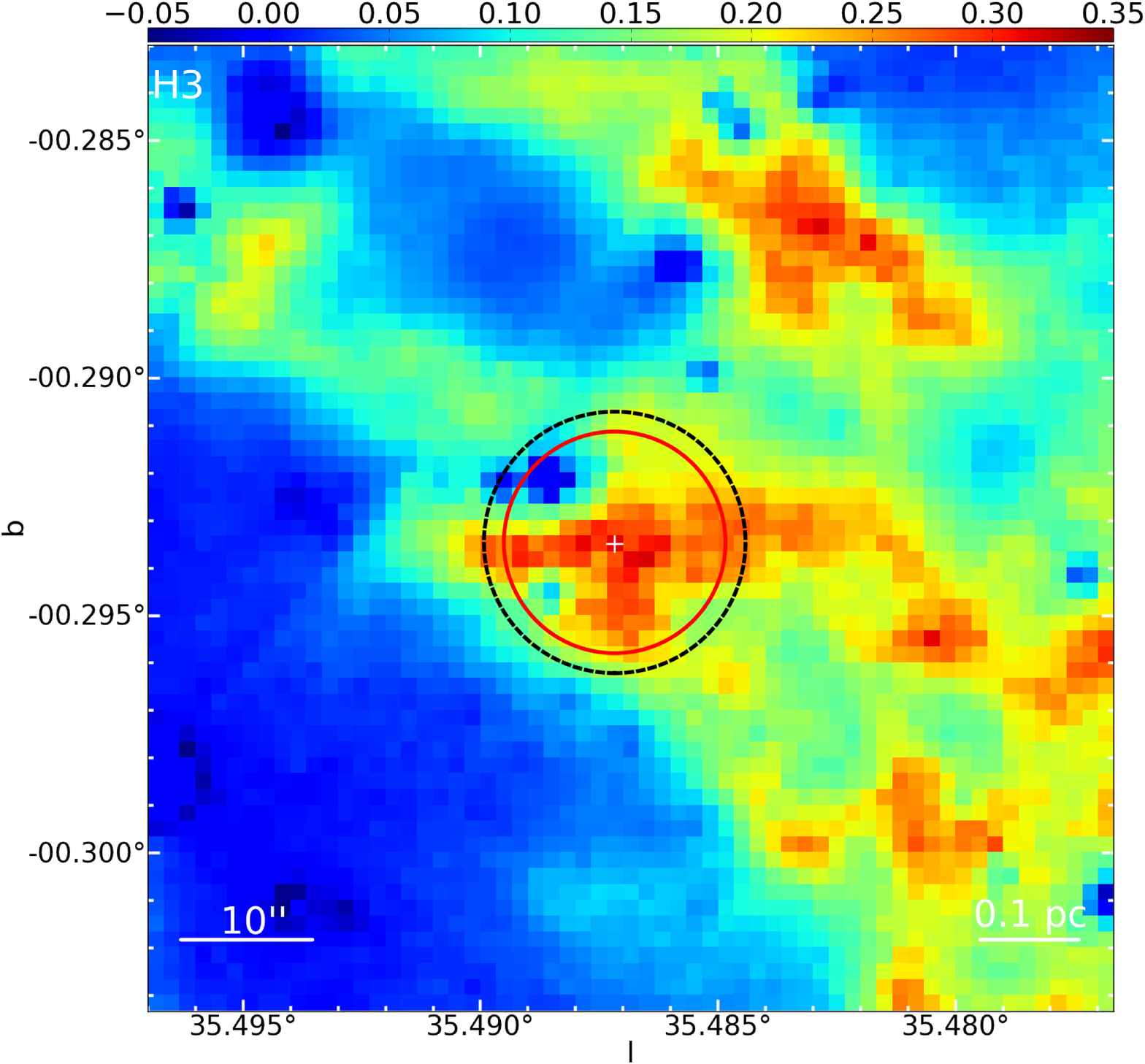} \\
\includegraphics[width=2.2in]{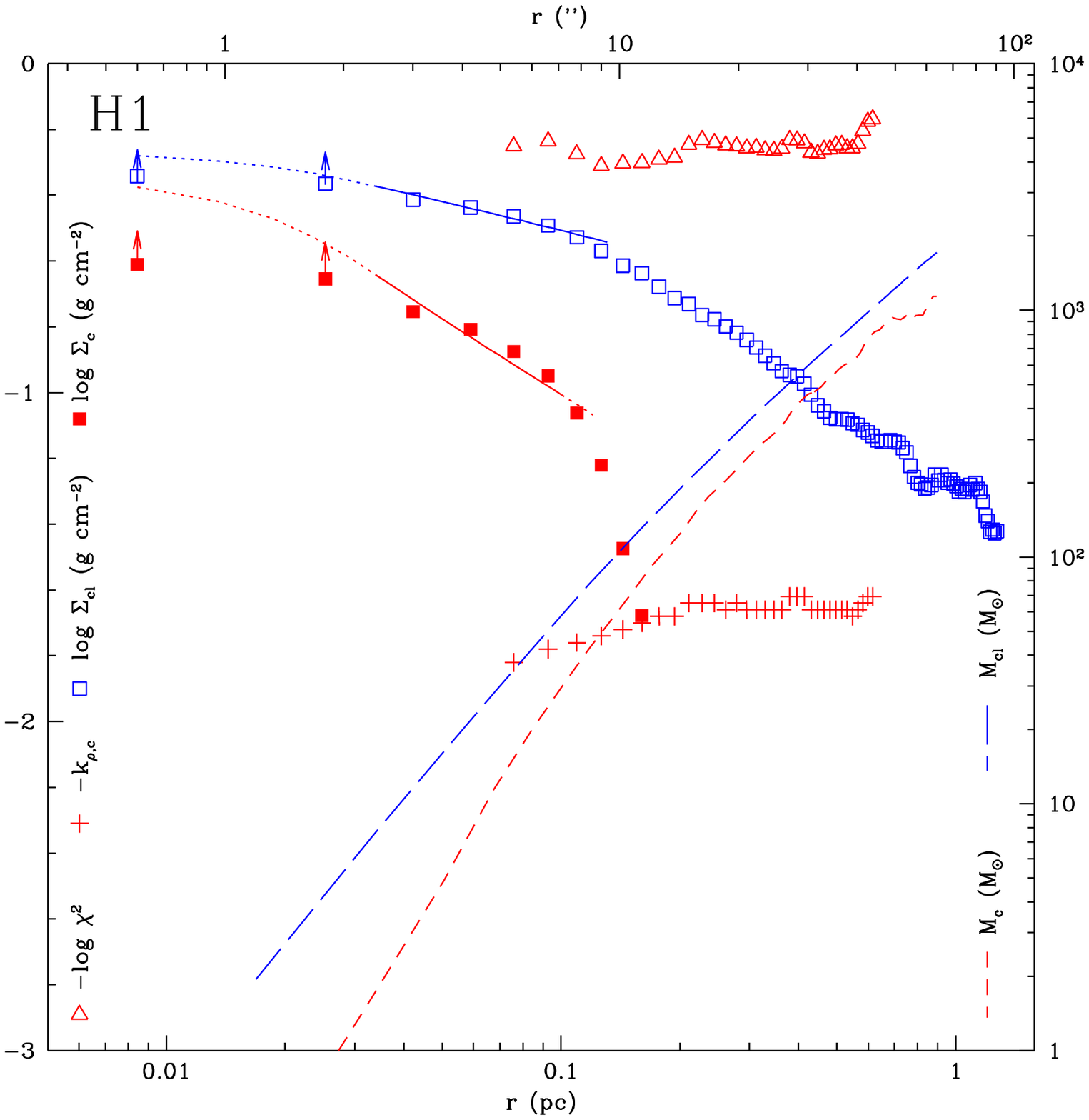} & \includegraphics[width=2.2in]{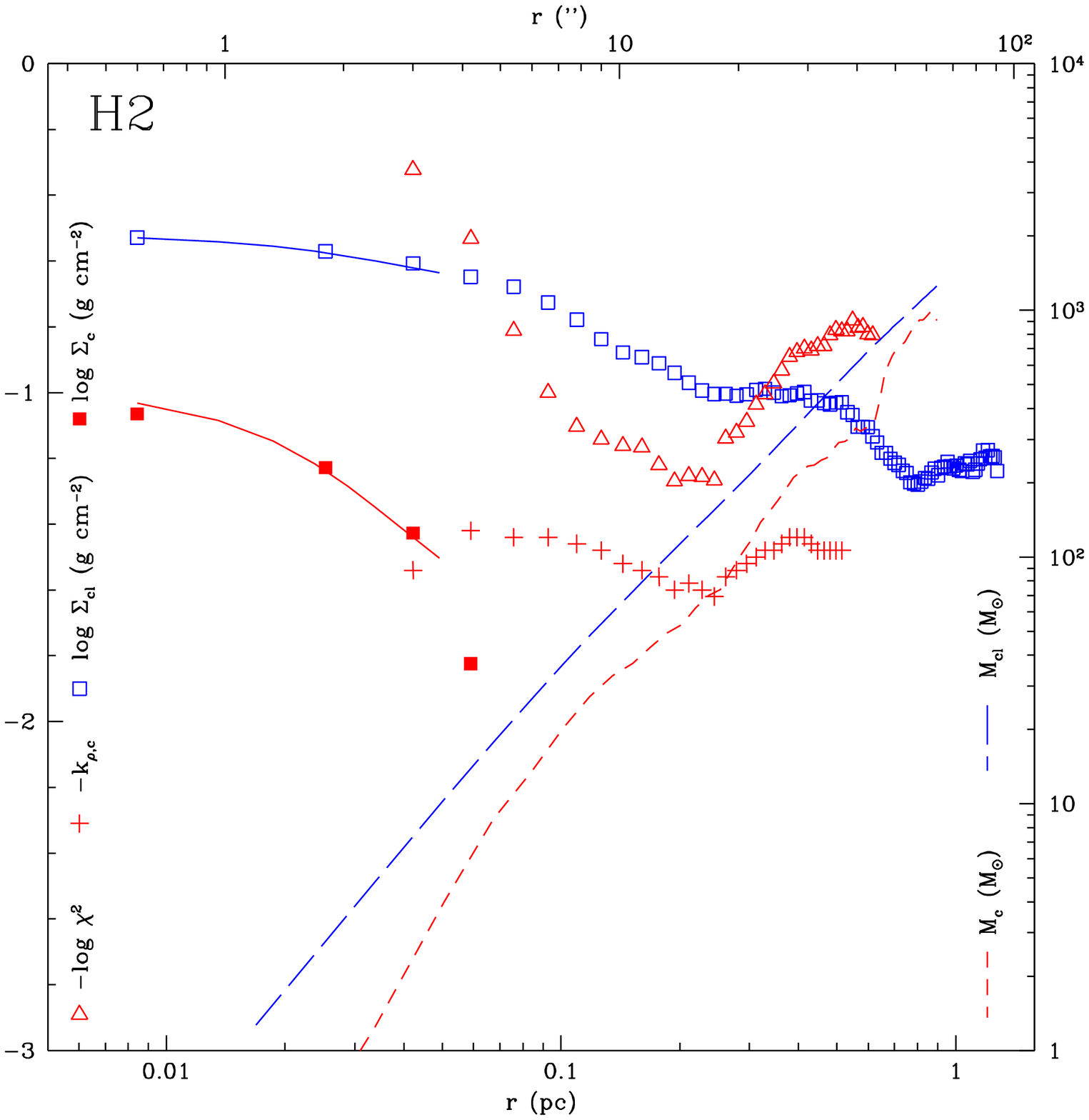} & \hspace{-0.1in} \includegraphics[width=2.2in]{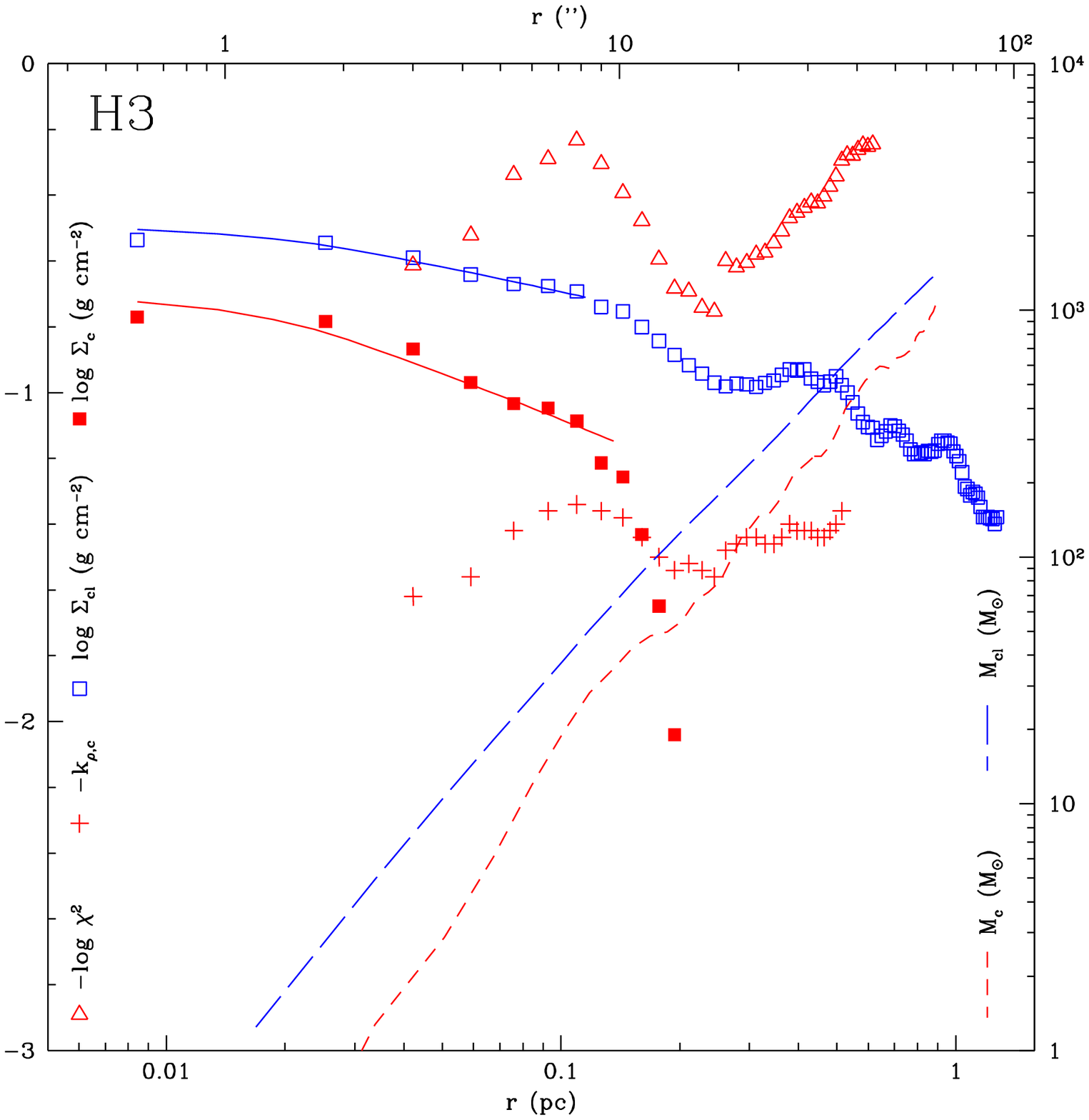}
\end{array}$
\end{center}
\vspace{-0.3in}
\caption{\footnotesize
Core G1, G2, G3, H1, H2, H3 $\Sigma$ maps (notation as Fig.~\ref{fig:coreA1}a) and azimuthally averaged radial profile figures (notation as Fig.~\ref{fig:coreA1}b).
\label{fig:cores6}
}
\end{figure*}

\begin{figure*}
\begin{center}$
\begin{array}{ccc}
\hspace{-0.0in} \includegraphics[width=2.15in]{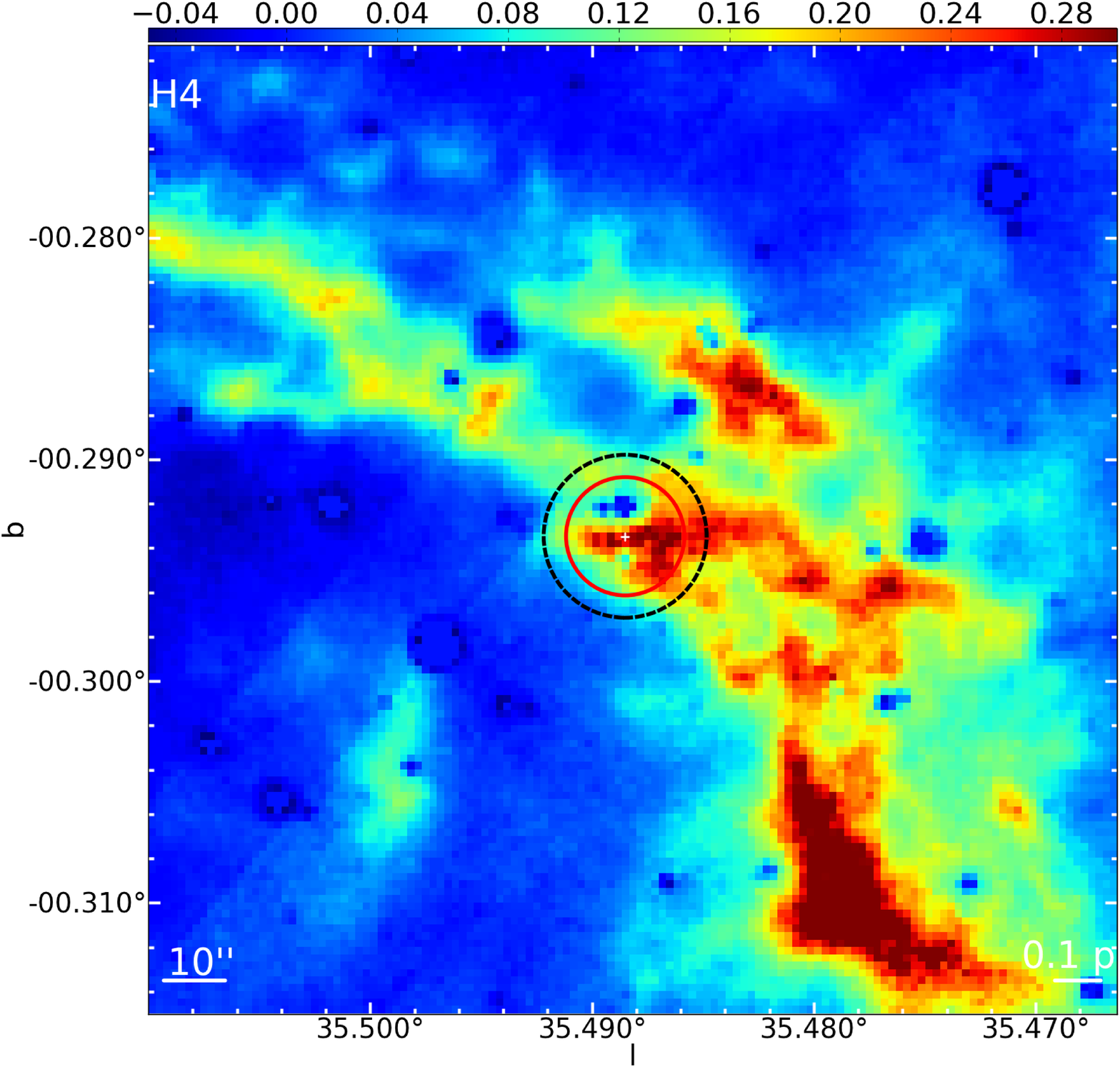} & \hspace{-0.2in} \includegraphics[width=2.15in]{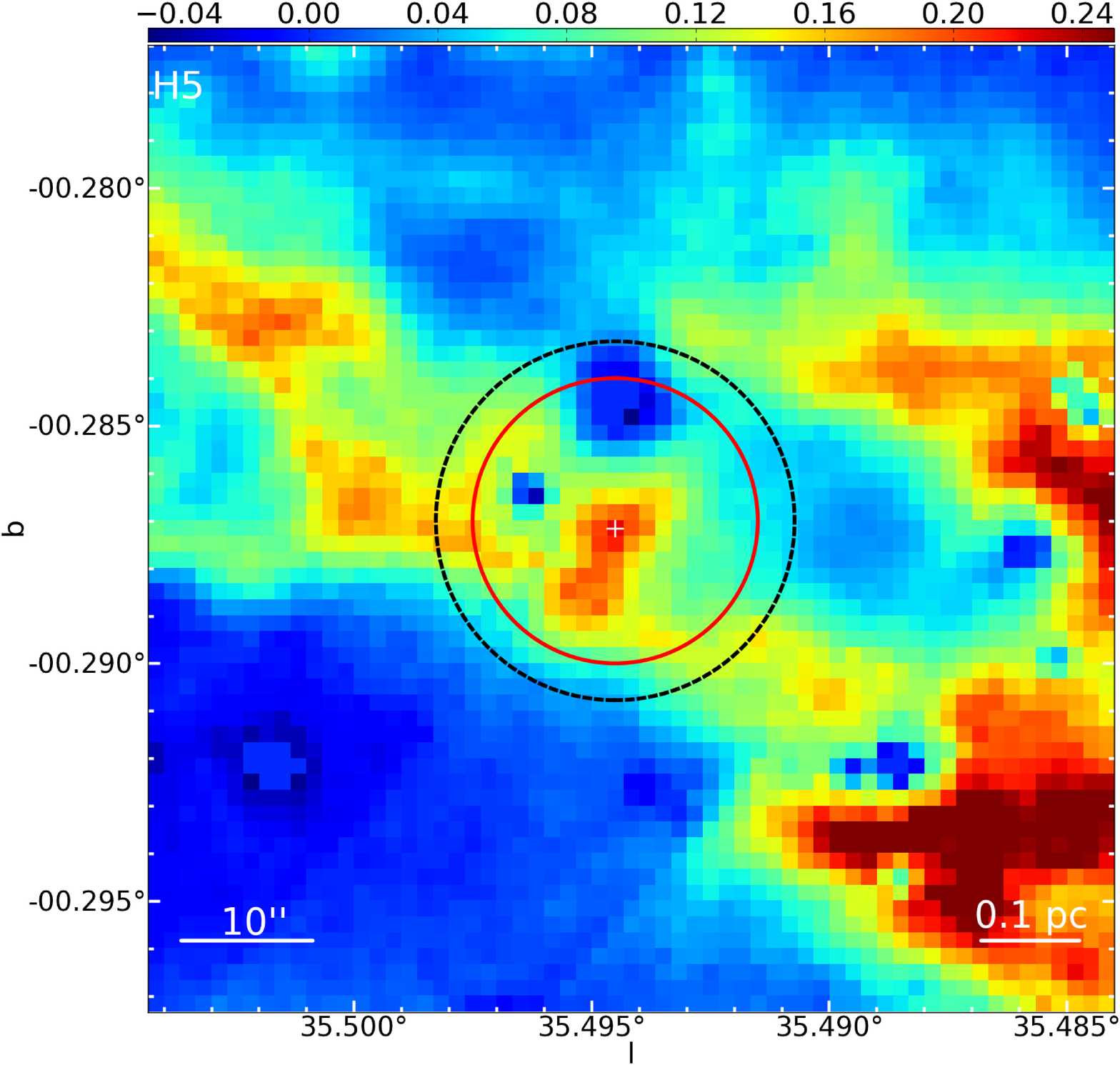} & \hspace{-0.3in} \includegraphics[width=2.15in]{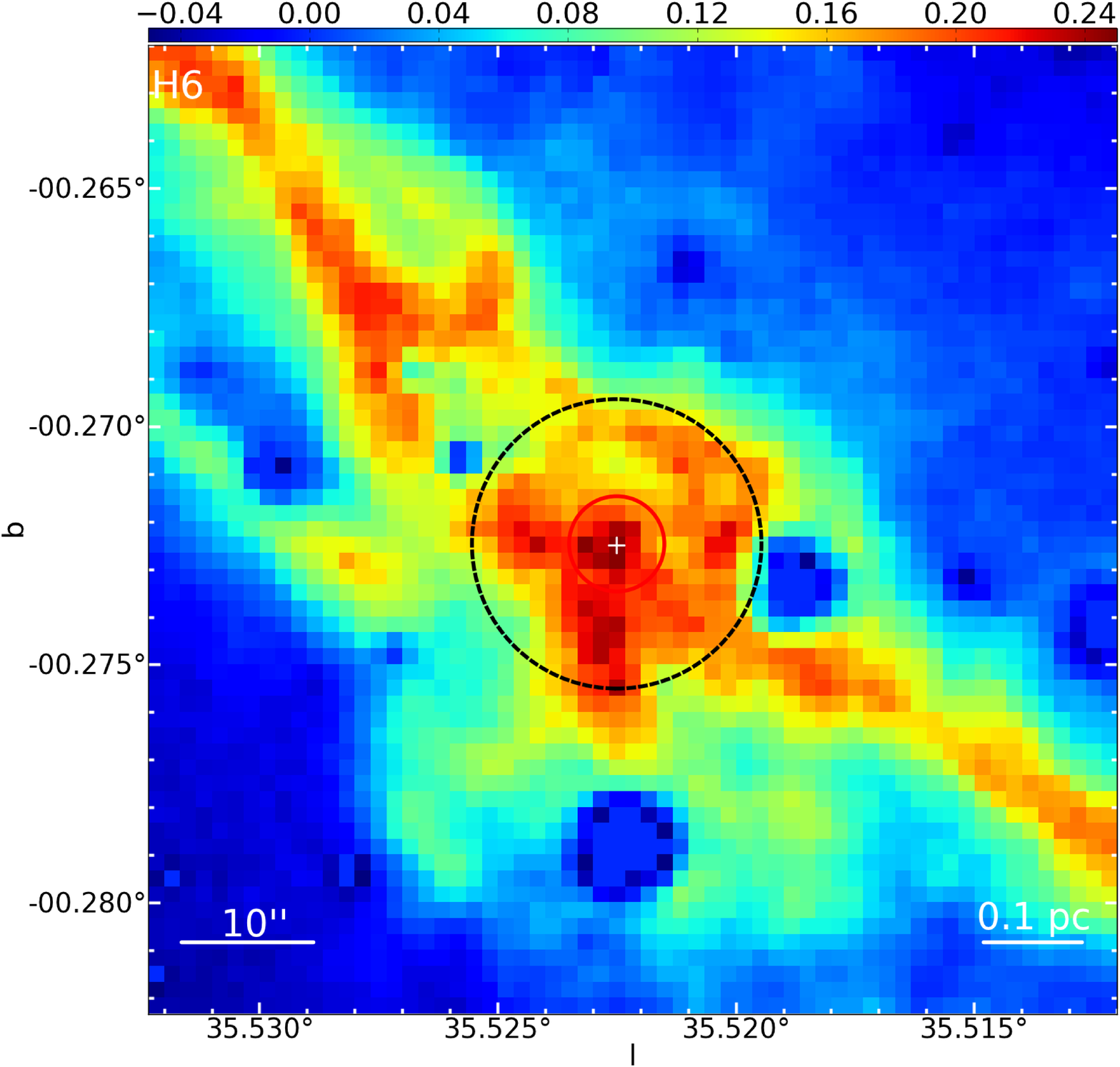} \\
\includegraphics[width=2.2in]{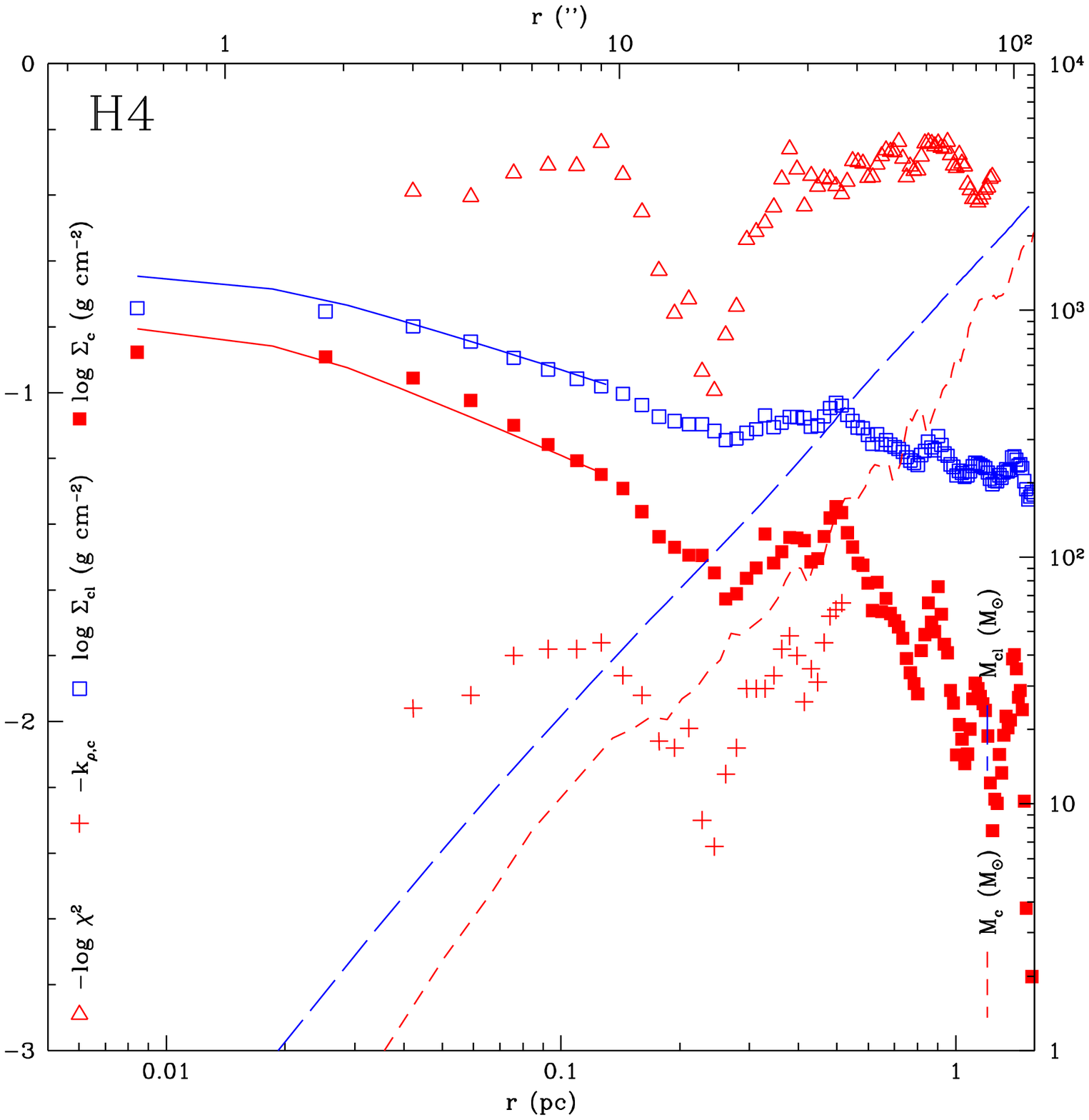} & \includegraphics[width=2.2in]{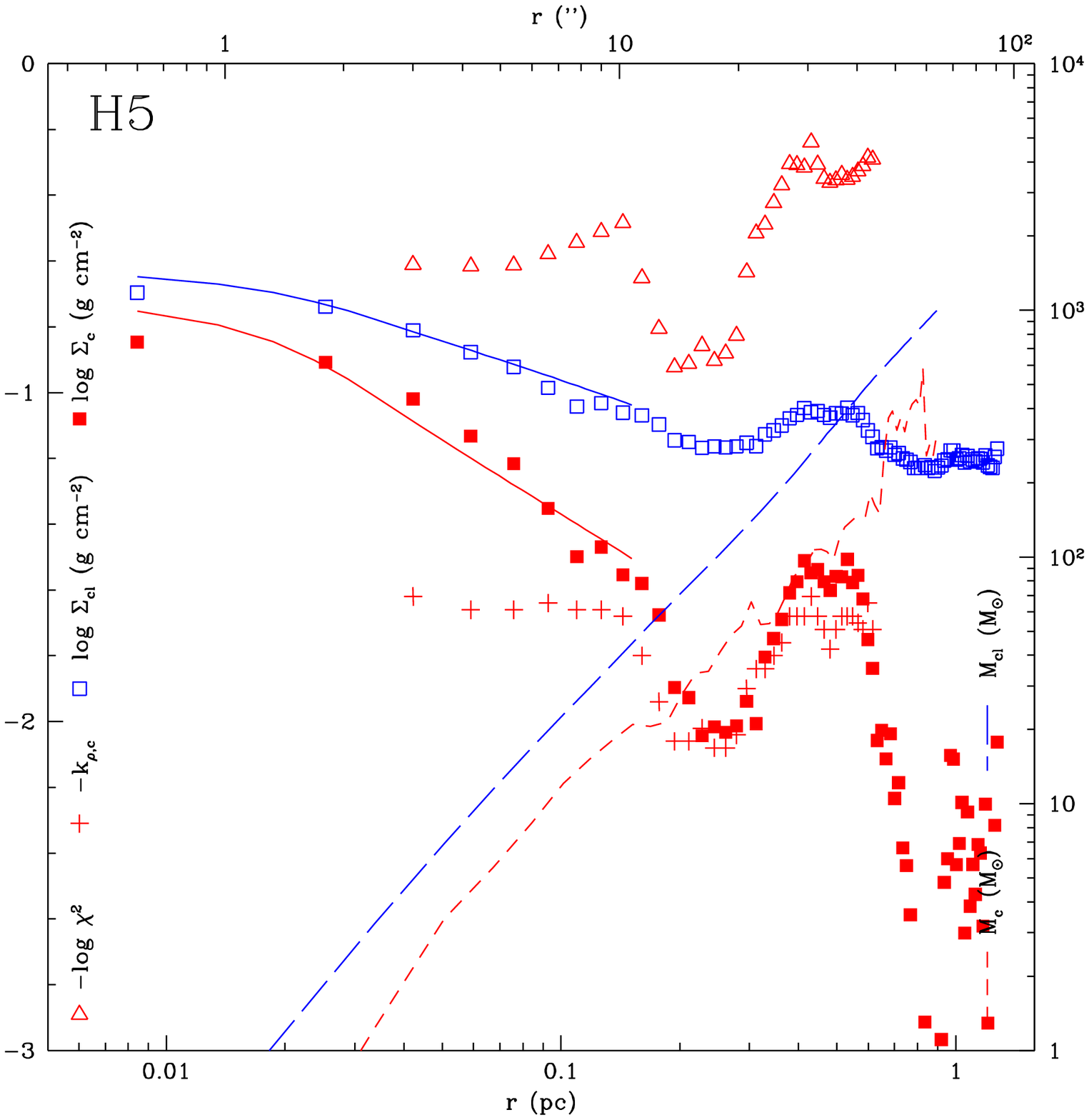} & \hspace{-0.1in} \includegraphics[width=2.2in]{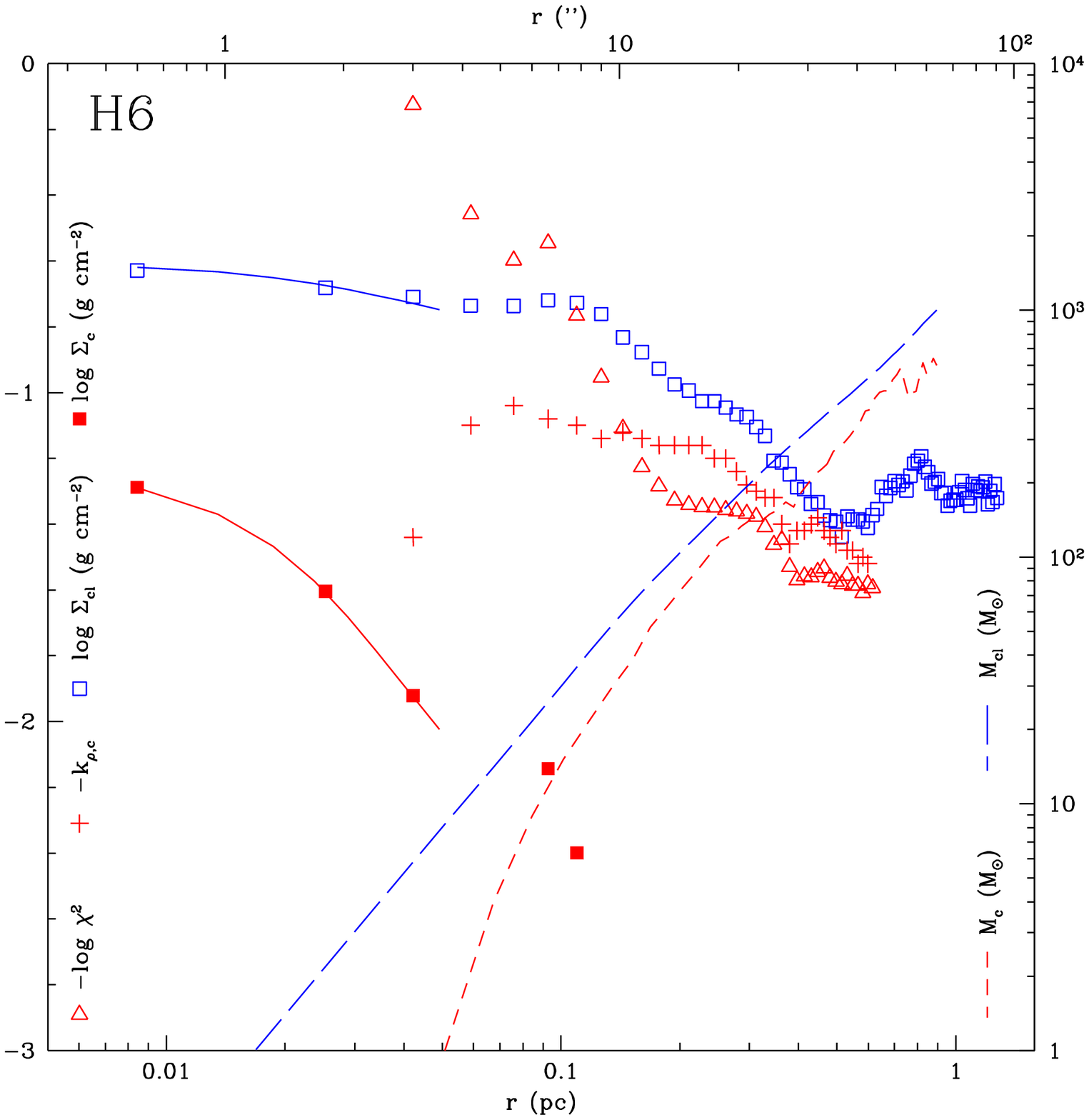} \\
\hspace{-0.0in} \includegraphics[width=2.15in]{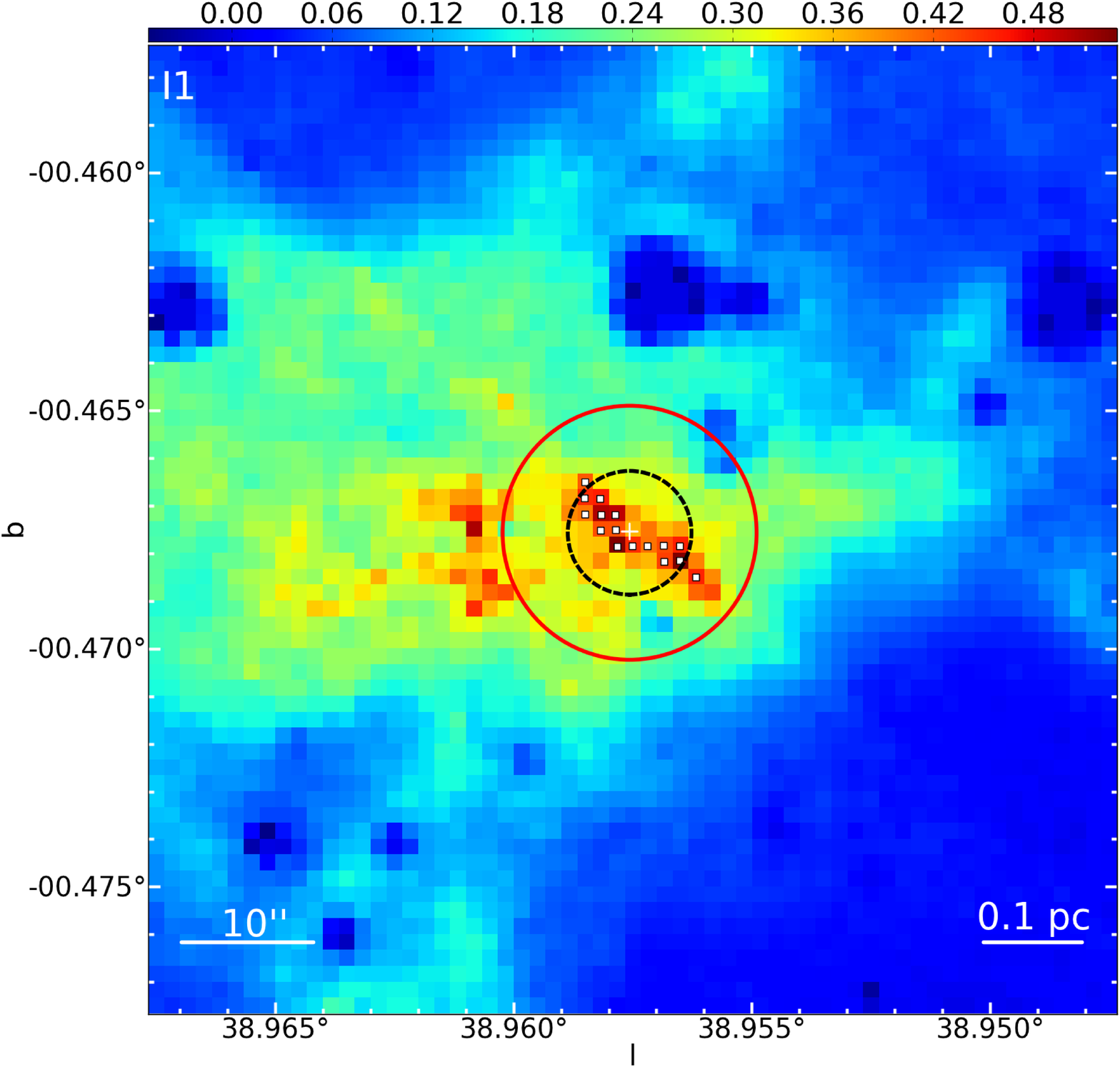} & \hspace{-0.2in} \includegraphics[width=2.15in]{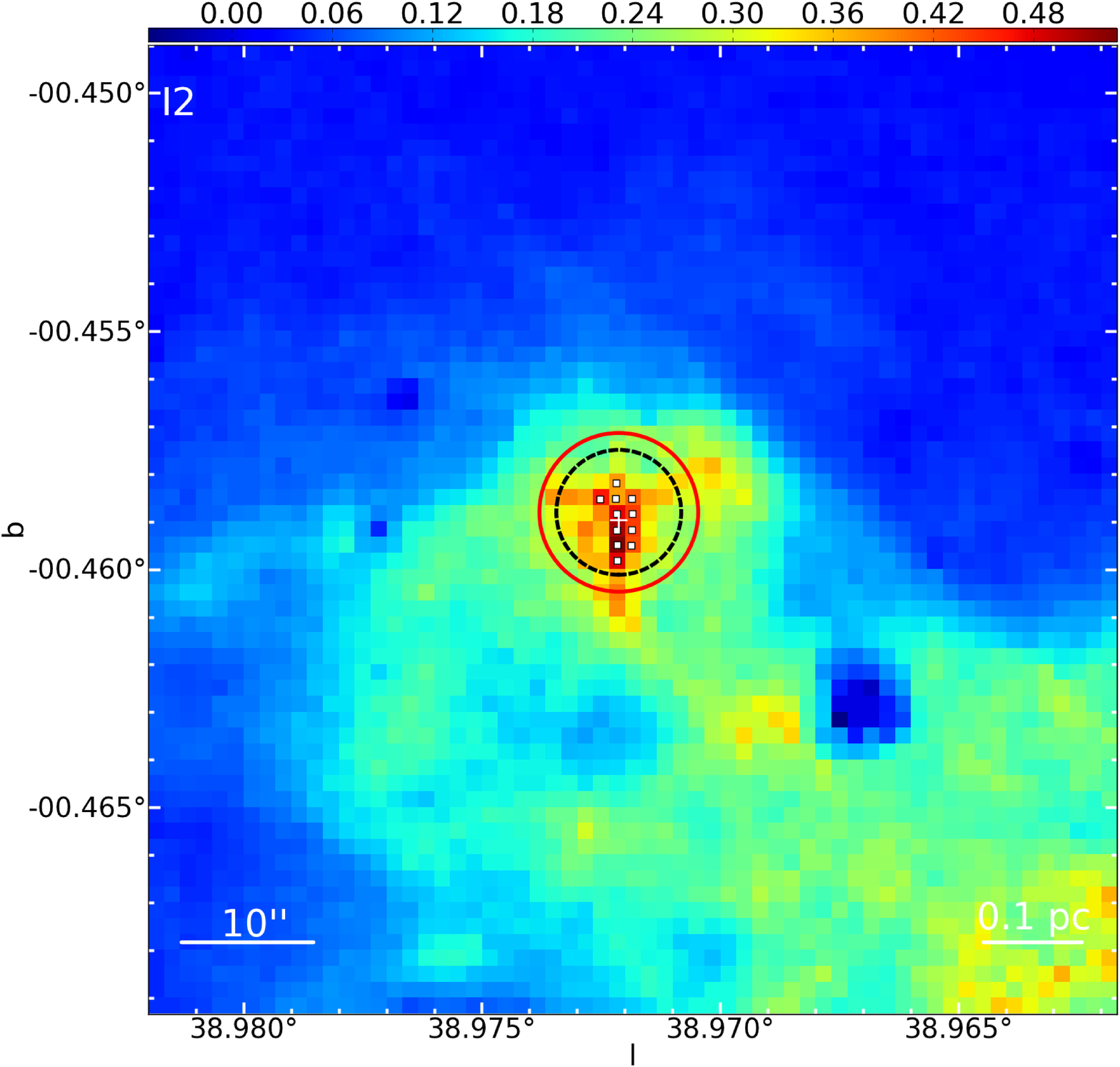} & \hspace{-0.3in} \includegraphics[width=2.15in]{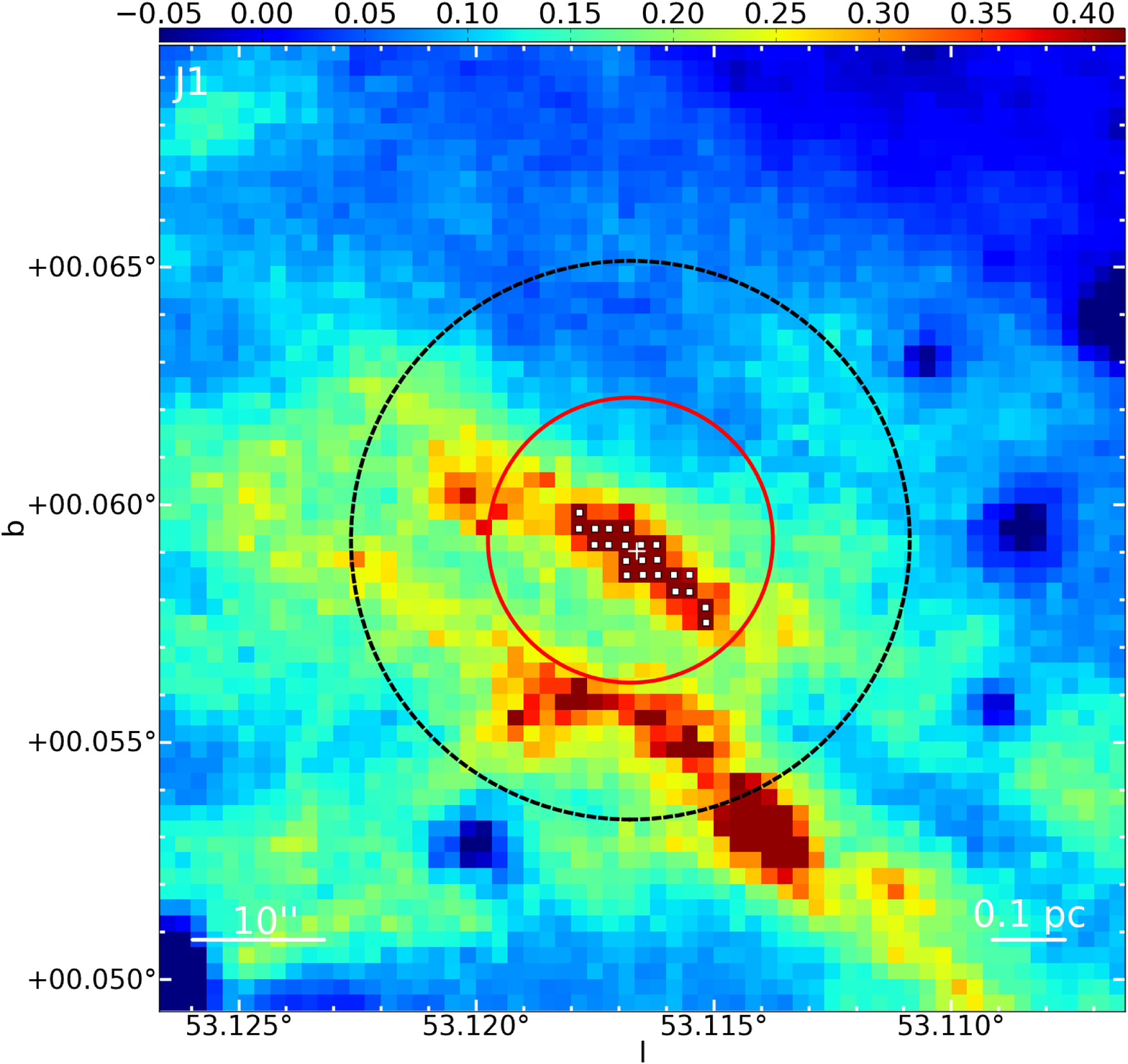} \\
\includegraphics[width=2.2in]{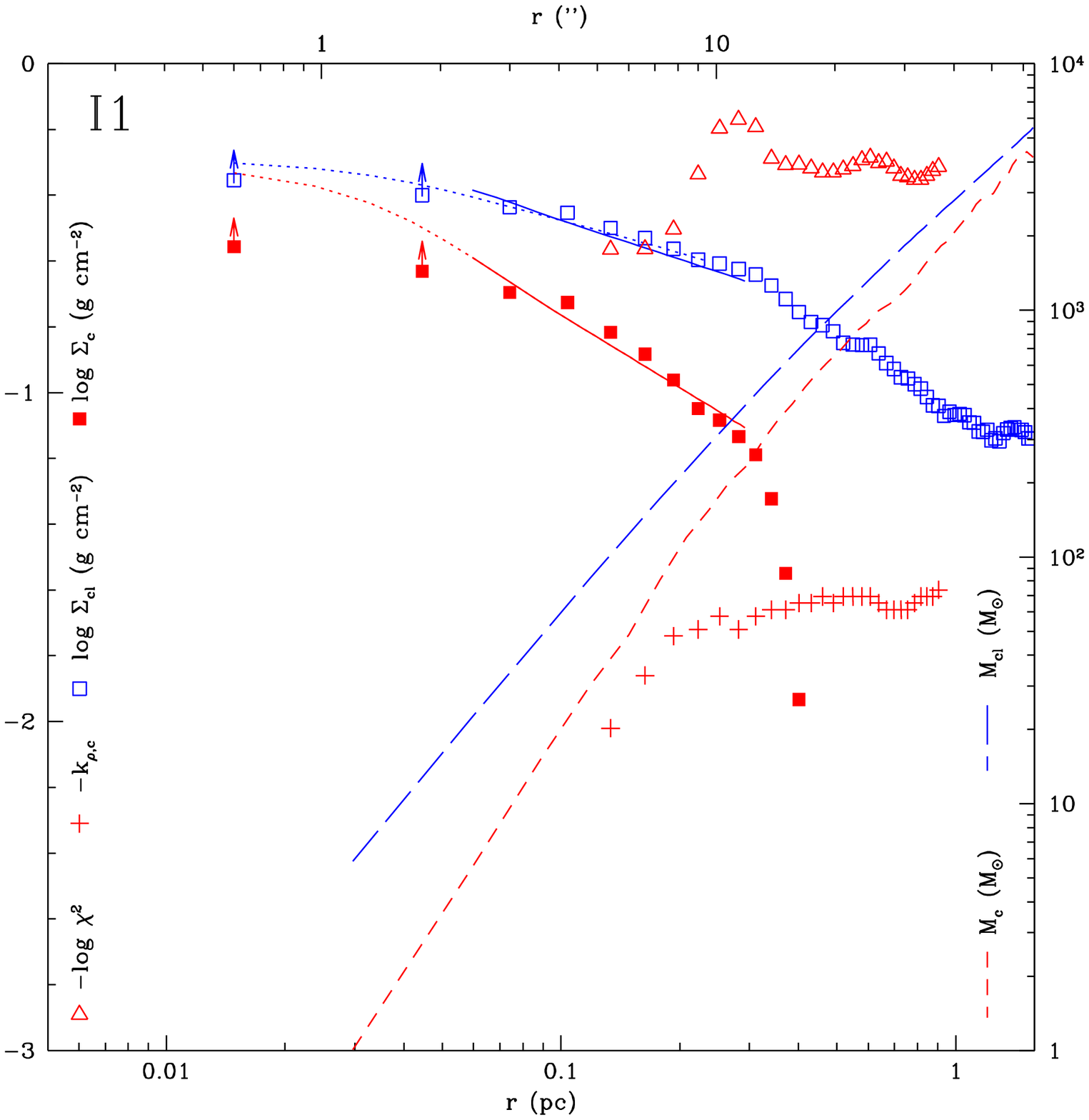} & \includegraphics[width=2.2in]{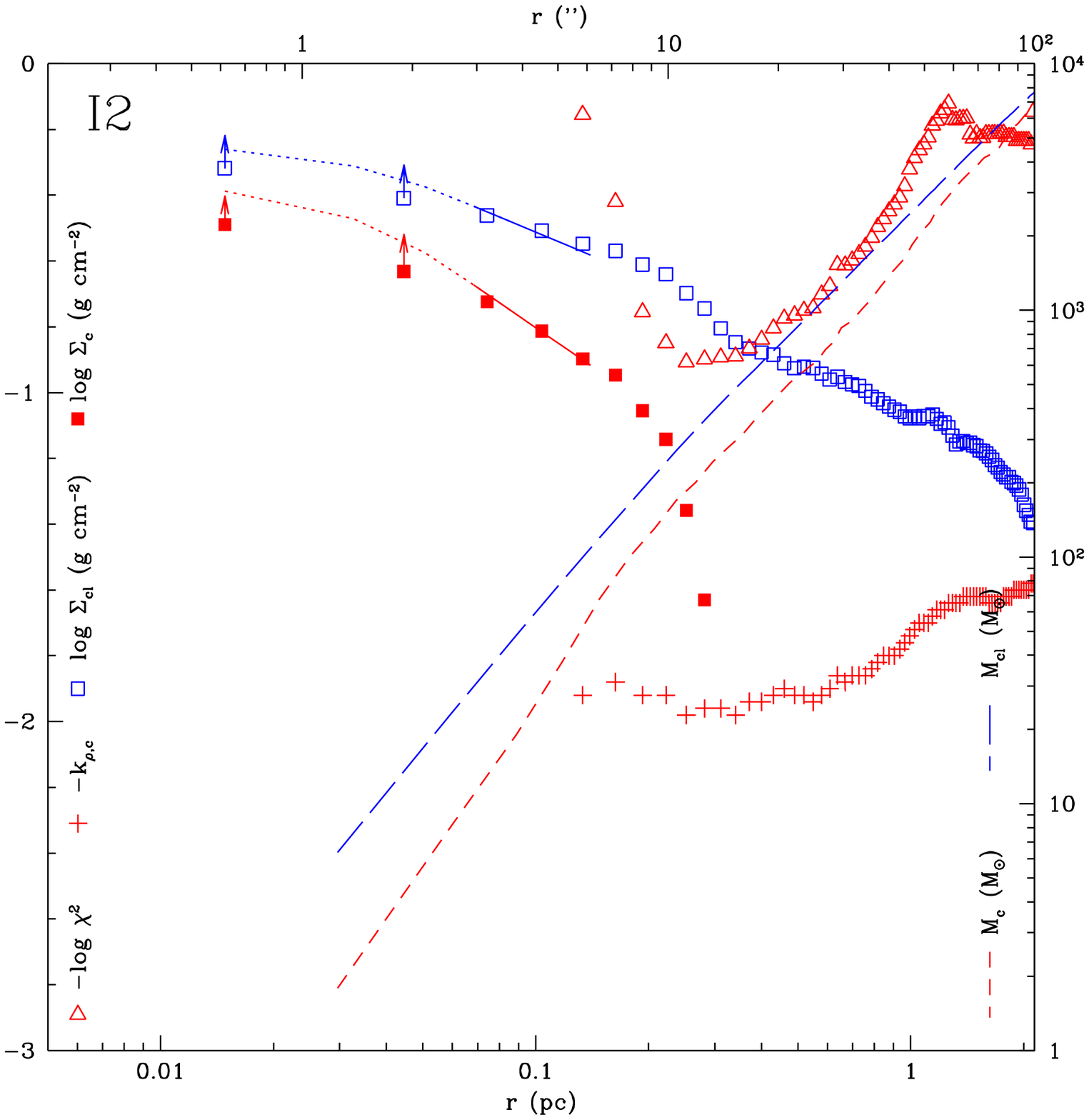} & \hspace{-0.1in} \includegraphics[width=2.2in]{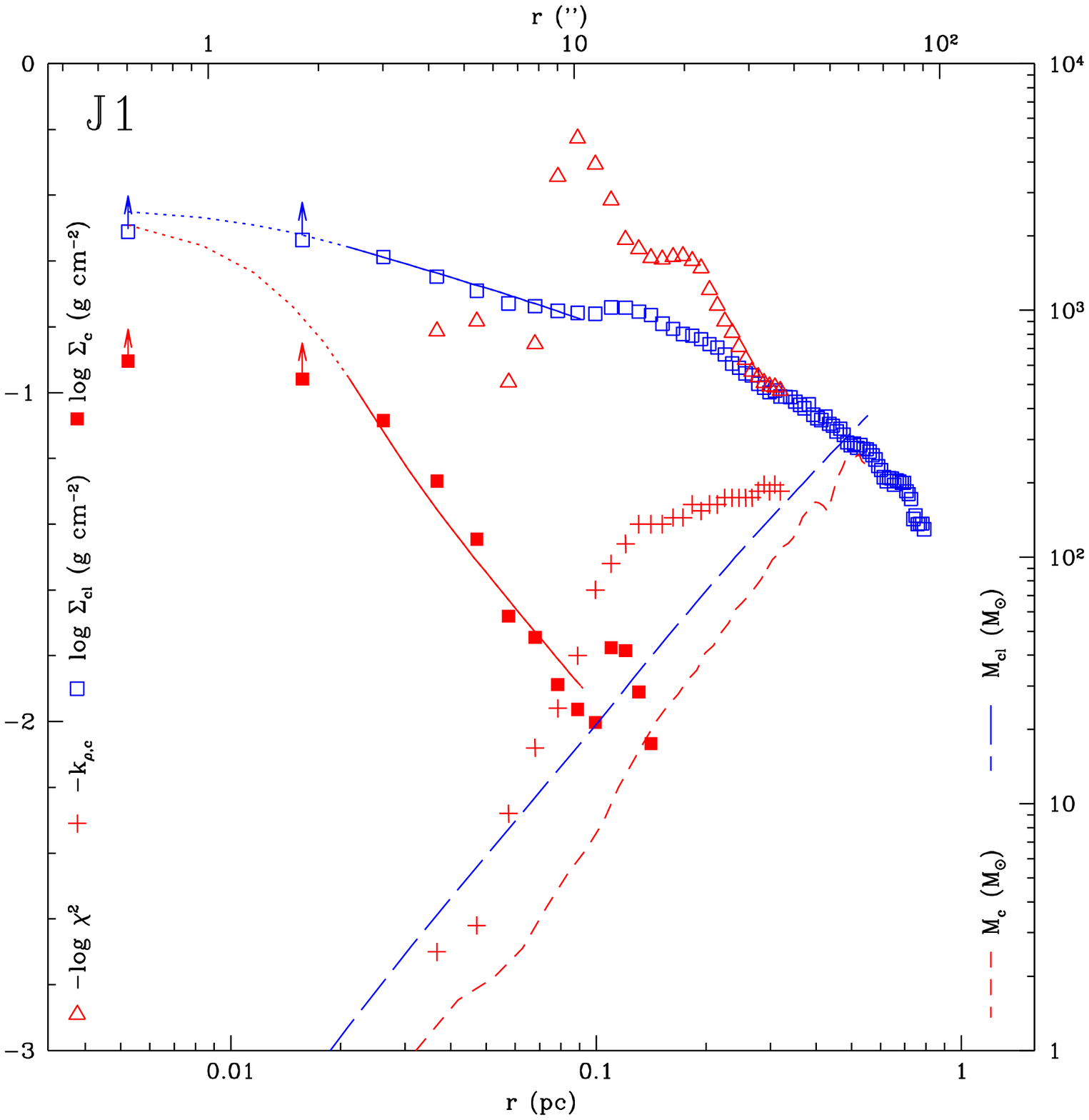}
\end{array}$
\end{center}
\vspace{-0.3in}
\caption{\footnotesize
Core H4, H5, H6, I1, I2, J1 $\Sigma$ maps (notation as Fig.~\ref{fig:coreA1}a) and azimuthally averaged radial profile figures (notation as Fig.~\ref{fig:coreA1}b).
\label{fig:cores7}
}
\end{figure*}

\begin{figure*}
\begin{center}$
\begin{array}{c}
\includegraphics[width=6.5in]{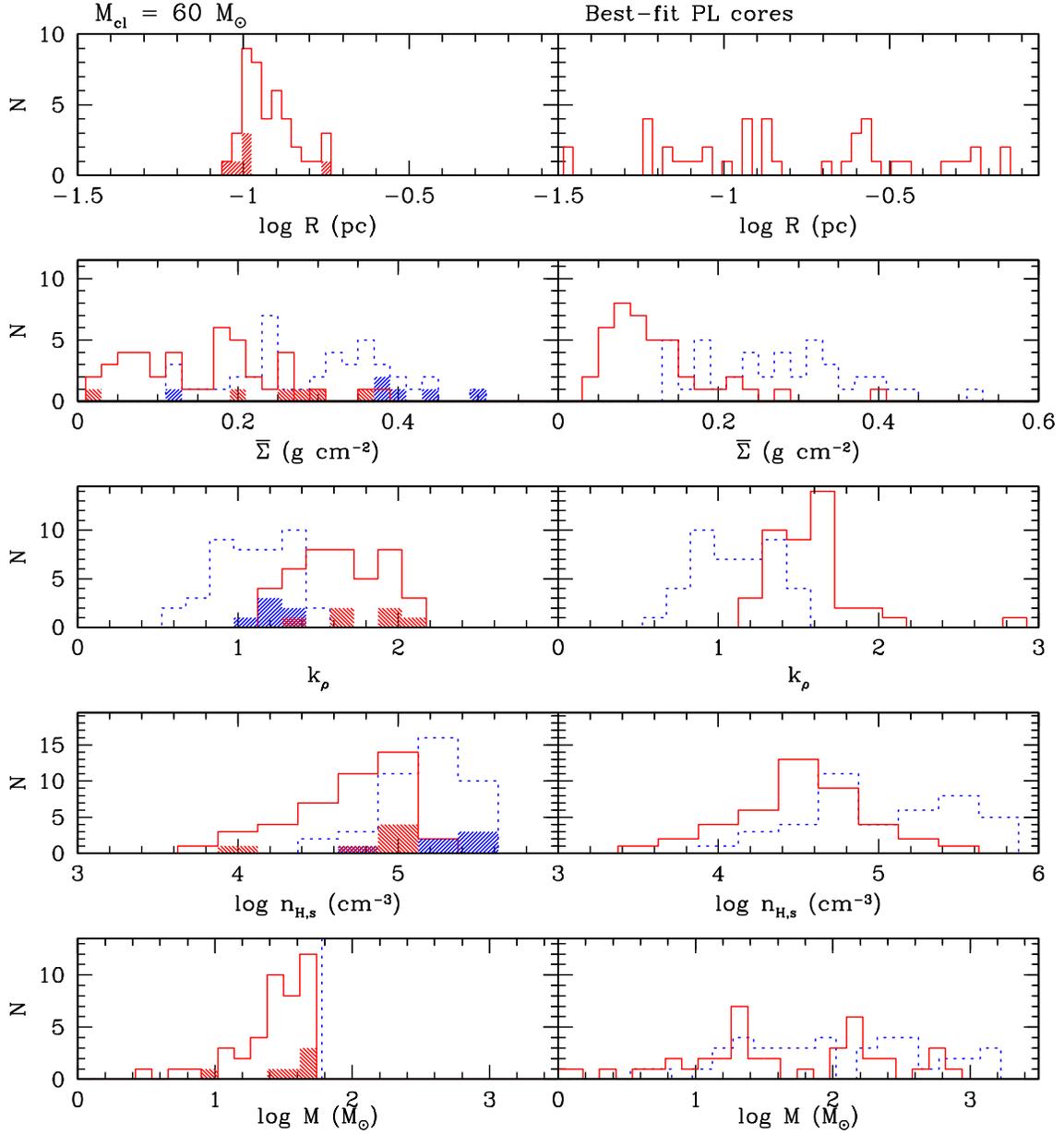}
\end{array}$
\end{center}
\caption{\footnotesize
{\it (a) Left column:} Distributions of properties of clumps and cores
at the scale where the observed total enclosed mass $M_{\rm
  cl}=60\:M_\odot$, which defines $R_{\rm cl}=R_c$. The first graph
shows the distribution of $R_{\rm cl}$ (red solid line) of the 42
cores enclosing this mass. The shaded subset shows the 5 cores (A1,
C2, H1, I1, I2) that have extended saturation in their centers on this scale. The
second panel down shows the mean mass surface density of the clumps,
$\bar{\Sigma}_{\rm cl}$ (blue dotted line; shaded subset as before),
and cores after clump envelope subtraction, $\bar{\Sigma}_{\rm c,PL}$
(red solid line), based on the power law fit. The third panel down
shows the distributions of $k_{\rm \rho, cl}$ (blue dotted line) and
$k_{\rho,c}$ (red solid line). The fourth panel shows the
distributions of $n_{\rm H,s,cl}$ (blue dotted line) and $n_{\rm
  H,s,c}$ (red solid line). The bottom panel shows the distribution of
$M_{\rm c,PL}$ (red solid line). The vertical blue dotted line shows
the $60\:M_\odot$ scale of the clump. {\it (b) Right column:} As for
(a) but now for the best-fit power law plus clump envelope models. The
only difference is that in the bottom panel, the clump mass, $M_{\rm
  cl}$, is now shown (blue dotted line).
}
\label{fig:hist}
\end{figure*}

The results for $M_c(r)$, $k_{\rho,c}(r)$ and $-{\rm log}\:\chi^2(r)$
are shown for Core A1 in Fig.~\ref{fig:coreA1}b. The location of the
peak value of $-\chi^2$ indicates the best-fitting power law (PL) core
radius, which occurs at $0.251\:{\rm pc}$ with a value of
$\chi^2=1.62$.  A circle of this best-fit core radius is shown in the
$\Sigma$ map of the core in Fig.~\ref{fig:coreA1}a. The total enclosed
mass at this scale is $M_{\rm cl}=303\:M_\odot$, the core mass is
$M_c=194\:M_\odot$, the mean core mass surface density is
$\bar{\Sigma}_c=0.204\:\gcc$ and the clump surrounding the core has
$\Sigma_{\rm cl,env}= \bar{\Sigma}_{\rm cl} - \bar{\Sigma}_c =
0.115\:\gcc$. The core mass based on integrating the power law profile
is $M_{\rm c,PL}=204\:M_\odot$, yielding a slightly higher mean mass
surface density of $\bar{\Sigma}_{\rm c, PL}=0.214\:\gcc$.

The best-fit total $\Sigma_{\rm cl}(r) = \Sigma_c(r)+\Sigma_{\rm cl,env}$ model profile is shown by the solid line in
Fig.~\ref{fig:coreA1}b (the dotted continuation in the inner region
indicates where annuli affected by saturation are not used in the
fitting). Figure~\ref{fig:coreA1}c shows the clump envelope subtracted
profile of $\Sigma_c(r)$, together with various projected power law
fits, including the best-fit value of $k_{\rho,c} = 1.88$. The parameters
of the best-fitting power law plus clump envelope model are listed in
Table 2. 

The distributions of $R_c$, $\bar{\Sigma}_{\rm cl}$ (which is the mean
total $\Sigma$ over the area of the core), $\bar{\Sigma}_{\rm c,PL}$, $k_{\rm
  \rho,cl}$, $k_{\rho,c}$, $n_{\rm H,s,cl}$, $n_{\rm H,s,c}$, $M_{\rm
  cl}$ and $M_{\rm c,PL}$ are shown in Figure~\ref{fig:hist}b and summarized in
Table 3. The values for each core are listed in
Table 2. 


It is important to note that these ``best-fit'' values may not
necessarily be the most accurate description of the core
structures. They are based on azimuthally-averaged quantities. The
$\Sigma$ map of a particular core should be inspected to gauge the
validity of this assumption. Also, the values of $\chi^2$ as a
function of radius should be checked to gauge the reasonableness and
uniqueness of the fit.

\begin{figure*}
\begin{center}$
\begin{array}{cc}
\includegraphics[width=6.0in]{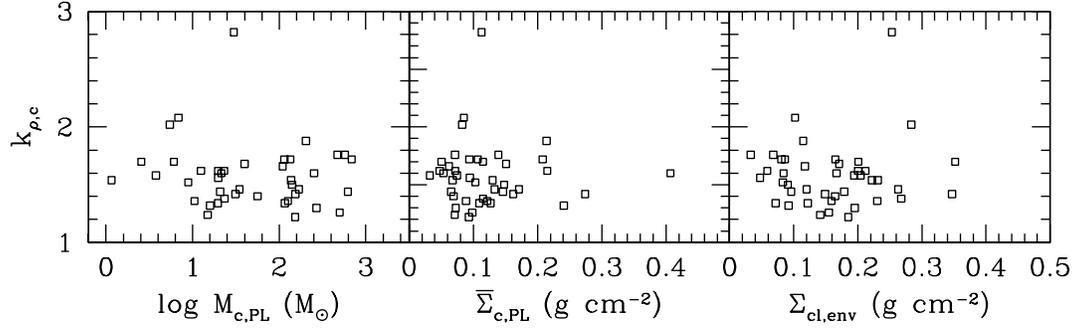}
\end{array}$
\vspace{-4.0in}
\caption{\footnotesize 
$k_{\rho,c}$ versus $M_{\rm c,PL}$, $\bar{\Sigma}_{\rm c,PL}$ and
  $\Sigma_{\rm cl,env}$ for the best-fit cores.
}
\label{fig:krho}
\end{center}

\end{figure*}

The radii and masses of the best-fit cores are generally, but not
always, larger than those at the $M_{\rm cl}=60\:M_\odot$ scale, and thus the volume densities are generally lower. The mean/median/dispersion values of
$k_{\rm \rho,cl}= 1.10/1.12/0.246$ and $k_{\rho,c}= 1.58/1.56/0.277$
are however very similar to those derived at the $M_{\rm
  cl}=60\:M_\odot$ scale, which suggests that the assumption by McKee
\& Tan (2002) and MT03 of a self-similar hierarchy of structure from
the clumps to core scales is a reasonable one. The fiducial value they
adopted of $k_\rho=1.5$ also is close to the average values found in
this sample. In Figure~\ref{fig:krho} we plot $k_{\rho,c}$ versus
$M_{\rm c,PL}$, $\bar{\Sigma}_{\rm c,PL}$ and $\Sigma_{\rm cl,env}$ for the best-fit cores. There are no apparent correlations of
$k_{\rho,c}$ with these properties.

\subsubsection{Best-fit Bonnor-Ebert Cores}

We perform a similar analysis as the power law plus constant envelope
fitting as a function of radius, but now using critical Bonnor-Ebert
profiles (varying the total effective sound speed, $c_s$ and surface
pressure, $P_0$) plus a constant envelope. See Dapp \& Basu (2009) for
more details about fitting Bonnor-Ebert profiles to column density
data.

The best-fitting model for Core A1 is shown in
Fig.~\ref{fig:coreA1}d. This has $R_c=0.670$~pc, $M_c=353\:M_\odot$,
$\bar{\Sigma}_c=0.0523\:\gcc$, $c_s=0.275\:{\rm km\:s^{-1}}$ and
$P_0/k=8.9\times 10^7 \:{\rm K\: cm^{-3}}$. However, the value of
$\chi^2=8.75$, which is significantly larger, i.e. worse, than the
best-fit power law plus clump envelope model fit (for which
$\chi^2=1.62$). Also the size of the Bonnor-Ebert fitted core is much
larger than the power law model: as can be see from
Fig.~\ref{fig:coreA1}a, on these larger scales the assumption of single
monolithic and azimuthally symmetric structure becomes less valid.

Carrying out the Bonnor-Ebert analysis for all 42 cores, we find the
fits are generally worse than for the power law models. The
best-fitting Bonnor-Ebert radii are typically larger than those of the
power law core models. For these reasons, we do not consider the
Bonnor-Ebert models further in our discussion.

\section{Discussion \& Conclusions}\label{S:discussion}
                                  
We have presented a new method to accurately probe mass surface
densities in the range $\sim 0.01$ to $\sim 0.5\:\gcc$ on arcsecond
scales in quiescent, infrared dark clouds, some of which are likely to
be the sites of future star formation. The method uses the small-scale
median filter method of background interpolation from regions around a
defined IRDC (BT09) and then estimates the level of foreground
emission by seeing if there are {\it independent, nearby, saturated
  cores} within the IRDC. If so, the foreground level is set equal to
that observed towards these saturated regions. The resulting $\Sigma$
measurements derived from this MIREX mapping depend on the assumed MIR
dust opacity per unit total mass, but do not depend on the dust
temperature, which is a distinct advantage over measurements based on
sub-mm/mm dust continuum emission.

Focusing on 42 core/clumps within 10 IRDCs, we have tried various
methods of characterizing their azimuthally-averaged structure. Our
preferred method, following the model of McKee \& Tan (2002, 2003),
involves fitting power law cores surrounded by a clump envelope, which
is assumed to have a constant value of $\Sigma$ that can be estimated
from the surrounding region. We have fitted these models as a function
of radius from the core center, deriving an overall best-fit, but also
presenting the full results of this radial characterization. The
typical value of the volume density power law index that best
describes the cores is $k_{\rho,c} \simeq 1.6$. This is close to the
fiducal value of 1.5 adopted by McKee \& Tan (2002, 2003), who based
their choice on previous measurements on the larger, $\sim$parsec,
scales of gas clumps. We find this power law index does not appear to
vary significantly with scale (i.e. between the $60\:M_\odot$ enclosed
mass scale and the best-fit power law plus clump scale), nor with
other core or clump properties, suggesting the presence of a
self-similar hierarchy of structure.

On the scale at which the total projected enclosed mass is
$60\:M_\odot$, the derived cores have about 50\% of this
mass. If massive star formation is to occur, then this is the material
that has a high probability of being incorporated into the massive
star.  These cores have typical radii of $\simeq 0.1$~pc, masses of
$\sim 30\:M_\odot$, mean mass surface densities of
$\bar{\Sigma}_c\simeq 0.15\:\gcc$ and surrounding clump mass surface
densities of similar values. If one regards
our method of clump envelope subtraction to be an overestimate, then
one can consider the typical properties of the clumps on these scales
as being representative of the gas that will form massive stars, i.e.,
with $M_{\rm cl}=60\:M_\odot$ and $\Sigma_{\rm cl}\simeq 0.3\:\gcc$.

The above values of $\Sigma_{\rm cl}$ are lower by factors of $\sim 3-7$ than
the fiducial value of $1\:\gcc$ considered by McKee \& Tan (2002,
2003). Note, their theoretical model is general and does not require a
particular value of $\Sigma_{\rm cl}$, so this difference does not require any
physical explanation. However, Krumholz \& McKee (2008) have proposed
massive star formation requires $\Sigma_{\rm cl}\gtrsim 1\:\gcc$,
based on a model in which fragmentation of massive cores is prevented
by radiative heating from surrounding lower-mass protostars. The high
value of $\Sigma_{\rm cl}$ is required so that the lower-mass protostars
accrete at high enough rates that they are luminous enough to
sufficiently heat the massive core.

Since massive star formation occurs relatively rarely, it may be that
it occurs preferentially in cores with higher values of $\Sigma_c$ and
$\Sigma_{\rm cl}$ than we have observed for the average of our sample,
which we note does show significant dispersion. However, we consider
this unlikely given that we have selected the highest $\Sigma$ regions
from 10~IRDCs that show some of the highest contrast against the
Galactic MIR background (selected from the larger sample of 38~IRDCs
studied by Rathborne et al. 2006). 

Saturation limits our $\Sigma$ maps to values of $\simeq
0.5\:{\rm g\:cm^{-2}}$ and this may be affecting our ability to find
the highest $\Sigma$ cores. However, we are excluding the saturated
regions when deriving core and clump density profiles and most cores
do not exhibit extensively saturated centers. We expect our choice of
MIR opacity per unit gas mass may be uncertain by $\sim 30\%$ (see
\S1), so this by itself is unlikely to explain the relatively low
values of $\Sigma$ that we are deriving compared to the Krumholz \&
McKee (2008) prediction.

Another possibility, is that these cores and clumps will evolve to
higher values of $\Sigma$ before massive star formation
occurs. Indeed, the fact that these are IR dark objects suggests that
they cannot yet be experiencing much radiative heating. However, there
is observational evidence for star formation activity in some of these
cores. For example, Y. Wang et al. (2006) observed water maser 
emission located 4.31\arcsec\ from the center of Core C1. K. Wang et al. (2011) have reported
protostellar outflows from Core C2.

\begin{figure*}
\begin{center}$
\begin{array}{c}
\includegraphics[width=6.5in]{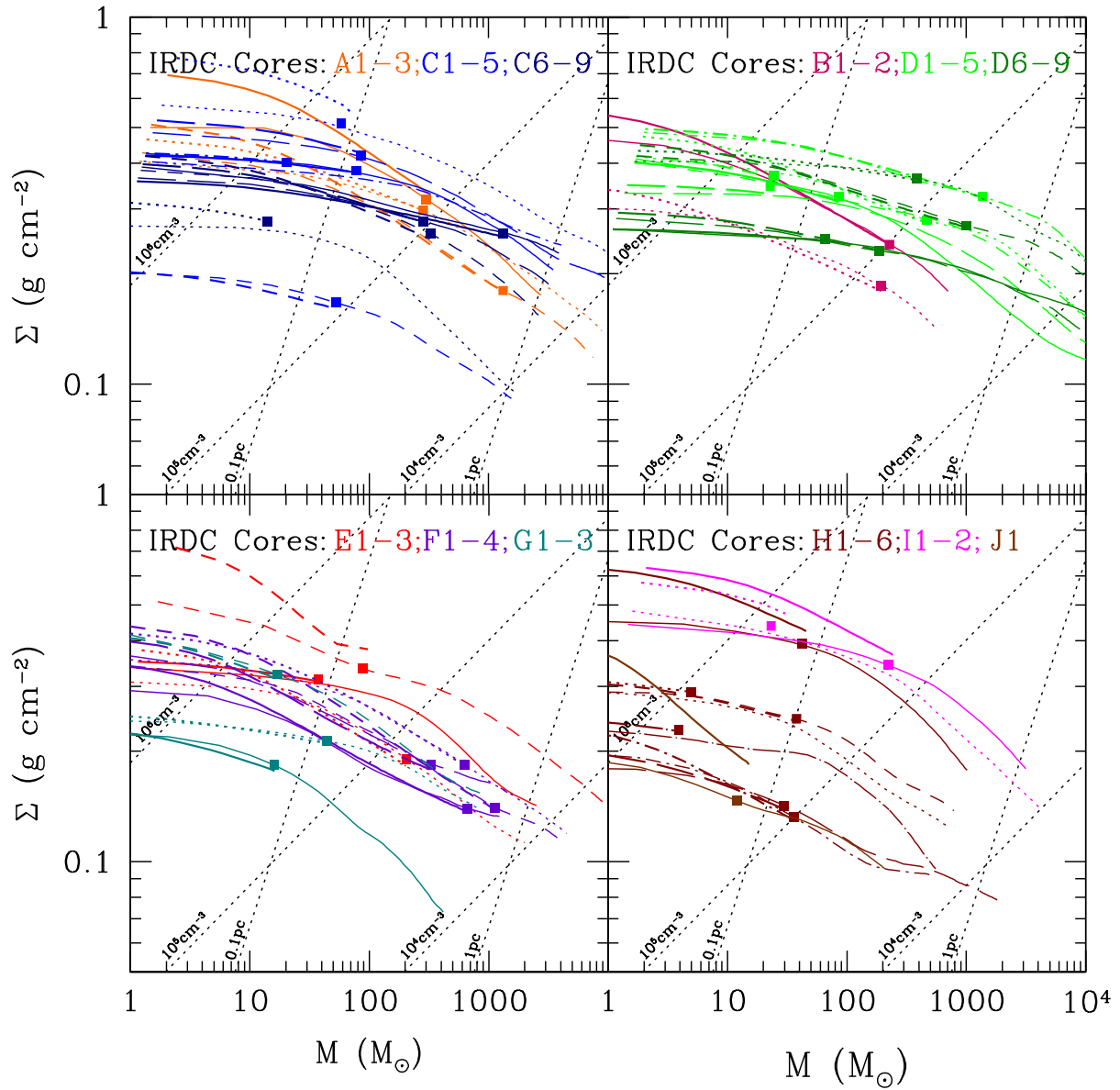}
\end{array}$
\end{center}
\caption{\footnotesize 
Mass surface density versus mass ($\Sigma - M$) diagram (including
lines of constant radial size and density [$n_{\rm H}$] for spherical clouds) for
the 42 IRDC core/clumps. For each range of core numbers, the total
mass surface density, $\Sigma_{\rm cl}$, of the cores are indicated by
color-coded thin solid, dotted, dashed, long-dashed, dot-dashed,
dot-long-dashed lines in order of increasing number (e.g. C1 (solid)
to C5 (dot-dashed); C6 (solid) to C9 (long-dashed)). Squares mark the
location of the best-fit cores.
Heavier lines extend inward from near the squares show $\Sigma_{\rm
  c,PL}+\bar{\Sigma}_{\rm cl,env}$, i.e. based on the fitted
power law density profile.
}
\label{fig:compare1}
\end{figure*}

\begin{figure*}
\begin{center}$
\begin{array}{c}
\includegraphics[width=6.5in]{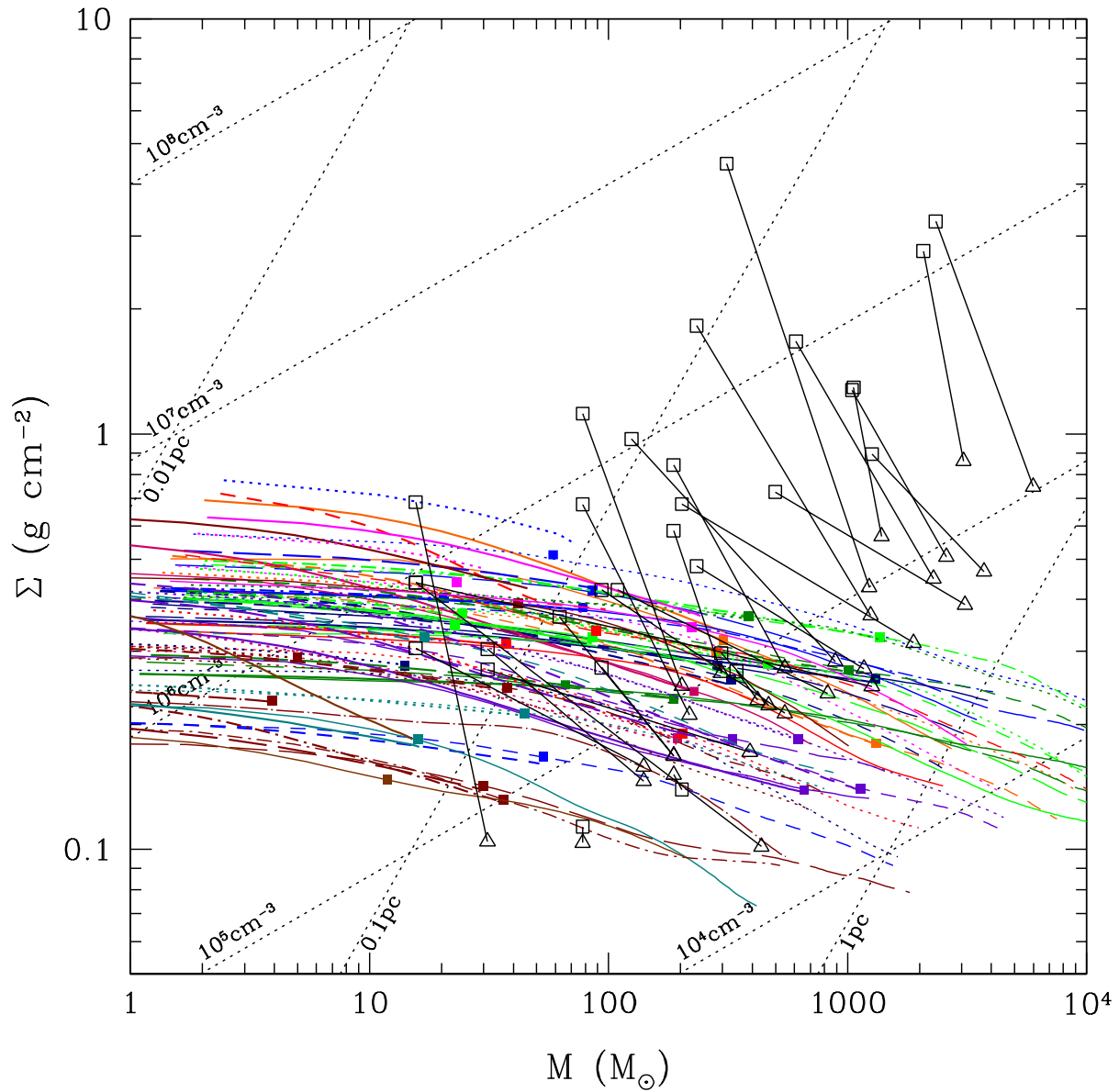}
\end{array}$
\end{center}
\caption{\footnotesize 
Same as Fig.~\ref{fig:compare1}, now combining all 42 cores together
(colored lines). The black symbols and lines show the masses of 31
actively star-forming core/clumps from Mueller et al. (2002), with
triangles indicating the masses above a density threshold of $n_{\rm
  H}\sim 3\times 10^4{\rm cm^{-3}}$ (we have scaled the masses by a
factor 1.56 to be consistent with our adopted gas-to-dust mass ratio)
and the squares indicating the masses inside the deconvolved source
size. Note, the properties of the clouds on this inner scale are not
directly resolved, but are inferred based on simple 1D radiative
transfer modeling. The IRDC cores/clumps overlap only with the
lower-$\Sigma$ range of the star-forming core/clump sample, perhaps indicating
there is an evolutionary growth in core/clump density as star formation proceeds.  
}
\label{fig:compare2}
\end{figure*}

To better compare our IRDC core/clump sample with more evolved
systems, in Figure~\ref{fig:compare1} we show the 42 core/clumps on
the $\Sigma$ versus $M$ diagram (following Tan 2007). Here, $\Sigma$
is measured from the total observed mass inside a given radial
distance from the core/clump center. In Figure~\ref{fig:compare2} we
compare these profiles to the properties of the 31 star-forming clumps
whose IR and sub-mm dust continuum emission was observed and modeled
by Mueller et al. (2002). Note that these properties depend on the
(1D) modeled temperature structure, dust emissivity (they used the same
Ossenkopf \& Henning (1994) dust model that we have adopted for our
MIREX maps) and gas-to-dust ratio (we have scaled Mueller et al.'s
masses by a factor 1.56 to be consistent with our adopted gas-to-dust
ratio). 

The IRDC cores/clumps overlap only with the lower-$\Sigma$ range of
the star-forming core/clump sample, perhaps indicating there is a
(physically plausible) evolutionary growth in core/clump density as
star formation proceeds. However, note that many (indeed most)
star-forming cores and clumps have $\Sigma<1\:\gcc$. Alternatively,
the lack of starless high $\Sigma$ core/clumps may be due to the
somewhat smaller volume of the Galaxy that we have probed with our nearby IRDC
sample, compared to the Mueller et al. star-forming core/clump sample.

Mueller et al. (2002) found density power law indices of $k_{\rm
  \rho,cl}=1.8\pm0.4$, slightly steeper than our derived values for
IRDC cores of $k_{\rho,c} \simeq 1.6$, but significantly steeper than
our value for clumps of $k_{\rm \rho,cl} \simeq 1.1$. Again, this
latter difference may indicate an evolution in cloud properties as
star formation proceeds.

Considering the above results, we suggest that the initial conditions of
local massive star formation in the Galaxy may be better characterized
with values of $\Sigma_{\rm cl}\simeq 0.2\:\gcc$ rather than
$1\:\gcc$, which would imply smaller accretion rates and longer
formation times that the fiducial values of MT03. The accretion rate
becomes
\begin{equation}
\label{mdothighmass}
\dot{m}_* = 1.37 \times 10^{-4}
\left(\frac{m_{*f}}{30\:{M_\odot}}\right)^{3/4} \left(\frac{\Sigma_{\rm cl}}{0.2\:\gcc}\right)^{3/4} 
\left(\frac{m_*}{m_{*f}}\right)^{0.5}~{M_\odot\:{\rm yr}^{-1}}
\end{equation}
for a core with $k_{\rho,c}=1.5$ and a star formation efficiency of 50\%, where $m_*$ is
the instantaneous protostellar mass and $m_{*f}$ is the final
protostellar mass. The star formation timescale becomes
\begin{equation}
\label{tsfhighmass}
t_{*f} = 4.31 \times 10^5 \left(\frac{m_{*f}}{30\:{M_\odot}}\right)^{1/4} 
\left(\frac{\Sigma}{0.2\:\gcc}\right)^{-3/4}~~~ {\rm yr}.
\end{equation}

In this case of massive star formation at relatively low values of
$\Sigma$, we expect fragmentation of the cores is prevented by
magnetic fields, i.e. if the core mass is equal to the magnetic
critical mass (Bertoldi \& McKee 1992)
\begin{equation}
\label{MB}
M_B = 1020 \left(\frac{R}{Z}\right)^2 \left(\frac{\bar{B}}{30\:{\rm \mu G}}\right)^{3} \left(\frac{\bar{n}_{\rm H}}{10^3\:{\rm cm^{-3}}}\right)^{-2}\:M_\odot,
\end{equation}
where $R$ and $Z$ are the major and minor axes of the core, $\bar{B}$ is the mean field strength in the core, and $\bar{n}_{\rm H}$ is the mean number density of H nuclei. Thus for
a core with $\bar{n}_{\rm H}=10^5\:{\rm cm^{-3}}$ and
$M_c= 100\:M_\odot$, typical of our sample, the condition $M_B=M_c$ requires a field strength
\begin{equation}
\label{Bfield}
\bar{B} = 300 \left(\frac{M_B}{100\:M_\odot}\right)^{1/3} \left(\frac{Z}{R}\right)^{2/3} \left(\frac{\bar{n}_{\rm H}}{10^5\:{\rm cm^{-3}}}\right)^{2/3}\:{\rm \mu G}.
\end{equation}
If cores have some significant magnetic support, then we expect
$R/Z>1$, perhaps $\sim 2$, so that the required field strength in
eq.(\ref{Bfield}) is then $190\:{\rm \mu G}$. Such field strengths are
similar to those observed in regions of active massive star formation
(e.g. Crutcher 2005). Indeed, Crutcher (2005) noted the observed mass
to flux ratios scattered about the critical value. Numerical
simulations of the collapse of marginally magnetically critical (rather than super critical,
e.g. Wang et al. 2010; Hennebelle et al. 2011) cores are required to
investigate this scenario for forming massive stars, and, more generally,
for explaining the high-mass tail of the initial mass function (Kunz \& Mouschovias 2009).

\acknowledgments MJB acknowledges support from a Sigma Xi Grant in Aid
of Research. JCT acknowledges support from NSF CAREER grant
AST-0645412; NASA Astrophysics Theory and Fundamental Physics grant
ATP09-0094; NASA Astrophysics Data Analysis Program ADAP10-0110 and a
Faculty Enhancement Opportunity grant from the University of Florida.
We thank Peter Barnes, Paola Caselli, Ed  
Churchwell, Francesco Fontani, Audra Hernandez, Jouni Kainulainen,  
Jens Kauffmann, Shuo Kong, Mark Krumholz, Chris McKee, Thushara  
Pillai, Sven Van Loo, Qizhou Zhang for helpful discussions.

\newpage

\clearpage
\mbox{~}

\newpage

\begin{figure*}
\begin{center}$
\begin{array}{c}
\includegraphics[width=6.5in]{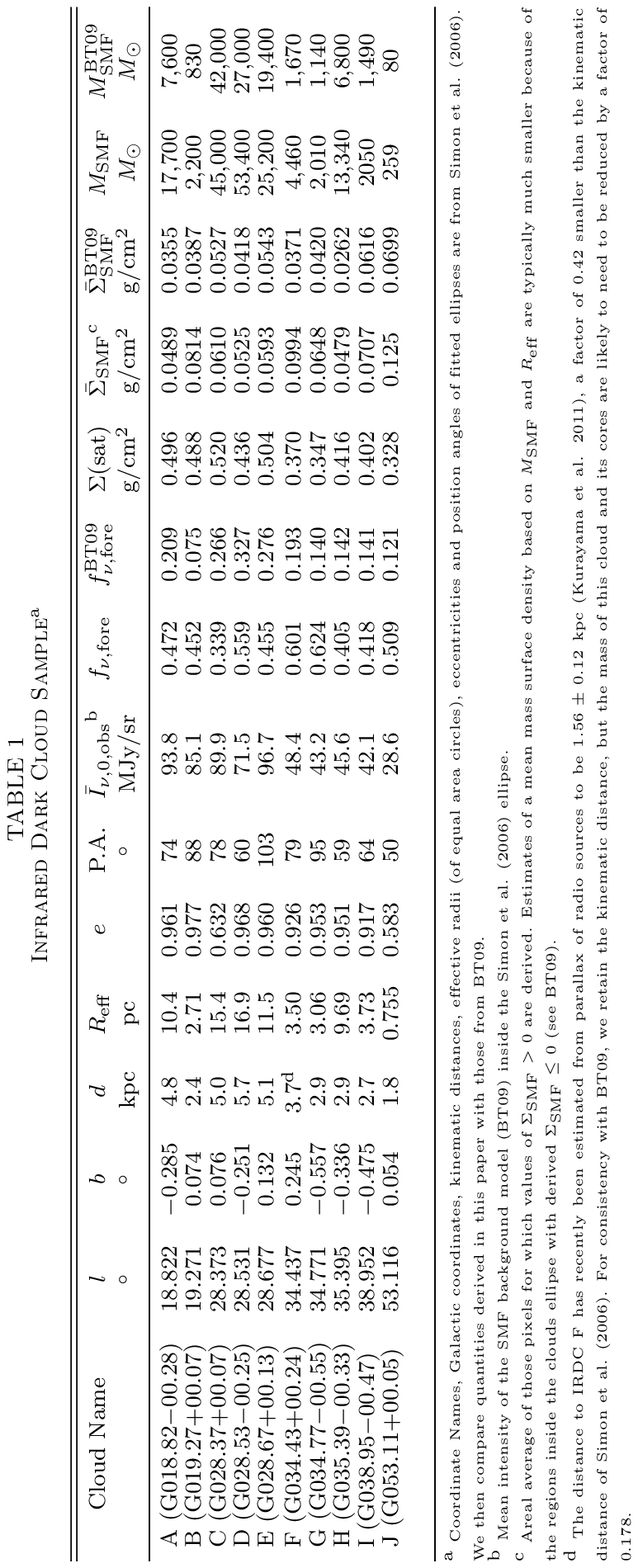}
\end{array}$
\end{center}
\end{figure*}
\begin{figure*}
\begin{center}$
\begin{array}{c}
\includegraphics[width=6.5in]{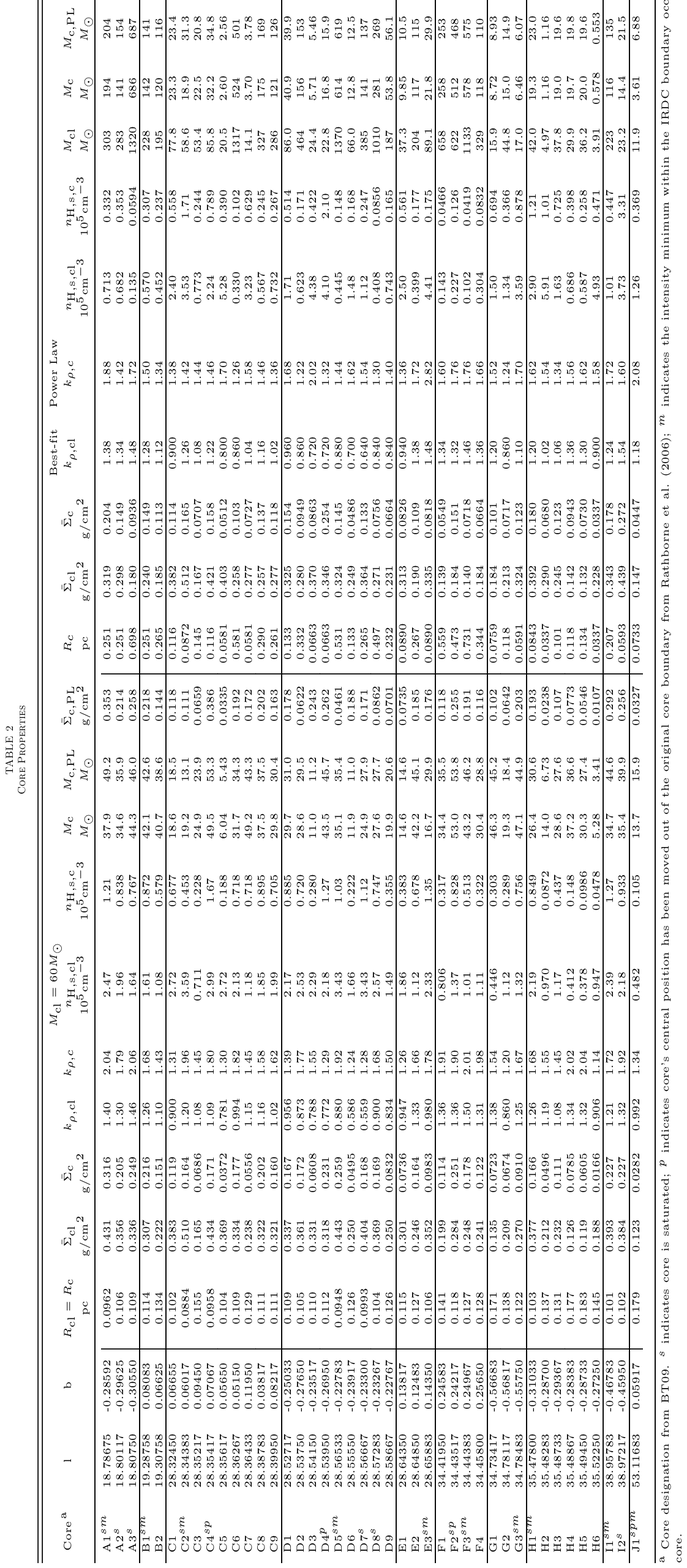}
\end{array}$
\end{center}
\end{figure*}
\begin{figure*}
\begin{center}$
\begin{array}{c}
\includegraphics[width=6.5in]{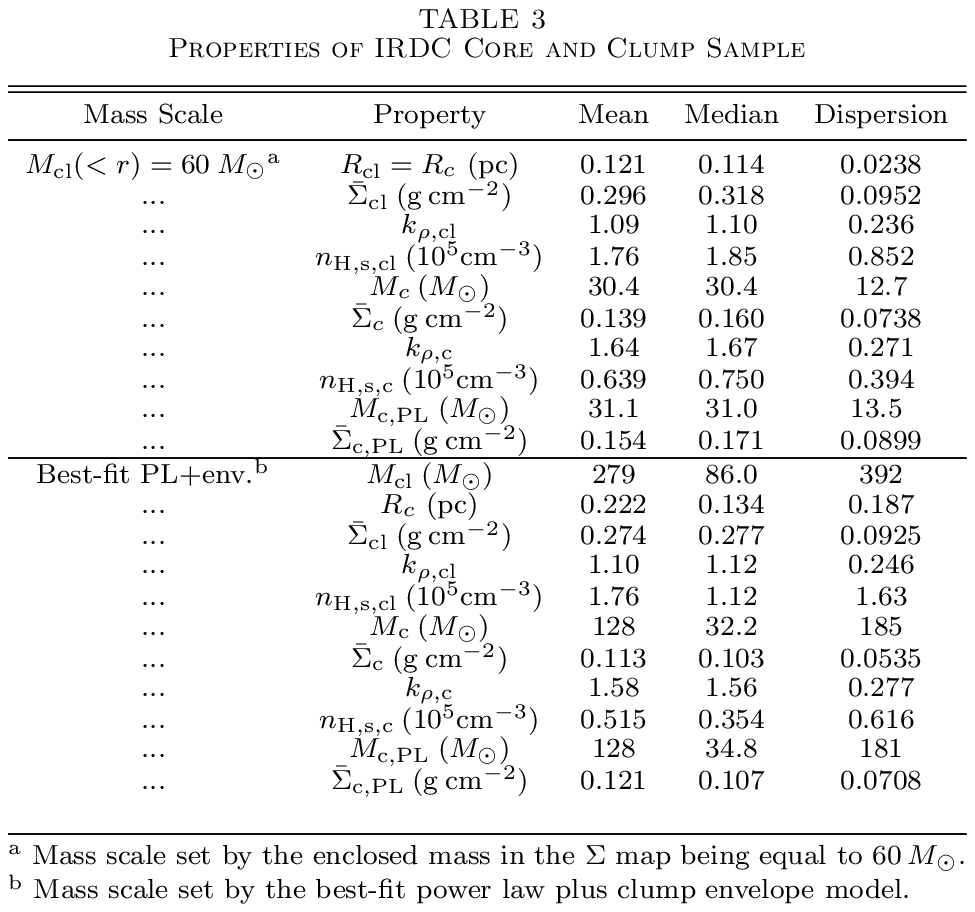}
\end{array}$
\end{center}
\end{figure*}

\end{document}